\documentclass[moor,nonblindrev]{informs1} 

\usepackage{times}
\usepackage{mathtools}
\usepackage{csquotes}
\usepackage{enumerate}
\usepackage[algoruled,boxed,lined]{algorithm2e}
\usepackage{bbm}

\newtheorem{fact}{Fact}
\newcommand{\tr}{\ensuremath{{\scriptscriptstyle\mathsf{T}}}}
\DeclareMathOperator*{\ropt}{\mathrm{ER}_\mathrm{OPT}}

\DeclareMathOperator*{\tini}{t_{\mathrm{ini}}}
\DeclareMathOperator*{\tinivalue}{2^8 \frac{(C+2)!^4}{\epsilon^6}}
\DeclareMathOperator*{\tinivalueinline}{2^8 (C+2)!^4/\epsilon^6}
\DeclareMathOperator*{\adj}{\mathrm{adj}}
\DeclareMathOperator*{\diag}{\mathrm{diag}}
\DeclarePairedDelimiter\floor{\lfloor}{\rfloor}
\DeclareMathOperator*{\maxdelta}{\frac{\epsilon^2}{4 \cdot (C+1)!}}
\DeclareMathOperator*{\maxdevmeangeneralcase}{\frac{\epsilon^3}{16 \cdot (C+2)!^2}}
\DeclareMathOperator*{\probamaxdevmeangeneralcase}{ 2^9 \frac{(C+3)!^4}{\epsilon^6 \cdot T^2}}
\DeclareMathOperator*{\probamaxdevmeangeneralcasewithouttime}{2^9 \frac{(C+3)!^4}{\epsilon^6}}

\newcommand{\norm}[1]{\left\|#1\right\|}
\newcommand{\norminduced}[1]{{\left\vert\kern-0.25ex\left\vert\kern-0.25ex\left\vert #1 
    \right\vert\kern-0.25ex\right\vert\kern-0.25ex\right\vert}}

\usepackage{natbib}
 \NatBibNumeric
 \def\bibfont{\small}%
 \bibpunct[, ]{[}{]}{,}{n}{}{,}%

\usepackage[colorlinks=true,breaklinks=true,bookmarks=true,urlcolor=blue,
     citecolor=blue,linkcolor=blue,bookmarksopen=false,draft=false]{hyperref}

\def\EMAIL#1{\href{mailto:#1}{#1}}
\def\URL#1{\href{#1}{#1}}         

\TheoremsNumberedThrough     

\EquationsNumberedThrough    

\makeatletter
\newcommand{\RemoveAlgoNumber}{\renewcommand{\fnum@algocf}{\AlCapSty{\AlCapFnt\algorithmcfname}}}
\newcommand{\RevertAlgoNumber}{\algocf@resetfnum}
\makeatother

\begin{document}

\RemoveAlgoNumber

\RUNAUTHOR{Flajolet and Jaillet}

\RUNTITLE{Logarithmic Regret Bounds for BwK}

\TITLE{Logarithmic Regret Bounds for Bandits with Knapsacks\footnote{Research funded in part by ONR, grants N00014-12-1-0033 and N00014-15-1-2083.}}

\ARTICLEAUTHORS{%
\AUTHOR{Arthur Flajolet}
\AFF{Operations Research Center \\ Massachusetts Institute of Technology \\ Cambridge, MA 02139 \\ \EMAIL{flajolet@mit.edu} \URL{}}
\AUTHOR{Patrick Jaillet}
\AFF{Department of Electrical Engineering and Computer Science \\ Laboratory for Information and Decision Systems \\ Operations Research Center \\ Massachusetts Institute of Technology \\ Cambridge, MA 02139 \\ \EMAIL{jaillet@mit.edu} \URL{}}
} 

\ABSTRACT{%
Optimal regret bounds for Multi-Armed Bandit problems are now well documented. They can be classified into two categories based on the growth rate with respect to the time horizon $T$: (i) small, distribution-dependent, bounds of order of magnitude $\ln(T)$ and (ii) robust, distribution-free, bounds of order of magnitude $\sqrt{T}$. The Bandits with Knapsacks model, an extension to the framework allowing to model resource consumption, lacks this clear-cut distinction. While several algorithms have been shown to achieve asymptotically optimal distribution-free bounds on regret, there has been little progress toward the development of small distribution-dependent regret bounds. We partially bridge the gap by designing a general-purpose algorithm with distribution-dependent regret bounds that are logarithmic in the initial endowments of resources in several important cases that cover many practical applications, including dynamic pricing with limited supply, bid optimization in online advertisement auctions, and dynamic procurement.
}%




\maketitle


%


\section{Introduction.}
\subsection{Motivation.}
Multi-Armed Bandit (MAB) is a benchmark model for repeated decision making in stochastic environments with very limited feedback on the outcomes of alternatives. In these circumstances, a decision maker must strive to find an overall optimal sequence of decisions while making as few suboptimal ones as possible when exploring the decision space in order to generate as much revenue as possible, a trade-off coined \emph{exploration-exploitation}. The original problem, first formulated in its predominant version in \citet{robbins1985some}, has spurred a new line of research that aims at introducing additional constraints that reflect more accurately the reality of the decision making process. \emph{Bandits with Knapsacks} (BwK), a model formulated in its most general form in \citet{badanidiyuru2013bandits}, fits into this framework and is characterized by the consumption of a limited supply of resources (e.g. time, money, and natural resources) that comes with every decision. This extension is motivated by a number of applications in electronic markets such as dynamic pricing with limited supply, see \citet{besbes2012blind} and \citet{babaioff2012dynamic}, online advertising, see \citet{slivkins2013dynamic}, online bid optimization for sponsored search auctions, see \citet{tranthanonlinebidoptimization2014}, and crowdsourcing, see \citet{badanidiyuru2012learning}. A unifying paradigm of online learning is to evaluate algorithms based on their regret performance. In the BwK theory, this performance criterion is expressed as the gap between the total payoff of an optimal oracle algorithm aware of how the rewards and the amounts of resource consumption are generated and the total payoff of the algorithm. Many approaches have been proposed to tackle the original MAB problem, where time is the only limited resource with a prescribed time horizon $T$, and the optimal regret bounds are now well documented. They can be classified into two categories with qualitatively different asymptotic growth rates. Many algorithms, such as UCB1, see \citet{auer2002finite}, Thompson sampling, see \citet{agrawal2011analysis}, and $\epsilon$-greedy, see \citet{auer2002finite}, achieve distribution-dependent, i.e. with constant factors that depend on the underlying unobserved distributions, asymptotic bounds on regret of order $\Theta(\ln(T))$, which is shown to be optimal in \citet{lai1985asymptotically}. While these results prove very satisfying in many settings, the downside is that the bounds can get arbitrarily large if a malicious opponent was to select the underlying distributions in an adversarial fashion. In contrast, algorithms such as Exp3, designed in \citet{auer2002nonstochastic}, achieve distribution-free bounds that can be computed in an online fashion, at the price of a less attractive growth rate $\Theta(\sqrt{T})$. The BwK theory lacks this clear-cut distinction. While provably optimal distribution-free bounds have recently been established, see \citet{agrawal2014bandits} and \citet{badanidiyuru2013bandits}, there has been little progress toward the development of asymptotically optimal distribution-dependent regret bounds. To bridge the gap, we introduce a template algorithm with proven regret bounds which are asymptotically logarithmic in the initial supply of each resource, in four important cases that cover a wide range of applications:
\begin{itemize}
	\item{Case 1, where there is a single limited resource other than time, which is not limited, and the amount of resource consumed as a result of making a decision is stochastic. Applications in online advertising, see \citet{tranthanangreedybudgetedMAB2010}, fit in this framework;}
	\item{Case 2, where there are arbitrarily many resources and the amounts of resources consumed as a result of making a decision are deterministic. Applications to network revenue management of perishable goods, see \citet{besbes2012blind}, shelf optimization of perishable goods, see \citet{graczova2014generalized}, and wireless sensor networks, see \citet{tran2012long}, fit in this framework;}
	\item{Case 3, where there are two limited resources, one of which is assumed to be time while the consumption of the other is stochastic, under a nondegeneracy condition. Typical applications include online bid optimization in sponsored search auctions, see \citet{tranthanonlinebidoptimization2014}, dynamic pricing with limited supply, see \citet{babaioff2012dynamic}, and dynamic procurement, see \citet{badanidiyuru2012learning};}
	\item{Case 4, where there are arbitrarily many resources, under a stronger nondegeneracy condition than for Case 3. Typical applications include dynamic ad allocation, see \citet{slivkins2013dynamic}, dynamic pricing of multiple products, see \citet{badanidiyuru2013bandits}, and network revenue management, see \citet{besbes2012blind}.}
\end{itemize}

\indent In terms of applicability and significance of the results, Case 3 is the most important case. Case 4 is the most general one but the analysis is more involved and requires stronger assumptions which makes it less attractive from a practical standpoint. The analysis is easier for Cases 1 and 2 but their modeling power is more limited.
\\
\indent In fast-paced environments, such as in ad auctions, the stochastic assumptions at the core of the BwK model are only valid for a short period of time but there are typically a large number of actions to be performed per second (e.g. submit a bid for a new ad auction). In these situations, the initial endowments of resources are thus typically large and logarithmic regret bounds can be significantly more attractive than distribution-free ones.

\subsection{Problem statement and contributions.}
At each time period $t \in \mathbb{N}$, a decision needs to be made among a predefined finite set of actions, represented by arms and labeled $k = 1, \cdots, K$. We denote by $a_t$ the arm pulled at time $t$. Pulling arm $k$ at time $t$ yields a random reward $r_{k, t} \in [0, 1]$ (after scaling) and incurs the consumption of $C \in \mathbb{N}$ different resource types by random amounts $c_{k, t}(1), \cdots, c_{k, t}(C) \in [0, 1]^C$ (after scaling). Note that time itself may or may not be a limited resource. At any time $t$ and for any arm $k$, the vector $(r_{k, t}, c_{k, t}(1), \cdots, c_{k, t}(C))$ is jointly drawn from a fixed probability distribution $\nu_k$ independently from the past. The rewards and the amounts of resource consumption can be arbitrarily correlated across arms. We denote by $(\mathcal{F}_t)_{t \in \mathbb{N}}$ the natural filtration generated by the rewards and the amounts of resource consumption revealed to the decision maker, i.e. $((r_{a_t, t}, c_{a_t, t}(1), \cdots, c_{a_t, t}(C)))_{t \in \mathbb{N}}$. The consumption of any resource $i \in \{1, \cdots, C \}$ is constrained by an initial budget $B(i) \in \mathbb{R}_+$. As a result, the decision maker can keep pulling arms only so long as he does not run out of any of the $C$ resources and the game ends at time period $\tau^*$, defined as:
\begin{equation}
	\label{eq-definition-stoppingtime}
	\tau^* = \min \{t \in \mathbb{N} \; | \; \exists i \in \{1, \cdots, C\},  \sum_{\tau=1}^t c_{a_\tau, \tau}(i) > B(i) \}.
\end{equation}
Note that $\tau^*$ is a stopping time with respect to $(\mathcal{F}_t)_{t \geq 1}$. When it comes to choosing which arm to pull next, the difficulty for the decision maker  lies in the fact that none of the underlying distributions, i.e. $(\nu_k)_{k=1, \cdots, K}$, are initially known. Furthermore, the only feedback provided to the decision maker upon pulling arm $a_t$ (but prior to selecting $a_{t+1}$) is $(r_{a_t, t}, c_{a_t, t}(1), \cdots, c_{a_t, t}(C))$, i.e. the decision maker does not observe the rewards that would have been obtained and the amounts of resources that would have been consumed as a result of pulling a different arm. The goal is to design a non-anticipating algorithm that, at any time $t$, selects $a_t$ based on the information acquired in the past so as to keep the pseudo regret defined as:
\begin{equation}
		\label{eq-general-def-regret}
		R_{B(1), \cdots, B(C)} = \ropt(B(1), \cdots, B(C)) - \mathbb{E}[ \sum_{t=1}^{ \tau^* - 1} r_{a_t, t}  ],
\end{equation}
as small as possible, where $\ropt(B(1), \cdots, B(C))$ is the maximum expected sum of rewards that can be obtained by a non-anticipating oracle algorithm that has knowledge of the underlying distributions. Here, an algorithm is said to be non-anticipating if the decision to pull a given arm does not depend on the future observations. We develop algorithms and establish distribution-dependent regret bounds, that hold for any choice of the unobserved underlying distributions $(\nu_k)_{k=1, \cdots, K}$, as well as distribution-independent regret bounds. This entails studying the asymptotic behavior of $R_{B(1), \cdots, B(C)}$ when all the budgets $(B(i))_{i=1, \cdots, C}$ go to infinity. In order to simplify the analysis, it is convenient to assume that the ratios $( B(i)/B(C))_{i=1, \cdots, C}$ are constants independent of any other relevant quantities and to denote $B(C)$ by $B$.
\begin{assumption}
	\label{assumption-budget-scale-linearly-wt-time}
	For any resource $i \in \{1, \cdots, C\}$, we have $B(i) = b(i) \cdot B$ for some fixed constant $b(i) \in (0, 1]$. Hence $b = \min\limits_{i=1, \cdots, C} b(i)$ is a positive quantity.
\end{assumption}
When time is a limited resource, we use the notation $T$ in place of $B$. Assumption \ref{assumption-budget-scale-linearly-wt-time} is widely used in the dynamic pricing literature where the inventory scales linearly with the time horizon, see \citet{besbes2012blind} and \citet{johnson2015online}. Assumption \ref{assumption-budget-scale-linearly-wt-time} will only prove useful when deriving distribution-dependent regret bounds and it can largely be relaxed, see Section \ref{sec-extensions} of the Appendix. \\
As the mean turns out to be an important statistics, we denote the mean reward and amounts of resource consumption by $\mu^r_{k}, \mu^{c}_{k}(1), \cdots, \mu^{c}_{k}(C)$ and their respective empirical estimates by $\bar{r}_{k, t}$, $\bar{c}_{k, t}(1), \cdots, \bar{c}_{k, t}(C)$. These estimates depend on the number of times each arm has been pulled by the decision maker up to, but not including, time $t$, which we write $n_{k, t}$. We end with a general assumption, which we use throughout the paper, meant to have the game end in finite time. 
\begin{assumption}
	\label{assumption-all-cost-non-zero}
	For any arm $k \in \{1, \cdots, K\}$, we have $\max\limits_{i=1, \cdots, C} \mu^{c}_{k}(i) > 0$.
\end{assumption}
Note that Assumption \ref{assumption-all-cost-non-zero} is automatically satisfied if time is a limited resource.


\paragraph{Contributions.}
We design an algorithm that runs in time polynomial in $K$ for which we establish $O( K^C \cdot \ln(B)/\Delta)$ (resp. $\sqrt{K^C \cdot B \cdot \ln(B)}$) distribution-dependent (resp. distribution-free) regret bounds, where $\Delta$ is a parameter that generalizes the optimality gap for the standard MAB problem. We establish these regret bounds in four cases of increasing difficulty making additional technical assumptions that become stronger as we make progress towards tackling the general case. We choose to present these intermediate cases since: (i) we get improved constant factors under weaker assumptions and (ii) they subsume many practical applications. Note that our distribution-dependent regret bounds scale as a polynomial function of $K$ of degree $C$, which may be unacceptable when the number of resources is large. We provide evidence that suggests that a linear dependence on $K$ can be achieved by tweaking the algorithm, at least in some particular cases of interest. Finally, we point out that the constant factors hidden in the $O$ notations are not scale-free, in the sense that jointly scaling down the amounts of resources consumed at each round along with their respective initial endowments worsens the bounds. As a consequence, initially scaling down the amounts of resource consumption in order to guarantee that they lie in $[0, 1]$ should be done with caution: the scaling factors should be as small as possible.

\subsection{Literature review.}
The Bandits with Knapsacks framework was first introduced in its full generality in \citet{badanidiyuru2013bandits}, but special cases had been studied before, see for example \citet{tranthanangreedybudgetedMAB2010}, \citet{ding2013multi}, and \citet{babaioff2012dynamic}. Since the standard MAB problem fits in the BwK framework, with time being the only scarce resource, the results listed in the introduction tend to suggest that regret bounds with logarithmic growth with respect to the budgets may be possible for BwK problems but very few such results are documented. When there are arbitrarily many resources and a time horizon, \citet{badanidiyuru2013bandits} and \citet{agrawal2014bandits} obtain $\tilde{O}(\sqrt{K \cdot T})$ distribution-free bounds on regret that hold on average as well as with high probability, where the $\tilde{O}$ notation hides logarithmic factors. These results were later extended to the contextual version of the problem in \citet{badanidiyuru2014resourcefulbandits} and \citet{agrawal2016efficient}. \citet{johnson2015online} extend Thompson sampling to tackle the general BwK problem and obtain distribution-dependent bounds on regret of order $\tilde{O}(\sqrt{T})$ (with an unspecified dependence on $K$) when time is a limited resource, under a nondegeneracy condition. \citet{trovobudgeted} develop algorithms for BwK problems with a continuum of arms and a single limited resource constrained by a budget $B$ and obtain $o(B)$ regret bounds. \citet{combes2015bandits} consider a closely related framework that allows to model any history-dependent constraint on the number of times any arm can be pulled and obtain $O(K \cdot \ln(T))$ regret bounds when time is a limited resource. However, the benchmark oracle algorithm  used in \citet{combes2015bandits} to define the regret is significantly weaker than the one considered here as it only has knowledge of the distributions of the rewards, as opposed to the joint distributions of the rewards and the amounts of resource consumption. \citet{babaioff2012dynamic} establish a $\Omega(\sqrt{T})$ distribution-dependent lower bound on regret for a dynamic pricing problem which can be cast as a BwK problem with a time horizon, a resource whose consumption is stochastic, and a continuum of arms. This lower bound does not apply here as we are considering finitely many arms and it is well known that the minimax regret can be exponentially smaller when we move from finitely many arms to uncountably many arms for the standard MAB problem, see \citet{kleinberg2003value}. \citet{tran2012knapsack} tackles BwK problems with a single limited resource whose consumption is deterministic and constrained by a global budget $B$ and obtain $O(K \cdot \ln(B))$ regret bounds. This result was later extended to the case of a stochastic resource in \citet{xia2015budgeted}. \citet{wu2015algorithms} study a contextual version of the BwK problem when there are two limited resources, one of which is assumed to be time while the consumption of the other is deterministic, and obtain $O(K \cdot \ln(T))$ regret bounds under a nondegeneracy condition. Logarithmic regret bounds are also derived in \citet{slivkins2013dynamic} for a dynamic ad allocation problem that can be cast as a BwK problem.

\paragraph{Organization.} The remainder of the paper is organized as follows. We present applications of the BwK model in Section \ref{sec-applications}. We expose the algorithmic ideas underlying our approach in Section \ref{sec-algorithmicideas} and apply theses ideas to Cases (1), (2), (3), and (4) in Sections \ref{sec-singlebudget}, \ref{sec-multiplebudgets}, \ref{sec-singlebudgettimehorizon}, and \ref{sec-stochastic-multiple-budget} respectively. We choose to discuss each case separately in a self-contained fashion so that readers can delve into the setting they are most interested in. This comes at the price of some overlap in the analysis. We relax some of the assumptions made in the course of proving the regret bounds and discuss extensions in Section \ref{sec-extensions} of the Appendix. To provide as much intuition as possible, the ideas and key technical steps are all included in the main body, sometimes through proof sketches, while the technical details are deferred to the Appendix.

\paragraph{Notations.} For a set $S$, $|S|$ denotes the cardinality of $S$ while $\mathbbm{1}_S$ is the indicator function of $S$. For a vector $x \in \mathbb{R}^n$ and $S$ a subset of $\{1, \cdots, n\}$, $x_S$ refers to the subvector $(x_i)_{i \in S}$. For a square matrix $A$, $\det(A)$  is the determinant of $A$ while $\adj(A)$ denotes its adjugate. For $x \in \mathbb{R}, x_+$ is the positive part of $x$. We use standard asymptotic notations such as $O(\cdot)$, $o(\cdot)$, $\Omega(\cdot)$, and $\Theta(\cdot)$.

\section{Applications.}
\label{sec-applications}
A number of applications of the BwK framework are documented in the literature. For the purpose of being self-contained, we review a few popular ones that satisfy our technical assumptions.

\subsection{Online advertising.}
\label{sec-applications-advertising}
\paragraph{Bid optimization in repeated second-price auctions.}
Consider a bidder participating in sealed second-price auctions who is willing to spend a budget $B$. This budget may be allocated only for a period of time (for the next $T$ auctions) or until it is completely exhausted. Rounds, indexed by $t \in \mathbb{N}$, correspond to auctions the bidder participates in. If the bid submitted by the bidder  for auction $t$ is larger than the highest bid submitted by the competitors, denoted by $m_t$, the bidder wins the auction, derives a private utility $v_t \in [0, 1]$ (whose monetary value is typically difficult to assess), and is charged $m_t$. Otherwise, $m_t$ is not revealed to the bidder and $v_t$ cannot be assessed. We consider a stochastic setting where the environment and the competitors are not fully adversarial: $((v_t, m_t))_{t \in \mathbb{N}}$ is assumed to be an i.i.d. stochastic process. The goal for the bidder is to design a strategy to maximize the expected total utility derived given that the bidder has selected a grid of bids to choose from $(b_1, \cdots, b_K)$ (e.g. $b_1 = \$0.10, \cdots, b_K = \$1)$. This is a BwK problem with two resources: time and money. Pulling arm $k$ at round $t$ corresponds to bidding $b_k$ in auction $t$, costs $c_{k, t} = m_t \cdot \mathbbm{1}_{b_k \geq m_t}$, and yields a reward $r_{k, t} = v_t \cdot \mathbbm{1}_{b_k \geq m_t}$. \citet{weed2015online} design bidding strategies for a variant of this problem where the bidder is not limited by a budget and $r_{k, t} = (v_t - m_t) \cdot \mathbbm{1}_{b_k \geq m_t}$. 
\\
\indent This model was first formalized in \citet{tranthanonlinebidoptimization2014} in the context of sponsored search auctions. In sponsored search auctions, advertisers can bid on keywords to have ads (typically in the form of a link followed by a text description) displayed alongside the search results of a web search engine. When a user types a search query, a set of relevant ads are selected and an auction is run in order to determine which ones will be displayed. The winning ads are allocated to ad slots based on the outcome of the auction and, in the prevailing cost-per-click pricing scheme, their owners get charged only if the user clicks on their ads. Because the auction is often a variant of a sealed second-price auction (e.g. a generalized second-price auction), very limited feedback is provided to the advertiser if the auction is lost. In addition, both the demand and the supply cannot be predicted ahead of time and are thus commonly modeled as random variables, see \citet{ghosh2009adaptive}. For these reasons, bidding repeatedly on a keyword can be formulated as a BwK problem. In particular, when the search engine has a single ad slot per query, this problem can be modeled as above: $B$ is the budget the advertiser is willing to spend on a predetermined keyword and rounds correspond to ad auctions the advertiser has been selected to participate in. If the advertiser wins the auction, his or her ad gets displayed and he or she derives a utility $v_t = \mathbbm{1}_{A_t}$, where $A_t$ is the event that the ad gets clicked on. The goal is to maximize the expected total number of clicks given the budget constraint. The advertiser may also be interested in optimizing the ad to be displayed, which will affect the probability of a click. In this case, the modeling is similar but arms correspond to pairs of bid values and ads.

\paragraph{Dynamic ad allocation.}
This problem was first modeled in the BwK framework in \citet{slivkins2013dynamic}. A publisher, i.e. the owner of a collection of websites where ads can be displayed, has previously agreed with $K$ advertisers, indexed by $k \in \{1, \cdots, K\}$, on a predetermined cost-per-click $p_k$. Additionally, advertiser $k$ is not willing to spend more than a prescribed budget, $B_k$, for a predetermined period of time (which corresponds to the next $T$ visits or rounds). Denote by $A^k_t$ the event that the ad provided by advertiser $k$ gets clicked on at round $t$. We consider a stochastic setting where the visitors are not fully adversarial: $(\mathbbm{1}_{A^k_t})_{t \in \mathbb{N}}$ is assumed to be an i.d.d. stochastic process for any advertiser $k$. The goal for the publisher is to maximize the total expected revenues  by choosing which ad to display at every round, i.e. every time somebody visits one of the websites. This situation can be modeled as a BwK problem with $K+1$ resources: time and money for each of the $K$ advertisers. Pulling arm $k \in \{1, \cdots, K\}$ at round $t$ corresponds to displaying the ad owned by advertiser $k$, incurs the costs $c_{k, t}(i) = p_k \cdot \mathbbm{1}_{A^k_t} \cdot \mathbbm{1}_{i = k}$ to advertiser $i \in \{1, \cdots, K\}$, and yields a revenue $r_{k, t} = p_k \cdot \mathbbm{1}_{A^k_t}$.

\subsection{Revenue management.}

\paragraph{Dynamic pricing with limited supply.}
This BwK model was first proposed in \citet{babaioff2012dynamic}. An agent has $B$ identical items to sell to $T$ potential customers that arrive sequentially. Customer $t \in \{1, \cdots, T\}$ is offered a take-it-or-leave-it price $p_t$ and purchases the item only if $p_t$ is no larger than his or her own valuation $v_t$, which is never disclosed. Customers are assumed to be non-strategic in the sense that their valuations are assumed to be drawn i.i.d. from a distribution unknown to the agent. The goal for the agent is to maximize the total expected revenues by offering prices among a predetermined list $(p_1, \cdots, p_K)$. This is a BwK problem with two resources: time and item inventory. Pulling arm $k \in \{1, \cdots, K\}$ at round $t$ corresponds to offering the price $p_k$, depletes the inventory of $c_{k, t} = \mathbbm{1}_{p_k \leq v_t}$ unit, and generates a revenue $r_{k, t} = p_k \cdot \mathbbm{1}_{p_k \leq v_t}$. \citet{badanidiyuru2013bandits} propose an extension where multiple units of $M$ different products may be offered to a customer, which then buys as many as needed of each kind in order to maximize his or her own utility function. In this case, the modeling is similar but arms correspond to vectors of dimension $2 M$ specifying the number of items offered along with the price tag for each product and there are $M+1$ resources: time and item inventory for each of the $M$ products.


\paragraph{Network revenue management.}
\subparagraph{Non-perishable goods.}
This is an extension of the dynamic pricing problem developed in \citet{besbes2012blind} which is particularly suited for applications in the online retailer industry, e.g. the online fashion sample sales industry, see \citet{johnson2015online}. Each product $m=1, \cdots, M$ is produced from a finite amount of $C$ different kinds of raw materials (which may be products themselves). Producing one unit of product $m \in \{1, \cdots, M\}$ consumes a deterministic amount of resource $i \in \{1, \cdots, C\}$ denoted by $c_{m}(i)$. Customer $t \in \{1, \cdots, T\}$ is offered a product $m_t \in \{1, \cdots, M\}$ along with a take-it-or-leave-it price $p^{m_t}_t$ and purchases it if his or her valuation $v^{m_t}_t$ is larger than $p^{m_t}_t$. Products are manufactured online as customers order them. We assume that $((v^1_t, \cdots, v^M_t))_{t \in \mathbb{N}}$ is an i.i.d. stochastic process with distribution unknown to the agent. This is a BwK problem with $C+1$ resources: time and the initial endowment of each resource. Given a predetermined list of arms $((m_k, p_k))_{k=1, \cdots, K}$, pulling arm $k$ at round $t$ corresponds to offering product $m_k$ at the price $p_k$, incurs the consumption of resource $i$ by an amount $c_{k, t}(i) = c_{m_k}(i) \cdot \mathbbm{1}_{p_k \leq v^{m_k}_t}$, and generates a revenue $r_{k, t} = p_k \cdot \mathbbm{1}_{p_k \leq v^{m_k}_t}$.

\subparagraph{Perishable goods.}
This is a variant of the last model developed for perishable goods, with applications in the food retail industry and the newspaper industry. At each time period $t \in \{1, \cdots, T\}$, a retailer chooses how many units $\lambda^m_t \in \mathbb{N}$ of product $m \in \{1, \cdots, M\}$ to manufacture along with a price offer for it $p^m_t$. At time $t$, the demand for product $m$ sold at the price $p$ is a random quantity denoted by $d^m_t(p)$. We assume that customers are non-strategic: for any vector of prices $(p_1, \cdots, p_m)$, $((d^1_t(p_1), \cdots d^M_t(p_M)))_{t \in \mathbb{N}}$ is an i.i.d. stochastic process with distribution unknown to the agent. Products perish at the end of each round irrespective of whether they have been purchased. Given a predetermined list of arms $((\lambda^1_k, p^1_k, \cdots, \lambda^M_k, p^M_k))_{k=1, \cdots, K}$, pulling arm $k$ at round $t$ corresponds to offering $\lambda^m_k$ units of product $m$ at the price $p^m_k$ for any $m \in \{1, \cdots, M\}$, incurs the consumption of resource $i$ by a deterministic amount $c_{k, t}(i) = \sum_{m=1}^M \lambda^m_k \cdot c_{m}(i)$ (where $c_m(i)$ is defined in the previous paragraph), and generates a revenue $r_{k, t} = \sum_{m=1}^M p^m_k \cdot \min(d^m_t(p^m_k), \lambda^m_k)$.

\paragraph{Shelf optimization for perishable goods.}
This is a variant of the model introduced in \citet{graczova2014generalized}. Consider a retailer who has an unlimited supply of $M$ different types of products. At each time period $t$, the retailer has to decide how many units, $\lambda^m_t$, of each product, $m \in \{1, \cdots, M\}$, to allocate to a promotion space given that at most $N$ items fit in the limited promotion space. Moreover, the retailer also has to decide on a price tag $p^m_t$ for each product $m$. All units of product $m \in \{1, \cdots, M\}$ perish by time period $T_m$ and the retailer is planning the allocation for the next $T$ time periods. At round $t$, the demand for product $m$ is a random quantity denoted by $d^m_t(p)$. Customers are non-strategic: for any vector of prices $(p_1, \cdots, p_m)$, $((d^1_t(p_1), \cdots d^M_t(p_M)))_{t \in \mathbb{N}}$ is an i.i.d. stochastic process with distribution unknown to the agent. This is a BwK problem with $M+1$ resources: time horizon and time after which each product perishes. Given a predetermined list of arms $((\lambda^1_k, p^1_k, \cdots, \lambda^M_k, p^M_k))_{k=1, \cdots, K}$ satisfying $\sum_{m=1}^M \lambda^m_k \leq K$ for any $k \in \{1, \cdots, K\}$, pulling arm $k$ at round $t$ corresponds to allocating $\lambda^m_k$ units of product $m$ to the promotion space for every $m \in \{1, \cdots, M\}$ with the respective price tags $(p^1_k, \cdots, p^M_k)$, incurs the consumption of resource $i$ by a deterministic amount $c_{k, t}(i) = 1$, and generates a revenue $r_{k, t} = \sum_{m=1}^M p^m_k \cdot \min(d^m_t(p^m_k), \lambda^m_k)$.

\subsection{Dynamic procurement.}
This problem was first studied in \citet{badanidiyuru2012learning}. Consider a buyer with a budget $B$ facing $T$ agents arriving sequentially, each interested in selling one good. Agent $t \in \{1, \cdots, T\}$ is offered a take-it-or-leave-it price, $p_t$, and makes a sell only if the value he or she attributes to the item, $v_t$, is no larger than $p_t$. We consider a stochastic setting where the sellers are not fully adversarial: $(v_t)_{t \in \mathbb{N}}$ is an i.i.d. stochastic process with distribution unknown to the buyer. The goal for the buyer is to maximize the total expected number of goods purchased by offering prices among a predetermined list $(p_1, \cdots, p_K)$. This is a BwK problem with two resources: time and money. Pulling arm $k$ at round $t$ corresponds to offering the price $p_k$, incurs a cost $c_{k, t} = p_k \cdot \mathbbm{1}_{p_k \geq v_t}$, and yields a reward $r_{k, t} = \mathbbm{1}_{p_k \geq v_t}$. It is also possible to model situations where the agents are selling multiple types of products and/or multiple units, in which case arms correspond to vectors specifying the number of units of each product required along with their respective prices, see \citet{badanidiyuru2013bandits}.\\
\indent Applications of this model to crowdsourcing platforms are described in \citet{badanidiyuru2012learning} and \citet{badanidiyuru2013bandits}. In this setting, agents correspond to workers that are willing to carry out microtasks which are submitted by buyers (called \enquote{requesters}) using a posted-price mechanism. Requesters are typically submitting large batches of jobs and can thus adjust the posted prices as they learn about the pool of workers.

\subsection{Wireless sensor networks.}
This is a variant of the model introduced in \citet{tran2012long}. Consider an agent collecting information using a network of wireless sensors powered by batteries. Activating sensor $k \in \{1, \cdots, K\}$ consumes some amount of energy, $c_k$, which is depleted from the sensor's initial battery level, $B_k$, and triggers a measurement providing a random amount of information (measured in bits), $r_{k, t}$, which is transmitted back to the agent. Sensors cannot harvest energy and the goal for the agent is to maximize the total expected amount of information collected over $T$ actions. This is a BwK problem with $K+1$ resources: time and the energy stored in the battery of each sensor. Pulling arm $k \in \{1, \cdots, K\}$ corresponds to activating sensor $k$, incurs the consumption of resource $i \in \{1, \cdots, K\}$ by a deterministic amount $c_{k, t} = c_k \cdot \mathbbm{1}_{k = i}$, and yields a random reward $r_{k, t}$.

\section{Algorithmic ideas.}
\label{sec-algorithmicideas} 
\subsection{Preliminaries.}
To handle the exploration-exploitation trade-off, an approach that has proved to be particularly successful hinges on the \emph{optimism in the face of uncertainty} paradigm. The idea is to consider all plausible scenarios consistent with the information collected so far and to select the decision that yields the most revenue among all the scenarios identified. Concentration inequalities are intrinsic to the paradigm as they enable the development of systematic closed form confidence intervals on the quantities of interest, which together define a set of plausible scenarios. We make repeated use of the following result.

\begin{lemma} Hoeffding's inequality
	\label{lemma-martingale-inequality}
	Consider $X_1, \cdots, X_n$ $n$ random variables with support in $[0, 1]$. 
	\\	
	If $\forall t \leq n \; \mathbb{E}[X_t \; | \; X_1, \cdots, X_{t-1} ] \leq \mu$, then $\mathbb{P}[X_1 + \cdots + X_n \geq n \mu + a] \leq \exp( - \frac{2 a^2}{n} ) \quad \forall a \geq 0$.
	\\
	If $\forall t \leq n \; \mathbb{E}[X_t \; | \; X_1, \cdots, X_{t-1} ] \geq \mu$, then $\mathbb{P}[X_1 + \cdots + X_n \leq n \mu - a] \leq \exp( - \frac{2 a^2}{n} ) \quad \forall a \geq 0$.
\end{lemma} \vspace{0.3cm}
\noindent
\citet{auer2002finite} follow the \emph{optimism in the face of uncertainty} paradigm to develop the Upper Confidence Bound algorithm (UCB1). UCB1 is based on the following observations: (i) the optimal strategy always consists in pulling the arm with the highest mean reward when time is the only limited resource, (ii) informally, Lemma \ref{lemma-martingale-inequality} shows that $\mu_k^r \in [\bar{r}_{k, t} - \epsilon_{k, t}, \bar{r}_{k, t} + \epsilon_{k, t}]$ at time $t$ with probability at least $1 - 2/t^3$ for $\epsilon_{k, t} = \sqrt{ 2 \ln(t) / n_{k, t}  }$, irrespective of the number of times arm $k$ has been pulled. Based on these observations, UCB1 always selects the arm with highest UCB index, i.e. $a_t \in \argmax_{k=1, \cdots, K} I_{k, t}$, where the UCB index of arm $k$ at time $t$ is defined as $I_{k, t} = \bar{r}_{k, t} + \epsilon_{k, t}$. The first term can be interpreted as an exploitation term, the ultimate goal being to maximize revenue, while the second term is an exploration term, the smaller $n_{k, t}$, the bigger it is. This fruitful paradigm go well beyond this special case and many extensions of UCB1 have been designed to tackle variants of the MAB problem, see for example \citet{slivkins2013dynamic}. \citet{agrawal2014bandits} embrace the same ideas to tackle BwK problems. The situation is more complex in this all-encompassing framework as the optimal oracle algorithm involves pulling several arms. In fact, finding the optimal pulling strategy given the knowledge of the underlying distributions is already a challenge in its own, see \citet{papadimitriou1999complexity} for a study of the computational complexity of similar problems. This raises the question of how to evaluate $\ropt(B(1), \cdots, B(C))$ in \eqref{eq-general-def-regret}. To overcome this issue, \citet{badanidiyuru2013bandits} upper bound the total expected payoff of any non-anticipating algorithm by the optimal value of a linear program, which is easier to compute.
\begin{lemma} Adapted from \citet{badanidiyuru2013bandits} \\
\label{lemma-general-bound-optimal-policy}
	The total expected payoff of any non-anticipating algorithm is no greater than $B$ times the optimal value of the linear program:
	\begin{equation}
		\label{eq-linear-program-general-upperbound-opt-strategy}
		\begin{aligned}
			& \sup_{ (\xi_k)_{k=1, \cdots, K} } 
			& & \sum_{k=1}^K \mu^r_k \cdot \xi_k \\
			& \text{subject to}
			& & \sum_{k=1}^K \mu^{c}_k(i) \cdot \xi_k \leq b(i), \quad i = 1, \cdots, C \\
			&
			& & \xi_k \geq 0, \quad k = 1, \cdots, K.
		\end{aligned}
	\end{equation}
	plus the constant term $\max\limits_{ \substack{ k=1, \cdots, K \\ i = 1, \cdots, C \\ \text{with } \mu^c_k(i)>0}} \frac{\mu^r_k}{\mu^{c}_k(i)}$.
\end{lemma} \vspace{0.3cm}
\noindent
The optimization problem \eqref{eq-linear-program-general-upperbound-opt-strategy} can be interpreted as follows. For any arm $k$, $B \cdot \xi_k$ corresponds to the expected number of times arm $k$ is pulled by the optimal algorithm. Hence, assuming we introduce a dummy arm $0$ which is equivalent to skipping the current round, $\xi_k$ can be interpreted as the probability of pulling arm $k$ at any round when there is a time horizon $T$. Observe that the constraints restrict the feasible set of expected number of pulls by imposing that the amounts of resources consumed are no greater than their respective budgets in expectations, as opposed to almost surely which would be a more stringent constraint. This explains why the optimal value of \eqref{eq-linear-program-general-upperbound-opt-strategy} is larger than the maximum achievable payoff. In this paper, we use standard linear programming notions such as the concept of a basis and a basic feasible solution. We refer to \citet{bertsimas1997introduction} for an introduction to linear programming. A pseudo-basis $x$ is described by two subsets $\mathcal{K}_x \subset \{1, \cdots, K\}$ and $\mathcal{C}_x \subset \{1, \cdots, C\}$ such that $|\mathcal{K}_x | = |\mathcal{C}_x |$. A pseudo-basis $x$ is a basis for \eqref{eq-linear-program-general-upperbound-opt-strategy} if the matrix $A_x = (\mu^c_k(i))_{(i,k) \in \mathcal{C}_x \times \mathcal{K}_x}$ is invertible. Furthermore, $x$ is said to be a feasible basis for \eqref{eq-linear-program-general-upperbound-opt-strategy} if the corresponding basic solution, denoted by $(\xi^x_k)_{k=1, \cdots, K}$ and determined by $\xi^x_k = 0$ for $k \not \in \mathcal{K}_x$ and $A_x \xi^x_{\mathcal{K}_x} = b_{\mathcal{C}_x}$ (where $b_{\mathcal{C}_x}$ is the subvector $(b(i))_{i \in \mathcal{C}_x}$), is feasible for \eqref{eq-linear-program-general-upperbound-opt-strategy}. When $x$ is a feasible basis for \eqref{eq-linear-program-general-upperbound-opt-strategy}, we denote by $\mathrm{obj}_{x} = \sum_{k=1}^K \mu^r_k \cdot \xi^{x}_k $ its objective function. From Lemma \ref{lemma-general-bound-optimal-policy}, we derive:
\begin{equation}
	\label{eq-simplified-general-upper-bound-on-regret}
		R_{B(1), \cdots, B(C)} \leq  B \cdot \mathrm{obj}_{x^*}   - \mathbb{E}[\sum_{t=1}^{\tau^*} r_{a_t, t}] + O(1),
\end{equation}
where $x^*$ is an optimal feasible basis for \eqref{eq-linear-program-general-upperbound-opt-strategy}. For mathematical convenience, we consider that the game carries on even if one of the resources is already exhausted so that $a_t$ is well defined for any $t \in \mathbb{N}$. Of course, the rewards obtained for $t \geq \tau^*$ are not taken into account in the decision maker's payoff when establishing regret bounds. 

\subsection{Solution methodology.}
Lemma \ref{lemma-general-bound-optimal-policy} also provides insight into designing algorithms. The idea is to incorporate confidence intervals on the mean rewards and the mean amounts of resource consumption into the offline optimization problem \eqref{eq-linear-program-general-upperbound-opt-strategy} and to base the decision upon the resulting optimal solution. There are several ways to carry out this task, each leading to a different algorithm. When there is a time horizon $T$, \citet{agrawal2014bandits} use high-probability lower (resp. upper) bounds on the mean amounts of resource consumption (resp. rewards) in place of the unknown mean values in \eqref{eq-linear-program-general-upperbound-opt-strategy} and pull an arm at random according to the resulting optimal distribution.  Specifically, at any round $t$, the authors suggest to compute an optimal solution $(\xi^*_{k, t})_{k=1, \cdots, K}$ to the linear program:
\begin{equation}
	\label{eq-linear-program-agrawal}
		\begin{aligned}
			& \sup_{ (\xi_k)_{k=1, \cdots, K} } 
			& & \sum_{k=1}^K ( \bar{r}_{k, t} + \epsilon_{k, t}) \cdot \xi_k \\
			& \text{subject to}
			& & \sum_{k=1}^K  (\bar{c}_{k, t}(i) - \epsilon_{k, t} ) \cdot \xi_k \leq (1- \gamma) \cdot b(i), \quad i = 1, \cdots, C-1 \\
			&
			& & \sum_{k=1}^K \xi_k \leq  1 \\
			&
			& & \xi_k \geq 0, \quad k = 1, \cdots, K,
		\end{aligned}
\end{equation}
for a well-chosen $\gamma \in (0, 1)$, and then to pull arm $k$ with probability $\xi^*_{k, t}$ or skip the round with probability $1 - \sum_{k=1}^K \xi^*_{k, t}$. If we relate this approach to UCB1, the intuition is clear: the idea is to be optimistic about both the rewards and the amounts of resource consumption. We argue that this approach cannot yield logarithmic regret bounds. First, because $\gamma$ has to be of order $1/\sqrt{T}$. Second, because, even if we were given an optimal solution to \eqref{eq-linear-program-general-upperbound-opt-strategy}, $(\xi^{x^*}_k)_{k=1, \cdots, K}$, before starting the game, consistently choosing which arm to pull at random according to this distribution at every round would incur regret $\Omega(\sqrt{T})$, as we next show. 
\begin{lemma}
\label{lemma-lower-bound-agrawal}
	For all the cases treated in this paper, pulling arm $k$ with probability $\xi^{x^*}_k$ at any round $t$ yields a regret of order $\Omega(\sqrt{T})$ unless pulling any arm in $\{ k \in \{1, \cdots, K \} \; | \; \xi^{x^*}_k > 0 \}$ incurs the same deterministic amount of resource consumption for all resources in $\mathcal{C}_{x^*}$ and for all rounds $t \in \mathbb{N}$.
	\proof{Proof.}
	We use the shorthand notation $\mathcal{K}_* = \{ k \in \{1, \cdots, K \} \; | \; \xi^{x^*}_k > 0 \}$. Observe that, for any resource $i \in \{1, \cdots, C-1\}$, we have:
	\begin{align*}
		T \cdot \mathrm{obj}_{x^*} - \mathbb{E}[\sum_{t=1}^{ \tau^*} r_{a_t, t}]
			& =  ( T - \mathbb{E}[\tau^*]) \cdot \mathrm{obj}_{x^*} \\
			& \geq \mathbb{E}[ ( \sum_{t=\tau^*}^{ T } c_{a_t, t}(i) + \sum_{t=1}^{ \tau^* -1 } c_{a_t, t}(i) - B(i) )_+ ] \cdot \mathrm{obj}_{x^*} + O(1) \\
			& = \mathbb{E}[ ( \sum_{t=1}^{ T } \{ c_{a_t, t}(i) - b(i) \} )_+ ] \cdot \mathrm{obj}_{x^*} + O(1),
	\end{align*}
	where the inequality is derived using $c_{a_t, t}(i) \leq 1$ for all rounds $t$ and $\sum_{t=1}^{ \tau^* -1 } c_{a_t, t}(i) \leq B(i)$. Since, for $i \in \mathcal{C}_{x^*}$, $(c_{a_t, t}(i))_{t \in \mathbb{N}}$ is an i.i.d. bounded stochastic process with mean $b(i)$, we have:
\begin{equation}
	\label{eq-lower-bound-agrawal}
	\mathbb{E}[ ( \sum_{t=1}^{ T } \{ c_{a_t, t}(i) - b(i) \} )_+ ] = \Omega(\sqrt{T}),
\end{equation}
provided that $c_{a_t, t}(i)$ has positive variance, which is true if there exists at least one arm $k \in \mathcal{K}_*$ such that $c_{k, t}(i)$ has positive variance or if there exist two arms $k, l \in \mathcal{K}_*$ such that $c_{k, t}(i)$ and $c_{l, t}(i)$ are not almost surely equal to the same deterministic value. Strictly speaking, this is not enough to conclude that $R_{B(1), \cdots, B(C-1), T} = \Omega(\sqrt{T})$ as $T \cdot \mathrm{obj}_{x^*}$ is only an upper bound on the maximum total expected payoff. However, in Sections \ref{sec-singlebudget}, \ref{sec-multiplebudgets}, \ref{sec-singlebudgettimehorizon}, and \ref{sec-stochastic-multiple-budget}, we show that there exists an algorithm that satisfies $T \cdot \mathrm{obj}_{x^*} - \mathbb{E}[\sum_{t=1}^{ \tau^*} r_{a_t, t}]= O(\ln(T))$ for all the cases considered in this paper. This result, together with \eqref{eq-lower-bound-agrawal} and Lemma \ref{lemma-general-bound-optimal-policy}, implies that the regret incurred when pulling arm $k$ with probability $\xi^{x^*}_k$ at any round is $\Omega(\sqrt{T})$. 
	\endproof
\end{lemma} \vspace{0.3cm}

\noindent The fundamental shortcoming of this approach is that it systematically leads us to plan to consume the same average amount of resource $i$ per round $b(i)$, for any resource $i=1, \cdots, C-1$, irrespective of whether we have significantly over- or under-consumed in the past. Based on this observation, a natural idea is to solve the linear program:
\begin{equation}
	\label{eq-linear-program-optimal-algorithm}
		\begin{aligned}
			& \sup_{ (\xi_k)_{k=1, \cdots, K} } 
			& & \sum_{k=1}^K ( \bar{r}_{k, t} + \epsilon_{k, t}) \cdot \xi_k \\
			& \text{subject to}
			& & \sum_{k=1}^K  (\bar{c}_{k, t}(i) - \epsilon_{k, t} ) \cdot \xi_k \leq (1- \gamma) \cdot b_t(i), \quad i = 1, \cdots, C-1 \\
			&
			& & \sum_{k=1}^K \xi_k \leq  1 \\
			&
			& & \xi_k \geq 0, \quad k = 1, \cdots, K,
		\end{aligned}
\end{equation}
instead of \eqref{eq-linear-program-agrawal}, where $b_t(i)$ denotes the ratio of the remaining amount of resource $i$ at time $t$ to the remaining time horizon, i.e. $T-t+1$. Bounding the regret incurred by this adaptive algorithm is, however, difficult from a theoretical standpoint. To address this issue, we propose the following family of algorithms, whose behaviors are similar to the adaptive algorithm but lend themselves to an easier analysis.

\vspace{0.3cm}
\begin{algorithm}[H]
\caption{UCB-Simplex}
\label{def-generic-algorithms}
Take $\lambda \geq 1$ and $(\eta_i)_{i=1, \cdots, C} \geq 0$ (these quantities will need to be carefully chosen). The algorithm is preceded by an initialization phase which consists in pulling each arm a given number of times, to be specified. For each subsequent time period $t$, proceed as follows. \\
	\textbf{Step-Simplex}: Find an optimal basis $x_t$ to the linear program:
	\begin{equation}
		\label{eq-algo-general-idea}
		\begin{aligned}
			& \sup_{ (\xi_k)_{k=1, \cdots, K} } 
			& & \sum_{k=1}^K (\bar{r}_{k, t} + \lambda \cdot \epsilon_{k, t}) \cdot \xi_k \\
			& \text{subject to}
			& & \sum_{k=1}^K ( \bar{c}_{k, t}(i) - \eta_i \cdot \epsilon_{k, t}) \cdot \xi_k \leq b(i), \quad i = 1, \cdots, C \\
			&
			& & \xi_k \geq 0, \quad k = 1, \cdots, K
		\end{aligned}
	\end{equation}
	Adapting the notations, $x_t$ is described by two subsets $\mathcal{K}_{x_t} \subset \{1, \cdots, K\}$ and $\mathcal{C}_{x_t} \subset \{1, \cdots, C\}$ such that $|\mathcal{K}_{x_t} | = |\mathcal{C}_{x_t} |$, the matrix $\bar{A}_{x_t, t} = (\bar{c}_{k,t}(i) - \eta_i \cdot \epsilon_{k, t})_{(i,k) \in \mathcal{C}_{x_t} \times \mathcal{K}_{x_t}}$, and the corresponding basic feasible solution $(\xi^{x_t}_{k, t})_{k=1, \cdots, K}$ determined by $\xi^{x_t}_{k,t} = 0$ for $k \not \in \mathcal{K}_{x_t}$ and $\bar{A}_{x_t, t} \xi^{x_t}_{\mathcal{K}_x, t} = b_{\mathcal{K}_{x_t}}$. 	 \\
	\textbf{Step-Load-Balance}: Identify the arms involved in the optimal basis, i.e. $\mathcal{K}_{x_t}$. There are at most $\min(K, C)$ such arms. Use a load balancing algorithm $\mathcal{A}_{x_t}$, to be specified, to determine which of these arms to pull.
\end{algorithm} \vspace{0.3cm}

\noindent  For all the cases considered in this paper, \eqref{eq-algo-general-idea} is always bounded and Step-Simplex is well defined. The Simplex algorithm is an obvious choice to carry out Step-Simplex, especially when $\eta_i = 0$ for any resource $i \in \{1, \cdots, C\}$, because, in this case, we only have to update one column of the constraint matrix per round which makes warm-starting properties attractive. However, note that this can also be done in time polynomial in $K$ and $C$, see \citet{grotschel2012geometric}. If we compare \eqref{eq-algo-general-idea} with \eqref{eq-linear-program-agrawal}, the idea remains to be overly optimistic but, as we will see, more about the rewards than the amounts of resource consumption through the exploration factor $\lambda$ which will typically be larger than $\eta_i$, thus transferring most of the burden of exploration from the constraints to the objective function. The details of Step-Load-Balance are purposefully left out and will be specified for each of the cases treated in this paper. When there is a time horizon $T$, the general idea is to determine, at any time period $t$ and for each resource $i=1, \cdots, C$, whether we have over- or under-consumed in the past and to perturb the probability distribution $(\xi^{x_t}_{k, t})_{k=1, \cdots, K}$ accordingly to get back on track.  \\
\indent The algorithm we propose is intrinsically tied to the existence of basic feasible optimal solutions to \eqref{eq-linear-program-general-upperbound-opt-strategy} and \eqref{eq-algo-general-idea}. We denote by $\mathcal{B}$ (resp. $\mathcal{B}_t$) the subset of bases of \eqref{eq-linear-program-general-upperbound-opt-strategy} (resp. \eqref{eq-algo-general-idea}) that are feasible for \eqref{eq-linear-program-general-upperbound-opt-strategy} (resp. \eqref{eq-algo-general-idea}). Step-Simplex can be interpreted as an extension of the index-based decision rule of UCB1. Indeed, Step-Simplex assigns an index $I_{x, t}$ to each basis $x \in \mathcal{B}_t$ and outputs $x_t \in \argmax_{x \in \mathcal{B}_t} I_{x, t}$, where $I_{x, t} = \mathrm{obj}_{x, t} + E_{x, t}$ with a clear separation (at least when $\eta_i =0$ for any resource $i$) between the exploitation term, $\mathrm{obj}_{x, t} = \sum_{k=1}^K \xi^{x}_{k, t} \cdot \bar{r}_{k, t}$, and the exploration term, $E_{x, t} = \lambda \cdot \sum_{k=1}^K \xi^{x}_{k, t} \cdot \epsilon_{k, t}$. Observe that, for $x \in \mathcal{B}_t$ that is also feasible for \eqref{eq-linear-program-general-upperbound-opt-strategy}, $(\xi^{x}_{k, t})_{k=1, \cdots, K}$ and $\mathrm{obj}_{x, t}$ are plug-in estimates of $(\xi^x_k)_{k=1, \cdots, K}$ and $\mathrm{obj}_{x}$ when $\eta_i = 0$ for any resource $i$. Also note that when $\lambda = 1$ and $\eta_i = 0$ for any resource $i$ and when time is the only limited resource, UCB-Simplex is identical to UCB1 as Step-Load-Balance is unambiguous in this special case, each basis involving a single arm. For any $x \in \mathcal{B}$, we define $\Delta_x = \mathrm{obj}_{x^*} - \mathrm{obj}_{x} \geq 0$ as the optimality gap. A feasible basis $x$ is said to be suboptimal if $\Delta_{x} > 0$. At any time $t$, $n_{x, t}$ denotes the number of times basis $x$ has been selected at Step-Simplex up to time $t$ while $n^x_{k, t}$ denotes the number of times arm $k$ has been pulled up to time $t$ when selecting $x$ at Step-Simplex. For all the cases treated in this paper, we will show that, under a nondegeneracy assumption, Step-Simplex guarantees that a suboptimal basis cannot be selected more than $O( \ln(B) )$ times on average, a result reminiscent of the regret analysis of UCB1 carried out in \citet{auer2002finite}. However, in stark contrast with the situation of a single limited resource, this is merely a prerequisite to establish a $O( \ln(B) )$ bound on regret. Indeed, a low regret algorithm must also balance the load between the arms as closely as possible to optimality. Hence, the choice of the load balancing algorithms $\mathcal{A}_x$ is crucial to obtain logarithmic regret bounds. 

\section{A single limited resource.}
\label{sec-singlebudget}
In this section, we tackle the case of a single resource whose consumption is limited by a global budget $B$, i.e. $C=1$ and $b(1) = 1$. To simplify the notations, we omit the indices identifying the resources as there is only one, i.e. we write $\mu^c_k$, $c_{k, t}$, $\bar{c}_{k, t}$, and $\eta$ as opposed to $\mu^{c}_k(1)$, $c_{k, t}(1)$, $\bar{c}_{k, t}(1)$, and $\eta_1$. We also use the shorthand $\epsilon = \min_{k=1, \cdots, K} \mu^c_k$. Recall that, under Assumption \ref{assumption-all-cost-non-zero}, $\epsilon$ is positive and a priori unknown to the decision maker. In order to derive logarithmic bounds, we will also need to assume that the decision maker knows an upper bound on the optimal value of \eqref{eq-linear-program-general-upperbound-opt-strategy}.
\begin{assumption}
	\label{assumption-simplying-assumption-single-budget}
	The decision maker knows $\kappa \geq \max\limits_{k=1, \cdots, K} \frac{\mu^r_k}{\mu^c_k}$ ahead of round $1$.
\end{assumption}
Assumption \ref{assumption-simplying-assumption-single-budget} is natural in repeated second-price auctions, as detailed in the last paragraph of this section. Moreover, note that if $\epsilon$ happens to be known ahead of round $1$ we can take $\kappa = 1/\epsilon$.
\paragraph{Specification of the algorithm.}
We implement UCB-Simplex with $\lambda = 1 + \kappa$ and $\eta = 0$. The initialization step consists in pulling each arm until the amount of resource consumed as a result of pulling that arm is non-zero. The purpose of this step is to have $\bar{c}_{k, t} > 0$ for all periods to come and for all arms. Step-Load-Balance is unambiguous here as basic feasible solutions involve a single arm. Hence, we identify a basis $x$ such that $\mathcal{K}_x = \{ k \}$ and $\mathcal{C}_x = \{1\}$ with the corresponding arm and write $x = k$ to simplify the notations. In particular, $k^* \in \{1, \cdots, K\}$ identifies an optimal arm in the sense defined in Section \ref{sec-algorithmicideas}. For any arm $k$, the exploration and exploitation terms defined in Section \ref{sec-algorithmicideas} specialize to:
$$
	\mathrm{obj}_{k, t} = \frac{ \bar{r}_{k, t} }{ \bar{c}_{k, t} } \; \text{and} \; E_{k, t} = ( 1 + \kappa ) \cdot \frac{ \epsilon_{k, t} }{ \bar{c}_{k, t}},
$$
while $\mathrm{obj}_{k} = \mu^r_k/\mu^c_k$, so that:
$$
	k^* \in \argmax_{k = 1, \cdots, K} \frac{\mu^r_k}{\mu^c_k}, \; a_t \in \argmax_{k = 1, \cdots, K} \frac{ \bar{r}_{k, t} + ( 1 + \kappa ) \cdot \epsilon_{k, t}  }{ \bar{c}_{k, t}}, \; \text{and} \; \Delta_k = \frac{\mu^r_{k^*}}{\mu^c_{k^*}} - \frac{\mu^r_k}{\mu^c_k}.
$$
We point out that, for the particular setting considered in this section, UCB-Simplex is almost identical to the fractional KUBE algorithm proposed in \citet{tran2012knapsack} to tackle the case of a single resource whose consumption is deterministic. It only differs by the presence of the scaling factor $1 + \kappa$ to favor exploration over exploitation, which becomes unnecessary when the amounts of resource consumed are deterministic, see Section \ref{sec-extensions} of the Appendix.
\paragraph{Regret analysis.}
We omit the initialization step in the theoretical analysis because the amount of resource consumed is $O(1)$ and the reward obtained is non-negative and not taken into account in the decision maker's total payoff. Moreover, the initialization step ends in finite time almost surely as a result of Assumption \ref{assumption-all-cost-non-zero}. First observe that \eqref{eq-simplified-general-upper-bound-on-regret} specializes to:
\begin{equation}
	\label{eq-simplified-upper-bound-on-regret}
		R_B \leq B \cdot \frac{\mu^r_{k^*}}{\mu^c_{k^*}} - \mathbb{E}[\sum_{t=1}^{\tau^*} r_{a_t, t}] + O(1).
\end{equation}
To bound the right-hand side, we start by estimating the expected time horizon.
\begin{lemma}
	\label{lemma-bound-stopping-time}
	For any non-anticipating algorithm, we have: $\mathbb{E}[\tau^*] \leq \frac{B + 1}{\epsilon}$.
	\proof{Sketch of proof.}
	By definition of $\tau^*$, we have $\sum_{t=1}^{\tau^* - 1} c_{a_t, t} \leq B$. Taking expectations on both sides yields $B \geq \mathbb{E}[ \sum_{t=1}^{\tau^*} \mu^c_{a_t}] - 1 \geq \mathbb{E}[\tau^*] \cdot \epsilon - 1$ by Assumption \ref{assumption-all-cost-non-zero}. Rearranging this last inequality yields the claim.
	\endproof
\end{lemma} \vspace{0.3cm}
\noindent The next result is crucial. Used in combination with Lemma \ref{lemma-bound-stopping-time}, it shows that any suboptimal arm is pulled at most $O(\ln(B))$ times in expectations, a well-known result for UCB1, see \citet{auer2002finite}. The proof is along the same lines as for UCB1, namely we assume that arm $k$ has already been pulled more than $\Theta(\ln(\tau^*)/(\Delta_k)^2)$ times and conclude that arm $k$ cannot be pulled more than a few more times, with the additional difficulty of having to deal with the random stopping time and the fact that the amount of resource consumed at each step is stochastic.

\begin{lemma}
	\label{lemma-bound-times-non-optimal-pulls}
	For any suboptimal arm $k$, we have:
	$$
		\mathbb{E}[n_{k, \tau^*}] \leq 2^6 (\frac{\lambda}{ \mu^c_k })^2 \cdot \frac{\mathbb{E}[ \ln(\tau^*) ]}{(\Delta_k)^2} + \frac{ 4 \pi^2 }{3 \epsilon^2}.
	$$
	\proof{Sketch of proof.}
		We use the shorthand notation $\beta_k =  2^5 (\lambda/\mu^c_k)^2 \cdot (1/\Delta_k)^2$. First observe that if we want to bound $\mathbb{E}[n_{k, \tau^*}]$, we may assume, without loss of generality, that arm $k$ has been pulled at least $\beta_k \cdot \ln(t)$ times at any time $t$ up to an additive term of $2 \beta_k \cdot \mathbb{E}[ \ln(\tau^*)]$ in the final inequality. We then just have to bound by a constant the probability that $k$ is selected at any time $t$ given that $n_{k, t} \geq \beta_k \cdot \ln(t)$. If $k$ is selected at time $t$, it must be that $k$ is optimal for \eqref{eq-algo-general-idea}, which, in particular, implies that $\mathrm{obj}_{k, t} + E_{k, t} \geq \mathrm{obj}_{k^*, t} + E_{k^*, t}$. This can only happen if either: (i) $\mathrm{obj}_{k, t}  \geq \mathrm{obj}_{k} + E_{k, t}$, i.e. the objective value of $k$ is overly optimistic, (ii) $\mathrm{obj}_{k^*, t}  \leq \mathrm{obj}_{k^*} - E_{k^*, t}$, i.e. the objective value of $k^*$ is overly pessimistic, or (iii) $\mathrm{obj}_{k^*} < \mathrm{obj}_{k} + 2 E_{k, t}$, i.e. the optimality gap of arm $k$ is small compared to its exploration factor. The probability of events (i) and (ii) can be bounded by $\sim 1/t^2$ in the same fashion, irrespective of how many times these arms have been pulled in the past. For example for event (i), this is because if $\bar{r}_{k, t} / \bar{c}_{k, t}  = \mathrm{obj}_{k, t} \geq \mathrm{obj}_{k} + E_{k, t} = \mu^r_k / \mu^c_k  + E_{k, t}$, then either (a) $\bar{r}_{k, t} \geq \mu^r_k + \epsilon_{k, t}$ or (b) $\bar{c}_{k, t} \leq \mu^c_k - \epsilon_{k, t}$ and both of these events have probability at most $\sim 1/t^2$ by Lemma \ref{lemma-martingale-inequality}. Indeed, if (a) and (b) do not hold, we have:
			\begin{align*}
				\frac{ \bar{r}_{k, t} }{ \bar{c}_{k, t} } - \frac{ \mu^r_k }{ \mu^c_k } 
					& = \frac{ ( \bar{r}_{k, t} - \mu^r_k) \mu^c_k + (\mu^c_k - \bar{c}_{k, t}) \mu^r_k }{ \bar{c}_{k, t} \cdot \mu^c_k } \\
					& <  \frac{ \epsilon_{k, t} }{ \bar{c}_{k, t} } + \frac{ \epsilon_{k, t} }{\bar{c}_{k, t}} \cdot  \frac{\mu^r_k}{\mu^c_k} \leq (1 + \kappa) \cdot \frac{ \epsilon_{k, t} }{ \bar{c}_{k, t} } = E_{k, t}.
			\end{align*}
			As for event (iii), observe that if $\mathrm{obj}_{k^*} < \mathrm{obj}_{k} + 2 E_{k, t}$ and $n_{k, t} \geq \beta_k \cdot \ln(t)$ then we have $\bar{c}_{k, t} \leq \mu^c_k/2$, which happens with  probability at most $\sim 1/t^2$ by Lemma \ref{lemma-martingale-inequality} given that arm $k$ has already been pulled at least $\sim \ln(t) / (\mu^k_c)^2$ times.
	\endproof
\end{lemma} \vspace{0.3cm}
\noindent Building on the last two results, we derive a distribution-dependent regret bound which improves upon the one derived in  \citet{xia2015budgeted}: the decision maker is only assumed to know $\kappa$, as opposed to a lower bound on $\epsilon$, ahead of round $1$. This is more natural in bidding applications as detailed in the last paragraph of this section. This bound generalizes the one obtained by \citet{auer2002finite} when time is the only scarce resource.

\begin{theorem}
	\label{lemma-log-B-regret-bound}
	We have:
	$$
		R_B \leq 2^6 \lambda^2 \cdot (\sum_{k \in \{1, \cdots, K \} \; | \; \Delta_k > 0} \; \frac{1}{ \mu^c_k \cdot \Delta_k}) \cdot \ln(\frac{B + 1}{ \epsilon }) + O(1).
	$$
	\proof{Sketch of proof.}
		We build upon \eqref{eq-simplified-upper-bound-on-regret}:
		\begin{align*}
			R_B 
			& \leq B \cdot \frac{ \mu^r_{k^*} }{\mu^c_{k^*}} - \mathbb{E}[\sum_{t=1}^{\tau^*} r_{a_t, t}] + O(1) \\
			& = B \cdot \frac{ \mu^r_{k^*} }{\mu^c_{k^*}} - \sum_{k=1}^K \mu^r_k \cdot \mathbb{E}[n_{k, \tau^*}] + O(1) \\
			& =  \frac{ \mu^r_{k^*} }{ \mu^c_{k^*} } \cdot ( B - \sum_{k \; | \; \Delta_k = 0} \mu^c_k \cdot \mathbb{E}[n_{k, \tau^*}] ) - \sum_{k \; | \; \Delta_k > 0} \mu^r_k \cdot \mathbb{E}[n_{k, \tau^*}] + O(1). 
		\end{align*}
	By definition of $\tau^*$, the resource is exhausted at time $\tau^*$, i.e. $B \leq \sum_{t=1}^{\tau^*} c_{a_t, t}$. Taking expectations on both sides yields $B \leq \sum_{k=1}^K \mu^c_k \cdot \mathbb{E}[n_{k, \tau^*}]$. Plugging this last inequality back into the regret bound, we get:
	\begin{align*}
		R_B 
			& \leq \sum_{k \; | \; \Delta_k > 0} \mu^c_k \cdot \Delta_k \cdot \mathbb{E}[n_{k, \tau^*}] + O(1).
	\end{align*}
	Using the upper bound of Lemma \ref{lemma-bound-stopping-time}, the concavity of the logarithmic function, and Lemma \ref{lemma-bound-times-non-optimal-pulls}, we derive:
	\begin{align*}
		R_B 
			& \leq  2^6 \lambda^2 \cdot (\sum_{k \; | \; \Delta_k > 0} \frac{1}{\mu^c_k \cdot  \Delta_k}) \cdot \ln(\frac{B + 1}{ \epsilon })  +  \frac{ 4 \pi^2 }{3 \epsilon^2} \cdot (\sum_{k \; | \; \Delta_k > 0} \mu^c_k \cdot \Delta_k) +  O(1) \nonumber 
	\end{align*}
	which yields the claim since $\Delta_k \leq \mu^r_{k^*} / \mu^c_{k^*} \leq \kappa$ and $\mu^c_k \leq 1$ for any arm $k$.
	\endproof
\end{theorem} \vspace{0.3cm}
	\noindent Observe that the set of optimal arms, namely $\argmax_{k=1, \cdots, K}  \mu^r_k / \mu^c_k$, does not depend on $B$ and that $\Delta_k$ is a constant independent of $B$ for any suboptimal arm. We conclude that $R_B = O( K \cdot \ln(B)/\Delta )$ with $\Delta = \min_{k \in \{1, \cdots, K \} \; | \; \Delta_k > 0} \Delta_k$. Interestingly, the algorithm we propose does not rely on $B$ to achieve this regret bound, much like what happens for UCB1 with the time horizon, see \citet{auer2002finite}. This result is optimal up to constant factors as the standard MAB problem is a special case of the framework considered in this section, see \citet{lai1985asymptotically} for a proof of a lower bound in this context. It is possible to improve the constant factors when the consumption of the resource is deterministic as we can take $\lambda = 1$ in this scenario and the resulting regret bound is scale-free, see Section \ref{sec-extensions} of the Appendix. Building on Theorem \ref{lemma-log-B-regret-bound}, we can also derive a near-optimal distribution-free regret bound in the same fashion as for UCB1.
\begin{theorem}
\label{lemma-sqrt-B-regret-bound}
We have:
$$
	R_B \leq 8 \lambda \cdot \sqrt{ K \cdot  \frac{B+1}{\epsilon} \cdot \ln(\frac{B+1}{ \epsilon } )} + O(1).
$$
\proof{Proof}
	To get the distribution-free bound, we start from the penultimate inequality derived in the proof sketch of Theorem \ref{lemma-log-B-regret-bound} and apply Lemma \ref{lemma-bound-times-non-optimal-pulls} only if $\Delta_k$ is big enough, noting that: 
	$$
		\sum_{k=1}^K \mathbb{E}[n_{k, \tau^*}] = \mathbb{E} [\tau^*] \leq (B +1)/\epsilon.
	$$ 
	Specifically, we have:
\begin{align*}
	R_B 
		& \leq \sup\limits_{ \substack{(n_1, \cdots, n_K) \geq 0 \\ \sum_{k=1}^K n_k \leq \frac{B+1}{\epsilon} }  } \{ \; \sum_{k \; | \; \Delta_k > 0} \min(\mu^c_k \cdot  \Delta_k \cdot n_k, 2^6 \lambda^2 \cdot \frac{\ln(\frac{B + 1}{ \epsilon })}{ \mu^c_k \cdot \Delta_k} + \frac{ 4 \pi^2 }{3 \epsilon^2} \cdot \mu^c_k \cdot \Delta_k ) \; \} + O(1) \\
		& \leq \sup\limits_{ \substack{(n_1, \cdots, n_K) \geq 0 \\ \sum_{k=1}^K n_k \leq \frac{B+1}{\epsilon}}  } \{ \; \sum_{k \; | \; \Delta_k > 0} \min(\mu^c_k \cdot  \Delta_k \cdot n_k, 2^6 \lambda^2 \cdot \frac{\ln(\frac{B + 1}{ \epsilon } )}{\mu^c_k \cdot \Delta_k} ) \; \} + K \cdot \frac{ 4 \pi^2 \kappa }{3 \epsilon^2}+ O(1) \\
		& \leq \sup\limits_{ \substack{(n_1, \cdots, n_K) \geq 0 \\ \sum_{k=1}^K n_k \leq \frac{B+1}{\epsilon} }  } \{ \; \sum_{k \; | \; \Delta_k > 0} \sqrt{2^6 \lambda^2 \cdot n_k \cdot \ln(\frac{B+1}{ \epsilon } ) }  \; \} + O(1) \\
		& \leq 8 \lambda \cdot \sqrt{ K \cdot  \frac{B+1}{\epsilon} \cdot \ln(\frac{B+1}{ \epsilon } )} + O(1),
\end{align*}
where the second inequality is obtained with $\Delta_k \leq \mu^r_{k^*} / \mu^c_{k^*} \leq \kappa$ and $\mu^c_k \leq 1$, the third inequality is derived by maximizing on $(\mu^c_k \cdot \Delta_k) \geq 0$ for all arms $k$, and the last inequality is obtained with the Cauchy$-$Schwarz inequality.
\endproof
\end{theorem} \vspace{0.3cm}
We conclude that $R_B = O( \sqrt{K \cdot B \cdot \ln(B)} )$, where the hidden constant factors are independent of the underlying distributions $(\nu_k)_{k=1, \cdots, K}$. 

\paragraph{Applications.}
Assumption \ref{assumption-simplying-assumption-single-budget} is natural for bidding in repeated second-price auctions when the auctioneer sets a reserve price $R$ (this is common practice in sponsored search auctions). Indeed, then we have:
\begin{align*}
	\mathbb{E}[c_{k, t}]
		 & = \mathbb{E}[m_t \cdot \mathbbm{1}_{b_k \geq m_t}] \\
		 & \geq R \cdot \mathbb{E}[\mathbbm{1}_{b_k \geq m_t}] \\
		 & \geq R \cdot \mathbb{E}[v_t \cdot \mathbbm{1}_{b_k \geq m_t}] = R \cdot \mathbb{E}[r_{k, t}],
\end{align*}
for any arm $k \in \{1, \cdots, K\}$ and Assumption \ref{assumption-simplying-assumption-single-budget} is satisfied with $\kappa = 1/R$.

\section{Arbitrarily many limited resources whose consumptions are deterministic.}
\label{sec-multiplebudgets}
In this section, we study the case of multiple limited resources when the amounts of resources consumed as a result of pulling an arm are deterministic and globally constrained by prescribed budgets $(B(i))_{i= 1, \cdots, C}$, where $C$ is the number of resources. Note that time need not be a constraint. Because the amounts of resources consumed are deterministic, we can substitute the notation $\mu^{c}_k(i)$ with $c_k(i)$ for any arm $k \in \{1, \cdots, K\}$ and any resource $i \in \{1, \cdots, C\}$. We point out that the stopping time need not be deterministic as the decision to select an arm at any round is based on the past realizations of the rewards. We define $\rho \leq \min(C, K)$ as the rank of the matrix $(c_k(i))_{ 1 \leq k \leq K,  1 \leq i \leq C  }$. 

\paragraph{Specification of the algorithm.}
We implement UCB-Simplex with an initialization step which consists in pulling each arm $\rho$ times. The motivation behind this step is mainly technical and is simply meant to have:
\begin{equation}
	\label{eq-ini-pull-deterministic}
	n_{k, t} \geq \rho + \sum_{x \in \mathcal{B} \; | \; k \in \mathcal{K}_x } n^x_{k, t} \quad \forall t \in \mathbb{N}, \forall k \in \{1, \cdots, K\}.
\end{equation} 
Compared to Section \ref{sec-singlebudget}, we choose to take $\lambda = 1$ and $\eta_i = 0$ for any $i \in \{1, \cdots, C\}$. As a result and since the amounts of resource consumption are deterministic, the exploration (resp. exploitation) terms defined in Section \ref{sec-algorithmicideas} specialize to $\mathrm{obj}_{x, t} = \sum_{k=1}^K \xi^{x}_k \cdot \bar{r}_{k, t}$ (resp. $E_{x, t} = \sum_{k=1}^K \xi^{x}_k \cdot \epsilon_{k, t}$). Compared to the case of a single resource, we are required to specify the load balancing algorithms involved in Step-Load-Balance of UCB-Simplex as a feasible basis selected at Step-Simplex may involve several arms. Although Step-Simplex will also need to be specified in Sections \ref{sec-singlebudgettimehorizon} and \ref{sec-stochastic-multiple-budget}, designing good load balancing algorithms is arguably easier here as the optimal load balance is known for each basis from the start. Nonetheless, one challenge remains: we can never identify the (possibly many) optimal bases of \eqref{eq-linear-program-general-upperbound-opt-strategy} with absolute certainty. As a result, any basis selected at Step-Simplex should be treated as potentially optimal when balancing the load between the arms involved in this basis, but this inevitably causes some interference issues as an arm may be involved in several bases, and worst, possibly several optimal bases. Therefore, one point that will appear to be of particular importance in the analysis is the use of load balancing algorithms that are decoupled from one another, in the sense that they do not rely on what happened when selecting other bases. More specifically, we use the following class of load balancing algorithms.

\vspace{0.3cm}
\begin{algorithm}[H] 
	\caption{Load balancing algorithm $\mathcal{A}_x$ for a feasible basis $x \in \mathcal{B}$}	
	\label{def-algorithm-deterministic-costs}
	If basis $x$ is selected at time $t$, pull any arm $k \in \mathcal{K}_x$ such that $n^x_{k, t} \leq n_{x, t} \cdot \frac{\xi^x_k}{\sum_{l=1}^K \xi^x_l}$.
\end{algorithm} \vspace{0.3cm}
\noindent The load balancing algorithms $\mathcal{A}_x$ thus defined are decoupled because, for each basis, the number of times an arm has been pulled when selecting any other basis is not taken into account. The following lemma shows that $\mathcal{A}_x$ is always well defined and guarantees that the ratios $(n^x_{k, t}/n^x_{l, t})_{k, l \in \mathcal{K}_x}$ remain close to the optimal ones $(\xi^x_k / \xi^x_l)_{k, l \in \mathcal{K}_x}$ at all times. 

\begin{lemma}
\label{lemma-step-2-well-defined-multiple-budget}
For any basis $x \in \mathcal{B}$, $\mathcal{A}_x$ is well defined and moreover, at any time $t$ and for any arm $k \in \mathcal{K}_x$, we have:
$$
	n_{x, t} \cdot \frac{\xi^x_k}{\sum_{l=1}^K \xi^x_l} - \rho \leq n^x_{k, t} \leq n_{x, t} \cdot \frac{\xi^x_k}{\sum_{l=1}^K \xi^x_l} + 1,
$$
while $n^x_{k, t} = 0$ for any arm $k \notin \mathcal{K}_x$.
\proof{Proof.}
	We need to show that there always exists an arm $k \in \mathcal{K}_{x}$ such that $n^x_{k, t} \leq n_{x, t} \cdot \xi^x_k / \sum_{l=1}^K \xi^x_l$. Suppose there is none, we have:
\begin{align*}
	n_{x, t} 
		& = \sum_{k \in \mathcal{K}_x} n^x_{k, t} > \sum_{k \in \mathcal{K}_x} n_{x, t} \cdot \frac{\xi^x_k}{\sum_{l=1}^K \xi^x_l} = n_{x, t},
\end{align*}
a contradiction. Moreover, we have, at any time $t$ and for any arm $k \in \mathcal{K}_x$, $n^x_{k, t} \leq n_{x, t} \cdot \xi^x_k/\sum_{l=1}^K \xi^x_l + 1$. Indeed, suppose otherwise and define $t^* \leq t$ as the last time arm $k$ was pulled, we have:
\begin{align*}
	n^x_{k, t^*} & = n^x_{k, t} - 1 > n_{x, t} \cdot \frac{\xi^x_k}{\sum_{l=1}^K \xi^x_l} \geq n_{x, t^*} \cdot \frac{\xi^x_k}{\sum_{l=1}^K \xi^x_l},
\end{align*}
which shows, by definition, that arm $k$ could not have been pulled at time $t^*$. We also derive as a byproduct that, at any time $t$ and for any arm $k \in \mathcal{K}_x$, $n_{x, t} \cdot \xi^x_k / \sum_{l=1}^K \xi^x_l - \rho \leq n^x_{k, t}$ since $n_{x, t} = \sum_{k \in \mathcal{K}_x} n^x_{k, t}$ and since a basis involves at most $\rho$ arms.
\endproof
\end{lemma} \vspace{0.3cm}
\noindent
Observe that the load balancing algorithms $\mathcal{A}_x$ run in time $O(1)$ but may require a memory storage capacity exponential in $C$ and polynomial in $K$, although always bounded by $O( B )$ (because we do not need to keep track of $n^x_{k, t}$ if $x$ has never been selected). In practice, only a few bases will be selected at Step-Simplex, so that a hash table is an appropriate data structure to store the sequences $(n^x_{k, t})_{k \in \mathcal{K}_x}$. In Section \ref{sec-extensions} of the Appendix, we introduce another class of load balancing algorithms that is both time and memory efficient while still guaranteeing logarithmic regret bounds (under an additional assumption) but no distribution-free regret bounds.

\paragraph{Regret Analysis.}
We use the shorthand notation:
$$
	\epsilon = \min\limits_{ \substack{ k=1, \cdots, K \\ i = 1, \cdots, C \\ \text{with } c_k(i)>0} } c_k(i).
$$ 
Note that $\epsilon <\infty$ under Assumption \ref{assumption-all-cost-non-zero}. We discard the initialization step in the theoretical study because the amounts of resources consumed are bounded by a constant and the total reward obtained is non-negative and not taken into account in the decision maker's total payoff. We again start by estimating the expected time horizon.

\begin{lemma}
	\label{lemma-bound-stopping-time-deterministic}
	For any non-anticipating algorithm, we have: $\mathbb{E}[\tau^*] \leq \frac{\sum_{i=1}^C b(i) \cdot B}{\epsilon}  + 1$.
	\proof{Proof.}
	By definition of $\tau^*$, we have $\sum_{t=1}^{\tau^* - 1} c_{a_t, t}(i) \leq B(i)$ for any resource $i \in \{1, \cdots, C\}$. Summing up these inequalities and using Assumption \ref{assumption-all-cost-non-zero} and the fact that $(c_{k, t}(i))_{t=1, \cdots, T}$ are deterministic, we get $(\tau^* - 1) \cdot \epsilon \leq \sum_{i=1}^C B(i)$. Taking expectations on both sides and using Assumption \ref{assumption-budget-scale-linearly-wt-time} yields the result.
	\endproof
\end{lemma} \vspace{0.3cm}
We follow by bounding the number of times any suboptimal basis can be selected at Step-Simplex in the same spirit as in Section \ref{sec-singlebudget}.
\begin{lemma}
	\label{lemma-bound-times-non-optimal-pulls-deterministic}
	For any suboptimal basis $x \in \mathcal{B}$, we have: 
	$$
		\mathbb{E}[n_{x, \tau^*}] \leq 16 \rho \cdot (\sum_{k=1}^K \xi^x_k)^2  \cdot \frac{\mathbb{E}[ \ln(\tau^*) ]}{ (\Delta_x)^2 } +  \rho \cdot \frac{\pi^2 }{3}. 
	$$
	\proof{Sketch of proof.}
		We use the shorthand notation $\beta_x =  8  \rho \cdot ( \sum_{k=1}^K \xi^x_k / \Delta_x )^2$. The proof is along the same lines as for Lemma \ref{lemma-bound-times-non-optimal-pulls}. First note that we may assume, without loss of generality, that $x$ has been selected at least $\beta_x \cdot \ln(t)$ times at any time $t$ up to an additive term of $2 \beta_x \cdot \mathbb{E}[\ln(\tau^*)]$ in the final inequality. We then just have to bound by a constant the probability that $x$ is selected at any time $t$ given that $n_{x, t} \geq \beta_x \cdot \ln(t)$. If $x$ is selected at time $t$, $x$ is an optimal basis to \eqref{eq-algo-general-idea}. Since the amounts of resources consumed are deterministic, $x^*$ is feasible to \eqref{eq-algo-general-idea} at time $t$, which implies that $\mathrm{obj}_{x, t} + E_{x, t} \geq \mathrm{obj}_{x^*, t} + E_{x^*, t}$. This can only happen if either: (i) $\mathrm{obj}_{x, t} \geq \mathrm{obj}_{x} + E_{x, t}$, (ii) $\mathrm{obj}_{x^*, t} \leq \mathrm{obj}_{x^*}  - E_{x^*, t}$, or (iii) $\mathrm{obj}_{x^*} < \mathrm{obj}_{x} + 2 E_{x, t}$. First note that (iii) is impossible because, assuming this is the case, we would have: 
					\begin{align*}
				\frac{\Delta_x}{2} 
					& < \sum_{k \in \mathcal{K}_x} \xi^x_k \cdot \sqrt{ \frac{2 \ln(t) }{ n_{k, t} } } \\
					& \leq \sum_{k \in \mathcal{K}_x} \xi^x_k \cdot \sqrt{ \frac{2 \ln(t) }{ \rho + n^x_{k, t} } } \\
					& \leq \sqrt{ \sum_{k \in \mathcal{K}_x} \xi^x_k } \cdot \sum_{k \in \mathcal{K}_x} \sqrt{ \xi^x_k } \cdot \sqrt{ \frac{ 2 \ln(t) }{ n_{x, t} }} \\
					& \leq \sqrt{\rho} \cdot \sum_{k \in \mathcal{K}_x} \xi^x_k \cdot \sqrt{ \frac{ 2 }{ \beta_x }} = \frac{\Delta_x}{2},
			\end{align*}
			where we use \eqref{eq-ini-pull-deterministic}, Lemma \ref{lemma-step-2-well-defined-multiple-budget} for each $k \in \mathcal{K}_x$, the Cauchy$-$Schwarz inequality, and the fact that a basis involves at most $\rho$ arms. Along the same lines as for Lemma \ref{lemma-bound-times-non-optimal-pulls}, the probability of events (i) and (ii) can be bounded by $\sim \rho/t^2$ in the same fashion, irrespective of how many times $x$ and $x^*$ have been selected in the past. For example for event (i), this is because if $\mathrm{obj}_{x, t} \geq \mathrm{obj}_{x} + E_{x, t}$, then there must exist $k \in \mathcal{K}_x$ such that $\bar{r}_{k, t} \geq \mu^r_k + \epsilon_{k, t}$, but any of these events have individual probability at most $\sim 1/t^2$ by Lemma \ref{lemma-martingale-inequality}. Indeed otherwise, if  $\bar{r}_{k, t} < \mu^r_k + \epsilon_{k, t}$ for all $k \in \mathcal{K}_x$, we would have:
				\begin{align*}
					\mathrm{obj}_{x, t} - \mathrm{obj}_{x} 
						& = \sum_{k \in \mathcal{K}_x} (\bar{r}_{k, t} - \mu^r_k ) \cdot \xi^x_k \\
						& < \sum_{k \in \mathcal{K}_x} \epsilon_{k, t} \cdot \xi^x_k = E_{x, t},
				\end{align*}
				a contradiction
	\endproof
\end{lemma} \vspace{0.3cm}
\noindent Lemma \ref{lemma-bound-times-non-optimal-pulls-deterministic} used in combination with Lemma \ref{lemma-bound-stopping-time-deterministic} shows that a suboptimal basis is selected at most $O(\ln(B))$ times. To establish the regret bound, what remains to be done is to lower bound the expected total payoff derived when selecting any of the optimal bases. This is more involved than in the case of a single limited resource because the load balancing step comes into play at this stage.

\begin{theorem}
		\label{lemma-log-B-regret-bound-deterministic}
		We have:
		$$
			R_{B(1), \cdots, B(C)} \leq 16 \frac{\rho \cdot \sum_{i=1}^C b(i) }{\epsilon} \cdot (\sum_{x \in \mathcal{B} \; | \; \Delta_x > 0} \frac{ 1 }{ \Delta_x } ) \cdot \ln( \frac{\sum_{i=1}^C b(i) \cdot B}{\epsilon}  + 1 ) + O(1).
		$$
		\proof{Sketch of proof.}
			The proof proceeds along the same lines as for Theorem \ref{lemma-log-B-regret-bound}. We build upon \eqref{eq-simplified-general-upper-bound-on-regret}:
	\begin{align*}
				R_{B(1), \cdots, B(C)} 
					& \leq B \cdot \sum_{k=1}^K \mu^r_k \cdot \xi^{x^*}_k - \mathbb{E}[\sum_{t=1}^{\tau^*} r_{a_t, t}] + O(1) \\
					& = B \cdot \sum_{k=1}^K \mu^r_k \cdot \xi^{x^*}_k - \sum_{x \in \mathcal{B}} \sum_{k=1}^K \mu^r_k \cdot \mathbb{E}[n^x_{k, \tau^*} ] + O(1).
	\end{align*}
	 Using the properties of the load balancing algorithm established in Lemma \ref{lemma-step-2-well-defined-multiple-budget}, we derive:
	\begin{align*}
			R_{B(1), \cdots, B(C)}  
					& \leq B \cdot  \sum_{k=1}^K \mu^r_k \cdot \xi^{x^*}_k - \sum_{x \in \mathcal{B}} \{ \frac{\mathbb{E}[ n_{x, \tau^*  }]}{\sum_{k=1}^K \xi^x_k} \cdot (\sum_{k=1}^K \mu^r_k \cdot \xi^x_k) \} + O(1) \\
					& = (\sum_{k=1}^K \mu^r_k \cdot \xi^{x^*}_k) \cdot ( B - \sum_{x \in \mathcal{B} \; | \; \Delta_x = 0} \frac{\mathbb{E}[ n_{x, \tau^*  }]}{\sum_{k=1}^K \xi^x_k} ) \\
					& - \sum_{x \in \mathcal{B} \; | \; \Delta_x > 0} \{ (\sum_{k=1}^K \mu^r_k \cdot \xi^x_k) \cdot  \frac{\mathbb{E}[ n_{x, \tau^*  }]}{\sum_{k=1}^K \xi^x_k} \} + O(1).
	\end{align*}
	Now observe that, by definition, at least one resource is exhausted at time $\tau^*$. Hence, there exists $i \in  \{1, \cdots, C\}$ such that:
	\begin{align*}
	B(i) 
		& \leq \sum_{x \in \mathcal{B}} \sum_{k \in \mathcal{K}_x} c_k(i) \cdot n^x_{k, \tau^*}  \\
		& \leq O(1) + \sum_{x \in \mathcal{B}} \frac{ n_{x, \tau^*  } }{ \sum_{k=1}^K \xi^x_k } \cdot \sum_{k \in \mathcal{K}_x} c_k(i) \cdot \xi^x_k   \\
		& \leq O(1) + b(i) \cdot \sum_{x \in \mathcal{B}} \frac{ n_{x, \tau^*  } }{\sum_{k=1}^K \xi^x_k}, 	
	\end{align*}
	where we use Lemma \ref{lemma-step-2-well-defined-multiple-budget} again and the fact that any basis $x \in \mathcal{B}$ satisfies all the constraints of \eqref{eq-linear-program-general-upperbound-opt-strategy}. We conclude that:
	$$
		\sum_{ x \in \mathcal{B} \; | \; \Delta_x = 0 } \frac{ n_{x, \tau^*}  }{ \sum_{k=1}^K \xi^x_k } \geq B - \sum_{ x \in \mathcal{B} \; | \; \Delta_x > 0 } \frac{ n_{x, \tau^*} }{ \sum_{k=1}^K \xi^x_k } + O(1).
	$$
	Taking expectations on both sides and plugging the result back into the regret bound yields:
	\begin{align*}
		R_{B(1), \cdots, B(C)}  
			& \leq \sum_{ x \in \mathcal{B} \; | \; \Delta_x > 0 } \frac{\Delta_x}{\sum_{k=1}^K \xi^x_k} \cdot \mathbb{E}[ n_{x, \tau^*  }] + O(1).
	\end{align*}
	Using Lemma \ref{lemma-bound-stopping-time-deterministic}, Lemma \ref{lemma-bound-times-non-optimal-pulls-deterministic}, and the concavity of the logarithmic function, we obtain:	
	\begin{align*}
		R_{B(1), \cdots, B(C)}  
			& \leq 16 \rho  \cdot (\sum_{x \in \mathcal{B} \; | \; \Delta_x > 0} \frac{ \sum_{k=1}^K \xi^x_k }{ \Delta_x } ) \cdot \ln( \frac{\sum_{i=1}^C b(i) \cdot B}{\epsilon}  + 1 )  \\
			& + \frac{\pi^2}{3} \rho \cdot (\sum_{x \in \mathcal{B} \; | \; \Delta_x > 0} \frac{\Delta_x}{\sum_{k=1}^K \xi^x_k} ) + O(1)
	\end{align*}
	which yields the claim since $\Delta_x \leq \sum_{k=1}^K \mu^r_k \cdot \xi^{x^*}_k \leq \sum_{i=1}^C b(i)/\epsilon$, $\sum_{k=1}^K \xi^{x}_k \geq b$, and $\sum_{k=1}^K \xi^x_k \leq \sum_{i=1}^C b(i)/\epsilon$.
		\endproof
\end{theorem} \vspace{0.3cm}
We point out that, if time is a limited resource with a time horizon $T$, we can also derive the (possibly better) regret bound:
$$
	R_{B(1), \cdots, B(C)} \leq 16 \rho \cdot (\sum_{x \in \mathcal{B} \; | \; \Delta_x > 0} \frac{ 1 }{ \Delta_x } ) \cdot \ln( T ) + O(1).
$$
Since the number of feasible bases to \eqref{eq-linear-program-general-upperbound-opt-strategy} is at most ${ K + \rho \choose K} \leq 2 K^{\rho}$, we get the distribution-dependent regret bound $O( \rho \cdot K^{\rho} \cdot \ln(B) /\Delta )$ where $\Delta = \min_{x \in \mathcal{B} \; | \; \Delta_x > 0} \Delta_x$. In Section \ref{sec-extensions} of the Appendix, we introduce an alternative class of load balancing algorithms which yields a better dependence on $K$ and $C$ with a regret bound of order $O(\rho^3 \cdot K \cdot \ln(B) / \Delta^2)$ provided that there is a unique optimal basis to \eqref{eq-linear-program-general-upperbound-opt-strategy}. Along the sames lines as in Section \ref{sec-singlebudget}, the distribution-dependent bound of Theorem \ref{lemma-log-B-regret-bound-deterministic} almost immediately implies a distribution-free one.

\begin{theorem}
		\label{lemma-sqrt-B-regret-bound-deterministic}
		We have:
		$$
			R_{B(1), \cdots, B(C)} \leq 4 \sqrt{ \rho \cdot |\mathcal{B}| \cdot (\frac{\sum_{i=1}^C b(i) \cdot B}{\epsilon}  + 1) \cdot \ln( \frac{\sum_{i=1}^C b(i) \cdot B}{\epsilon}  + 1 )  } + O(1).
		$$
		\proof{Sketch of proof.}
			The proof is along the same lines as for the case of a single limited resource, we start from the penultimate inequality derived in the proof sketch of Theorem \ref{lemma-log-B-regret-bound-deterministic} and apply Lemma \ref{lemma-bound-times-non-optimal-pulls-deterministic} only if $\Delta_x$ is big enough, taking into account the fact that $\sum_{x \in \mathcal{B}} \mathbb{E}[n_{x, \tau^*}] \leq \mathbb{E} [\tau^*] \leq \sum_{i=1}^C b(i) \cdot B/\epsilon + 1$.
		\endproof
\end{theorem} \vspace{0.3cm}
We conclude that $R_{B(1), \cdots, B(C)} = O( \sqrt{\rho \cdot K^\rho \cdot B \cdot \ln(B)} )$, where the hidden constant factors are independent of the underlying distributions $(\nu_k)_{k=1, \cdots, K}$. If time is a limited resource, we can also derive the (possibly better) regret bound:
\begin{align*}
	R_{B(1), \cdots, B(C)} \leq 4 \sqrt{ \rho \cdot |\mathcal{B}| \cdot T \cdot \ln( T )  } + O(1).
\end{align*}
In any case, we stress that the dependence on $K$ and $C$ is not optimal since \citet{badanidiyuru2013bandits} and \citet{agrawal2014bandits} obtain a $\tilde{O}(\sqrt{K \cdot B})$ bound on regret, where the $\tilde{O}$ notation hides factors logarithmic in $B$. Observe that the regret bounds derived in this section do not vanish with $b$. This can be remedied by strengthening Assumption \ref{assumption-all-cost-non-zero}, additionally assuming that $c_{k, t}(i) > 0$ for any arm $k \in \{1, \cdots, K\}$ and resource $i \in \{1, \cdots, C\}$. In this situation, we can refine the analysis and substitute $\sum_{i=1}^C b(i)$ with $b$ in the regret bounds of Propositions \ref{lemma-log-B-regret-bound-deterministic} and \ref{lemma-sqrt-B-regret-bound-deterministic} which become scale-free. 

\paragraph{Applications.}
BwK problems where the amounts of resources consumed as a result of pulling an arm are deterministic find applications in network revenue management of perishable goods, shelf optimization of perishable goods, and wireless sensor networks, as detailed in Section \ref{sec-applications}. 

\section{A time horizon and another limited resource.}
\label{sec-singlebudgettimehorizon}
In this section, we investigate the case of two limited resources, one of which is assumed to be time, with a time horizon $T$, while the consumption of the other is stochastic and constrained by a global budget $B$. To simplify the notations, we omit the indices identifying the resources since the second limited resource is time and we write $\mu^c_k$, $c_{k, t}$, $\bar{c}_{k, t}$, $B$, and $T$ as opposed to $\mu^{c}_k(1)$, $c_{k, t}(1)$, $\bar{c}_{k, t}(1)$, $B(1)$, and $B(2)$. Moreover, we refer to resource $i = 1$ as \enquote{the} resource. Observe that, in the particular setting considered in this section, $\tau^* = \min( \tau(B), T+1)$ with $\tau(B) =  \min \{t \in \mathbb{N} \; | \; \sum_{\tau=1}^t c_{a_\tau, \tau} > B \}$. Note that the budget constraint is not limiting if $B \geq T$, in which case the problem reduces to the standard MAB problem. Hence, without loss of generality under Assumption \ref{assumption-budget-scale-linearly-wt-time}, we assume that the budget scales linearly with time, i.e. $B = b \cdot T$ for a fixed constant $b \in (0, 1)$, and we study the asymptotic regime $T \rightarrow \infty$. Motivated by technical considerations, we make two additional assumptions for the particular setting considered in this section that are perfectly reasonable in many applications, such as in repeated second-price auctions, dynamic procurement, and dynamic pricing, as detailed in the last paragraph of this section.
\begin{assumption}
\label{assumption-cost-bounds-rewards-budget-and-time-horizon}
There exists $\sigma > 0$ such that $\mu^r_{k} \leq \sigma \cdot \mu^c_{k}$ for any arm $k \in \{1, \cdots, K\}$.
\end{assumption}
\begin{assumption}
	\label{assumption-simplying-assumption-upperboundknown-budget-and-time-horizon}
	The decision maker knows $\kappa > 0$ such that:
	$$
		|\mu^r_k - \mu^r_l| \leq \kappa \cdot | \mu^c_k - \mu^c_l | \quad  \forall (k, l) \in \{1, \cdots, K\}^2,
	$$	
	ahead of round $1$.
\end{assumption}
Note that $\sigma$, as opposed to $\kappa$, is not assumed to be known to the decision maker. Assumption \ref{assumption-cost-bounds-rewards-budget-and-time-horizon} is relatively weak and is always satisfied in practical applications. In particular, note that if $\mu^c_{k} > 0$ for all arms $k \in \{1, \cdots, K\}$, we can take $\sigma = 1 / \min_{k=1, \cdots, K} \mu^c_k$.	

\paragraph{Specification of the algorithm.}
	We implement UCB-Simplex with $\lambda = 1 + 2  \kappa$, $\eta_1 = 1$, $\eta_2 = 0$, and an initialization step which consists in pulling each arm once. Because the amount of resource consumed at each round is a random variable, a feasible basis for \eqref{eq-algo-general-idea} may not be feasible for \eqref{eq-linear-program-general-upperbound-opt-strategy} and conversely. In particular, $x^*$ may not be feasible for \eqref{eq-algo-general-idea}, thus effectively preventing it from being selected at Step-Simplex, and an infeasible basis for \eqref{eq-linear-program-general-upperbound-opt-strategy} may be selected instead. This is in contrast to the situation studied in Section \ref{sec-multiplebudgets} and this motivates the choice $\eta_1 > 0$ to guarantee that any feasible solution to \eqref{eq-linear-program-general-upperbound-opt-strategy} will be feasible to \eqref{eq-algo-general-idea} with high probability at any round $t$. 
\\
\indent Just like in Section \ref{sec-multiplebudgets}, we need to specify Step-Load-Balance because a basis selected at Step-Simplex may involve up to two arms. To simplify the presentation, we introduce a dummy arm $k=0$ with reward $0$ and resource consumption $0$ (pulling this arm corresponds to skipping the round) and $K$ dummy arms $k = K+1, \cdots, 2 K$ with reward identical to arm $K-k$ but resource consumption $1$ so that any basis involving a single arm can be mapped to an \enquote{artificial} one involving two arms. Note, however, that we do not introduce new variables $\xi_k$ in \eqref{eq-algo-general-idea} for these arms as they are only used for mathematical convenience in Step-Load-Balance once a basis has been selected at Step-Simplex. Specifically, if a basis $x_t$ involving a single arm determined by $\mathcal{K}_{x_t} = \{ k_t \}$ and $\mathcal{C}_{x_t} = \{1\}$ (resp. $\mathcal{C}_{x_t} = \{2\}$) is selected at Step-Simplex, we map it to the basis $x'_t$ determined by $\mathcal{K}_{x'_t} = \{0, k_t\}$ (resp. $\{k_t, K + k_t\}$) and $\mathcal{C}_{x'_t} = \{1, 2\}$. We then use a load balancing algorithm specific to this basis, denoted by $\mathcal{A}_{ x_t }$, to determine which of the two arms in $\mathcal{K}_{x'_t}$ to pull. Similarly as in Section \ref{sec-multiplebudgets}, using load balancing algorithms that are decoupled from one another is crucial because the decision maker can never identify the optimal bases with absolute certainty. This implies that each basis should be treated as potentially optimal when balancing the load between the arms, but this inevitably causes interference issues as an arm may be involved in several bases. Compared to Section \ref{sec-multiplebudgets}, we face an additional challenge when designing the load balancing algorithms: the optimal load balances are initially unknown to the decision maker. It turns out that we can still approximately achieve the unknown optimal load balances by enforcing that, at any round $t$, the total amount of resource consumed remains close to the pacing target $b \cdot t$ with high probability, as precisely described below.

\vspace{0.3cm}
\begin{algorithm}[H]
		\label{def-pulling-strategy-time-horizon}
		\caption{Load balancing algorithm $\mathcal{A}_{x}$ for any basis $x$}		
		For any time period $t$, define $b_{x, t}$ as the total amount of resource consumed when selecting $x$ in the past $t-1$ rounds. Suppose that $x$ is selected at time $t$. Without loss of generality, write $\mathcal{K}_x = \{k, l\}$ with $\bar{c}_{k, t} - \epsilon_{k, t} \geq \bar{c}_{l, t} - \epsilon_{l, t}$. Pull arm $k$ if $b_{x, t} \leq n_{x, t} \cdot b$ and pull arm $l$ otherwise.
\end{algorithm} \vspace{0.3cm}
\noindent Observe that a basis $x$ with $\mathcal{K}_x = \{k, l\}$ is feasible for \eqref{eq-linear-program-general-upperbound-opt-strategy} if either $\mu^c_{k} > b > \mu^c_{l}$ or $\mu^c_{l} > b > \mu^c_{k}$. Assuming we are in the first situation, the exploration and exploitation terms defined in Section \ref{sec-algorithmicideas} specialize to:
$$
	\mathrm{obj}_{x, t} =  \xi^{x}_{l, t} \cdot \bar{r}_{l, t} + \xi^{x}_{k, t} \cdot \bar{r}_{k, t} \text{ and } E_{x, t} = \lambda \cdot ( \xi^{x}_{l, t} \cdot \epsilon_{l, t} + \xi^{x}_{k, t} \cdot \epsilon_{k, t})
$$
with:
$$
	\xi^{x}_{l, t} =  \frac{ (\bar{c}_{k, t} - \epsilon_{k, t}) - b }{ (\bar{c}_{k, t} - \epsilon_{k, t}) - (\bar{c}_{l, t} - \epsilon_{l, t}) } \text{ and } \xi^{x}_{k, t} = \frac{ b - (\bar{c}_{l, t} - \epsilon_{l, t}) }{ (\bar{c}_{k, t} - \epsilon_{k, t})  - (\bar{c}_{l, t} - \epsilon_{l, t}) },
$$
provided that $\bar{c}_{k, t} - \epsilon_{k, t} > b > \bar{c}_{l, t} - \epsilon_{l, t}$. Moreover, their offline counterparts are given by:
$$
	\mathrm{obj}_{x} =  \xi^{x}_{l} \cdot \mu^r_{l} + \xi^{x}_{k} \cdot \mu^r_{k}, \; \xi^{x}_{l} = \frac{ \mu^c_{k} - b }{ \mu^c_{k} - \mu^c_{l}}, \; \text{and } \xi^{x}_{k} = \frac{ b - \mu^c_{l} }{ \mu^c_{k} - \mu^c_{l}}.
$$
\paragraph{Regret Analysis.}
We start by pointing out that, in degenerate scenarios, using the linear relaxation \eqref{eq-linear-program-general-upperbound-opt-strategy} as an upper bound on $\ropt(B, T)$ already dooms us to $\Omega(\sqrt{T})$ regret bounds. Precisely, if there exists a unique optimal basis $x^*$ to \eqref{eq-linear-program-general-upperbound-opt-strategy} that happens to be degenerate, i.e. $\mathcal{K}_{x^*} = \{k^*\}$ (pre-mapping) with $\mu^c_{k^*} = b$, then, in most cases, $T \cdot \mathrm{obj}_{x^*} \geq \ropt(B, T) + \Omega(\sqrt{T})$ as shown below. 
\begin{lemma}
	\label{lemma-non-degeneracy-assumption-is-necessary-budget-and-time-horizon}
	If there exists $k^* \in \{1, \cdots, K\}$ such that: (i) the i.i.d. process $(c_{k^*, t})_{t \in \mathbb{N}}$ has positive variance, (ii) $\mu^c_{k^*} = b$, and (iii) $(\xi_k)_{k=1, \cdots, K}$ determined by $\xi_{k^*} = 1$ and $\xi_k = 0$ for $k \neq k^*$ is the unique optimal solution to \eqref{eq-linear-program-general-upperbound-opt-strategy}, then there exists a subsequence of $( \frac{ T \cdot \mathrm{obj}_{x^*} - \ropt(B, T)  }{\sqrt{T}} )_{T \in \mathbb{N}}$ that does not converge to $0$.
	\proof{Sketch of proof.}
		For any time horizon $T \in \mathbb{N}$ and any arm $k \in \{1, \cdots, K\}$, we denote by $n^{\mathrm{opt}}_{k, T}$ the expected number of times arm $k$ is pulled by the optimal non-anticipating algorithm  when the time horizon is $T$ and the budget is $B = b \cdot T$. We expect that consistently pulling arm $k^*$ is near-optimal. Unfortunately, this is also nothing more than an i.i.d. strategy which implies, along the same lines as in Lemma \ref{lemma-lower-bound-agrawal}, that $\mathbb{E}[\tau^*] = T - \Omega(\sqrt{T})$ so that the total expected payoff is $\mathbb{E}[ \tau^* ] \cdot \mu^r_{k^*} = T \cdot \mathrm{obj}_{x^*} - \Omega(\sqrt{T})$. To formalize these ideas, we study two cases: $T - n^{\mathrm{opt}}_{k^*, T} = \Omega(\sqrt{T})$ (Case A) and $T - n^{\mathrm{opt}}_{k^*, T} = o(\sqrt{T})$ (Case B) and we show that $\ropt(B, T) = T \cdot \mathrm{obj}_{x^*} - \Omega(\sqrt{T})$ in both cases. In Case A, this is because the optimal value of \eqref{eq-linear-program-general-upperbound-opt-strategy} remains an upper bound on the maximum total expected payoff if we add the constraint $\xi_{k^*} \leq n^{\mathrm{opt}}_{k^*, T}/T$ to the linear program \eqref{eq-linear-program-general-upperbound-opt-strategy} by definition of $n^{\mathrm{opt}}_{k^*, T}$. Since the constraint $\xi_{k^*} \leq 1$ is binding for \eqref{eq-linear-program-general-upperbound-opt-strategy}, the optimal value of this new linear program can be shown to be smaller than $\mathrm{obj}_{x^*} - \Omega( (T - n^{\mathrm{opt}}_{k^*, T})/T)$ (by strong duality and strict complementary slackness). In Case B, up to an additive term of order $o(\sqrt{T})$ in the final bound, the optimal non-anticipating algorithm is equivalent to consistently pulling arm $k^*$, which is an i.i.d. strategy so the study is very similar to that of Lemma \ref{lemma-lower-bound-agrawal}.
	\endproof
\end{lemma} \vspace{0.3cm}
 Dealing with these degenerate scenarios thus calls for a completely different approach than the one taken on in the BwK literature and we choose instead to rule them out in such a way that there can be no degenerate optimal basis to \eqref{eq-linear-program-general-upperbound-opt-strategy}. 
\begin{assumption}
	\label{assumption-simplying-assumption-analysis-budget-and-time-horizon}
	We have $|\mu^c_k - b| > 0 $ for any arm $k \in \{1, \cdots, K\}$.
\end{assumption}
\noindent
We use the shorthand notation $\epsilon = \min_{k=1, \cdots, K} |\mu^c_k - b|$. Assumption \ref{assumption-simplying-assumption-analysis-budget-and-time-horizon} is equivalent to assuming that any basis for \eqref{eq-linear-program-general-upperbound-opt-strategy} is non-degenerate. This assumption can be relaxed to some extent at the price of more technicalities. However, in light of Lemma \ref{lemma-non-degeneracy-assumption-is-necessary-budget-and-time-horizon}, the minimal assumption is that there is no degenerate optimal basis to \eqref{eq-linear-program-general-upperbound-opt-strategy}. As a final remark, we stress that Assumption \ref{assumption-simplying-assumption-analysis-budget-and-time-horizon} is only necessary to carry out the analysis but Step-Simplex can be implemented in any case as $\epsilon$ is not assumed to be known to the decision maker. \\
\indent We are now ready to establish regret bounds. Without loss of generality, we can assume that any pseudo-basis for \eqref{eq-linear-program-general-upperbound-opt-strategy} involves two arms, one of which may be a dummy arm introduced in the specification of the algorithm detailed above.
As stressed at the beginning of this section, UCB-Simplex may sometimes select an infeasible basis or even a pseudo-basis $x$ with $\det(A_x) = 0$ (i.e. such that $\mu^c_k = \mu^c_l$ assuming $\mathcal{K}_x = \{k, l\}$). Interestingly the load balancing algorithm plays a crucial role to guarantee that this does not happen very often. 
\begin{lemma}
	\label{lemma-bound-times-infeasible-pulls-budget-time-horizon}
	For any basis $x \notin \mathcal{B}$, we have:
	$$
		\mathbb{E}[n_{x, T}] \leq \frac{2^6}{\epsilon^3} \cdot \ln( T ) +  \frac{ 10 \pi^2 }{3 \epsilon^2}.
	$$ 
	The same inequality holds if $x$ is a pseudo-basis but not a basis for \eqref{eq-linear-program-general-upperbound-opt-strategy}.
	\proof{Proof.}
		We use the shorthand notation $\beta_x = 2^5/\epsilon^3$. Without loss of generality, we can assume that $\mathcal{K}_x = \{k, l\}$ with $\mu^c_k, \mu^c_l > b$ (the situation is symmetric if the reverse inequality holds). Along the same lines as in Lemma \ref{lemma-bound-times-non-optimal-pulls}, we only have to bound by a constant the probability that $x$ is selected at any round $t$ given that $x$ has already been selected at least $\beta_x \cdot \ln(t)$ times. If $x$ is selected at round $t$ and $n_{x, t} \geq \beta_x \cdot \ln(t)$, then $b_{x, t}$ must be larger than $n_{x, t} \cdot b$ by at least a margin of $\sim 1/\epsilon^2 \cdot \ln(t)$ with high probability given that $\mu^c_k, \mu^c_l > b$. Moreover, at least one arm, say $k$, has been pulled at least $\sim 1/\epsilon^3 \cdot \ln(t)$ times and, as a result, $\bar{c}_{k, \tau} - \epsilon_{k, \tau} \geq b$ with high probability for the last $s \sim 1/\epsilon^2 \cdot \ln(t)$ rounds $\tau = \tau_1, \cdots, \tau_s$ where $x$ was selected. This implies that arm $l$ must have been pulled at least $\sim 1/\epsilon^2 \cdot \ln(t)$ times already by definition of the load balancing algorithm but then we have $\bar{c}_{l, t} - \epsilon_{l, t} \geq b$ with high probability and $x$ cannot be feasible for \eqref{eq-algo-general-idea} at time $t$ with high probability.
	\endproof
\end{lemma} \vspace{0.3cm}
What remains to be done is to: (i) show that suboptimal bases are selected at most $O(\ln(T))$ times and (ii) lower bound the expected total payoff derived when selecting any of the optimal bases. The major difficulty lies in the fact that the amounts of resource consumed, the rewards obtained, and the stopping time are correlated in a non-trivial way through the budget constraint and the decisions made in the past. This makes it difficult to study the expected total payoff derived when selecting optimal bases independently from the amounts of resource consumed and the rewards obtained when selecting suboptimal ones. However, a key point is that, by design, the pulling decision made at Step-Load-Balance is based solely on the past history associated with the basis selected at Step-Simplex because the load balancing algorithms are decoupled. For this reason, the analysis proceeds in two steps irrespective of the number of optimal bases. In a first step, we show that, for any basis $x$ for \eqref{eq-linear-program-general-upperbound-opt-strategy}, the amount of resource consumed per round when selecting $x$ remains close to the pacing target $b$ with high probability. This enables us to show that the ratios $(\mathbb{E}[n^x_{k, T}]/\mathbb{E}[n^x_{l, T}])_{k, l \in \mathcal{K}_x}$ are close to the optimal ones $(\xi^x_k/\xi^x_l)_{k, l \in \mathcal{K}_x}$, as precisely stated below.  
	\begin{lemma}
		\label{lemma-load-balance-budget-time-horizon}
		For any basis $x \in \mathcal{B}$ and time period $t$, we have:
		$$
		\mathbb{P}[|b_{x, t} - n_{x, t} \cdot b| \geq u + (\frac{4}{\epsilon})^2 \cdot \ln(t)] \leq \frac{4}{ \epsilon^2} \cdot \exp(- \epsilon^2 \cdot u ) + \frac{8}{\epsilon^2 \cdot t^2} \quad \forall u \geq 1,
		$$
		which, in particular, implies that:
		\begin{equation}
			\label{eq-lower-bound-pulls-budget-time-horizon}
				\begin{aligned}
					& \mathbb{E}[n^x_{k, T}] \geq  \xi^x_k \cdot \mathbb{E}[n_{x, T}] - 13/\epsilon^5 - 16/\epsilon^3 \cdot \ln(T) \\
					& \mathbb{E}[n^x_{l, T}] \geq  \xi^x_l \cdot \mathbb{E}[n_{x, T}] - 13/\epsilon^5 - 16/\epsilon^3 \cdot \ln(T).
				\end{aligned}
		\end{equation}		
		\proof{Sketch of proof.}
				Without loss of generality, we can assume that $\mathcal{K}_x = \{k, l\}$ with $\mu^c_k > b > \mu^c_l$. Observe that, if the decision maker knew that $\mu^c_k > b > \mu^c_l$ ahead of round $1$, he would always pull the \enquote{correct} arm in order not to deviate from the pacing target $n_{x, t} \cdot b$ and  $|b_{x, t} - n_{x, t} \cdot b|$ would remain small with high probability given Assumption \ref{assumption-simplying-assumption-analysis-budget-and-time-horizon}. However, because this information is not available	 ahead of round $1$, the decision maker is led to pull the incorrect arm when arm $k$ and $l$ are swapped, in the sense that $\bar{c}_{k, t} - \epsilon_{k, t} \leq \bar{c}_{l, t} - \epsilon_{l, t}$. Fortunately, at any time $t$, there could have been at most $1/\epsilon^2 \cdot \ln(t)$ swaps with probability at least $\sim 1 - 1/t^2$ given Assumption \ref{assumption-simplying-assumption-analysis-budget-and-time-horizon}. To derive \eqref{eq-lower-bound-pulls-budget-time-horizon}, we use: $|\mathbb{E}[ b_{x, T} - n_{x, T} \cdot b ]| \leq \int_0^T \mathbb{P}[|b_{x, T} - n_{x, T} \cdot b| \geq u] \mathrm{d}u$ and $\mathbb{E}[ b_{x, T} ] = \mu^k_c \cdot \mathbb{E}[n^x_{k, T}] + \mu^l_c \cdot \mathbb{E}[n^x_{l, T}]$.
		\endproof				
	\end{lemma} \vspace{0.3cm}

\noindent The next step is to show, just like in Section \ref{sec-multiplebudgets}, that any suboptimal feasible basis is selected at most $O(\ln( T ))$ times on average. Interestingly, the choice of the load balancing algorithm plays a minor role in the proof. Any load balancing algorithm that pulls each arm involved in a basis at least a constant fraction of the time this basis is selected does enforce this property.

\begin{lemma}
	\label{lemma-bound-times-non-optimal-pulls-budget-time-horizon}
	For any suboptimal basis $x \in \mathcal{B}$, we have:
	$$
		\mathbb{E}[n_{x, T}] \leq 2^9  \frac{\lambda^2}{\epsilon^3} \cdot \frac{\ln(T)}{(\Delta_x)^2} + \frac{10 \pi^2}{\epsilon^2}.
	$$
	\proof{Sketch of proof.}
			We use the shorthand notation $\beta_x = 2^8/\epsilon^3 \cdot (\lambda/\Delta_x)^2$. Without loss of generality, we can assume that $\mathcal{K}_{x^*} = \{k^*, l^*\}$ with $\mu^c_{k^*} > b > \mu^c_{l^*}$ and $\mathcal{K}_x = \{k, l\}$ with $\mu^c_k > b > \mu^c_l$. Along the same lines as in Lemma \ref{lemma-bound-times-non-optimal-pulls}, we only have to bound by a constant the probability that $x$ is selected at any round $t$ given that $x$ has already been selected at least $\beta_x \cdot \ln(t)$ times. If $x$ is selected at time $t$, $x$ is optimal for \eqref{eq-algo-general-idea}. Observe that $(\xi^{x^*}_k)_{k = 1, \cdots, K}$ is a feasible solution to \eqref{eq-algo-general-idea} when $\bar{c}_{k^*, t} - \epsilon_{k^*, t} \leq \mu^c_{k^*}$ and $\bar{c}_{l^*, t} - \epsilon_{l^*, t} \leq \mu^c_{l^*}$, which happens with probability at least $\sim 1 - 1/t^2$. As a result, $\mathrm{obj}_{x, t} + E_{x, t} \geq \mathrm{obj}_{x^*}$ when additionally $\bar{r}_{k^*, t} + \epsilon_{k^*, t} \geq \mu^r_{k^*}$ and $\bar{r}_{l^*, t} + \epsilon_{l^*, t} \geq \mu^r_{l^*}$, which also happens with probability at least $\sim 1 - 1/t^2$. If $\mathrm{obj}_{x, t} + E_{x, t} \geq \mathrm{obj}_{x^*}$ then we have either (i) $\mathrm{obj}_{x, t} \geq \mathrm{obj}_{x} + E_{x, t}$ or (ii) $\mathrm{obj}_{x^*} < \mathrm{obj}_{x} + 2 E_{x, t}$. Observe that (ii) can only happen with probability at most $\sim 1/t^2$ given that $n_{x, t} \geq \beta_x \cdot \ln(t)$ because (ii) implies that either $n_{l, t} \leq 8 (\lambda/\Delta_x)^2 \cdot \ln(t)$ or $n_{k, t} \leq 8 (\lambda/\Delta_x)^2 \cdot \ln(t)$ but the load balancing algorithm guarantees that each arm is pulled a fraction of the time $x$ is selected (using Lemma \ref{lemma-load-balance-budget-time-horizon}). As for (i), if $\mathrm{obj}_{x, t} \geq \mathrm{obj}_{x} + E_{x, t}$, then, using Assumption \ref{assumption-cost-bounds-rewards-budget-and-time-horizon}, either $\bar{r}_{k, t} \geq \mu^r_k + \epsilon_{k, t}$, $\bar{c}_{k, t} \notin [\mu^c_k - \epsilon_{k, t}, \mu^c_k + \epsilon_{k, t}]$, $\bar{r}_{l, t} \geq \mu^r_l + \epsilon_{l, t}$, or $\bar{c}_{l, t} \notin [\mu^c_l - \epsilon_{l, t}, \mu^c_l + \epsilon_{l, t}]$ but all of these events have individual probability at most $\sim 1/t^2$ by Lemma \ref{lemma-martingale-inequality}.
	\endproof
\end{lemma} \vspace{0.3cm}
\noindent	
		In a last step, we show, using Lemma \ref{lemma-load-balance-budget-time-horizon}, that, at the cost of an additive logarithmic term in the regret bound, we may assume that the game lasts exactly $T$ rounds. This enables us to combine Lemmas \ref{lemma-bound-times-infeasible-pulls-budget-time-horizon}, \ref{lemma-load-balance-budget-time-horizon}, and \ref{lemma-bound-times-non-optimal-pulls-budget-time-horizon} to establish a distribution-dependent regret bound.
	\begin{theorem}
		\label{lemma-log-B-regret-bound-time-horizon}
		We have:
		$$
			R_{B, T} \leq 2^9 \frac{\lambda^2}{\epsilon^3} \cdot (\sum_{x \in \mathcal{B} \; | \; \Delta_x > 0}  \frac{1}{\Delta_x}) \cdot \ln(T) + O(\frac{K^2 \cdot \sigma}{\epsilon^3} \cdot \ln(T)),
		$$
		where the $O$ notation hides universal constant factors.
		\proof{Sketch of proof.}
			We build upon \eqref{eq-simplified-general-upper-bound-on-regret}:
		\begin{align*}
				R_{B, T} 
					& \leq T \cdot \sum_{k=1}^K \mu^r_k \cdot \xi^{x^*}_k - \mathbb{E}[\sum_{t=1}^{\tau^*} r_{a_t, t}] + O(1) \\
					& \leq T \cdot \sum_{k=1}^K \mu^r_k \cdot \xi^{x^*}_k - \mathbb{E}[\sum_{t=1}^{T} r_{a_t, t}] + \sigma \cdot \mathbb{E}[ (\sum_{t=1}^{T} c_{a_t, t} - B)_+]  + O(1),
		\end{align*}
		where we use Assumption \ref{assumption-cost-bounds-rewards-budget-and-time-horizon} for the second inequality. Moreover:
		\begin{align*}
		\mathbb{E}[ (\sum_{t=1}^{T} c_{a_t, t} - B)_+] 
			& \leq \sum_{x \in \mathcal{B}} \mathbb{E}[ |b_{x, T} - n_{x, T} \cdot b |] + \sum_{x \notin \mathcal{B}} \mathbb{E}[n_{x, T}]  \\
			& + \sum_{ \substack{ x \text{ pseudo-basis for } \eqref{eq-linear-program-general-upperbound-opt-strategy} \\ \text{ with } \det(A_x) = 0 }} \mathbb{E}[n_{x, T}]  = O(\frac{K^2}{\epsilon^3} \ln(T)),
	\end{align*}
	using Lemmas \ref{lemma-bound-times-infeasible-pulls-budget-time-horizon} and \ref{lemma-load-balance-budget-time-horizon}. Plugging this last inequality back into the regret bound yields:
	\begin{align*}
		R_{B, T}
					& \leq T \cdot \sum_{k=1}^K \mu^r_k \cdot \xi^{x^*}_k - \mathbb{E}[\sum_{t=1}^{T} r_{a_t, t}] + O(\frac{K^2 \cdot \sigma}{\epsilon^3} \ln(T))  \\			
					& \leq T \cdot \sum_{k=1}^K \mu^r_k \cdot \xi^{x^*}_k  - \sum_{x \in \mathcal{B}} \sum_{k=1}^K \mu^r_k \cdot \mathbb{E}[n^x_{k, T} ] + O(\frac{K^2 \cdot \sigma}{\epsilon^3} \ln(T))  \\
					& \leq T \cdot \sum_{k=1}^K \mu^r_k \cdot \xi^{x^*}_k  - \sum_{x \in \mathcal{B}} (\sum_{k=1}^K \mu^r_k \cdot \xi^x_k) \cdot \mathbb{E}[n_{x, T} ] + O(\frac{K^2 \cdot \sigma}{\epsilon^3} \ln(T))  \\
					& = \sum_{k=1}^K \mu^r_k \cdot \xi^{x^*}_k \cdot ( T - \sum_{x \in \mathcal{B} \; | \; \Delta_x = 0} \mathbb{E}[n_{x, T}] ) - \sum_{x \in \mathcal{B} \; | \; \Delta_x > 0} (\sum_{k=1}^K \mu^r_k \cdot \xi^x_k) \cdot \mathbb{E}[n_{x, T} ] + O(\frac{K^2 \cdot \sigma}{\epsilon^3} \ln(T))  \\
					& = \sum_{k=1}^K \mu^r_k \cdot \xi^{x^*}_k \cdot (\sum_{x \in \mathcal{B} \; | \; \Delta_x > 0} \mathbb{E}[n_{x, T}] + \sum_{x \notin \mathcal{B}}  \mathbb{E}[n_{x, T}] + \sum_{ \substack{ x \text{ pseudo-basis for } \eqref{eq-linear-program-general-upperbound-opt-strategy} \\ \text{ with } \det(A_x) = 0 } } \mathbb{E}[n_{x, T}])  \\
					& - \sum_{x \in \mathcal{B} \; | \; \Delta_x > 0} (\sum_{k=1}^K \mu^r_k \cdot \xi^x_k) \cdot \mathbb{E}[n_{x, T} ]  + O(\frac{K^2 \cdot \sigma}{\epsilon^3} \ln(T))  \\
					& \leq \sum_{x \in \mathcal{B} \; | \; \Delta_x > 0}  \Delta_x \cdot \mathbb{E}[n_{x, T} ] + O(\frac{K^2 \cdot \sigma}{\epsilon^3} \ln(T)) \\	
					& \leq 2^9  \frac{\lambda^2}{\epsilon^3} \cdot (\sum_{x \in \mathcal{B} \; | \; \Delta_x > 0}  \frac{1}{\Delta_x}) \cdot \ln(T) + O(\frac{K^2 \cdot \sigma}{\epsilon^3} \ln(T)), 
	\end{align*}
	where we use Lemma \ref{lemma-load-balance-budget-time-horizon}	for the third inequality, Lemma \ref{lemma-bound-times-infeasible-pulls-budget-time-horizon} along with $\sum_{k=1}^K \mu^r_k \cdot \xi^{x^*}_k \leq \sum_{k=1}^K \xi^{x^*}_k \leq 1$ for the fourth inequality, and Lemma \ref{lemma-bound-times-non-optimal-pulls-budget-time-horizon} for the last inequality.
		\endproof
	\end{theorem} \vspace{0.3cm}
	\noindent
	Since there are at most $2 K^2$ feasible bases, we get the regret bound $O( K^2 \cdot ( 1/\Delta + \sigma/\epsilon^3) \cdot \ln(T))$, where $\Delta = \min_{x \in \mathcal{B} \; | \; \Delta_x > 0} \Delta_x$. Along the sames lines as in Sections \ref{sec-singlebudget} and \ref{sec-multiplebudgets}, pushing the analysis further almost immediately yields a distribution-free regret bound.
\begin{theorem}
		\label{lemma-sqrt-B-regret-bound-time-horizon}
		We have:
		$$
			R_{B, T} \leq 2^5 \frac{ \lambda }{\epsilon^{3/2}} \cdot \sqrt{ |\mathcal{B}| \cdot T \cdot \ln( T )  } + O(\frac{K^2 \cdot \sigma}{\epsilon^3} \ln(T)),
		$$
		where the $O$ notation hides universal constant factors.
		\proof{Sketch of proof.}
			The proof is along the same lines as for Theorems \ref{lemma-sqrt-B-regret-bound} and \ref{lemma-sqrt-B-regret-bound-deterministic}, we start from the penultimate inequality derived in the proof sketch of Theorem \ref{lemma-log-B-regret-bound-time-horizon} and apply Lemma \ref{lemma-bound-times-non-optimal-pulls-budget-time-horizon} only if $\Delta_x$ is big enough, taking into account the fact that $\sum_{x \in \mathcal{B}} 	\mathbb{E}[n_{x, T}] \leq T$.
		\endproof
\end{theorem} \vspace{0.3cm}
We conclude that $R_{B, T} = O( \sqrt{  K^2 \cdot T \cdot \ln(T)} )$, where the hidden factors are independent of the underlying distributions $(\nu_k)_{k=1, \cdots, K}$. Just like in Section \ref{sec-multiplebudgets}, we stress that the dependence on $K$ is not optimal since \citet{badanidiyuru2013bandits} and \citet{agrawal2014bandits} obtain a $\tilde{O}(\sqrt{K \cdot T})$ bound on regret, where the $\tilde{O}$ notation hides factors logarithmic in $T$. Observe that the regret bounds derived in Theorems \ref{lemma-log-B-regret-bound-time-horizon} and \ref{lemma-sqrt-B-regret-bound-time-horizon} do not vanish with $b$, which is not the expected behavior. This is a shortcoming of the analysis that can easily be remedied when $\min_{k=1, \cdots, K} \mu^c_k > 0$ provided that instead of pulling the dummy arm $0$ we always pull the other arm involved in the basis (i.e. we never skip rounds). Note that not skipping rounds can only improve the regret bounds derived in Theorems \ref{lemma-log-B-regret-bound-time-horizon} and \ref{lemma-sqrt-B-regret-bound-time-horizon}: arm $0$ was introduced only in order to harmonize the notations for mathematical convenience. 
\begin{theorem}
	\label{lemma-regret-bound-b-small-time-horizon}
	Relax Assumption \ref{assumption-simplying-assumption-analysis-budget-and-time-horizon} and redefine $\epsilon = \min_{k=1, \cdots, K} \mu^c_k$. Suppose that $b \leq \epsilon/2$ and that we never skip rounds, then we have:
	$$
			R_{B, T} \leq 2^{12} \frac{\lambda^2}{\epsilon^3} \cdot (\sum_{x \in \mathcal{B} \; | \; \Delta_x > 0} \frac{1}{\Delta_{x}})  \cdot \ln( \frac{B+1}{\epsilon} ) + O( \frac{K^2 \cdot \kappa}{\epsilon^3} \cdot \ln( \frac{B+1}{\epsilon} ) )
	$$
	and 
	$$
		R_{B, T} \leq 2^6 \frac{\lambda}{\epsilon^{3/2}} \cdot \sqrt{ K \cdot \frac{B+1}{\epsilon}  \cdot \ln( \frac{B+1}{\epsilon} ) } + O(\frac{K^2 \cdot \kappa}{\epsilon^3} \cdot \ln( \frac{B+1}{\epsilon} )),
	$$
	where the $O$ notations hide universal constant factors.
\end{theorem} \vspace{0.3cm}


\paragraph{Applications.} 
Similarly, as in the case of a single resource, Assumptions \ref{assumption-cost-bounds-rewards-budget-and-time-horizon} and \ref{assumption-simplying-assumption-upperboundknown-budget-and-time-horizon} are natural when bidding in repeated second-price auctions if the auctioneer sets a reserve price $R$ (which is common practice in sponsored search auctions). Indeed, we have:
\begin{align*}
	|\mathbb{E}[c_{k, t}] - \mathbb{E}[c_{l, t}]|
		 & = \mathbb{E}[m_t \cdot \mathbbm{1}_{b_k \geq m_t > b_l}] \\
		 & \geq R \cdot \mathbb{E}[\mathbbm{1}_{b_k \geq m_t > b_l}] \\
		 & \geq R \cdot \mathbb{E}[v_t \cdot \mathbbm{1}_{b_k \geq m_t > b_l}] = R \cdot |\mathbb{E}[r_{k, t}] - \mathbb{E}[r_{l, t}]|,
\end{align*}
for any pair of arms $(k, l) \in \{1, \cdots, K\}$ with $b_k \geq b_l$. Hence, Assumption \ref{assumption-cost-bounds-rewards-budget-and-time-horizon} (resp. \ref{assumption-simplying-assumption-upperboundknown-budget-and-time-horizon}) is satisfied with $\sigma = 1/R$ (resp. $\kappa = 1/R$). \\
\indent In dynamic procurement, Assumptions \ref{assumption-cost-bounds-rewards-budget-and-time-horizon} and \ref{assumption-simplying-assumption-upperboundknown-budget-and-time-horizon} are satisfied provided that the agents are not willing to sell their goods for less than a known price $P$. Indeed, in this case, pulling any arm $k$ associated with a price $p_k \leq P$ is always suboptimal and we have:
\begin{align*}
	|\mathbb{E}[c_{k, t}] - \mathbb{E}[c_{l, t}]|
		 & = p_k \cdot \mathbb{P}[p_k \geq v_t] - p_l \cdot \mathbb{P}[p_l \geq v_t] \\
		 & \geq p_k \cdot \mathbb{P}[p_k \geq v_t > p_l] \\
		 & \geq P \cdot \mathbb{P}[p_k \geq v_t > p_l] = P \cdot |\mathbb{E}[r_{k, t}] - \mathbb{E}[r_{l, t}]|,
\end{align*}
for any pair of arms $(k, l) \in \{1, \cdots, K\}$ with $p_k \geq p_l \geq P$. Hence, Assumption \ref{assumption-cost-bounds-rewards-budget-and-time-horizon} (resp. \ref{assumption-simplying-assumption-upperboundknown-budget-and-time-horizon}) is satisfied with $\sigma = 1/P$ (resp. $\kappa = 1/P$). \\
\indent In dynamic pricing, Assumptions \ref{assumption-cost-bounds-rewards-budget-and-time-horizon} and \ref{assumption-simplying-assumption-upperboundknown-budget-and-time-horizon} are satisfied if the distribution of valuations has a positive probability density function $f(\cdot)$. Indeed, in this case, we have:
\begin{align*}
	|\mathbb{E}[r_{k, t}] - \mathbb{E}[r_{l, t}]| 
		 & = |p_l \cdot \mathbb{P}[ p_l \leq v_t ] - p_k \cdot \mathbb{P}[ p_k \leq v_t] |   \\
		 & = |p_l \cdot \mathbb{P}[ p_l \leq v_t < p_k] + (p_l - p_k) \cdot \mathbb{P}[ p_k \leq v_t] |  \\
		 & \leq \max_{r=1, \cdots, K} p_r \cdot \mathbb{P}[ p_l \leq v_t < p_k ] + |p_k - p_l| \\
		 & \leq (\max_{r=1, \cdots, K} p_r + \frac{1}{\inf f(\cdot)}) \cdot \mathbb{P}[ p_l \leq v_t < p_k ] \\
		 & = (\max_{r=1, \cdots, K} p_r + \frac{1}{\inf f(\cdot)}) \cdot |\mathbb{E}[r_{k, t}] - \mathbb{E}[r_{l, t}]|,
\end{align*}
for any pair of arms $(k, l) \in \{1, \cdots, K\}$ with $k \geq l$. Hence, Assumption \ref{assumption-cost-bounds-rewards-budget-and-time-horizon} (resp. \ref{assumption-simplying-assumption-upperboundknown-budget-and-time-horizon}) is satisfied with $\sigma = \max_{k=1, \cdots, K} p_k + 1/\inf f(\cdot)$ (resp. $\kappa = \max_{k=1, \cdots, K} p_k + 1/\inf f(\cdot)$).



\section{Arbitrarily many limited resources.}
\label{sec-stochastic-multiple-budget}	
In this section, we tackle the general case of arbitrarily many limited resources. Additionally, we assume that one of them is time, with index $i = C$, but this assumption is almost without loss of generality, as detailed at the end of this section. To simplify the presentation, we consider the regime $K \geq C$, which is the most common in applications. This implies that $|\mathcal{K}_x| = |\mathcal{C}_x| \leq C$ for any pseudo-basis $x$. We also use the shorthand notation $\bar{A}_{t} = (\bar{c}_{k,t}(i))_{(i,k) \in \{1, \cdots, C\} \times \{1, \cdots, K\}}$ at any round $t$. For similar reasons as in Section \ref{sec-singlebudgettimehorizon}, we are led to make two additional assumptions which are discussed in the last paragraph of this section.
\begin{assumption}
\label{assumption-cost-bounds-rewards-general-case}
There exists $\sigma > 0$ such that $r_{k, t} \leq \sigma \cdot \min\limits_{i=1, \cdots, C} c_{k, t}(i)$ for any arm $k \in \{1, \cdots, K\}$ and for any round $t \in \mathbb{N}$.
\end{assumption}
Note that Assumption \ref{assumption-cost-bounds-rewards-general-case} is stronger than Assumption \ref{assumption-cost-bounds-rewards-budget-and-time-horizon} given that the amounts of resources consumed at each round have to dominate the rewards almost surely, as opposed to on average. Assumption \ref{assumption-cost-bounds-rewards-general-case} is not necessarily satisfied in all applications but it simplifies the analysis and can be relaxed at the price of an additive term of order $O(\ln^2(T))$ in the final regret bounds, see the last paragraph of this section.
\begin{assumption}
	\label{assumption-simplying-assumption-general case}
	There exists $\epsilon > 0$, known to the decision maker ahead of round $1$, such that every basis $x$ for \eqref{eq-linear-program-general-upperbound-opt-strategy} is $\epsilon$-non-degenerate for \eqref{eq-linear-program-general-upperbound-opt-strategy} and satisfy $|\det(A_x)| \geq \epsilon$. 
\end{assumption}
Without loss of generality, we assume that $\epsilon \leq 1$. Observe that Assumption \ref{assumption-simplying-assumption-general case} generalizes Assumption \ref{assumption-simplying-assumption-analysis-budget-and-time-horizon} but is more restrictive because $\epsilon$ is assumed to be known to the decision maker initially. Just like in Section \ref{sec-singlebudgettimehorizon}, this assumption can be relaxed to a large extent. For instance, if $\epsilon$ is initially unknown, taking $\epsilon$ as a vanishing function of $T$ yields the same asymptotic regret bounds. However, note that Lemma \ref{lemma-non-degeneracy-assumption-is-necessary-budget-and-time-horizon} carries over to this more general setting and, as a result, the minimal assumption we need to get logarithmic rates is that any optimal basis for \eqref{eq-linear-program-general-upperbound-opt-strategy} is non-degenerate.

\paragraph{Specification of the algorithm.}
We implement UCB-Simplex with $\lambda = 1 + 2 (C+1)!^2/\epsilon$, $\eta_i = 0$ for any $i \in \{1, \cdots, C\}$, and an initialization step which consists in pulling each arm $\tinivalueinline \cdot \ln(T)$ times in order to get i.i.d. samples. Hence, Step-Simplex is run for the first time after round $\tini = K \cdot \tinivalueinline \cdot \ln(T)$. Compared to Section \ref{sec-singlebudgettimehorizon}, the initialization step plays as a substitute for the choice $\eta_i > 0$ which was meant to incentivize exploration. This significantly simplifies the analysis but the downside is that $\epsilon$ has to be known initially. Similarly, as in Section \ref{sec-singlebudgettimehorizon}, we introduce a dummy arm which corresponds to skipping the round (i.e. pulling this arm yields a reward $0$ and does not consume any resource) so that any basis can be mapped to one for which the time constraint is always binding, i.e. without loss of generality we assume that $C \in \mathcal{C}_x$ for any pseudo-basis $x$. Following the ideas developed in Section \ref{sec-singlebudgettimehorizon}, we design load balancing algorithms for any basis $x$ that pull arms in order to guarantee that, at any round $t$, the total amount of resource $i$ consumed remains close to the target $t \cdot b(i)$ with high probability for any resource $i \in \mathcal{C}_x$. This is more involved that in Section \ref{sec-singlebudgettimehorizon} since we need to enforce this property for multiple resources but, as we show in this section, this can be done by perturbing the probability distribution solution to \eqref{eq-algo-general-idea} taking into account whether we have over- or under-consumed in the past for each binding resource $i \in \mathcal{C}_{x_t}$.

\vspace{0.3cm}
\begin{algorithm}[H]
	\caption{Load balancing algorithm $\mathcal{A}_{x}$ for any basis $x$}
		\label{def-pulling-strategy-general-case}
		For any time period $t > \tini$ and $i \in \mathcal{C}_x - \{C\}$, define $b_{x, t}(i)$ as the total amount of resource $i$ consumed when selecting basis $x$ in the past $t-1$ rounds. Suppose that basis $x$ is selected at time $t$ and define the vector $e^x_t$ by $e^x_{C, t} = 0$ and $e^x_{i, t} = -1$ (resp. $e^x_{i, t} = 1$) if $b_{x, t}(i) \geq n_{x, t} \cdot b(i)$ (resp. $b_{x, t}(i) < n_{x, t} \cdot b(i)$) for any $i \in \mathcal{C}_x - \{C\}$. Since $x$ is selected at round $t$, $\bar{A}_{x, t}$ is invertible and we can define, for any $\delta \geq 0$, $p^x_{k, t}(\delta) = (\bar{A}_{x, t}^{-1} (b_{\mathcal{C}_x} + \delta \cdot e^x_t ))_k$ for $k \in \mathcal{K}_x$ and $p^x_{k, t}(\delta) = 0$ otherwise, which together define the probability distribution $p^x_t(\delta) = (p^x_{k, t}(\delta))_{k \in \{1, \cdots, K\}}$. Define 
		$$
			\delta^*_{x, t} = \max\limits_{ \substack{\delta \geq 0 \\ (\bar{A}_{t} p^x_t(\delta))_i \leq b(i), i \notin \mathcal{C}_x \\ p^x_t(\delta) \geq 0 } } \delta
		$$ 
		and $p^x_t = p^x_{t}(\delta^*_{x, t})$. Note that $\delta^*_{x, t}$ is well defined as $x$ must be feasible for \eqref{eq-algo-general-idea} if it is selected at Step-Simplex. Pull an arm at random according to the distribution $p^x_t$.
\end{algorithm} \vspace{0.3cm}
\noindent Observe that the load balancing algorithms generalize the ones designed in Section \ref{sec-singlebudgettimehorizon} (up to the change $\eta_i = 0$). Indeed, when there is a single limited resource other than time, the probability distribution $p^x_t$ is a Dirac supported at the arm with smallest (resp. largest) empirical cost when $b_{x, t} \geq n_{x, t} \cdot b$ (resp. $b_{x, t} < n_{x, t} \cdot b$). Similarly, as in Section \ref{sec-multiplebudgets}, the load balancing algorithms $\mathcal{A}_x$ may require a memory storage capacity exponential in $C$ and polynomial in $K$, but, in practice, we expect that only a few bases will be selected at Step-Simplex, so that a hash table is an appropriate data structure to store the sequences $(b_{x, t}(i))_{i \in \mathcal{C}_x}$. Note, however, that the load balancing algorithms are computationally efficient because $p^x_t$ can be computed in $O(C^2)$ running time if $\bar{A}_{x, t}^{-1}$ is available once we have computed an optimal basic feasible solution to \eqref{eq-algo-general-idea}, which is the case if we use the revised simplex algorithm. 

\paragraph{Regret analysis.}
The regret analysis follows the same recipe as in Section \ref{sec-singlebudgettimehorizon} but the proofs are more technical and are thus deferred to Section \ref{sec-proofs-general-case} of the Appendix. First, we show that the initialization step guarantees that infeasible bases or pseudo-bases $x$ with $\det(A_x) = 0$ cannot be selected more than $O(\ln(T))$ times on average at Step-Simplex.

\begin{lemma}
	\label{lemma-bound-times-infeasible-pulls-general-case}
			For any basis $x \notin \mathcal{B}$, we have: 
			$$
				\mathbb{E}[n_{x, T}] \leq \probamaxdevmeangeneralcasewithouttime.
			$$
			The same inequality holds if $x$ is a pseudo-basis but not a basis for \eqref{eq-linear-program-general-upperbound-opt-strategy}.
\end{lemma} \vspace{0.3cm}
The next step is to show that the load balancing algorithms guarantee that, for any basis $x$, the amount of resource $i \in \mathcal{C}_x$ (resp. $i \notin \mathcal{C}_x$) consumed per round when selecting $x$ remains close to (resp. below) the pacing target $b(i)$ with high probability. This enables us to show that the ratios $(\mathbb{E}[n^x_{k, T}]/\mathbb{E}[n^x_{l, T}])_{k, l \in \mathcal{K}_x}$ are close to the optimal ones $(\xi^x_k/\xi^x_l)_{k, l \in \mathcal{K}_x}$.

\begin{lemma}
		\label{lemma-load-balance-general-case}
		For any feasible basis $x$ and time period $t$, we have:
		\begin{equation}
			\label{eq-load-balance-general-case-eq1}
			\mathbb{P}[|b_{x, t}(i) - n_{x, t} \cdot b(i)| \geq u ] \leq 2^5 \frac{(C+1)!^2}{\epsilon^4} \cdot \exp(- u \cdot (\maxdelta)^2) + \probamaxdevmeangeneralcasewithouttime \cdot \frac{1}{T} \quad \forall u \geq 1,		
		\end{equation}
		for any resource $i \in \mathcal{C}_x$ while
		\begin{equation}
			\label{eq-load-balance-general-case-eq2}
				\mathbb{P}[b_{x, t}(i) - n_{x, t} \cdot b(i) \geq 2^8  \frac{(C+3)!^3}{\epsilon^6} \cdot \ln(T) ] \leq 2^{10} \frac{(C+4)!^4}{\epsilon^6 \cdot T},
		\end{equation}
		for any resource $i \notin \mathcal{C}_x$. In particular, this implies that:
		\begin{equation}
			\label{eq-lower-bound-pulls-general-case}
				\begin{aligned}
					& \mathbb{E}[n^x_{k, T}] \geq  \xi^x_k \cdot \mathbb{E}[n_{x, T}] - 2^{10}  \frac{(C+3)!^4 }{\epsilon^9},
				\end{aligned}
		\end{equation}				
		for any arm $k \in \mathcal{K}_x$.
\end{lemma} \vspace{0.3cm}

\noindent Next, we show that a suboptimal basis cannot be selected more than $O(\ln(T))$ times on average at Step-Simplex. Just like in Section \ref{sec-singlebudgettimehorizon}, the exact definition of the load balancing algorithms has little impact on the result: we only need to know that, for any feasible basis $x$, each arm $k \in \mathcal{K}_x$ is pulled at least a fraction of the time $x$ is selected with high probability.
\begin{lemma}
	\label{lemma-bound-times-non-optimal-pulls-general-case}
	For any suboptimal basis $x \in \mathcal{B}$, we have: 
	$$
		\mathbb{E}[n_{x, T}] \leq 2^{10} \frac{(C+3)!^3 \cdot \lambda^2 }{\epsilon^6} \cdot \frac{\ln(T)}{(\Delta_x)^2} + 2^{11} \frac{(C+4)!^4}{\epsilon^6}.
	$$
\end{lemma} \vspace{0.3cm}
We are now ready to derive both distribution-dependent and distribution-independent regret bounds.
\begin{theorem}
	We have:
	\label{lemma-log-B-regret-bound-general-case}
	$$
		R_{B(1), \cdots, B(C-1), T} \leq 2^{10} \frac{(C+3)!^3 \cdot \lambda^2}{\epsilon^6} \cdot (\sum_{x \in \mathcal{B} \; | \; \Delta_x > 0} \frac{1}{\Delta_x}) \cdot \ln(T) + O(\frac{\sigma \cdot |\mathcal{B}| \cdot (C+3)!^4 }{\epsilon^6} \cdot \ln(T)),
	$$
	where the $O$ notation hides universal constant factors.
\end{theorem} \vspace{0.3cm}

\begin{theorem}
		\label{lemma-sqrt-B-regret-bound-general-case}
		We have:
		$$
			R_{B(1), \cdots, B(C-1), T} \leq 2^5 \frac{(C+3)!^{2} \cdot \lambda}{\epsilon^3} \cdot \sqrt{ |\mathcal{B}| \cdot T \cdot \ln( T )  } + O(\frac{\sigma \cdot |\mathcal{B}| \cdot (C+3)!^4 }{\epsilon^6} \cdot \ln(T)),
		$$
		where the $O$ notation hides universal constant factors.
\end{theorem} \vspace{0.3cm}
Since the number of feasible bases is at most $2 K^C$, we get the distribution-dependent regret bound $R_{B(1), \cdots, B(C-1), T} = O( K^C \cdot (C+3)!^4 / \epsilon^6 \cdot ( \lambda^2/\Delta + \sigma) \cdot \ln(T))$ where $\Delta = \min_{x \in \mathcal{B} \; | \; \Delta_x > 0} \Delta_x$ and the distribution-independent bound $R_{B(1), \cdots, B(C-1), T} = O( (C+3)!^{2} \cdot \lambda / \epsilon^3 \cdot \sqrt{K^C \cdot T \cdot \ln(T)} )$. We stress that the dependence on $K$ and $C$ is not optimal since \citet{agrawal2014bandits} obtain a $\tilde{O}(\sqrt{K \cdot T})$ distribution-independent bound on regret, where the $\tilde{O}$ notation hides factors logarithmic in $T$. Just like in Section \ref{sec-singlebudgettimehorizon}, we can also derive regret bounds that vanish with $b$ under the assumption that pulling any arm incurs some positive amount of resource consumption in expectations for all resources, but this requires a minor tweak of the algorithm.
\begin{theorem}
\label{lemma-regret-bound-b-small-general-case}
Suppose that:
$$
	\epsilon \leq \min\limits_{ \substack{i = 1, \cdots, C-1 \\ k = 1, \cdots, K } } \mu^c_k(i)
$$ 
and that $b \leq \epsilon$. If the decision maker artificially constrains himself or herself to a time horizon $\tilde{T} = b \cdot T / \epsilon \leq T$, then the regret bounds derived in Theorems \ref{lemma-log-B-regret-bound-general-case} and \ref{lemma-sqrt-B-regret-bound-general-case} hold with $T$ substituted with $\tilde{T}$.
\end{theorem} \vspace{0.3cm}
Similarly, if the decision maker is not constrained by a time horizon, artificially constraining himself or herself to a time horizon $\tilde{T} = \min_{i=1, \cdots, C} B(i)/ \epsilon$ yields the regret bounds derived in Theorems \ref{lemma-log-B-regret-bound-general-case} and \ref{lemma-sqrt-B-regret-bound-general-case} with $T$ substituted with $\tilde{T}$.


\paragraph{Applications.}
In dynamic pricing and online advertising applications, Assumption \ref{assumption-cost-bounds-rewards-general-case} is usually not satisfied as pulling an arm typically incurs the consumption of only a few resources. We can relax this assumption but this comes at the price of an additive term of order $O(\ln^2(T))$ in the final regret bounds.
\begin{theorem}
	\label{lemma-relax-assumption-cost-dominate-rewards-general-case}
If Assumption \ref{assumption-cost-bounds-rewards-general-case} is not satisfied, the regret bounds derived in Theorems \ref{lemma-log-B-regret-bound-general-case} and \ref{lemma-sqrt-B-regret-bound-general-case} hold with $\sigma = 0$ up to an additive term of order $O(  \frac{(C+4)!^4 \cdot |\mathcal{B}|^2}{b \cdot \epsilon^6} \cdot  \ln^2(T) )$.
\end{theorem} \vspace{0.3cm}
As for Assumption \ref{assumption-simplying-assumption-general case}, the existence of degenerate optimal bases to \eqref{eq-linear-program-general-upperbound-opt-strategy} is determined by a complex interplay between the mean rewards and the mean amounts of resource consumption. However, we point out that the set of parameters $(\mu^r_k)_{k=1, \cdots, K}$ and $(\mu^c_k(i))_{k=1, \cdots, K, i=1, \cdots, C}$ that satisfy these conditions has Lebesgue measure $0$, hence such an event is unlikely to occur in practice. Additionally, while $\epsilon$ is typically not known in applications, taking $\epsilon$ as a vanishing function of $T$ yields the same asymptotic regret bounds. 


\section{Concluding remark.}
In this paper, we develop an algorithm with a $O(K^C \cdot \ln(B) / \Delta)$ distribution-dependent bound on regret, where $\Delta$ is a parameter that generalizes the optimality gap for the standard MAB problem. It is however unclear whether the dependence on $K$ is optimal. Extensions discussed in Section \ref{sec-extensions} of the Appendix suggest that it may be possible to achieve a linear dependence on $K$ but this calls for the development of more efficient load balancing algorithms.

%
%
%
\begin{APPENDICES}

\section{Extensions.}
\label{sec-extensions}
\subsection{Improving the multiplicative factors in the regret bounds.}

\subsubsection{A single limited resource whose consumption is deterministic.}
\label{sec-ext-single-budget}
If the amounts of resource consumed are deterministic, we can substitute the notation $\mu^{c}_k$ for $c_k$. Moreover, we can take $\lambda = 1$ and, going through the analysis of Lemma \ref{lemma-bound-times-non-optimal-pulls}, we can slightly refine the regret bound. Specifically, we have $\mathbb{E}[n_{k, \tau^*}] \leq \frac{16}{(c_k)^2} \cdot \frac{\mathbb{E}[\ln(\tau^*)]}{(\Delta_k)^2} + \frac{\pi^2}{3}$, for any arm $k$ such that $\Delta_k > 0$. Moreover, $\tau^* \leq B/\epsilon + 1$ in this setting since:
\begin{align*}
	B
		& \geq \sum_{t=1}^{\tau^*-1} c_{a_t, t} \geq (\tau^*-1) \cdot \epsilon,
\end{align*}
by definition of $\tau^*$. As a result, the regret bound derived in Theorem \ref{lemma-log-B-regret-bound} turns into:
$$
	R_B \leq 16 (\sum_{k \; | \; \Delta_k > 0} \frac{1}{c_k \cdot \Delta_k}) \cdot \ln(\frac{B}{\epsilon} + 1) + O(1),
$$
which is identical (up to universal constant factors) to the bound obtained by \citet{tran2012knapsack}. Note that this bound is scale-free.

\subsubsection{Arbitrarily many limited resources whose consumptions are deterministic.}
\label{sec-ext-deterministic}
We propose another load balancing algorithm that couples bases together. This is key to get a better dependence on $K$ because, otherwise, we have to study each basis independently from the other ones.

\vspace{0.3cm}
\begin{algorithm}[H]
	\label{def-algorithm-deterministic-costs-alternative}
	\caption{Load balancing algorithm $\mathcal{A}_x$ for a feasible basis $x \in \mathcal{B}$.}
	If $x$ is selected at time $t$, pull any arm $a_t \in \argmax\limits_{k \in \mathcal{K}_x} \frac{\xi^{x}_k}{ n_{k, t} }$.
\end{algorithm} \vspace{0.3cm}

\noindent Observe that this load balancing algorithm is computationally efficient with a $O(K)$ runtime (once we have computed an optimal basic feasible solution to \eqref{eq-algo-general-idea}) and requires $O(K)$ memory space. The shortcoming of this approach is that, if there are multiple optimal bases to \eqref{eq-linear-program-general-upperbound-opt-strategy}, the optimal load balance for each optimal basis will not be preserved since we take into account the number of times we have pulled each arm when selecting any  basis (for which we strive to enforce different ratios). Hence, the following assumption will be required for the analysis.

\begin{assumption}
	There is a unique optimal basis to \eqref{eq-linear-program-general-upperbound-opt-strategy}.
\end{assumption}

\paragraph{Regret Analysis.}
All the proofs are deferred to Section \ref{sec-proof-extensions}. We start by  bounding, for each arm $k$, the number of times this arm can be pulled when selecting any of the suboptimal bases. This is in stark contrast with the analysis carried out in Section \ref{sec-multiplebudgets} where we bound the number of times each suboptimal basis has been selected. 
\begin{lemma}
	\label{lemma-bound-times-non-optimal-pulls-deterministic-alternaterule}
	For any arm $k \in \{1, \cdots, K\}$, we have:
	$$
		\mathbb{E}[ \sum_{ x \in \mathcal{B} \; | \; k \in \mathcal{K}_x, \; x \neq x^* }  n^x_{k, \tau^*} ] \leq 16 \frac{ \rho \cdot (\sum_{i=1}^C b(i))^2}{ \epsilon^2 } \cdot \frac{\mathbb{E}[ \ln(\tau^*) ]}{(\Delta_k)^2} + K \cdot \frac{\pi^2 }{3},
	$$
	where $\Delta_k = \min_{ x \in \mathcal{B} \; | \; k \in \mathcal{K}_x, \; x \neq x^*} \Delta_x$.
\end{lemma} \vspace{0.3cm}
In contrast to Section \ref{sec-multiplebudgets}, we can only guarantee that the ratios $(n^x_{k, t}/n^x_{l, t})_{k, l \in \mathcal{K}_x}$ remain close to the optimal ones $(\xi^x_k / \xi^x_l)_{k, l \in \mathcal{K}_x}$ at all times for the optimal basis $x = x^*$. This will not allow us to derive distribution-free regret bounds for this particular class of load balancing algorithms.

\begin{lemma}
\label{lemma-step-2-well-defined-multiple-budget-alternaterule}
At any time $t$ and for any arm $k \in \mathcal{K}_{x^*}$, we have:
\begin{equation}
	\label{eq-deterministic-alternaterule-nb-pull-arm-lowerbound}
	n_{k, t} \geq n_{x^*, t} \cdot \frac{\xi^{x^*}_k}{\sum_{l=1}^K \xi^{x^*}_l} - \rho \cdot (\sum_{x \in \mathcal{B}, x \neq x^*} n_{x, t} + 1)
\end{equation}
and 
\begin{equation}
	\label{eq-deterministic-alternaterule-nb-pull-arm-upperbound}
		n_{k, t} \leq n_{x^*, t} \cdot \frac{\xi^{x^*}_k}{\sum_{l=1}^K \xi^{x^*}_l} + \sum_{x \in \mathcal{B}, x \neq x^*} n_{x, t} + 1.
\end{equation}
\end{lemma} \vspace{0.3cm}
Bringing everything together, we are now ready to establish regret bounds.

\begin{theorem} 
	\label{lemma-log-B-regret-bound-deterministic-alternate}
	We have:
		\begin{align*}
			R_{B(1), \cdots, B(C)}
				& \leq 32 \frac{ \rho^3 \cdot (\sum_{i=1}^C b(i))^3 }{ \epsilon^3 \cdot b}  \cdot (\sum_{k=1}^K \frac{1}{(\Delta_k)^2} ) \cdot \ln( \frac{ \sum_{i=1}^C b(i) \cdot B}{\epsilon}  + 1 ) + O(1),
		\end{align*}
		where the $O$ notation hides universal constant factors.
\end{theorem} \vspace{0.3cm}
We derive a distribution-dependent regret bound of order $O( \rho^3 \cdot K \cdot \frac{\ln(B)}{\Delta^2})$ where $\Delta = \min_{x \in \mathcal{B} \; | \; \Delta_x > 0}$ $\Delta_x$ but no non-trivial distribution-free regret bound.

\subsubsection{Arbitrarily many limited resources.}
A straightforward extension of the load balancing algorithm developed in the case of deterministic resource consumption in Section \ref{sec-ext-deterministic} guarantees that the total number of times any suboptimal basis is pulled is of order $O(K \cdot \ln(T))$. However, in contrast to Section \ref{sec-ext-deterministic}, this is not enough to get logarithmic regret bounds as $\xi^{x}_{k, t}$ fluctuates around the optimal load balance $\xi^x_{k, t}$ with a magnitude of order at least $\sim 1/\sqrt{t}$, and, as a result, the ratios $(\mathbb{E}[n^x_{k, T}]/\mathbb{E}[n^x_{l, T}])_{k, l \in \mathcal{K}_x}$ might be very different from the optimal ones $(\xi^x_k/\xi^x_l)_{k, l \in \mathcal{K}_x}$.

\vspace{0.3cm}
\begin{algorithm}[H]
	\caption{Load balancing algorithm $\mathcal{A}_x$ for a feasible basis $x \in \mathcal{B}$.}
	\label{def-algorithm-general-case-alternative}
	If $x$ is selected at time $t$, pull any arm $a_t \in \argmax\limits_{k \in \mathcal{K}_x} \frac{\xi^{x}_{k, t} }{n_{k, t}}$.
\end{algorithm} \vspace{0.3cm}

\begin{lemma}
	\label{lemma-bound-times-non-optimal-pulls-general-case-alternative}
	For any arm $k \in \{1, \cdots, K\}$, we have:
	$$
		\mathbb{E}[ \sum_{ x \in \mathcal{B} \; | \; k \in \mathcal{K}_x, \; \Delta_x > 0}  n^x_{k, T} ] \leq 16 C \cdot \lambda^2 \cdot \frac{ \ln(T) }{(\Delta_k)^2} + 2^{12}  \frac{K \cdot (C+3)!^2}{\epsilon^6},
	$$
	where $\Delta_k = \min_{ x \in \mathcal{B} \; | \; k \in \mathcal{K}_x, \; \Delta_x > 0} \Delta_x$.
\end{lemma} \vspace{0.3cm}

\subsection{Relaxing Assumption \ref{assumption-budget-scale-linearly-wt-time}.}
The regret bounds obtained in Sections \ref{sec-multiplebudgets}, \ref{sec-singlebudgettimehorizon}, and  \ref{sec-stochastic-multiple-budget} can  be extended when the ratios converge as opposed to being fixed, as precisely stated below, but this requires slightly more work.
\begin{assumption}
\label{assumption-budget-ratios-converge}
For any resource $i \in \{1, \cdots, C\}$, the ratio $B(i)/B(C)$ converges to a finite value $b(i) \in (0, 1]$. Moreover, $b = \min\limits_{i=1, \cdots, C} b(i)$ is a positive quantity.
\end{assumption}
To state the results, we need to redefine some notations and to work with the linear program:
	\begin{equation}
		\label{eq-linear-program-general-upperbound-opt-strategy-determistic-when-relaxing-budgetscalinglinearly}
		\begin{aligned}
			& \sup_{ (\xi_k)_{k=1, \cdots, K} } 
			& & \sum_{k=1}^K \mu^r_k \cdot \xi_k \\
			& \text{subject to}
			& & \sum_{k=1}^K \mu^c_k(i) \cdot \xi_k \leq \frac{B(i)}{B(C)}, \quad i = 1, \cdots, C \\
			&
			& & \xi_k \geq 0, \quad k = 1, \cdots, K.
		\end{aligned}
	\end{equation}
We redefine $\mathcal{B}$ as the set of bases that are feasible to \eqref{eq-linear-program-general-upperbound-opt-strategy-determistic-when-relaxing-budgetscalinglinearly} and, for $x \in \mathcal{B}$, $\Delta_x$ is redefined as the optimality gap of $x$ with respect to \eqref{eq-linear-program-general-upperbound-opt-strategy-determistic-when-relaxing-budgetscalinglinearly}. We also redefine $\mathcal{O} = \{ x \in \mathcal{B} \; | \; \Delta_x = 0 \}$ as the set of optimal bases to \eqref{eq-linear-program-general-upperbound-opt-strategy-determistic-when-relaxing-budgetscalinglinearly}. Moreover, we define $\mathcal{B}_{\infty}$ (resp. $\mathcal{O}_{\infty}$) as the set of feasible (resp. optimal) bases to \eqref{eq-linear-program-general-upperbound-opt-strategy} and, for $x \in \mathcal{B}_\infty$, $\Delta^\infty_x$ is the optimality gap of $x$ with respect to \eqref{eq-linear-program-general-upperbound-opt-strategy}. Our algorithm remains the same provided that we substitute $b(i)$ with $B(i)/B(C)$ for any resource $i \in \{1, \cdots, C\}$. Specifically, Step-Simplex consists in solving:
	\begin{equation}
		\label{eq-algo-general-idea-determistic-when-relaxing-budgetscalinglinearly}
		\begin{aligned}
			& \sup_{ (\xi_k)_{k=1, \cdots, K} } 
			& & \sum_{k=1}^K (\bar{r}_{k, t} + \lambda \cdot \epsilon_{k, t}) \cdot \xi_k \\
			& \text{subject to}
			& & \sum_{k=1}^K (\bar{c}_{k, t}(i) - \eta_i \cdot \epsilon_{k, t}) \cdot \xi_k \leq \frac{B(i)}{B(C)}, \quad i = 1, \cdots, C \\
			&
			& & \xi_k \geq 0, \quad k = 1, \cdots, K
		\end{aligned}
	\end{equation}
and Step-Load-Balance is identical up to the substitution of $b(i)$ with $B(i)/B(C)$.
\paragraph{Regret Analysis.}
As it turns out, the logarithmic regret bounds established in Theorems \ref{lemma-log-B-regret-bound-deterministic}, \ref{lemma-log-B-regret-bound-time-horizon}, and \ref{lemma-log-B-regret-bound-general-case} do not always extend when Assumption \ref{assumption-budget-scale-linearly-wt-time} is relaxed even though these bounds appear to be very similar to the one derived in Theorem \ref{lemma-log-B-regret-bound} when there is a single limited resource. The fundamental difference is that the set of optimal bases may not converge while it is always invariant in the case of a single limited resource. Typically, the ratios $(B(i)/B(C))_{i = 1, \cdots, C}$ may oscillate around $(b(i))_{i=1, \cdots, C}$ in such a way that there exist two optimal bases for \eqref{eq-linear-program-general-upperbound-opt-strategy} while there is a unique optimal basis for this same optimization problem whenever the right-hand side of the inequality constraints is slightly perturbed around this limit. This alternately causes one of these two bases to be slightly suboptimal, a situation difficult to identify and to cope with for the decision maker. Nevertheless, this difficulty does not arise in several situations of interest which generalize Assumption \ref{assumption-budget-scale-linearly-wt-time}, as precisely stated below. The proofs are deferred to Section \ref{sec-proof-extensions}.

\subparagraph{Arbitrarily many limited resources whose consumptions are deterministic.}
\begin{theorem}
	\label{lemma-log-B-regret-bound-deterministic-when-relaxing-budgetscalinglinearly}	
	Suppose that Assumption \ref{assumption-budget-ratios-converge} holds. If there exists a unique optimal basis to \eqref{eq-linear-program-general-upperbound-opt-strategy} or if $B(i)/B(C) - b(i) = O( \ln(B(C))/B(C))$ for all resources $i \in \{1, \cdots, C - 1\}$ then, we have:
	$$
		R_{B(1), \cdots, B(C)} = O( \frac{\rho \cdot \sum_{i=1}^C b(i)}{\epsilon \cdot b} \cdot (\sum_{x \in \mathcal{B}_\infty \; | \; \Delta^\infty_x > 0} \; \frac{ 1 }{ \Delta^\infty_x }) \cdot \ln( \frac{\sum_{i=1}^C b(i) \cdot B(C)}{\epsilon} + 1) + |\mathcal{O}_\infty| \cdot \frac{\ln(B(C))}{\epsilon \cdot b}),
	$$
	where the $O$ notation hides universal constant factors.
\end{theorem} \vspace{0.3cm}

\subparagraph{A time horizon and another limited resource.}
\begin{theorem}
	\label{lemma-log-B-regret-bound-time-horizon-when-relaxing-budgetscalinglinearly}	
	Suppose that Assumptions \ref{assumption-cost-bounds-rewards-budget-and-time-horizon}, \ref{assumption-simplying-assumption-upperboundknown-budget-and-time-horizon}, and \ref{assumption-simplying-assumption-analysis-budget-and-time-horizon} hold and that the ratio $B/T$ converges to $b \in (0, 1]$. If there exists a unique optimal basis to \eqref{eq-linear-program-general-upperbound-opt-strategy} or if $B/T - b = O( \ln(T)/T)$, then, we have:
	$$
		R_{B, T} = O(\frac{\lambda^2}{\epsilon^3} \cdot (\sum_{x \in \mathcal{B}_\infty \; | \; \Delta^\infty_x > 0}  \frac{1}{\Delta^\infty_x}) \cdot \ln(T) + \frac{K^2 \cdot \sigma}{\epsilon^3} \cdot \ln(T)),
	$$
	where the $O$ notation hides universal constant factors.
\end{theorem} \vspace{0.3cm}

\subparagraph{Arbitrarily many limited resources with a time horizon.}

\begin{theorem}
	\label{lemma-log-B-regret-bound-general-case-when-relaxing-budgetscalinglinearly}	
	Suppose that Assumptions \ref{assumption-cost-bounds-rewards-general-case}, \ref{assumption-simplying-assumption-general case}, and \ref{assumption-budget-ratios-converge} hold. If there exists a unique optimal basis to \eqref{eq-linear-program-general-upperbound-opt-strategy} or if $B(i)/T - b(i) = O(\ln(T)/T)$ for all resources $i \in \{1, \cdots, C-1\}$, then, we have:
	$$
		R_{B(1), \cdots, B(C-1), T} = O(\frac{(C+3)!^3 \cdot \lambda^2}{\epsilon^6} \cdot (\sum_{x \in \mathcal{B}_\infty \; | \; \Delta^\infty_x > 0} \frac{1}{\Delta^\infty_x}) \cdot \ln(T) + \frac{ \sigma \cdot |\mathcal{B}_\infty| \cdot (C+3)!^4 }{\epsilon^6} \cdot \ln(T)),
	$$
	where the $O$ notation hides universal constant factors.
\end{theorem} \vspace{0.3cm}

\section{Proofs for Section \ref{sec-algorithmicideas}.}

\subsection{Proof of Lemma \ref{lemma-general-bound-optimal-policy}.}
The proof can be found in \citet{badanidiyuru2013bandits}. For the sake of completeness, we reproduce it here. The optimization problem \eqref{eq-linear-program-general-upperbound-opt-strategy} is a linear program whose dual reads:
	\begin{equation}
	\label{eq-dual-linear-program-general-upperbound-opt-strategy}
		\begin{aligned}
		& \inf_{ (\zeta_i)_{i =1, \cdots, C} } 
		& & \sum_{i=1}^C b(i) \cdot \zeta_i \\
		& \text{subject to}
		& & \sum_{i=1}^C \mu^{c}_k(i) \cdot \zeta_i \geq \mu^r_k,  \quad k = 1, \cdots, K \\
		&
		& & \zeta_i \geq 0, \quad i = 1, \cdots, C.
		\end{aligned}
	\end{equation}
	Observe that \eqref{eq-linear-program-general-upperbound-opt-strategy} is feasible therefore \eqref{eq-linear-program-general-upperbound-opt-strategy} and \eqref{eq-dual-linear-program-general-upperbound-opt-strategy} have the same optimal value. Note that \eqref{eq-linear-program-general-upperbound-opt-strategy} is bounded under Assumption \ref{assumption-all-cost-non-zero} as $\xi_k \in[0, b(i)/\mu^{c}_k(i)]$ for any feasible point and any resource $i \in \{1, \cdots, C\}$ such that $\mu^c_k(i) > 0$. Hence, \eqref{eq-dual-linear-program-general-upperbound-opt-strategy} has an optimal basic feasible solution $(\zeta^*_1, \cdots, \zeta^*_C)$. Consider any non-anticipating algorithm. Let $Z_t$ be the sum of the total payoff accumulated in rounds $1$ to $t$ plus the \enquote{cost} of the remaining resources, i.e. $Z_t = \sum_{\tau = 1}^t r_{a_\tau, \tau} + \sum_{i=1}^C \zeta^*_i \cdot ( B(i) - \sum_{\tau = 1}^t c_{a_\tau, \tau}(i) )$. Observe that $(Z_t)_t$ is a supermartingale with respect to the filtration $(\mathcal{F}_t)_t$ as $\mathbb{E}[Z_t \; | \; \mathcal{F}_{t-1}] = \mathbb{E}[ \mu^r_{a_t} - \sum_{i=1}^C \zeta^*_i \cdot \mu^{c}_{a_t}(i) \; | \; \mathcal{F}_{t-1}]  + Z_{t-1} \leq Z_{t-1}$ since $(\zeta^*_1, \cdots, \zeta^*_C)$ is feasible for \eqref{eq-dual-linear-program-general-upperbound-opt-strategy}. Moreover, note that $(Z_t)_t$ has bounded increments since $|Z_t - Z_{t-1}| = | r_{a_t, t} - \sum_{i=1}^C \zeta^*_i \cdot c_{a_t, t}(i) | \leq 1 + \sum_{i=1}^C \zeta^*_i  < \infty$. We also have $\mathbb{E}[\tau^*] < \infty$ as:
	\begin{align*}
		\mathbb{E}[\tau^*] 
			& = \sum_{t=1}^{\infty} \mathbb{P}[\tau^* \geq t] \\
			& \leq \sum_{t=1}^{\infty} \mathbb{P}[\sum_{\tau=1}^{t-1} c_{a_{\tau}, \tau}(i) \leq B(i), i = 1, \cdots, C] \\
			& \leq 1 + \sum_{t=1}^{\infty} \mathbb{P}[\sum_{\tau=1}^{t} \sum_{i=1}^C c_{a_{\tau}, \tau}(i) \leq t \cdot \epsilon - (t \cdot \epsilon - \sum_{i=1}^C B(i) ) ] \\
			& \leq ( \frac{\sum_{i=1}^C B(i)}{\epsilon} + 2) + \sum_{t \geq \frac{\sum_{i=1}^C B(i)}{\epsilon}}^{\infty} \exp( - \frac{2 (t \cdot \epsilon - \sum_{i=1}^C B(i))^2}{t} ) \\
			& < \infty,
	\end{align*}
where the third inequality results from an application of Lemma \ref{lemma-martingale-inequality} and 
$$
	\epsilon = \min\limits_{ \substack{ k=1, \cdots, K \\ i = 1, \cdots, C \\ \text{with } \mu^c_k(i)>0} } \mu^c_k(i).
$$ 
By Doob's optional stopping theorem, $\mathbb{E}[Z_{\tau^*}] \leq \mathbb{E}[Z_0] = \sum_{i=1}^C \zeta^*_i \cdot B(i)$. Observe that: 
\begin{align*}
	\mathbb{E}[Z_{\tau^*}] 
		& = \mathbb{E}[r_{a_{\tau^*}, \tau^*} - \sum_{i=1}^C \zeta^*_i \cdot c_{a_{\tau^*}, \tau^*}(i)  + Z_{\tau^*-1} ] \\
		& \geq \mathbb{E}[- \sum_{i=1}^C \zeta^*_i + \sum_{t=1}^{\tau^* - 1} r_{a_t, t}].
\end{align*}
Using Assumption \ref{assumption-all-cost-non-zero} and since $(\zeta^*_i)_{i=1, \cdots, C}$ is a basic feasible solution, for every $i \in \{1, \cdots, C\}$ such that $\zeta^*_i > 0$ there must exist $k \in \{1, \cdots, K\}$ such that $\zeta^*_i \leq \mu^r_k/\mu^c_k(i)$ with $\mu^c_k(i) >0$.
	We get:
	$$ \mathbb{E}[Z_{\tau^*}] \geq  \mathbb{E}[ \sum_{t=1}^{\tau^* - 1} r_{a_t, t}] - \max\limits_{ \substack{ k=1, \cdots, K \\ i = 1, \cdots, C \\ \text{with } \mu^c_k(i)>0}} \frac{\mu^r_k}{\mu^{c}_k(i)}
	$$
	and finally:
	\begin{align*}
				\mathbb{E}[ \sum_{t=1}^{\tau^* - 1} r_{a_t, t}] 
					& \leq \sum_{i=1}^C \zeta^*_i \cdot B(i)  + \max\limits_{ \substack{ k=1, \cdots, K \\ i = 1, \cdots, C \\ \text{with } \mu^c_k(i)>0}} \frac{\mu^r_k}{\mu^{c}_k(i)} \\
					& = B \cdot \sum_{i=1}^C \zeta^*_i \cdot b(i) + \max\limits_{ \substack{ k=1, \cdots, K \\ i = 1, \cdots, C \\ \text{with } \mu^c_k(i)>0}} \frac{\mu^r_k}{\mu^{c}_k(i)}.
	\end{align*}
	By strong duality, $\sum_{i=1}^C \zeta^*_i \cdot b(i) $ is also the optimal value of \eqref{eq-linear-program-general-upperbound-opt-strategy}.

\section{Proofs for Section \ref{sec-singlebudget}.}

\subsection{Proof of Lemma \ref{lemma-bound-stopping-time}.}
	By definition of $\tau^*$, we have $\sum_{t=1}^{\tau^* - 1} c_{a_t, t} \leq B$. Taking expectations on both sides yields: 	
	\begin{align*}
		B 
			& \geq \mathbb{E}[\sum_{t=1}^{\tau^* - 1} c_{a_t, t}] \\
			& \geq \mathbb{E}[\sum_{t=1}^{\tau^*} c_{a_t, t}] - 1 \\
			& = \sum_{t=1}^{\infty} \mathbb{E}[I_{\tau^* \geq t} \cdot c_{a_t, t}] - 1 \\
			& = \sum_{t=1}^{\infty} \mathbb{E}[I_{\tau^* \geq t} \cdot \mathbb{E}[c_{a_t, t} \; | \; \mathcal{F}_{t-1}]] - 1 \\
			& = \sum_{t=1}^{\infty} \mathbb{E}[I_{\tau^* \geq t} \cdot \mu^c_{a_t}] - 1\\
			& \geq \sum_{t=1}^{\infty} \mathbb{E}[I_{\tau^* \geq t} \cdot \epsilon] - 1 \\
			& = \mathbb{E}[\tau^*] \cdot \epsilon - 1,
	\end{align*}
	where we use the fact that $c_{k, t} \leq 1$ for all arms $k$ to derive the second inequality, the fact that $\tau^*$ is a stopping time for the second equality, the fact that $a_t$ is deterministically determined by the past, i.e. $a_t \in \mathcal{F}_{t-1}$, for the third equality and Assumption \ref{assumption-all-cost-non-zero} for the third inequality. We conclude that $\mathbb{E}[\tau^*] \leq \frac{B+1}{\epsilon}$.
	
\subsection{Proof of Lemma \ref{lemma-bound-times-non-optimal-pulls}.}

		We break down the analysis in a series of facts. Consider any arm $k$ such that $\Delta_k > 0$. We use the shorthand notation $\beta_k =  2^5 (\frac{\lambda}{ \mu^c_k })^2 \cdot (\frac{1}{\Delta_k})^2$.
		\begin{fact}
			\label{fact-wlog-assume-n-big}
			\begin{equation}
				\label{eq-wlog-assume-n-big}
				\mathbb{E}[n_{k, \tau^*}] \leq 2 \beta_k \cdot \mathbb{E}[\ln( \tau^* )] +  \mathbb{E}[\sum_{t=1}^{\tau^*}  I_{a_t = k} \cdot I_{n_{k, t} \geq \beta_k \ln(t)} ].
			\end{equation}
			\proof{Proof.}
				Define the random variable $T_k = \beta_k \cdot \ln( \tau^* )$. We have:
				\begin{align*}
					\mathbb{E}[n_{k, \tau^*}] 
						& = \mathbb{E}[n_{k, \tau^*} \cdot I_{n_{k, \tau^*} < T_k}] + \mathbb{E}[n_{k, \tau^*} \cdot I_{n_{k, \tau^*} \geq T_k}] \\
						& \leq \beta_k \cdot \mathbb{E}[\ln( \tau^* )] + \mathbb{E}[n_{k, \tau^*} \cdot I_{n_{k, \tau^*} \geq T_k}].
				\end{align*}
				Define $T^*_k$ as the first time $t$ such that $n_{k, t} \geq T_k$ and $T^*_k = \infty$ if no such $t$ exists. We have:
				\begin{align*}
				\mathbb{E}[n_{k, \tau^*} \cdot I_{n_{k, \tau^*} \geq T_k}] 
					& = \mathbb{E}[\sum_{t=1}^{\tau^*} I_{a_t = k} \cdot I_{n_{k, \tau^*} \geq T_k} ] \\
					& = \mathbb{E}[\sum_{t=1}^{T^*_k - 1} I_{a_t = k} \cdot I_{n_{k, \tau^*} \geq T_k} ] + \mathbb{E}[\sum_{t=T^*_k}^{\tau^*} I_{a_t = k} \cdot I_{n_{k, \tau^*} \geq T_k} ] \\
					& \leq \mathbb{E}[ n_{k, T^*_k - 1}  \cdot I_{n_{k, \tau^*} \geq T_k} ] + \mathbb{E}[\sum_{t=T^*_k}^{\tau^*}  I_{a_t = k} \cdot I_{n_{k, t} \geq T_k} ] \\
					& \leq \beta_k \cdot \mathbb{E}[\ln( \tau^* )]  + \mathbb{E}[\sum_{t=1}^{\tau^*}  I_{a_t = k} \cdot I_{n_{k, t} \geq \beta_k \ln(t)} ],
				\end{align*}
				since, by definition of $T^*_k$, $n_{k, T^*_k - 1} \leq T_k$ if $T^*_k$ if finite, which is always true if $n_{k, \tau^*} \geq T_k$ (the sequence $(n_{k, t})_t$ is non-decreasing and $\tau^*$ is finite almost surely as a byproduct of Lemma \ref{lemma-bound-stopping-time}). Conversely, $n_{k, t} \geq T_k \geq \beta_k \ln(t)$  for $t \in \{T^*_k, \cdots, \tau^*\}$. Wrapping up, we obtain:
				$$
					\mathbb{E}[n_{k, \tau^*}] \leq 2 \beta_k \cdot \mathbb{E}[\ln( \tau^* )] +  \mathbb{E}[\sum_{t=1}^{\tau^*}  I_{a_t = k} \cdot I_{n_{k, t} \geq \beta_k \ln(t)} ].
				$$
			\endproof
		\end{fact} \vspace{0.3cm}		
		\noindent Fact \ref{fact-wlog-assume-n-big} enables us to assume that arm $k$ has been pulled at least $\beta_k \ln(t)$ times out of the last $t$ time periods. The remainder of this proof is dedicated to show that the second term of the right-hand side of \eqref{eq-wlog-assume-n-big} can be bounded by a constant. Let us first rewrite this term:
		\begin{align}
			\mathbb{E}[\sum_{t=1}^{\tau^*}  I_{a_t = k} \cdot I_{n_{k, t} \geq \beta_k \ln(t)} ] 
				& \leq \mathbb{E}[\sum_{t=1}^{\tau^*}  I_{ \mathrm{obj}_{k, t} + E_{k, t} \geq \mathrm{obj}_{k^*, t} + E_{k^*, t} } \cdot I_{n_{k, t} \geq \beta_k \ln(t)} ] \nonumber \\
				& \leq  
					\mathbb{E}[\sum_{t=1}^{\tau^*} I_{ \mathrm{obj}_{k, t}  \geq \mathrm{obj}_{k} + E_{k, t} } ] \label{eq-first-term-upper-bound-only-B}\\
				& + \mathbb{E}[\sum_{t=1}^{\tau^*} I_{ \mathrm{obj}_{k^*, t}  \leq \mathrm{obj}_{k^*} - E_{k^*, t} } ] \label{eq-second-term-upper-bound-only-B} \\
				& + \mathbb{E}[\sum_{t=1}^{\tau^*} I_{ \mathrm{obj}_{k^*} < \mathrm{obj}_{k} + 2 E_{k, t} } \cdot I_{n_{k, t} \geq \beta_k \ln(t)} ]. \label{eq-third-term-upper-bound-only-B}
		\end{align}
		To derive this last inequality, simply observe that if $\mathrm{obj}_{k, t}  < \mathrm{obj}_{k} + E_{k, t}$ and $\mathrm{obj}_{k^*, t} > \mathrm{obj}_{k^*} - E_{k^*, t}$ while $\mathrm{obj}_{k, t} + E_{k, t} \geq \mathrm{obj}_{k^*, t} + E_{k^*, t}$, it must be that $\mathrm{obj}_{k^*} < \mathrm{obj}_{k} + 2 E_{k, t}$. Let us study \eqref{eq-first-term-upper-bound-only-B}, \eqref{eq-second-term-upper-bound-only-B}, and \eqref{eq-third-term-upper-bound-only-B} separately. 		
	
		\begin{fact}
			\label{fact-bound-last-term-is-zero}
			$$
				\mathbb{E}[\sum_{t=1}^{\tau^*} I_{ \mathrm{obj}_{k^*} < \mathrm{obj}_{k} + 2 E_{k, t} } \cdot I_{n_{k, t} \geq \beta_k \ln(t)} ] \leq \frac{ 2 \pi^2 }{3 \epsilon^2}.
			$$
			\proof{Proof.}
				Observe that when both $n_{k, t} \geq \beta_k \ln(t)$ and $\mathrm{obj}_{k^*} < \mathrm{obj}_{k} + 2 E_{k, t}$, we have:
				\begin{align*}
						\frac{\Delta_k}{2}
							& < E_{k, t} \\
							& \leq \frac{ \lambda }{ \bar{c}_{k, t} } \cdot \sqrt{ \frac{2}{ \beta_k } }\\
							& \leq \frac{ \mu^c_k }{ 2 \bar{c}_{k, t} } \cdot \frac{\Delta_k}{2},
					\end{align*}		
				by definition of $\beta_k$. This implies that $\bar{c}_{k, t} \leq \mu^c_k /2$. Thus:
				\begin{align*}
				\mathbb{E}[\sum_{t=1}^{\tau^*} I_{ \mathrm{obj}_{k^*} < \mathrm{obj}_{k} + 2 E_{k, t} } \cdot I_{n_{k, t} \geq \beta_k \ln(t)} ] 
					& \leq \mathbb{E}[\sum_{t=1}^{\tau^*} I_{  \bar{c}_{k, t}  < \mu^c_k/2 } \cdot I_{n_{k, t} \geq \beta_k \ln(t)} ]
			\end{align*}			
			We upper bound this last term using the concentration inequalities of Lemma \ref{lemma-martingale-inequality} observing that:
			\begin{align*}
				\mathbb{E}[\sum_{t=1}^{\infty}  I_{\bar{c}_{k, t} < \mu^c_k/2} \cdot I_{n_{k, t} \geq \beta_k \ln(t)}]
					& = \sum_{t=1}^{\infty} \mathbb{P}[\bar{c}_{k, t} < \frac{ \mu^c_k }{2} \; ; \; n_{k, t} \geq \beta_k \ln(t)] \\
					& \leq \sum_{t=1}^{\infty} \sum_{s = \beta_k \ln(t)}^t \mathbb{P}[\bar{c}_{k, t} < \mu_k^c - \frac{ \mu^c_k }{2} \; ; \; n_{k, t} = s].
			\end{align*}						
			Denote by $t_1, \cdots, t_s$ the first $s$ random times at which arm $k$ is pulled (these random variables are finite almost surely). We have:
			$$
				\mathbb{P}[\bar{c}_{k, t} < \mu_k^c - \frac{ \mu^c_k }{2} \; ; \; n_{k, t} = s] 
					\leq \mathbb{P}[ \sum_{l = 1}^s c_{k, t_l} < s \cdot \mu_k^c - s \cdot \frac{ \mu^c_k }{2}].
			$$
			Observe that, for any $l \leq s$:
			\begin{align*}
				\mathbb{E}[c_{k, t_l} \; | \; c_{k, t_1}, \cdots, c_{k, t_{l-1}}] 
					& = \mathbb{E}[ \sum_{\tau=1}^\infty I_{t_l = \tau} \cdot \mathbb{E}[c_{k, \tau} \; | \;  \mathcal{F}_{\tau-1}] \; | \; c_{k, t_1}, \cdots, c_{k, t_{l-1}}]  \\
					& = \mathbb{E}[ \sum_{\tau=1}^\infty I_{t_l = \tau} \cdot \mu^c_{k} \; | \; c_{k, t_1}, \cdots, c_{k, t_{l-1}}]  \\
					& = \mu^c_{k},
			\end{align*}						
			since the algorithm is not randomized ($\{ t_l = \tau \} \in \mathcal{F}_{\tau-1}$) and using the tower property. Hence, we can apply Lemma \ref{lemma-martingale-inequality} to get:
			\begin{align*}
			 \sum_{t=1}^{\infty} \mathbb{P}[\bar{c}_{k, t} < \frac{ \mu^c_k }{2} \; ; \; n_{k, t} \geq \beta_k \ln(t)] 
					& \leq \sum_{t=1}^{\infty} \sum_{s = \beta_k \ln(t)}^\infty  \exp(- s \cdot \frac{ (\mu^c_k) ^2}{2}) \\
					& \leq \sum_{t=1}^{\infty} \frac{\exp(- \frac{( \mu^c_k )^2}{2} \cdot \beta_k \ln(t))}{ 1 - \exp(- \frac{(\mu^c_k)^2}{2})} \\
					& \leq \frac{1}{1 - \exp(-\frac{( \mu^c_k )^2}{2})} \sum_{t=1}^\infty \frac{1}{t^2} \\
					& \leq \frac{ 2 \pi^2 }{3 ( \mu^c_k )^2} \\
					& \leq \frac{ 2 \pi^2 }{3 \epsilon^2},
			\end{align*}			
			where we use the fact that $\beta_k \geq 2^5 (\frac{1 + \kappa}{\mu^c_k })^2 \cdot (\frac{ \mu^c_{k^*} }{\mu^r_{k^*}})^2 \geq 2^5 (\frac{1 + \frac{1}{\kappa}}{\mu^c_k  })^2 \geq \frac{4}{(\mu^c_k )^2}$ for the third inequality (using Assumption \ref{assumption-simplying-assumption-single-budget}), the fact that $\exp(-x) \leq 1 - \frac{x}{2}$ for $x \in [0, 1]$ for the fourth inequality, and Assumption \ref{assumption-all-cost-non-zero} for the last inequality. 
			\endproof						
			
		\end{fact} \vspace{0.3cm}	
		\noindent
	 	Let us now elaborate on \eqref{eq-first-term-upper-bound-only-B}.
	 	
	 	\begin{fact}
	 		\label{fact-study-bound-first-term}
	 		$$
	 			\mathbb{E}[\sum_{t=1}^{\tau^*} I_{ \mathrm{obj}_{k, t} \geq \mathrm{obj}_{k} + E_{k, t} } ] \leq \frac{\pi^2}{3}.
	 		$$
		\proof{Proof.}
			Note that if $\bar{r}_{k, t} / \bar{c}_{k, t}  = \mathrm{obj}_{k, t} \geq \mathrm{obj}_{k} + E_{k, t} = \mu^r_k / \mu^c_k  + E_{k, t}$, then either $\bar{r}_{k, t} \geq \mu^r_k + \epsilon_{k, t}$ or $\bar{c}_{k, t} \leq \mu^c_k - \epsilon_{k, t}$, otherwise we would have:
			\begin{align*}
				\frac{ \bar{r}_{k, t} }{ \bar{c}_{k, t} } - \frac{ \mu^r_k }{ \mu^c_k } 
					& = \frac{ ( \bar{r}_{k, t} - \mu^r_k) \mu^c_k + (\mu^c_k - \bar{c}_{k, t}) \mu^r_k }{ \bar{c}_{k, t} \cdot \mu^c_k } \\
					& <  \frac{ \epsilon_{k, t} }{ \bar{c}_{k, t} } + \frac{ \epsilon_{k, t} }{\bar{c}_{k, t}} \cdot  \frac{\mu^r_k}{\mu^c_k} \\
					& \leq (1 + \kappa) \cdot \frac{ \epsilon_{k, t} }{ \bar{c}_{k, t} } \\
					&  = E_{k, t},
			\end{align*}
			a contradiction. Therefore:
			
			\begin{align*}
				\mathbb{E}[\sum_{t=1}^{\tau^*} I_{ \mathrm{obj}_{k, t} \geq \mathrm{obj}_{k} + E_{k, t} } ]
					& \leq \sum_{t=1}^{\infty}  \mathbb{P}[\bar{r}_{k, t} \geq \mu^r_k + \epsilon_{k, t}] + \mathbb{P}[\bar{c}_{k, t} \leq \mu^c_k - \epsilon_{k, t}] \\
					& \leq \sum_{t=1}^{\infty}  \sum_{s=1}^t \mathbb{P}[\bar{r}_{k, t} \geq \mu^r_k + \sqrt{ \frac{2 \ln(t)}{s} } \; ; \; n_{k, t} = s] \\
					& + \sum_{t=1}^{\infty}  \sum_{s=1}^t \mathbb{P}[\bar{c}_{k, t} \leq \mu^c_k - \sqrt{ \frac{2 \ln(t)}{s} } \; ; \; n_{k, t} = s] \\
					& = \sum_{t=1}^{\infty}  \sum_{s=1}^t \mathbb{P}[\sum_{l = 1}^s r_{k, t_l} \geq s \cdot \mu^r_k + \sqrt{ s \cdot 2 \ln(t) } \; ; \; n_{k, t} = s] \\
					& + \sum_{t=1}^{\infty}  \sum_{s=1}^t \mathbb{P}[\sum_{l = 1}^s c_{k, t_l} \leq s \cdot \mu^c_k - \sqrt{ s \cdot 2 \ln(t) } \; ; \; n_{k, t} = s] \\
					& \leq \sum_{t=1}^{\infty} \sum_{s=1}^t 2 \exp( - 4 \ln(t) ) \\					
					& = \frac{\pi^2}{3},
			\end{align*}
			where the random variables $(t_l)_l$ are defined similarly as in the proof of Fact \ref{fact-bound-last-term-is-zero} and the last inequality results from an application of Lemma \ref{lemma-martingale-inequality}.
		\endproof			 			
	 	\end{fact} \vspace{0.3cm}
		\noindent
		What remains to be done is to bound \eqref{eq-second-term-upper-bound-only-B}.
		
		\begin{fact}
			\label{fact-bound-second-term}
			$$				
				\mathbb{E}[\sum_{t=1}^{\tau^*} I_{ \mathrm{obj}_{k^*, t}  \leq \mathrm{obj}_{k^*} - E_{k^*, t} } ] \leq \frac{\pi^2}{3}.
			$$			
			\proof{Proof.}
				We proceed along the same lines as in the proof of Fact \ref{fact-study-bound-first-term}. As a matter of fact, the situation is perfectly symmetric because, in the course of proving Fact \ref{fact-study-bound-first-term}, we do not rely on the fact that we have pulled arm $k$ more than $\beta_k \ln(t)$ times at any time $t$. If $ \bar{r}_{k^*, t} / \bar{c}_{k^*, t}  = \mathrm{obj}_{k^*, t} \leq \mathrm{obj}_{k^*} - E_{k^*, t} =  \mu^r_{k^*} / \mu^c_{k^*}  -  E_{k^*, t}$, then we have either $ \bar{r}_{k^*, t} \leq \mu^r_{k^* } - \epsilon_{k^*, t}$ or $\bar{c}_{k^*, t} \geq \mu^c_{k^*} + \epsilon_{k^*, t}$, otherwise we would have:
			\begin{align*}
				\frac{ \bar{r}_{k^*, t} }{ \bar{c}_{k^*, t} } - \frac{ \mu^r_{k^*} }{ \mu^c_{k^*} } 
					& = \frac{ ( \bar{r}_{k^*, t} - \mu^r_{k^*}) \mu^c_{k^*} + (\mu^c_{k^*} - \bar{c}_{k^*, t}) \mu^r_{k^*} }{ \bar{c}_{k^*, t} \cdot \mu^c_{k^*} } \\
					& > - \frac{ \epsilon_{k^*, t} }{ \bar{c}_{k^*, t} } - \frac{ \epsilon_{k^*, t} }{\bar{c}_{k^*, t}} \cdot \frac{\mu^r_{k^*}}{\mu^c_{k^*}}   \\
					& \geq - ( 1 + \kappa) \cdot \frac{ \epsilon_{k^*, t} }{ \bar{c}_{k^*, t} }  \\
					& = - E_{k^*, t},
			\end{align*}
			a contradiction. Therefore:
			\begin{align*}	
					\mathbb{E}[\sum_{t=1}^{\tau^*} I_{ \mathrm{obj}_{k^*, t} \leq \mathrm{obj}_{k^*} - E_{k^*, t} } ]
						& \leq \mathbb{E}[\sum_{t=1}^{\infty} I_{ \bar{r}_{k^*, t} \leq \mu^r_{k^*} - \epsilon_{k, t} } +  I_{ \bar{c}_{k^*, t} \geq \mu^c_{k^*} + \epsilon_{k, t} } ]  \\
						& \leq \sum_{t=1}^\infty \sum_{s=1}^t \mathbb{P}[\bar{r}_{k^*, t} \leq \mu^r_{k^*} - \sqrt{ \frac{2 \ln(t)}{s} } \; ; \; n_{k^*, t} = s ] \\
						& +  \sum_{t=1}^\infty \sum_{s=1}^t \mathbb{P}[\bar{c}_{k^*, t} \geq \mu^c_{k^*} + \sqrt{ \frac{2 \ln(t)}{s} } \; ; \; n_{k^*, t} = s] \\
						& \leq \sum_{t=1}^\infty \sum_{s=1}^t \frac{2}{t^4} \\
						& = \frac{\pi^2}{3},
			\end{align*} 
			where the third inequality is obtained using Lemma \ref{lemma-martingale-inequality} in the same fashion as in Fact \ref{fact-study-bound-first-term}.
			\endproof
		\end{fact} \vspace{0.3cm}		
		\noindent
		We conclude:
	$$
		\mathbb{E}[n_{k, \tau^*}] \leq 2 \beta_k \cdot \mathbb{E}[ \ln(\tau^*) ] + \frac{ 4 \pi^2 }{3 \epsilon^2}.
	$$

\subsection{Proof of Theorem \ref{lemma-log-B-regret-bound}.}
First observe that:
\begin{align*}
	\mathbb{E}[ \sum_{t=1}^{\tau^*} r_{a_t, t} ] 
		& = \sum_{t=1}^{\infty} \mathbb{E}[I_{\tau^* \geq t} \cdot \mathbb{E}[r_{a_t, t} \; | \; \mathcal{F}_{t-1}]] \\
		& = \sum_{t=1}^{\infty} \mathbb{E}[I_{\tau^* \geq t} \cdot \mu^r_{a_t}] \\
		& = \sum_{t=1}^{\infty} \sum_{k=1}^K \mu^r_k \cdot \mathbb{E}[I_{\tau^* \geq t} \cdot I_{a_t = k}] \\
		& = \sum_{k=1}^K \mu^r_k \cdot \mathbb{E}[\sum_{t=1}^{\infty} I_{\tau^* \geq t} \cdot I_{a_t = k}]  \\
		& = \sum_{k=1}^K \mu^r_k \cdot \mathbb{E}[n_{k, \tau^*}],
\end{align*}
since $\tau^*$ is a stopping time. Plugging this equality into \eqref{eq-simplified-upper-bound-on-regret} yields:
		\begin{align*}
			R_B 
			& \leq B \cdot \frac{ \mu^r_{k^*} }{\mu^c_{k^*}} - \sum_{k=1}^K \mu^r_k \cdot \mathbb{E}[n_{k, \tau^*}] + O(1) \\
			& =  \frac{ \mu^r_{k^*} }{ \mu^c_{k^*} } \cdot ( B - \sum_{k \; | \; \Delta_k = 0} \mu^c_k \cdot \mathbb{E}[n_{k, \tau^*}] ) - \sum_{k \; | \; \Delta_k > 0} \mu^r_k \cdot \mathbb{E}[n_{k, \tau^*}] + O(1). 
		\end{align*}
	Along the same lines as for the rewards, we can show that $\mathbb{E}[ \sum_{t=1}^{\tau^*} c_{a_t, t} ] =  \sum_{k=1}^K \mu^c_k \cdot \mathbb{E}[n_{k, \tau^*}]$. By definition of $\tau^*$, we have $B \leq \sum_{t=1}^{\tau^*} c_{a_t, t}$ almost surely. Taking expectations on both sides yields:
	\begin{align*}
		B 
			& \leq \sum_{k=1}^K \mu^c_k \cdot \mathbb{E}[n_{k, \tau^*}] \\
			& = \sum_{k \; | \; \Delta_k = 0} \mu^c_k \cdot \mathbb{E}[n_{k, \tau^*}] + \sum_{k \; | \; \Delta_k > 0} \mu^c_k \cdot \mathbb{E}[n_{k, \tau^*}].
	\end{align*}
	Plugging this inequality back into the regret bound, we get:
	\begin{align}
		R_B 
			& \leq  \sum_{k \; | \; \Delta_k > 0} ( \frac{ \mu^r_{k^*} }{ \mu^c_{k^*} } \cdot \mu^c_k  - \mu^r_k )\cdot \mathbb{E}[n_{k, \tau^*}] + O(1) \nonumber \\
			& =  \sum_{k \; | \; \Delta_k > 0} \mu^c_k \cdot \Delta_k \cdot \mathbb{E}[n_{k, \tau^*}] + O(1). \label{eq-proof-log-B-start-here-sqrt-B}
	\end{align}
	Using the upper bound of Lemma \ref{lemma-bound-stopping-time}, the concavity of the logarithmic function, and Lemma \ref{lemma-bound-times-non-optimal-pulls}, we derive:
	\begin{align*}
		R_B 
			& \leq  2^6 \lambda^2 \cdot (\sum_{k \; | \; \Delta_k > 0} \frac{1}{\mu^c_k \cdot  \Delta_k}) \cdot \ln(\frac{B + 1}{ \epsilon })  +  \frac{ 4 \pi^2 }{3 \epsilon^2} \cdot (\sum_{k \; | \; \Delta_k > 0} \mu^c_k \cdot \Delta_k) +  O(1) \\
			& \leq 2^6 \lambda^2 \cdot (\sum_{k \; | \; \Delta_k > 0} \frac{1}{ \mu^c_k \cdot  \Delta_k }) \cdot \ln(\frac{B + 1}{ \epsilon }) + K \cdot \frac{ 4 \pi^2 \kappa }{3 \epsilon^2} + O(1), 
	\end{align*}
	since $\Delta_k \leq  \mu^r_{k^*} / \mu^c_{k^*}  \leq \kappa$ and $\mu^c_k \leq 1$ for any arm $k$.

\section{Proofs for Section \ref{sec-multiplebudgets}.}

\subsection{Proof of Lemma \ref{lemma-bound-times-non-optimal-pulls-deterministic}.}
Consider any suboptimal basis $x \in \mathcal{B}$. The proof is along the same lines as for Lemma \ref{lemma-bound-times-non-optimal-pulls} and follows the exact same steps. We use the shorthand notation $\beta_x =  8  \rho \cdot ( \frac{\sum_{k=1}^K \xi^x_k}{ \Delta_x} )^2$.
		\begin{fact}
			\begin{equation}
				\label{eq-wlog-assume-n-big-deterministic}
				\mathbb{E}[n_{x, \tau^*}] \leq 2 \beta_x \cdot \mathbb{E}[\ln( \tau^* )] + \mathbb{E}[\sum_{t=1}^{ \tau^* }  I_{x_t = x} \cdot I_{n_{x, t} \geq \beta_x \ln(t)} ].
			\end{equation}		
		\end{fact} \vspace{0.3cm}		
		\noindent
		We omit the proof as it is analogous to the proof of Fact \ref{fact-wlog-assume-n-big}. As in Lemma \ref{lemma-bound-times-non-optimal-pulls}, we break down the second term in three terms and bound each of them by a constant:
		
		\begin{align}
		\mathbb{E}[\sum_{t=1}^{ \tau^* }  I_{x_t = x} \cdot I_{n_{x, t} \geq \beta_x \ln(t)} ] 
			& =	 \mathbb{E}[\sum_{t=1}^{ \tau^* }  I_{ \mathrm{obj}_{x, t} + E_{x, t} \geq \mathrm{obj}_{x^*, t} + E_{x^*, t}} \cdot I_{n_{x, t} \geq \beta_x \ln(t)} ] \nonumber \\
			& \leq  
				\mathbb{E}[\sum_{t=1}^{ \tau^* } I_{ \mathrm{obj}_{x, t} \geq \mathrm{obj}_{x} + E_{x, t} } ] \label{eq-first-term-upper-bound-deterministic} \\
				& + \mathbb{E}[\sum_{t=1}^{ \tau^* } I_{ \mathrm{obj}_{x^*, t} \leq \mathrm{obj}_{x^*}  - E_{x^*, t} } ] \label{eq-second-term-upper-bound-deterministic} \\
				& + \mathbb{E}[\sum_{t=1}^{ \tau^* } I_{\mathrm{obj}_{x^*} < \mathrm{obj}_{x} + 2 E_{x, t} } \cdot I_{n_{x, t} \geq \beta_x \ln(t)} ]. \label{eq-third-term-upper-bound-deterministic}
		\end{align}

		\begin{fact}
			\label{fact-bound-last-term-is-small-deterministic}
			$$
				\mathbb{E}[\sum_{t=1}^{ \tau^* } I_{\mathrm{obj}_{x^*} < \mathrm{obj}_{x} + 2 E_{x, t} } \cdot I_{n_{x, t} \geq \beta_x \ln(t)} ] = 0.
			$$
		\proof{Proof.}
			If $\mathrm{obj}_{x^*} < \mathrm{obj}_{x} + 2 E_{x, t}$, we get:
			\begin{align*}
				\frac{\Delta_x}{2} 
					& < \sum_{k \in \mathcal{K}_x} \xi^x_k \cdot \sqrt{ \frac{2 \ln(t) }{ n_{k, t} } } \\
					& \leq \sum_{k \in \mathcal{K}_x} \xi^x_k \cdot \sqrt{ \frac{2 \ln(t) }{ \rho + n^x_{k, t} } } \\
					& \leq \sqrt{ \sum_{k \in \mathcal{K}_x} \xi^x_k } \cdot \sum_{k \in \mathcal{K}_x} \sqrt{ \xi^x_k } \cdot \sqrt{ \frac{ 2 \ln(t) }{ n_{x, t} }},
			\end{align*}
			where we use \eqref{eq-ini-pull-deterministic} and Lemma \ref{lemma-step-2-well-defined-multiple-budget} for each $k \in \mathcal{K}_x$ such that $\xi^x_k \neq 0$ (otherwise, if $\xi^x_k = 0$, the inequality is trivial). This implies:
			\begin{align*}
				n_{x, t} 
					& < 8  \rho \cdot ( \frac{\sum_{k=1}^K \xi^x_k}{ \Delta_x} )^2 \cdot \ln(t) \\
					& = \beta_x \cdot \ln(t),
			\end{align*}
			using the Cauchy$-$Schwarz inequality and the fact that a basis involves at most $\rho$ arms. 
			
		\endproof
		\end{fact} \vspace{0.3cm}

		\begin{fact}
			\label{fact-study-bound-first-term-deterministic}
			$$
				\mathbb{E}[\sum_{t=1}^{\tau^*} I_{ \mathrm{obj}_{x, t} \geq \mathrm{obj}_{x} + E_{x, t} } ] \leq \rho \cdot \frac{\pi^2}{6}.
			$$
			\proof{Proof.}
				If $\mathrm{obj}_{x, t} \geq \mathrm{obj}_{x} + E_{x, t}$, there must exist $k \in \mathcal{K}_x$ such that $\bar{r}_{k, t} \geq \mu^r_k + \epsilon_{k, t}$, otherwise:
				\begin{align*}
					\mathrm{obj}_{x, t} - \mathrm{obj}_{x} 
						& = \sum_{k \in \mathcal{K}_x} (\bar{r}_{k, t} - \mu^r_k ) \cdot \xi^x_k \\
						& < \sum_{k \in \mathcal{K}_x} \epsilon_{k, t} \cdot \xi^x_k \\
						& = E_{x, t},
				\end{align*}
				where the inequality is strict because there must exist $l \in \mathcal{K}_x$ such that $\xi^x_l > 0$ as a result of Assumption \ref{assumption-all-cost-non-zero} (at least one resource constraint is binding for a feasible basis to \eqref{eq-linear-program-general-upperbound-opt-strategy} aside from the basis $\tilde{x}$ associated with $\mathcal{K}_{\tilde{x}} = \emptyset$). We obtain:
				\begin{align*}
				\mathbb{E}[\sum_{t=1}^{\tau^*}  I_{ \mathrm{obj}_{x, t} \geq \mathrm{obj}_{x} + E_{x, t} } ]
					& \leq \sum_{k \in \mathcal{K}_x} \sum_{t=1}^{\infty}  \mathbb{P}[ \bar{r}_{k, t} \geq \mu^r_k + \epsilon_{k, t} ] \\
					& \leq \rho \cdot \frac{\pi^2}{6},
				\end{align*}
				where the last inequality is derived along the same lines as in the proof of Fact \ref{fact-study-bound-first-term}.
			\endproof					
		\end{fact} \vspace{0.3cm}		
		
		\begin{fact}
			\label{fact-bound-second-term-deterministic}
			$$
				\mathbb{E}[\sum_{t=1}^{ \tau^* } I_{ \mathrm{obj}_{x^*, t} \leq \mathrm{obj}_{x^*} - E_{x^*, t} } ]  \leq \rho \cdot \frac{\pi^2}{6}.
			$$
			\proof{Proof.}
				Similar to Fact \ref{fact-study-bound-first-term-deterministic}.
			\endproof
		\end{fact} \vspace{0.3cm}

\subsection{Proof of Theorem \ref{lemma-log-B-regret-bound-deterministic}.}
The proof proceeds along the same lines as for Theorem \ref{lemma-log-B-regret-bound}. We build upon \eqref{eq-simplified-general-upper-bound-on-regret}:
	\begin{align*}
				R_{B(1), \cdots, B(C)} 
					& \leq B \cdot \sum_{k=1}^K \mu^r_k \cdot \xi^{x^*}_k - \mathbb{E}[\sum_{t=1}^{\tau^*} r_{a_t, t}] + O(1) \\
					& = B \cdot  \sum_{k=1}^K \mu^r_k \cdot \xi^{x^*}_k  - \sum_{t=1}^\infty \mathbb{E}[I_{\tau^* \geq t} \cdot \sum_{k=1}^K \sum_{x \in \mathcal{B}} r_{k, t} \cdot I_{x_t = x, a_t = k}] + O(1) \\
					& = B \cdot  \sum_{k=1}^K \mu^r_k \cdot \xi^{x^*}_k  - \sum_{t=1}^\infty \mathbb{E}[I_{\tau^* \geq t} \cdot \sum_{k=1}^K \sum_{x \in \mathcal{B}} I_{x_t = x, a_t = k} \cdot \mathbb{E}[r_{k, t} \; | \; \mathcal{F}_{t-1} ]] + O(1) \\
					& = B \cdot  \sum_{k=1}^K \mu^r_k \cdot \xi^{x^*}_k  - \sum_{x \in \mathcal{B}} \sum_{k=1}^K \mu^r_k \cdot \mathbb{E}[ \sum_{t=1}^{\tau^*} I_{x_t = x, a_t = k}] + O(1) \\
					& = B \cdot \sum_{k=1}^K \mu^r_k \cdot \xi^{x^*}_k - \sum_{x \in \mathcal{B}} \sum_{k=1}^K \mu^r_k \cdot \mathbb{E}[n^x_{k, \tau^*} ] + O(1),
	\end{align*}
	where we use the fact that $x_t$ and $a_t$ are determined by the events of the first $t-1$ rounds and that $\tau^*$ is a stopping time. Using the properties of the load balancing algorithm established in Lemma \ref{lemma-step-2-well-defined-multiple-budget}, we derive:
	\begin{align*}
			R_{B(1), \cdots, B(C)}  
					& \leq B \cdot \sum_{k=1}^K \mu^r_k \cdot \xi^{x^*}_k - \sum_{x \in \mathcal{B}} \sum_{k \in \mathcal{K}_x} \{ \mu^r_k \cdot \frac{ \xi^x_k }{\sum_{l=1}^K \xi^x_l} \cdot \mathbb{E}[ n_{x, \tau^*} ] - \rho \} + O(1) \\
					& = B \cdot  \sum_{k=1}^K \mu^r_k \cdot \xi^{x^*}_k - \sum_{x \in \mathcal{B}} \{ \frac{\mathbb{E}[ n_{x, \tau^*  }]}{\sum_{k=1}^K \xi^x_k} \cdot (\sum_{k=1}^K \mu^r_k \cdot \xi^x_k) - (\rho)^2 \} + O(1) \\
					& = (\sum_{k=1}^K \mu^r_k \cdot \xi^{x^*}_k) \cdot ( B - \sum_{x \in \mathcal{B} \; | \; \Delta_x = 0} \frac{\mathbb{E}[ n_{x, \tau^*  }]}{\sum_{k=1}^K \xi^x_k} ) \\
					& - \sum_{x \in \mathcal{B} \; | \; \Delta_x > 0} \{ (\sum_{k=1}^K \mu^r_k \cdot \xi^x_k) \cdot  \frac{\mathbb{E}[ n_{x, \tau^*  }]}{\sum_{k=1}^K \xi^x_k} \} + O(1).
	\end{align*}
	Now observe that, by definition, at least one resource is exhausted at time $\tau^*$. Hence, there exists $i \in  \{1, \cdots, C\}$ such that the following holds almost surely:
	\begin{align*}
	B(i) 
		& \leq \sum_{x \in \mathcal{B}} \sum_{k \in \mathcal{K}_x} c_k(i) \cdot n^x_{k, \tau^*}  \\
		& \leq \sum_{x \in \mathcal{B}} \sum_{k \in \mathcal{K}_x} [ c_k(i) \cdot ( \frac{ \xi^x_k  }{\sum_{l=1}^K \xi^x_l} \cdot n_{x, \tau^*} + 1 ) ] \\
		& = | \mathcal{B} | \cdot \rho + \sum_{x \in \mathcal{B}} \frac{ n_{x, \tau^*  } }{ \sum_{k=1}^K \xi^x_k } \cdot \sum_{k \in \mathcal{K}_x} c_k(i) \cdot \xi^x_k   \\
		& \leq | \mathcal{B} | \cdot \rho + b(i) \cdot \sum_{x \in \mathcal{B}} \frac{ n_{x, \tau^*  } }{\sum_{k=1}^K \xi^x_k}, 	
	\end{align*}
	where we use Lemma \ref{lemma-step-2-well-defined-multiple-budget} again and the fact that any basis $x \in \mathcal{B}$ satisfies all the constraints of \eqref{eq-linear-program-general-upperbound-opt-strategy}. We conclude that the inequality:
	$$
		\sum_{ x \in \mathcal{B} \; | \; \Delta_x = 0 } \frac{ n_{x, \tau^*}  }{ \sum_{k=1}^K \xi^x_k } \geq B - \sum_{ x \in \mathcal{B} \; | \; \Delta_x > 0 } \frac{ n_{x, \tau^*} }{ \sum_{k=1}^K \xi^x_k } - \frac{| \mathcal{B} | \cdot \rho }{b}
	$$
	holds almost surely. Taking expectations on both sides and plugging the result back into the regret bound yields:
	\begin{align}
		R_{B(1), \cdots, B(C)}  
			& \leq \sum_{ x \in \mathcal{B} \; | \; \Delta_x > 0 } \frac{(\sum_{k=1}^K \mu^r_k \cdot \xi^{x^*}_k - \sum_{k=1}^K \mu^r_k \cdot \xi^x_k)}{ \sum_{k=1}^K \xi^x_k } \cdot \mathbb{E}[ n_{x, \tau^*  }] \\
			& + (\sum_{k=1}^K \mu^r_k \cdot \xi^{x^*}_k) \cdot \frac{| \mathcal{B} | \cdot \rho}{b} + O(1) \nonumber \\
			& \leq \sum_{ x \in \mathcal{B} \; | \; \Delta_x > 0 } \frac{\Delta_x}{\sum_{k=1}^K \xi^x_k} \cdot \mathbb{E}[ n_{x, \tau^*  }] + O(1), \label{eq-proof-log-B-start-here-sqrt-B-deterministic}
	\end{align}
	where we use the fact that:
	\begin{equation}
		\label{eq-proof-log-B-bound-deltax-deterministic}
		\begin{aligned}
			\sum_{k=1}^K \mu^r_k \cdot \xi^{x^*}_k 
				& \leq \sum_{k=1}^K  \frac{\sum_{i=1}^C c_k(i) }{\epsilon} \cdot \xi^{x^*}_k  \\
				& = \frac{1}{\epsilon} \cdot \sum_{i=1}^C \sum_{k=1}^K  c_k(i) \cdot \xi^{x^*}_k  \\
				& \leq \frac{\sum_{i=1}^C b(i)}{\epsilon},
		\end{aligned}
	\end{equation}
	using Assumption \ref{assumption-all-cost-non-zero} and the fact that $x^*$ is a feasible basis to \eqref{eq-linear-program-general-upperbound-opt-strategy}. 
	Using Lemma \ref{lemma-bound-stopping-time-deterministic}, Lemma \ref{lemma-bound-times-non-optimal-pulls-deterministic}, and the concavity of the logarithmic function, we obtain:	
	\begin{align*}
		R_{B(1), \cdots, B(C)}  
			& \leq 16 \rho  \cdot (\sum_{x \in \mathcal{B} \; | \; \Delta_x > 0} \frac{ \sum_{k=1}^K \xi^x_k }{ \Delta_x } ) \cdot \ln( \frac{ \sum_{i=1}^C b(i) \cdot B}{\epsilon}  + 1 )  \\
			& + \frac{\pi^2}{3} \rho \cdot (\sum_{x \in \mathcal{B} \; | \; \Delta_x > 0} \frac{\Delta_x}{\sum_{k=1}^K \xi^x_k} ) + O(1) \\
			& \leq 16 \frac{\rho \cdot \sum_{i=1}^C b(i)}{\epsilon} \cdot (\sum_{x \in \mathcal{B} \; | \; \Delta_x > 0} \frac{ 1 }{ \Delta_x } ) \cdot \ln( \frac{\sum_{i=1}^C b(i) \cdot B}{\epsilon}  + 1  ) + O(1).
	\end{align*}
	To derive this last inequality, we use: (i) $\Delta_x \leq \sum_{k=1}^K \mu^r_k \cdot \xi^{x^*}_k \leq \sum_{i=1}^C b(i) / \epsilon$ (see \eqref{eq-proof-log-B-bound-deltax-deterministic}), (ii) the fact that, for any basis $x \in \mathcal{B}$, at least one of the first $C$ inequalities is binding in \eqref{eq-linear-program-general-upperbound-opt-strategy}, which implies that there exists $i \in \{1, \cdots, C\}$ such that:
	\begin{align*}
		\sum_{k=1}^K \xi^{x}_k 
			& \geq \sum_{k=1}^K c_k(i) \cdot \xi^{x}_k \\
			& = b(i) \\
			& \geq b,
	\end{align*}
	and (iii) the inequality:
	\begin{align*}
		\sum_{k=1}^K \xi^x_k 
			& \leq \sum_{k=1}^K  \frac{\sum_{i=1}^C c_k(i) }{\epsilon} \cdot \xi^{x}_k  \\
			& = \frac{1}{\epsilon} \cdot \sum_{i=1}^C \sum_{k=1}^K  c_k(i) \cdot \xi^{x}_k  \\
			& \leq \frac{\sum_{i=1}^C b(i)}{\epsilon},
	\end{align*}
	for any basis $x \in \mathcal{B}$. \\
	As a side note, observe a possibly better regret bound is given by:
	\begin{align*}
		R_{B(1), \cdots, B(C)}  
			& \leq 16 \rho \cdot (\sum_{x \in \mathcal{B} \; | \; \Delta_x > 0} \frac{ 1 }{ \Delta_x } ) \cdot \ln( T  ) + O(1),
	\end{align*}
	if time is a limited resource since, in this case, $\tau^* \leq T$ and the constraint $\sum_{k=1}^K \xi^x_k \leq 1$ is part of \eqref{eq-linear-program-general-upperbound-opt-strategy}.
	
\subsection{Proof of Theorem \ref{lemma-sqrt-B-regret-bound-deterministic}.}
Along the same lines as for the case of a single limited resource, we start from inequality \eqref{eq-proof-log-B-start-here-sqrt-B-deterministic} derived in the proof of Theorem \ref{lemma-log-B-regret-bound-deterministic} and apply Lemma \ref{lemma-bound-times-non-optimal-pulls-deterministic} only if $\Delta_x$ is big enough, taking into account the fact that:
$$
	\sum_{x \in \mathcal{B}} \mathbb{E}[n_{x, \tau^*}] \leq \mathbb{E} [\tau^*] \leq \frac{\sum_{i=1}^C b(i) \cdot B}{\epsilon}  + 1.
$$
Specifically, we have: 
\begin{align*}
	& R_{B(1), \cdots, B(C)}  \\
		& \leq \sup\limits_{ \substack{(n_x)_{x \in \mathcal{B}} \geq 0 \\ \sum_{x \in \mathcal{B}} n_x \leq \frac{\sum_{i=1}^C b(i) \cdot B}{\epsilon}  + 1 }  } \{ \; \sum_{x \in \mathcal{B} \; | \; \Delta_x > 0} \min( \frac{\Delta_x}{\sum_{k=1}^K \xi^x_k} \cdot n_x, \\
		& \quad \quad \quad 16 \rho \cdot \frac{\sum_{k=1}^K \xi^x_k}{\Delta_x} \cdot \ln( \frac{\sum_{i=1}^C b(i) \cdot B}{\epsilon}  + 1 ) + \frac{\pi^2}{3}  \rho \cdot \frac{\Delta_x}{\sum_{k=1}^K \xi^x_k} ) \; \} + O(1) \\
		& \leq \sup\limits_{ \substack{(n_x)_{x \in \mathcal{B}} \geq 0 \\ \sum_{x \in \mathcal{B}} n_x \leq \frac{\sum_{i=1}^C b(i) \cdot B}{\epsilon}  + 1 }  } \{ \; \sum_{x \in \mathcal{B} \; | \; \Delta_x > 0} \min( \frac{\Delta_x}{\sum_{k=1}^K \xi^x_k} \cdot n_x, \\ 
		& \quad \quad \quad 16 \rho \cdot \frac{\sum_{k=1}^K \xi^x_k}{\Delta_x} \cdot \ln( \frac{\sum_{i=1}^C b(i) \cdot B}{\epsilon}  + 1 ) ) \; \} +  \frac{\pi^2}{3} \frac{|\mathcal{B}| \cdot \rho}{\epsilon}  + O(1) \\
		& \leq \sup\limits_{ \substack{(n_x)_{x \in \mathcal{B}} \geq 0 \\ \sum_{x \in \mathcal{B}} n_x \leq \frac{\sum_{i=1}^C b(i) \cdot B}{\epsilon}  + 1 }  } \{ \; \sum_{x \in \mathcal{B} \; | \; \Delta_x > 0} \sqrt{ 16 \rho \cdot \ln(  \frac{\sum_{i=1}^C b(i) \cdot B}{\epsilon}  + 1 ) \cdot  n_x} \; \} + O(1) \\
		& \leq 4 \sqrt{\rho \cdot \ln(  \frac{\sum_{i=1}^C b(i) \cdot B}{\epsilon}  + 1 )  } \cdot \sup\limits_{ \substack{(n_x)_{x \in \mathcal{B}} \geq 0 \\ \sum_{x \in \mathcal{B}} n_x \leq \frac{\sum_{i=1}^C b(i) \cdot B}{\epsilon}  + 1 }  } \{ \; \sum_{x \in \mathcal{B} \; | \; \Delta_x > 0} \sqrt{n_x} \; \} + O(1) \\
		& \leq 4 \sqrt{ \rho \cdot |\mathcal{B}| \cdot (\frac{\sum_{i=1}^C b(i) \cdot B}{\epsilon}  + 1) \cdot \ln( \frac{\sum_{i=1}^C b(i) \cdot B}{\epsilon}  + 1 )  } + O(1),
\end{align*}
where we use $\Delta_x \leq \sum_{i=1}^C b(i) / \epsilon$ and $\sum_{k=1}^K \xi^x_k \geq b$ for the second inequality (see the end of the proof of Theorem \ref{lemma-log-B-regret-bound-deterministic}), we maximize over  $\Delta_x / \sum_{k=1}^K \xi^x_k \geq 0$ for each $x \in \mathcal{B}$ to derive the third inequality, and we use Cauchy-Schwartz for the last inequality. 

\section{Proofs for Section \ref{sec-singlebudgettimehorizon}.}

\subsection{Proof of Lemma \ref{lemma-non-degeneracy-assumption-is-necessary-budget-and-time-horizon}.}
For any  $T \in \mathbb{N}$ and any arm $k \in \{1, \cdots, K\}$, we denote by $n^{\mathrm{opt}}_{k, T}$ the expected number of times that arm $k$ is pulled by the optimal non-anticipating algorithm (which is characterized by a high-dimensional dynamic program) when the time horizon is $T$ and the budget is $B = b \cdot T$. We prove the claim in two steps. First, we show that if $T - n^{\mathrm{opt}}_{k^*, T} = \Omega(\sqrt{T})$ (Case A) or $T - n^{\mathrm{opt}}_{k^*, T} = o(\sqrt{T})$ (Case B) then $\ropt(B, T) = T \cdot \mathrm{obj}_{x^*} - \Omega(\sqrt{T})$. This is enough to establish the result because if $T - n^{\mathrm{opt}}_{k^*, T} \neq \Omega(\sqrt{T})$ then we can extract a subsequence of $(T - n^{\mathrm{opt}}_{k^*, T})/\sqrt{T}$ that converges to $0$ and we can conclude with Case B.

\paragraph{Case A: $T - n^{\mathrm{opt}}_{k^*, T}= \Omega(\sqrt{T})$.}
	Consider the linear program:
	\begin{equation}
		\label{eq-proof-nondegeneracy-required-budget-and-time-horizon}
		\begin{aligned}
			& \sup_{ (\xi_k)_{k=1, \cdots, K} } 
			& & \sum_{k=1}^K \mu^r_k \cdot \xi_k \\
			& \text{subject to}
			& & \sum_{k=1}^K \mu^{c}_k \cdot \xi_k \leq b \\
			&
			& & \sum_{k=1}^K  \xi_k \leq 1 \\
			&
			& & \xi_{k^*} \leq \Gamma \\
			&
			& & \xi_k \geq 0, \quad k = 1, \cdots, K
		\end{aligned}
	\end{equation}
	parametrized by $\Gamma$ and its dual:
	\begin{equation}
		\label{eq-proof-nondegeneracy-required-budget-and-time-horizon-dual}
		\begin{aligned}
		& \inf_{ (\zeta_1, \zeta_2, \zeta_3)} 
		& & b \cdot \zeta_1 + \zeta_2 + \Gamma \cdot \zeta_3 \\
		& \text{subject to}
		& &  \mu^c_k \cdot \zeta_1 + \zeta_2 \geq \mu^r_k,  \quad k \neq k^* \\
		&
		& &  \mu^c_{k^*} \cdot \zeta_1 + \zeta_2 + \zeta_3 \geq \mu^r_{k^*} \\
		&
		& & \zeta_1, \zeta_2, \zeta_3 \geq 0.
		\end{aligned}
	\end{equation}
	Since the vector $(\xi_k)_{k=1, \cdots, K}$ determined by $\xi_{k^*} = 1$ and $\xi_k = 0$ for $k \neq k^*$ is the only optimal solution to \eqref{eq-proof-nondegeneracy-required-budget-and-time-horizon} when $\Gamma = 1$ (by assumption), we can find a strictly complementary optimal solution to the dual \eqref{eq-proof-nondegeneracy-required-budget-and-time-horizon-dual} $\zeta^*_1, \zeta^*_2, \zeta^*_3 > 0$. Moreover, by definition of $n^{\mathrm{opt}}_{k^*, T}$, we can show, along the same lines as in the proof of Lemma \ref{lemma-general-bound-optimal-policy}, that $\ropt(B, T)$ is no larger than $T$ times the optimal value of \eqref{eq-proof-nondegeneracy-required-budget-and-time-horizon} when $\Gamma = n^{\mathrm{opt}}_{k^*, T}/T$ (up to a constant additive term of order $O(1)$). By weak duality, and since  $(\zeta^*_1, \zeta^*_2, \zeta^*_3)$ is feasible for \eqref{eq-proof-nondegeneracy-required-budget-and-time-horizon-dual} for any $\Gamma$, this implies:
	\begin{align*}
		\ropt(B, T) 
			& \leq T \cdot (b \cdot \zeta^*_1 + \zeta^*_2 + \frac{ n^{\mathrm{opt}}_{k^*, T} }{T} \cdot \zeta^*_3) + O(1) \\
			& \leq T \cdot (b \cdot  \zeta^*_1 + \zeta^*_2 + \zeta^*_3 - \frac{T - n^{\mathrm{opt}}_{k^*, T} }{T} \cdot \zeta^*_3) + O(1) \\
			& \leq T \cdot \mathrm{obj}_{x^*} - \Omega(\sqrt{T}),
	\end{align*}
	where we use the fact that $b \cdot \zeta^*_1 + \zeta^*_2 + \zeta^*_3$ is the optimal value of \eqref{eq-proof-nondegeneracy-required-budget-and-time-horizon} when $\Gamma = 1$ by strong duality (both \eqref{eq-proof-nondegeneracy-required-budget-and-time-horizon} and \eqref{eq-proof-nondegeneracy-required-budget-and-time-horizon-dual} are feasible) and  $\zeta^*_3 > 0$.

\paragraph{Case B: $T - n^{\mathrm{opt}}_{k^*, T} = o(\sqrt{T})$.}
First observe that since the vector $(\xi_k)_{k=1, \cdots, K}$ determined by $\xi_{k^*} = 1$ and $\xi_k = 0$ for $k \neq k^*$ is the only optimal solution to \eqref{eq-linear-program-general-upperbound-opt-strategy}, it must be that $\mu^r_{k^*} > 0$ (since $0$ is a feasible solution to \eqref{eq-linear-program-general-upperbound-opt-strategy} with objective value $0$). For any $t \in \mathbb{N}$, denote by $a_t$ the arm pulled by the optimal non-anticipating algorithm at time $t$ and define $\tau^*_T$ as the corresponding stopping time when the time horizon is $T$. We have:
\begin{align*}
	\ropt(B, T) 
		& = \mathbb{E}[\sum_{t=1}^{\tau^*_T-1} r_{a_t, t} ] \\
		& = \sum_{k=1}^K \mu^r_k \cdot n^{\mathrm{opt}}_{k, T}  \\
		& \leq \sum_{k \neq k^*} n^{\mathrm{opt}}_{k, T}  + \mu^r_{k^*} \cdot n^{\mathrm{opt}}_{k^*, T} \\
		& \leq (T - n^{\mathrm{opt}}_{k^*, T}) + \mu^r_{k^*} \cdot (\mathbb{E}[\tau^*_T] - 1) \\
		& = T \cdot \mu^r_{k^*} - \mu^r_{k^*} \cdot (T - \mathbb{E}[\tau^*_T] + 1) + o(\sqrt{T}) \\
		& = T \cdot \mathrm{obj}_{x^*} - \mu^r_{k^*} \cdot \mathbb{E}[\sum_{t=\tau^*_T}^{T} 1 ] + o(\sqrt{T}) \\
		& \leq T \cdot \mathrm{obj}_{x^*} - \mu^r_{k^*} \cdot \mathbb{E}[ ( \sum_{t=\tau^*_T}^{ T } c_{k^*, t} + \sum_{t=1}^{ \tau^*_T -1 } c_{a_t, t} - B )_+ ] \\
		& \leq T \cdot \mathrm{obj}_{x^*} - \mu^r_{k^*} \cdot \Big( \mathbb{E}[ (\sum_{t=1}^{T} \{c_{k^*, t} - b\})_+] -  \sum_{k \neq k^*} n^{\mathrm{opt}}_{k, T} \Big) + o(\sqrt{T}) \\
		& = T \cdot \mathrm{obj}_{x^*} - \mu^r_{k^*} \cdot \mathbb{E}[(\sum_{t=1}^{T} \{c_{k^*, t} - b\})_+ ] + o(\sqrt{T}).
\end{align*}
 The first inequality is obtained using the fact that the rewards are bounded by $1$. The second inequality is obtained using the fact that $\sum_{k=1}^K n^{\mathrm{opt}}_{k, T} = \mathbb{E}[\tau^*_T] - 1 \leq T$. The third inequality is obtained along the same lines as in the proof of Lemma \ref{lemma-lower-bound-agrawal}, using $\sum_{t=1}^{ \tau^*_T -1 } c_{a_t, t} \leq B$ by definition of $\tau^*_T$. We use the inequality $(y+z)_+ \geq y_+ - |z|$ (true for any $(y, z) \in \mathbb{R}^2$) and the fact the amount of resource consumed at any step is no larger than $1$ for the fourth inequality. Since $(c_{k^*, t} - b)_{t \in \mathbb{N}}$ is an i.i.d. zero-mean bounded stochastic process with positive variance, $\frac{1}{\sqrt{T}} \cdot \mathbb{E}[ (\sum_{t=1}^{T} \{c_{k, t} - b\})_+]$ converges to a positive value and we conclude:
\begin{align*}
	\ropt(B, T) 
		& \leq T \cdot \mathrm{obj}_{x^*} - \Omega(\sqrt{T}),
\end{align*}
since $\mu^r_{k^*} > 0$.


\subsection{Proof of Lemma \ref{lemma-bound-times-infeasible-pulls-budget-time-horizon}.}
		Consider $x$ either an infeasible basis to \eqref{eq-linear-program-general-upperbound-opt-strategy} or a pseudo-basis for \eqref{eq-linear-program-general-upperbound-opt-strategy}. Without loss of generality, we can assume that $x$ involves two arms (one of which may be a dummy arm introduced in the specification of the algorithm given in Section \ref{sec-singlebudgettimehorizon}) and that $\mathcal{K}_x = \{k, l\}$ with $\mu^c_k, \mu^c_l > b$ (the situation is symmetric if the reverse inequality holds). Defining $\beta_x = 32 / \epsilon^3$, we have:
\begin{align*}
	\mathbb{E}[n_{x, T}] 
		& \leq 2 \beta_x \cdot \ln( T ) +  \mathbb{E}[\sum_{t=1}^{T}  I_{ x_t = x } \cdot I_{n_{x, t} \geq \beta_x \ln(t)}] \\
		& \leq 2 \beta_x \cdot \ln( T ) +  \mathbb{E}[\sum_{t=1}^{T}  I_{ x_t = x } \cdot I_{ n_{x, t} \geq \beta_x \ln(t) } \cdot I_{ b_{x, t} \geq n_{x, t} \cdot ( b + \epsilon/2 )  }   ] \\
		& + \sum_{t=1}^{T}  \sum_{s = \beta_x \ln(T)}^t \mathbb{P}[ b_{x, t} < s \cdot ( b + \epsilon ) - s \cdot \epsilon/2, n_{x, t} = s ] \\
		& \leq 2 \beta_x \cdot \ln( T ) +  \mathbb{E}[\sum_{t=1}^{T}  I_{ x_t = x } \cdot I_{ n_{x, t} \geq \beta_x \ln(t) } \cdot I_{ b_{x, t} \geq n_{x, t} \cdot ( b + \epsilon/2 )  }   ] + \sum_{t=1}^{T}  \sum_{s = \beta_x \ln(T)}^\infty \exp(- s \frac{\epsilon^2}{2}) \\
		& \leq 2 \beta_x \cdot \ln( T ) +  \mathbb{E}[\sum_{t=1}^{T}  I_{ x_t = x } \cdot I_{ n_{x, t} \geq \beta_x \ln(t) } \cdot I_{ b_{x, t} \geq n_{x, t} \cdot b + \epsilon \beta_x /4 \cdot  \ln(t)  }   ] + \frac{ 2 \pi^2 }{3 \epsilon^2}.
\end{align*}
The first inequality is derived along the same lines as in Fact \ref{fact-wlog-assume-n-big}. The third inequality is obtained by observing that, as a result of Assumption \ref{assumption-simplying-assumption-analysis-budget-and-time-horizon}, the average amount of resource consumed any time basis $x$ is selected at Step-Simplex is at least $b + \epsilon$ no matter which of arm $k$ or $l$ is pulled. Finally, we use the same bounds as in Fact \ref{fact-bound-last-term-is-zero} for the last two inequalities. Observe that if $x$ is selected at time $t$, either $\bar{c}_{k, t} - \epsilon_{k, t} \leq b$ or $\bar{c}_{l, t} - \epsilon_{l, t} \leq b$, otherwise $x$ would have been infeasible for \eqref{eq-algo-general-idea}. Moreover, if $n_{x, t} \geq \beta_x \ln(T)$, then we have either $n^x_{k, t} \geq \beta_x/2 \ln(T)$ or $n^x_{l, t} \geq \beta_x/2 \ln(T)$ since there are only two arms in $\mathcal{K}_x$. By symmetry, we study the first situation and look at:

	\begin{align*}
		 \mathbb{E}[ & \sum_{t=1}^{T}  I_{ x_t = x } \cdot I_{ n^x_{k, t} \geq \beta_x/2 \cdot \ln(t) } \cdot I_{ b_{x, t} \geq n_{x, t} \cdot b + \epsilon \beta_x /4 \cdot \ln(t)  } ] \\
		 & \leq \mathbb{E}[ \sum_{t=1}^{T}  I_{ x_t = x } \cdot I_{ n^x_{k, t} \geq \beta_x/2 \cdot \ln(t) } \cdot I_{ b_{x, t} \geq n_{x, t} \cdot b + \epsilon \beta_x /4 \cdot \ln(t)  } \cdot  I_{ \bar{c}_{k, \tau_q} - \epsilon_{k, \tau_q} \geq b, \; q = q_{t}-\epsilon \beta_x /4 \ln(t), \cdots, q_t} ] \\
		 & + \sum_{t=1}^{T}  \sum_{\tau = 1}^t \sum_{s = \beta_x/4 \cdot \ln(t)}^t \mathbb{P}[ \bar{c}_{k, \tau} < b + \frac{\epsilon}{2}, n_{k, \tau} = s ] \\
		 & \leq \mathbb{E}[ \sum_{t=1}^{T}  I_{ x_t = x } \cdot I_{ n^x_{k, t} \geq \beta_x/2 \cdot \ln(t) } \cdot I_{ b_{x, t} \geq n_{x, t} \cdot b + \epsilon \beta_x /4 \cdot \ln(t)  } \cdot I_{ \bar{c}_{k, \tau_q} - \epsilon_{k, \tau_q}  \geq b, \; q = q_{t}-\epsilon \beta_x /4 \ln(t), \cdots, q_t} ] \\
		 & + \sum_{t=1}^{T}  \sum_{\tau = 1}^t \sum_{s = \beta_x/4 \cdot \ln(t)}^\infty \exp(- s \cdot \frac{\epsilon^2}{2}) \\
		 & \leq \mathbb{E}[ \sum_{t=1}^{T}  I_{ x_t = x } \cdot I_{ n^x_{k, t} \geq \beta_x/2 \cdot \ln(t) } \cdot I_{ b_{x, t} \geq n_{x, t} \cdot b + \epsilon \beta_x /4 \cdot \ln(t)  } \cdot I_{ \bar{c}_{k, \tau_q} - \epsilon_{k, \tau_q} \geq b, \; q = q_{t}-\epsilon \beta_x /4 \ln(t), \cdots, q_t} ] \\
		 & + \frac{ 2 \pi^2 }{3 \epsilon^2},
	\end{align*}
	where $(\tau_q)_{q \in \mathbb{N}}$ denote the random times at which basis $x$ is selected and, for a time $t$ at which basis $x$ is selected, $q_t$ denotes the index $q \in \mathbb{N}$ such that $\tau_q = t$. The first inequality is a consequence of $n^{x}_{k, \tau_q} = n^{x}_{k, t} - (q_t - q) \geq n^{x}_{k, t}  - \epsilon \beta_x /4 \ln(t) \geq \beta_x /4 \ln(t)$ for $q = q_{t} - \epsilon \beta_x /4 \ln(t), \cdots, q_t$ and $n_{k, \tau_q} \geq n^x_{k, \tau_q}$, which implies $\epsilon_{k, \tau_q} \leq \epsilon/2$. We use the same bounds as in Fact \ref{fact-bound-last-term-is-zero} for the last two inequalities. Now observe that, for any $q \in \{ q_{t}-\epsilon \beta_x /4 \ln(t), \cdots, q_t \}$, we have $\bar{c}_{l, \tau_q} - \epsilon_{l, \tau_q} \leq b$ since $\bar{c}_{k, \tau_q} - \epsilon_{k, \tau_q} \geq b$ and since $x$ is feasible basis to \eqref{eq-algo-general-idea} at time $\tau_q$ (by definition). This implies that, for any $q \in \{ q_{t}-\epsilon \beta_x /4 \ln(t), \cdots, q_t \}$, arm $l$ was pulled at time $\tau_q$ by definition of the load balancing algorithm since the amount of resource consumed at any round cannot be larger than $1$ and $b_{x, \tau_q} \geq b_{x, t} - (q_t - q) \geq n_{x, t} \cdot b + \epsilon \beta_x /4 \cdot \ln(t) - (q_t -q) \geq n_{x, t} \cdot b \geq n_{x, \tau_q} b$. Hence, we get:
	\begin{align*}
		\mathbb{E}[ & \sum_{t=1}^{T}  I_{ x_t = x } \cdot I_{ n^x_{k, t} \geq \beta_x/2 \cdot \ln(t) } \cdot I_{ b_{x, t} \geq n_{x, t} \cdot b + \epsilon \beta_x /4 \cdot \ln(t)  } ] \\
		& \leq \mathbb{E}[ \sum_{t=1}^{T} I_{ n^x_{l, t} \geq \epsilon \beta_x / 4 \cdot \ln(t) } \cdot I_{\bar{c}_{l, t} - \epsilon_{l, t} \leq b}] + \frac{ 2 \pi^2 }{3 \epsilon^2} \\
		 & \leq \sum_{t=1}^{T}  \mathbb{P}[ \bar{c}_{l, t} \leq b + \frac{\epsilon}{2}, n^x_{l, t} \geq \epsilon \beta_x / 4 \cdot \ln(t) ] + \frac{ 2 \pi^2 }{3 \epsilon^2}  \\
		 & \leq \sum_{t=1}^{T} \sum_{s=\epsilon \beta_x / 4 \cdot \ln(t)}^t \mathbb{P}[ \bar{c}_{l, t} \leq b + \frac{\epsilon}{2}, n_{l, t} = s] + \frac{ 2 \pi^2 }{3 \epsilon^2} \\
		 & \leq \sum_{t=1}^{T}  \sum_{s=\epsilon \beta_x / 4 \cdot \ln(t)}^\infty \exp(- s \cdot \frac{\epsilon^2}{2}) + \frac{ 2 \pi^2 }{3 \epsilon^2} \\
		 & \leq \frac{ 4 \pi^2 }{3 \epsilon^2}.
	\end{align*}
Bringing everything together, we derive:
\begin{align*}
	\mathbb{E}[n_{x, T}] \leq \frac{2^6}{\epsilon^3} \cdot \ln( T ) +  \frac{ 10 \pi^2 }{3 \epsilon^2}.
\end{align*}

\subsection{Proof of Lemma \ref{lemma-load-balance-budget-time-horizon}.}
			Without loss of generality, we can assume that $x$ involves two arms (one of which may be a dummy arm introduced in the specification of the algorithm given in Section \ref{sec-singlebudgettimehorizon}) and that $\mathcal{K}_x = \{k, l\}$ with $\mu^c_{k} > b > \mu^c_{l}$. We say that a \enquote{swap} occurred at time $\tau$ if basis $x$ was selected at time $\tau$ and $\bar{c}_{k, \tau} - \epsilon_{k, \tau} \leq b \leq \bar{c}_{l, \tau} - \epsilon_{l, \tau}$. We define $n^{\mathrm{swap}}_{x, t}$ as the total number of swaps that have occurred before time $t$, i.e. $n^{\mathrm{swap}}_{x, t} = \sum_{\tau=1}^{t-1} I_{x_\tau = x} \cdot I_{\bar{c}_{k, \tau} - \epsilon_{k, \tau} \leq b \leq \bar{c}_{l, \tau} - \epsilon_{l, \tau}}$. Consider $u \geq 1$ and define $\gamma_x = (4/\epsilon)^2$. First note that:
			\begin{align*}
				\mathbb{P}[n^{\mathrm{swap}}_{x, t} \geq \gamma_x \ln(t)] 
					& \leq  \sum_{q=\gamma_x \ln(t)}^t \mathbb{P}[ \bar{c}_{k, \tau_q} - \epsilon_{k, \tau_q} \leq b \leq \bar{c}_{l, \tau_q} - \epsilon_{l, \tau_q} ] \\
					& \leq  \sum_{q=\gamma_x \ln(t)}^t \mathbb{P}[ \bar{c}_{k, \tau_q} \leq b + \frac{\epsilon}{2}, n_{k, \tau_q} \geq \frac{\gamma_x}{2} \ln(t)] \\
					& +  \sum_{q=\gamma_x \ln(t)}^t  \mathbb{P}[ b - \frac{\epsilon}{2} \leq \bar{c}_{l, \tau_q}, n_{l, \tau_q} \geq \frac{\gamma_x}{2} \ln(t) ] \\
					& \leq 2 \sum_{q=1}^t \sum_{s=\gamma_x/2 \cdot \ln(t)}^\infty \exp(- s \cdot \frac{\epsilon^2}{2}) \\
					& \leq \frac{ 8 }{ \epsilon^2 \cdot t^2},
			\end{align*}
			where $(\tau_q)_{q \in \mathbb{N}}$ are defined as the times at which basis $x$ is selected. The first inequality is derived observing that if $n^{\mathrm{swap}}_{x, t} \geq \gamma_x \ln(t)$ then it must be that basis $x$ was selected for the $q$th time, for some $q \geq \gamma_x \ln(t)$, and that we had $\bar{c}_{k, \tau_q} - \epsilon_{k, \tau_q} \leq b \leq \bar{c}_{l, \tau_q} - \epsilon_{l, \tau_q}$. To obtain the second inequality, we observe that, at any time $\tau$, at least one of arm $k$ and $l$ must have been pulled $n_{x, \tau}/2$ times and that $\epsilon_{k, \tau} \leq \epsilon/2$ when $n_{k, \tau} \geq \gamma_x/2 \ln(t)$ (a similar inequality holds for arm $l$). The last two inequalities are derived in the same fashion as in Lemma \ref{lemma-bound-times-infeasible-pulls-budget-time-horizon}. This yields:
			\begin{align*}
				\mathbb{P}[ & |b_{x, t} - n_{x, t} \cdot b | \geq u + \gamma_x \ln(t)] \\
					& \leq \mathbb{P}[|b_{x, t} - n_{x, t} \cdot b | \geq u + \gamma_x \ln(t) \; ; \; n^{\mathrm{swap}}_{x, t} \leq \gamma_x \ln(t)]  + \mathbb{P}[n^{\mathrm{swap}}_{x, t} \geq \gamma_x \ln(t)] \\ 
					& \leq \mathbb{P}[|b_{x, t} - n_{x, t} \cdot b | \geq u + \gamma_x \ln(t) \; ; \; n^{\mathrm{swap}}_{x, t} \leq \gamma_x \ln(t)]  + \frac{ 8 }{ \epsilon^2 \cdot t^2}.
			\end{align*}
			Note that, by definition of the load balancing algorithm, we are led to pull arm $k$ (resp. arm $l$) at time $\tau_q$ if the budget spent so far when selecting basis $x$, denoted by $b_{x, \tau_q}$, is below (resp. above) the \enquote{target} of $n_{x, \tau_q} \cdot b$ assuming there is no \enquote{swap} at time $\tau_q$ (i.e. $\bar{c}_{k, \tau_q} - \epsilon_{k, \tau_q}  \geq \bar{c}_{l, \tau_q} - \epsilon_{l, \tau_q}$). Hence, if $b_{x, t} - n_{x, t} \cdot b \geq u + \gamma_x \ln(t)$ and $n^{\mathrm{swap}}_{x, t} \leq \gamma_x \ln(t)$, we must have been pulling arm $l$ for at least $s \geq \floor*{u}$ rounds $t_1 \leq \cdots \leq t_s \leq t - 1$ where basis $x$ was selected since the last time, denoted by $t_0$, where basis $x$ was selected and the budget was below the target, i.e. $b_{x, t_0} \leq n_{x, t_0} \cdot b$ (because the amounts of resource consumed at each round are almost surely bounded by $1$). Moreover, we have:
			\begin{align*}
				\sum_{\tau = t_0+1}^{t-1} I_{x_\tau = x} \cdot & (c_{k, \tau} \cdot I_{\bar{c}_{k, \tau} - \epsilon_{k, \tau} \geq \bar{c}_{l, \tau} - \epsilon_{l, \tau}} + c_{l, \tau} \cdot I_{\bar{c}_{k, \tau} - \epsilon_{k, \tau} < \bar{c}_{l, \tau} - \epsilon_{l, \tau}}) \\
					& = b_{x, t} - b_{x, t_0+1}  \\
					& \geq ( n_{x, t} - n_{x, t_0} ) \cdot b + u - 1 + \gamma_x \ln(t) \\
					& \geq  s  \cdot b + u - 1 + \gamma_x \ln(t).
			\end{align*}
			This implies:
			$$
				\sum_{q=1}^s c_{l, t_q} \geq s \cdot b + u - 1
			$$
			since $\sum_{\tau = t_0+1}^{t-1} I_{x_\tau = x} \cdot I_{\bar{c}_{k, \tau}  - \epsilon_{k, \tau} < b} \cdot c_{k, \tau}  \leq n^{\mathrm{swap}}_{x, t} \leq \gamma_x \ln(t)$. Hence, if $u \geq 1$:
\begin{align*}
		\mathbb{P}[b_{x, t} & - n_{x, t} \cdot b \geq u + \gamma_x \ln(t) \; ; \; n^{\mathrm{swap}}_{x, t} \leq \gamma_x \ln(t)] \\
			& \leq \sum_{s=\floor*{u}}^t \mathbb{P}[\sum_{q=1}^s c_{l, t_q} \geq s \cdot b + u - 1] \\
			& = \sum_{s=\floor*{u}}^t \mathbb{P}[\sum_{q=1}^s c_{l, t_q} \geq s \cdot \mu^c_l + s \cdot (b - \mu^c_l)] \\
			& \leq \sum_{s=\floor*{u}}^t \mathbb{P}[\sum_{q=1}^s c_{l, t_q} \geq s \cdot \mu^c_l + s \cdot \epsilon] \\
			& \leq \sum_{s=\floor*{u}}^t \exp(- 2 \epsilon^2 \cdot s ) \\
			& \leq \frac{ \exp(- 2 \epsilon^2 \cdot \floor*{u} ) }{ 1 - \exp(- 2 \epsilon^2) } \\
			& \leq \frac{2}{ \epsilon^2} \cdot \exp(- \epsilon^2 \cdot u ),
\end{align*}		
where we use Lemma \ref{lemma-martingale-inequality} for the third inequality and the fact that $\exp(- 2 v) \leq 1 - v/2$ for $v \in [0, 1]$ for the last inequality. With a similar argument, we conclude:
$$
	\mathbb{P}[|b_{x, t} - n_{x, t}| \cdot b \geq u + \gamma_x \ln(t) \; ; \; n^{\mathrm{swap}}_{x, t} \leq \gamma_x \ln(t)] \leq \frac{4}{ \epsilon^2} \cdot \exp(- \epsilon^2 \cdot u ).
$$
This last result enables us to show that:
	\begin{align*}
		| & \mathbb{E}[b_{x, T}] - \mathbb{E}[n_{x, T}] \cdot b | \\
			& \leq \mathbb{E}[|b_{x, T} - n_{x, T} \cdot b|] \\
			& = \int_0^T \mathbb{P}[ |b_{x, T} - n_{x, T} \cdot b | \geq u ] \mathrm{d}u \\
			& \leq \int_0^T \mathbb{P}[ |b_{x, T} - n_{x, T} \cdot b | \geq u \; ; \; n^{\mathrm{swap}}_{x, T} \leq \gamma_x \ln(T)] \mathrm{d}u + T \cdot \mathbb{P}[n^{\mathrm{swap}}_{x, T} \geq \gamma_x \ln(T)] \\
			& \leq \int_0^T \mathbb{P}[ |b_{x, T} - n_{x, T} \cdot b | \geq u + 1 + \gamma_x \ln(T) \; ; \; n^{\mathrm{swap}}_{x, T} \leq \gamma_x \ln(T)] \mathrm{d}u + 1 + \gamma_x \ln(T) + \frac{8}{\epsilon^2} \\
			& \leq  \frac{4}{\epsilon^2} \cdot \int_0^T \exp(- \epsilon^2 \cdot u ) \mathrm{d}u + 1 + \gamma_x \ln(T) + \frac{8}{\epsilon^2} \\	
			& = \frac{13}{\epsilon^4} + (\frac{4}{\epsilon})^2 \ln(T).
	\end{align*}
	We get:
	$$
		\mathbb{E}[n^x_{k, T}] \cdot \mu^c_k + \mathbb{E}[n^x_{l, T}] \cdot \mu^c_l = \mathbb{E}[b_{x, T}] \geq \mathbb{E}[n_{x, T}] \cdot b - \frac{13}{\epsilon^4} - (\frac{4}{\epsilon})^2 \ln(T),
	$$	
	which, in combination with $\mathbb{E}[n^x_{k, T}] + \mathbb{E}[n^x_{l, T}] = \mathbb{E}[n_{x, T}]$, shows that:
	\begin{align*}
		\mathbb{E}[n^x_{k, T}] 
			& \geq ( \frac{ b - \mu^c_l }{ \mu^c_k - \mu^c_l }  ) \cdot \mathbb{E}[n_{x, T}] -  \frac{13}{\epsilon^4 \cdot (\mu^c_k - \mu^c_l)} - \frac{4^2}{\epsilon^2 \cdot (\mu^c_k - \mu^c_l)} \ln(T)\\
			& \geq \xi^x_k \cdot \mathbb{E}[n_{x, T}] - \frac{13}{\epsilon^5} - \frac{16}{\epsilon^3} \ln(T).
	\end{align*}
	Symmetrically, we get:
	$$
		\mathbb{E}[n^x_{l, T}] \geq \xi^x_l \cdot \mathbb{E}[n_{x, T}] - \frac{13}{\epsilon^5} - \frac{16}{\epsilon^3} \ln(T).
	$$

\subsection{Proof of Lemma \ref{lemma-bound-times-non-optimal-pulls-budget-time-horizon}.}

		Consider any suboptimal basis $x \in \mathcal{B}$. We use the shorthand notation $\beta_x = \frac{2^8}{\epsilon^3} \cdot (\frac{\lambda}{\Delta_x})^2$. Without loss of generality, we can assume that both $x$ and $x^*$ involve two arms (one of which may be a dummy arm introduced in the specification of the algorithm given in Section \ref{sec-singlebudgettimehorizon}) and that $\mathcal{K}_{x^*} = \{k^*, l^*\}$ with $\mu^c_{k^*} > b > \mu^c_{l^*}$ and $\mathcal{K}_x = \{k, l\}$ with $\mu^c_k > b > \mu^c_l$. The proof is along the same lines as for Lemmas \ref{lemma-bound-times-non-optimal-pulls} and \ref{lemma-bound-times-non-optimal-pulls-deterministic}. We break down the analysis in a series of facts where we emphasize the main differences. We start off with an inequality analogous to Fact \ref{fact-wlog-assume-n-big}. 
	\begin{align*}
	\mathbb{E}[n_{x, T}] 
		& \leq 2 \beta_x \cdot \ln( T ) + \mathbb{E}[\sum_{t=1}^{T}  I_{x_t = x} \cdot I_{n_{x, t} \geq \beta_x \ln(t)} ] \\	
			& \leq 2 \beta_x \cdot \ln( T )  + \mathbb{E}[ \sum_{t=1}^{T}  I_{x_t = x} \cdot I_{n_{x, t} \geq \beta_x \ln(t)} \cdot I_{ \bar{c}_{k^*, t} - \epsilon_{k^*, t} \leq \mu^c_{k^*}  }  \cdot  I_{ \bar{c}_{l^*, t} - \epsilon_{l^*, t} \leq \mu^c_{l^*}  } ] \\
			& + \sum_{t=1}^T \mathbb{P}[\bar{c}_{l^*, t}  > \mu^c_{l^*} + \epsilon_{l^*, t}] + \mathbb{P}[\bar{c}_{k^*, t} > \mu^c_{k^*} + \epsilon_{k^*, t}] \\
			& \leq 2 \beta_x \cdot \ln( T )  + \mathbb{E}[ \sum_{t=1}^{T}  I_{x_t = x} \cdot I_{n_{x, t} \geq \beta_x \ln(t)} \cdot I_{ \bar{c}_{k^*, t} - \epsilon_{k^*, t} \leq \mu^c_{k^*}  } \cdot  I_{ \bar{c}_{l^*, t} - \epsilon_{l^*, t} \leq \mu^c_{l^*}  } ] + \frac{\pi^2}{3} \\
			& \leq 2 \beta_x \cdot \ln( T )  + \mathbb{E}[ \sum_{t=1}^{T}  I_{x_t = x} \cdot  I_{ \mathrm{obj}_{x, t} + E_{x, t} \geq \sum_{k \in \{k^*, l^*\}} (\bar{r}_{k, t} + \lambda \epsilon_{k, t}) \cdot \xi^{x^*}_k } \cdot I_{n_{x, t} \geq \beta_x \ln(t)} ] + \frac{\pi^2}{3} \\
			& \leq 2 \beta_x \cdot \ln( T )  + \mathbb{E}[ \sum_{t=1}^{T}  I_{x_t = x}  \cdot  I_{ \mathrm{obj}_{x, t} + E_{x, t} \geq \mathrm{obj}_{x^*} }   \cdot I_{n_{x, t} \geq \beta_x \ln(t)} ] \\
			& + \sum_{t=1}^T \mathbb{P}[\bar{r}_{l^*, t}  < \mu^r_{l^*} - \epsilon_{l^*, t}] + \mathbb{P}[\bar{r}_{k^*, t} < \mu^r_{k^*} - \epsilon_{k^*, t}] + \frac{\pi^2}{3} \\
			& \leq 2 \beta_x \cdot \ln( T )  + \mathbb{E}[ \sum_{t=1}^{T}  I_{x_t = x} \cdot I_{ \mathrm{obj}_{x, t} + E_{x, t} \geq \mathrm{obj}_{x^*} }   \cdot I_{n_{x, t} \geq \beta_x \ln(t)} ] + \frac{2 \pi^2}{3}.
 	\end{align*}
		The first inequality is derived in the same fashion as in Fact \ref{fact-wlog-assume-n-big} substituting $k$ with $x$. The third and last inequalities are obtained using Lemma \ref{lemma-martingale-inequality} in the same fashion as in Fact \ref{fact-study-bound-first-term}. The fourth inequality is obtained by observing that (i) if $x_t = x$ then $x_t$ is optimal for \eqref{eq-algo-general-idea} and (ii) $(\xi^*_k)_{k=1, \cdots, K}$ is feasible for \eqref{eq-algo-general-idea} if $\bar{c}_{l^*, t} - \epsilon_{l^*, t} \leq \mu^c_{l^*}$ and $\bar{c}_{k^*, t} - \epsilon_{k^*, t} \leq \mu^c_{k^*}$. The fifth inequality results from $\lambda \geq 1$ and $\mathrm{obj}_{x^*} = \sum_{k \in \{k^*, l^*\}} \mu^r_k \cdot \xi^{x^*}_k$.  The second term in the last upper bound can be broken down in two terms similarly as in Lemmas \ref{lemma-bound-times-non-optimal-pulls} and \ref{lemma-bound-times-non-optimal-pulls-deterministic}:
		\begin{align}
		\mathbb{E}[ & \sum_{t=1}^{T}  I_{x_t = x} \cdot I_{ \mathrm{obj}_{x, t} + E_{x, t} \geq \mathrm{obj}_{x^*} }   \cdot I_{n_{x, t} \geq \beta_x \ln(t)} ] \nonumber \\
			& \leq \mathbb{E}[\sum_{t=1}^{T} I_{x_t \in \mathcal{B}_t} \cdot I_{ \mathrm{obj}_{x, t} \geq \mathrm{obj}_{x} + E_{x, t} } \cdot I_{n_{x, t} \geq \beta_x \ln(t)} ] \label{eq-first-term-upper-bound-time-horizon} \\
				& + \mathbb{E}[\sum_{t=1}^{T} I_{x_t  \in \mathcal{B}_t } \cdot I_{\mathrm{obj}_{x^*} \leq \mathrm{obj}_{x} + 2 E_{x, t} }  \cdot I_{n_{x, t} \geq \beta_x \ln(t)}]. \label{eq-third-term-upper-bound-time-horizon}
		\end{align}
		We carefully study each term separately.
	
		\begin{fact}
			\label{fact-bound-last-term-is-small-time-horizon}
			\begin{align*}
				\mathbb{E}[ & \sum_{t=1}^{T} I_{x \in \mathcal{B}_t} \cdot I_{\mathrm{obj}_{x^*} \leq \mathrm{obj}_{x} + 2 E_{x, t} } \cdot I_{n_{x, t} \geq \beta_x \ln(t)} ] \leq \frac{6 \pi^2}{\epsilon^2}.
			\end{align*}
		\proof{Proof.}
			Using the shorthand notations $\alpha_x = 8  (\frac{\lambda}{\Delta_x})^2$ and $\gamma_x = (\frac{4}{\epsilon})^2$, we have:
			\begin{align*}
				\mathbb{E}[ & \sum_{t=1}^{T} I_{x \in \mathcal{B}_t} \cdot I_{\mathrm{obj}_{x^*} \leq \mathrm{obj}_{x} + 2 E_{x, t} }  \cdot I_{n_{x, t} \geq \beta_x \ln(t)} ] \\
					& \leq \mathbb{E}[\sum_{t=1}^{T} I_{ \Delta_x \leq 2 \lambda \cdot \max( \epsilon_{k, t}, \epsilon_{l, t} ) } \cdot I_{n_{x, t} \geq \beta_x \ln(t)} ] \\
					& = \mathbb{E}[\sum_{t=1}^{T} I_{ \min(n_{k, t}, n_{l, t}) \leq \alpha_x \ln(t) } \cdot I_{n_{x, t} \geq \beta_{x} \ln(t)} ] \\
					& \leq \sum_{t=1}^{T} \mathbb{P}[n_{l, t} \leq \alpha_x \ln(t) \; ; \; n_{x, t} \geq  \beta_{x} \ln(t)]  + \sum_{t=1}^{T} \mathbb{P}[n_{k, t} \leq \alpha_x \ln(t) \; ; \; n_{x, t} \geq  \beta_{x} \ln(t)].
			\end{align*}
			The first inequality is derived using $E_{x, t} = \lambda \cdot ( \xi^{x}_{k, t} \cdot \epsilon_{k, t} + \xi^{x}_{l, t} \cdot \epsilon_{l, t})$ and $\xi^{x}_{k, t} + \xi^{x}_{l, t} \leq 1$ (this is imposed as a constraint in \eqref{eq-algo-general-idea}). Observe now that $\alpha_x/\beta_x$ is a constant factor independent of $\Delta_x$. Thus, we just have to show that if $x$ has been selected at least $\beta_{x} \ln(t)$ times, then both $k$ and $l$ have been pulled at least a constant fraction of the time with high probability. This is the only time the load balancing algorithm comes into play in the proof of Lemma \ref{lemma-bound-times-non-optimal-pulls-budget-time-horizon}. We study the first term and we conclude the study by symmetry. We have:
\begin{align*}
	\mathbb{P}[ & n_{l, t} \leq \alpha_x \ln(t)  \; ; \; n_{x, t} \geq \beta_{x} \ln(t)] \\
		& \leq \mathbb{P}[n_{l, t} \leq \alpha_x \ln(t) \; ; \; n_{x, t} \geq \beta_{x} \ln(t) \; ; \; \sum_{\tau=1}^t I_{x_\tau = x} \cdot I_{a_\tau = k} \cdot c_{k, \tau} \geq (b + \epsilon/2) \cdot n^x_{k, t}] \\ 
		& + \sum_{s = 4/\epsilon^2 \ln(t)}^t \mathbb{P}[\sum_{\tau=1}^t I_{x_\tau = x} \cdot I_{a_\tau = k} \cdot c_{k, \tau} \leq (b + \epsilon/2) \cdot s \; ; \; n^x_{k, t} = s] \\ 
		& \leq \mathbb{P}[n_{l, t} \leq \alpha_x \ln(t) \; ; \; n_{x, t} \geq \beta_{x} \ln(t)  \; ; \; b_{x, t} - n_{x, t} \cdot b \geq \epsilon/2 \cdot n^x_{k, t} - n^x_{l, t} ] \\
		& + \sum_{s = 4/\epsilon^2 \ln(t)}^t \exp( - s \cdot \frac{\epsilon^2}{2}) \\
		& \leq  \mathbb{P}[ b_{x, t} - n_{x, t} \cdot b \geq 2 \gamma_x \ln(t)] + (\frac{2}{\epsilon \cdot t})^2 \\
		& \leq \frac{ 16 }{ \epsilon^2 \cdot t^2}.
\end{align*}			
The first inequality is obtained observing that if $n_{l, t} \leq \alpha_x \ln(t)$ and $n_{x, t} \geq \beta_{x} \ln(t)$, we have: 
$$
	n^x_{k, t} = n_{x, t} - n^x_{l, t} \geq (\frac{2^8}{\epsilon^3}  - 8) \cdot (\frac{\lambda}{\Delta_x})^2 \cdot \ln(t) \geq \frac{4}{\epsilon^2} \cdot \ln(t)
$$ 
because $\lambda \geq 1$ and $\Delta_x \leq \mathrm{obj}_{x^*} = \sum_{k=1}^K \mu^r_k \cdot \xi^{x^*}_{k} \leq \sum_{k=1}^K \xi^{x^*}_{k} \leq 1$ since $x^*$ is a feasible basis to \eqref{eq-linear-program-general-upperbound-opt-strategy}. To derive the second inequality, we use Lemma \ref{lemma-martingale-inequality} for the second term and remark that:
$$
	b_{x, t} - n_{x, t} \cdot b \geq  (\sum_{\tau=1}^t I_{x_\tau = x} \cdot I_{a_\tau = k} \cdot c_{k, \tau} -  n^{x}_{k, t} \cdot b) - n^x_{l, t}
$$ 
since $b \leq 1$. The third inequality is derived using: 
$$
	\epsilon/2 \cdot n^x_{k, t} - n^x_{l, t} \geq \epsilon/2 \cdot n_{x, t} - 2 \cdot n^x_{l, t} \geq (\frac{2^7}{\epsilon^2} - 16) \cdot \ln(t) \geq 2 \gamma_x \ln(t)
$$ 
and the last inequality is obtained with Lemma \ref{lemma-load-balance-budget-time-horizon}.

		\endproof				
		
		\end{fact} \vspace{0.3cm}			
		
		\begin{fact}
			\label{fact-study-bound-first-term-time-horizon}
			$$
				\mathbb{E}[\sum_{t=1}^{T} I_{x_t  \in \mathcal{B}_t }  \cdot I_{ \mathrm{obj}_{x, t}  \geq \mathrm{obj}_{x} + E_{x, t} } \cdot I_{n_{x, t} \geq \beta_x \ln(t)}  ] \leq \frac{ 3 \pi^2 }{\epsilon^2}.
			$$
			\proof{Proof.}
			First observe that:
			\begin{align*}
				\mathbb{E}[ & \sum_{t=1}^{T} I_{x_t  \in \mathcal{B}_t }  \cdot  I_{ \mathrm{obj}_{x, t} \geq \mathrm{obj}_{x} + E_{x, t} } \cdot I_{n_{x, t} \geq \beta_x \ln(t)} ]  \\
				& \leq \mathbb{E}[\sum_{t=1}^{T} I_{x_t  \in \mathcal{B}_t }  \cdot  I_{ \mathrm{obj}_{x, t} \geq \mathrm{obj}_{x} + E_{x, t} } \cdot I_{n_{k, t} \geq \beta_x/2 \cdot \ln(t)} ]  	 \\
				& + \mathbb{E}[\sum_{t=1}^{T} I_{x_t  \in \mathcal{B}_t }  \cdot  I_{ \mathrm{obj}_{x, t} \geq \mathrm{obj}_{x} + E_{x, t} }  \cdot I_{n_{l, t} \geq \beta_x/2 \cdot \ln(t)} ] \\ 
				& \leq \mathbb{E}[\sum_{t=1}^{T} I_{x_t  \in \mathcal{B}_t }  \cdot  I_{ \mathrm{obj}_{x, t} \geq \mathrm{obj}_{x} + E_{x, t} } \cdot I_{ \bar{c}_{k, t} - \epsilon_{k, t} > b} ]  	 \\
				& + \mathbb{E}[\sum_{t=1}^{T} I_{x_t  \in \mathcal{B}_t }  \cdot  I_{ \mathrm{obj}_{x, t} \geq \mathrm{obj}_{x} + E_{x, t} } \cdot I_{ \bar{c}_{l, t} - \epsilon_{l, t} < b} ] \\ 
				& + \sum_{t=1}^T \sum_{s=\beta_x/2 \cdot \ln(t)}^t  \mathbb{P}[ \bar{c}_{k, t} - \epsilon_{k, t} \leq b, n_{k, t} = s]  + \mathbb{P}[ \bar{c}_{l, t} - \epsilon_{l, t} \geq  b, n_{l, t} = s] \\
				& \leq 2 \cdot \mathbb{E}[\sum_{t=1}^{T} I_{ \mathrm{obj}_{x, t} \geq \mathrm{obj}_{x} + E_{x, t} } \cdot I_{ \bar{c}_{k, t} - \epsilon_{k, t} \geq b \geq \bar{c}_{l, t} - \epsilon_{l, t} }  \cdot I_{\bar{c}_{k, t} - \epsilon_{k, t} > \bar{c}_{l, t} - \epsilon_{l, t}}]  \\
				& + \sum_{t=1}^T \sum_{s=\beta_x/2 \cdot \ln(t)}^t \mathbb{P}[ \bar{c}_{k, t} \leq b + \epsilon/2, n_{k, t} = s] + \mathbb{P}[ \bar{c}_{l, t}  \geq  b + \epsilon/2, n_{l, t} = s] \\
				& \leq 2 \cdot \mathbb{E}[\sum_{t=1}^{T} I_{ \mathrm{obj}_{x, t} \geq \mathrm{obj}_{x} + E_{x, t} } \cdot I_{ \bar{c}_{k, t} - \epsilon_{k, t} \geq b \geq \bar{c}_{l, t} - \epsilon_{l, t} } \cdot I_{\bar{c}_{k, t} - \epsilon_{k, t} > \bar{c}_{l, t} - \epsilon_{l, t}}]  \\
				& + 2 \cdot \sum_{t=1}^T \sum_{s=\beta_x/2 \cdot \ln(t)}^t \exp(- s \frac{\epsilon^2}{2}) \\ 
				& \leq 2 \cdot \mathbb{E}[\sum_{t=1}^{T} I_{ \mathrm{obj}_{x, t} \geq \mathrm{obj}_{x} + E_{x, t} } \cdot I_{ \bar{c}_{k, t} - \epsilon_{k, t} \geq b \geq \bar{c}_{l, t} - \epsilon_{l, t} }  \cdot I_{\bar{c}_{k, t} - \epsilon_{k, t} > \bar{c}_{l, t} - \epsilon_{l, t}}]  + \frac{ 4 \pi^2 }{3 \epsilon^2}.
			\end{align*}								
			 The third inequality is obtained by observing that $\epsilon_{k, t}, \epsilon_{l, t} \leq \frac{\epsilon}{2}$ for $n_{k, t}, n_{l, t} \geq \frac{\beta_x}{2} \ln(t)$ (because $\lambda \geq 1$ and $\Delta_x \leq \mathrm{obj}_{x^*} = \sum_{k=1}^K \mu^r_k \cdot \xi^{x^*}_{k} \leq \sum_{k=1}^K \xi^{x^*}_{k} \leq 1$) and that, if $x_t \in \mathcal{B}_t$ and (for example) $\bar{c}_{l, t} - \epsilon_{l, t} < b$, it must be that $\bar{c}_{k, t} - \epsilon_{k, t} \geq b$. The last two inequalities are obtained in the same fashion as in Lemma \ref{lemma-bound-times-infeasible-pulls-budget-time-horizon} observing that $\beta_x/2 \geq 4/\epsilon^2$. At this point, the key observation is that if $\mathrm{obj}_{x, t} \geq \mathrm{obj}_{x} + E_{x, t} $, $\bar{c}_{k, t} - \epsilon_{k, t} \geq b \geq \bar{c}_{l, t} - \epsilon_{l, t}$, and $\bar{c}_{k, t} - \epsilon_{k, t} > \bar{c}_{l, t} - \epsilon_{l, t}$, at least one of the following six events occurs: $\{ \bar{r}_{k, t} \geq \mu^r_k + \epsilon_{k, t} \}$, $\{ \bar{r}_{l, t} \geq \mu^r_l + \epsilon_{l, t} \}$, $\{ \bar{c}_{k, t} \leq \mu^c_k - \epsilon_{k, t} \}$, $\{ \bar{c}_{k, t} \geq \mu^c_k + \epsilon_{k, t} \}$, $\{ \bar{c}_{l, t} \leq \mu^c_l - \epsilon_{l, t} \}$ or $\{ \bar{c}_{l, t} \geq \mu^c_l + \epsilon_{l, t} \}$. Otherwise, using the shorthand notations $\tilde{c}_k = \bar{c}_{k, t} - \epsilon_{k, t}$ and $\tilde{c}_l = \bar{c}_{l, t} - \epsilon_{l, t}$, we have:
				\begin{align*}
					\mathrm{obj}_{x, t} &  - \mathrm{obj}_{x} \\
						& = [ \frac{ \tilde{c}_k - b }{ \tilde{c}_k - \tilde{c}_l} \cdot \bar{r}_{l, t}
							+ \frac{ b - \tilde{c}_l}{ \tilde{c}_k - \tilde{c}_l} \cdot \bar{r}_{k, t} ] 
						    - [ \frac{ \mu^c_{k} - b }{ \mu^c_{k} - \mu^c_{l} } \cdot \mu^r_{l}
							+ \frac{ b - \mu^c_{l} }{ \mu^c_{k} - \mu^c_{l} } \cdot \mu^r_{k} ] \\
						& < [ \frac{ \tilde{c}_k - b }{ \tilde{c}_k - \tilde{c}_l} \cdot (\mu^r_l + \epsilon_{l, t}) 
							+ \frac{ b - \tilde{c}_l}{ \tilde{c}_k - \tilde{c}_l} \cdot (\mu^r_k + \epsilon_{k, t}) ] 
						 - [ \frac{ \mu^c_{k} - b }{ \mu^c_{k} - \mu^c_{l} } \cdot \mu^r_{l}
							+ \frac{ b - \mu^c_{l} }{ \mu^c_{k} - \mu^c_{l} } \cdot \mu^r_{k} ] \\
						& = \frac{1}{\lambda} \cdot E_{x, t} 
							+ (\mu^r_k - \mu^r_l) \cdot [ \frac{ b - \tilde{c}_l}{ \tilde{c}_k - \tilde{c}_l} -  \frac{ b - \mu^c_{l} }{ \mu^c_{k} - \mu^c_{l} } ] \\
						& = \frac{1}{\lambda} \cdot E_{x, t} 
						 + \frac{(\mu^r_k - \mu^r_l)}{(\tilde{c}_k - \tilde{c}_l) \cdot (\mu^c_k - \mu^c_l) } \cdot [ (\mu^c_k - b) ( \mu^c_l - \tilde{c}_l ) + ( b - \mu^c_l ) ( \mu^c_k - \tilde{c}_k ) ] \\
						& \leq \frac{1}{\lambda} \cdot E_{x, t} 
						 +  \frac{\kappa}{\tilde{c}_k - \tilde{c}_l} \cdot |(\mu^c_k - b) ( \mu^c_l - \tilde{c}_l ) + ( b - \mu^c_l ) ( \mu^c_k - \tilde{c}_k ) | \\
						& \leq \frac{1}{\lambda} \cdot E_{x, t}
							+ 2 \frac{\kappa}{\tilde{c}_k - \tilde{c}_l } \cdot [ (\tilde{c}_k - b) \cdot \epsilon_{l, t} + (b - \tilde{c}_l) \cdot \epsilon_{k, t}    ] \\
						& = \frac{1}{\lambda} \cdot E_{x, t} + \frac{2 \kappa}{\lambda} \cdot E_{x, t} \\
						& = E_{x, t},
				\end{align*}
				a contradiction. The first inequality is strict because either $\tilde{c}_k > b$ or $\tilde{c}_l < b$. The second inequality is derived using Assumption \ref{assumption-simplying-assumption-upperboundknown-budget-and-time-horizon}. The third inequality is derived from the observation that the expression $(\mu^c_k - b) ( \mu^c_l - \tilde{c}_l ) + ( b - \mu^c_l ) ( \mu^c_k - \tilde{c}_k )$ is a linear function of $(\mu^c_k, \mu^c_l)$ (since the cross term $\mu^c_k \cdot \mu^c_l$ cancels out) so that $|(\mu^c_k - b) ( \mu^c_l - \tilde{c}_l ) + ( b - \mu^c_l ) ( \mu^c_k - \tilde{c}_k ) |$ is convex in $(\mu^c_k, \mu^c_l)$ and the maximum of this expression over the polyhedron $[\bar{c}_{k, t} - \epsilon_{k, t}, \bar{c}_{k, t} + \epsilon_{k, t}] \times [\bar{c}_{l, t} - \epsilon_{l, t}, \bar{c}_{l, t} + \epsilon_{l, t}]$ is attained at an extreme point. We obtain:
				\begin{align*}
				\mathbb{E}[ & \sum_{t=1}^{T} I_{x \in \mathcal{B}_t} I_{ \mathrm{obj}_{x, t} \geq \mathrm{obj}_{x} + E_{x, t} } \cdot I_{n_{x, t} \geq \beta_x \ln(t)}  ] \\
					& \leq \sum_{t=1}^{\infty}  \mathbb{P}[ \bar{r}_{k, t} \geq \mu^r_k + \epsilon_{k, t} ]  + \mathbb{P}[ \bar{r}_{l, t} \geq \mu^r_l + \epsilon_{l, t}] \\
					& + \sum_{t=1}^{\infty} \mathbb{P}[ \bar{c}_{l, t} \geq \mu^c_l + \epsilon_{l, t} ] + \mathbb{P}[ \bar{c}_{k, t} \geq \mu^c_k + \epsilon_{k, t}] \\
					& + \sum_{t=1}^{\infty}  \mathbb{P}[ \bar{c}_{k, t} \leq \mu^c_k - \epsilon_{k, t} ] + \mathbb{P}[ \bar{c}_{l, t} \leq \mu^c_l - \epsilon_{l, t} ] + \frac{ 4 \pi^2 }{3 \epsilon^2} \\
					& \leq \pi^2  + \frac{ 4 \pi^2 }{3 \epsilon^2} \\
					& \leq \frac{ 3 \pi^2 }{\epsilon^2},
				\end{align*}
				using the same argument as in Fact \ref{fact-study-bound-first-term}.
			\endproof					
		\end{fact} \vspace{0.3cm}

\subsection{Proof of Theorem \ref{lemma-log-B-regret-bound-time-horizon}.}
	We build upon \eqref{eq-simplified-general-upper-bound-on-regret}:
	\begin{align*}
				R_{B, T} 
					& \leq T \cdot \sum_{k=1}^K \mu^r_k \cdot \xi^{x^*}_k - \mathbb{E}[\sum_{t=1}^{\tau^*} r_{a_t, t}] + O(1) \\
					& = T \cdot \sum_{k=1}^K \mu^r_k \cdot \xi^{x^*}_k - \mathbb{E}[\sum_{t=1}^{T} r_{a_t, t}] + \mathbb{E}[\sum_{t=\tau^*+1}^{T} r_{a_t, t}]  + O(1)  \\					
					& \leq T \cdot \sum_{k=1}^K \mu^r_k \cdot \xi^{x^*}_k - \mathbb{E}[\sum_{t=1}^{T} r_{a_t, t}] + \sigma \cdot \mathbb{E}[\sum_{t=\tau^*+1}^{T} c_{a_t, t}]  + O(1)  \\
					& \leq T \cdot \sum_{k=1}^K \mu^r_k \cdot \xi^{x^*}_k - \mathbb{E}[\sum_{t=1}^{T} r_{a_t, t}] + \sigma \cdot \mathbb{E}[ (\sum_{t=1}^{T} c_{a_t, t} - B)_+]  + O(1).
	\end{align*}
	The second inequality is a consequence of Assumption \ref{assumption-cost-bounds-rewards-budget-and-time-horizon}:
	\begin{align*}
		\mathbb{E}[\sum_{t=\tau^*+1}^{T} c_{a_t, t}]
			& = \mathbb{E}[\sum_{t=1}^{T} c_{a_t, t}] - \mathbb{E}[\sum_{t=1}^{\tau^*} c_{a_t, t}] \\
			& = \mathbb{E}[\sum_{t=1}^{T}  \mathbb{E}[c_{a_t, t} \; | \; \mathcal{F}_{t-1}]] - \mathbb{E}[\sum_{t=1}^{\infty}  I_{\tau^* \geq t} \cdot  \mathbb{E}[c_{a_t, t} \; | \; \mathcal{F}_{t-1}]] \\ 
			& = \mathbb{E}[\sum_{t=1}^{T}  \mu^c_{a_t} ] - \mathbb{E}[\sum_{t=1}^{\infty} I_{\tau^* \geq t} \cdot \mu^c_{a_t}] \\ 
			& = \mathbb{E}[\sum_{t=\tau^*+1}^{T}  \mu^c_{a_t} ] \\
			& \geq \frac{1}{\sigma} \cdot \mathbb{E}[\sum_{t=\tau^*+1}^{T} \mu^r_{a_t}] = \frac{1}{\sigma} \cdot \mathbb{E}[\sum_{t=\tau^*+1}^{T} r_{a_t, t}],
	\end{align*}
	since $\tau^*$ is a stopping time. To derive the third inequality, observe that if $\tau^* = T + 1$, we have:
	$$
		\sum_{t=\tau^*+1}^{T} c_{a_t, t} = 0 \leq (\sum_{t=1}^{T} c_{a_t, t} - B)_+,
	$$ 
	while if $\tau^* < T + 1$ we have run out of resources before round $T$, i.e. $\sum_{t=1}^{\tau^*} c_{a_t, t} \geq B$, which implies:
	\begin{align*}
		\sum_{t=\tau^*+1}^{T} c_{a_t, t} 
			& \leq \sum_{t=\tau^*+1}^{T} c_{a_t, t} + \sum_{t=1}^{\tau^*} c_{a_t, t} -  B \\
			& \leq (\sum_{t=1}^{T} c_{a_t, t} - B)_+.
	\end{align*}
	Now observe that:
	\begin{align*}
		\mathbb{E}[ (\sum_{t=1}^{T} c_{a_t, t} - B)_+] 
			& =  \mathbb{E}[ ( \sum_{x \text{ pseudo-basis for } \eqref{eq-linear-program-general-upperbound-opt-strategy} } \{ b_{x, T} - n_{x, T} \cdot b \} )_+ ]  \\
			& \leq \sum_{x \in \mathcal{B}} \mathbb{E}[ |b_{x, T} - n_{x, T} \cdot b |] + \sum_{x \notin \mathcal{B}} \mathbb{E}[n_{x, T}] + \sum_{ \substack{ x \text{ pseudo-basis for } \eqref{eq-linear-program-general-upperbound-opt-strategy} \\ \text{ with } \det(A_x) = 0}} \mathbb{E}[n_{x, T}] \\
			& = O(\frac{K^2}{\epsilon^3} \ln(T)),
	\end{align*}	
	where we use the fact that $c_{k, t} \leq 1$ at any time $t$ and for all arms $k$ for the first inequality and Lemma \ref{lemma-bound-times-infeasible-pulls-budget-time-horizon} along with the proof of Lemma \ref{lemma-load-balance-budget-time-horizon} for the last equality. Plugging this last inequality back into the regret bound yields:
	\begin{align}
		R_{B, T}
					& \leq T \cdot \sum_{k=1}^K \mu^r_k \cdot \xi^{x^*}_k - \mathbb{E}[\sum_{t=1}^{T} r_{a_t, t}] + O(\frac{K^2 \cdot \sigma }{\epsilon^3} \ln(T)) \nonumber \\			
					& \leq T \cdot \sum_{k=1}^K \mu^r_k \cdot \xi^{x^*}_k  - \sum_{x \in \mathcal{B}} \sum_{k=1}^K \mu^r_k \cdot \mathbb{E}[n^x_{k, T} ] + O(\frac{K^2 \cdot \sigma }{\epsilon^3} \ln(T)) \nonumber \\
					& \leq T \cdot \sum_{k=1}^K \mu^r_k \cdot \xi^{x^*}_k  - \sum_{x \in \mathcal{B}} (\sum_{k=1}^K \mu^r_k \cdot \xi^x_k) \cdot \mathbb{E}[n_{x, T} ] + O(\frac{K^2 \cdot \sigma }{\epsilon^3} \ln(T)) \nonumber \\
					& = \sum_{k=1}^K \mu^r_k \cdot \xi^{x^*}_k \cdot ( T - \sum_{x \in \mathcal{B} \; | \; \Delta_x = 0} \mathbb{E}[n_{x, T}] ) - \sum_{x \in \mathcal{B} \; | \; \Delta_x > 0} (\sum_{k=1}^K \mu^r_k \cdot \xi^x_k) \cdot \mathbb{E}[n_{x, T} ] + O(\frac{K^2 \cdot \sigma }{\epsilon^3} \ln(T)) \nonumber \\
					& = \sum_{k=1}^K \mu^r_k \cdot \xi^{x^*}_k \cdot ( \sum_{x \in \mathcal{B} \; | \; \Delta_x > 0} \mathbb{E}[n_{x, T}] + \sum_{x \notin \mathcal{B}}  \mathbb{E}[n_{x, T}] + \sum_{ \substack{ x \text{ pseudo-basis for } \eqref{eq-linear-program-general-upperbound-opt-strategy} \\ \text{ with } \det(A_x) = 0}} \mathbb{E}[n_{x, T}]) \nonumber \\
					& - \sum_{x \in \mathcal{B} \; | \; \Delta_x > 0} (\sum_{k=1}^K \mu^r_k \cdot \xi^x_k) \cdot \mathbb{E}[n_{x, T} ]  + O(\frac{K^2 \cdot \sigma }{\epsilon^3} \ln(T)) \nonumber \\
					& \leq \sum_{x \in \mathcal{B} \; | \; \Delta_x > 0}  \Delta_x \cdot \mathbb{E}[n_{x, T} ] + O(\frac{K^2 \cdot \sigma }{\epsilon^3} \ln(T)) \label{eq-proof-log-B-start-here-sqrt-B-time-horizon} \\	
					& \leq 2^9  \frac{\lambda^2}{\epsilon^3} \cdot (\sum_{x \in \mathcal{B} \; | \; \Delta_x > 0}  \frac{1}{\Delta_x}) \cdot \ln(T) + O(\frac{K^2 \cdot \sigma }{\epsilon^3} \ln(T)), \nonumber
	\end{align}
	where we use Lemma \ref{lemma-load-balance-budget-time-horizon}	for the third inequality, Lemma \ref{lemma-bound-times-infeasible-pulls-budget-time-horizon} along with $\sum_{k=1}^K \mu^r_k \cdot \xi^{x^*}_k \leq \sum_{k=1}^K \xi^{x^*}_k \leq 1$ for the fourth inequality, and Lemma \ref{lemma-bound-times-non-optimal-pulls-budget-time-horizon} for the last inequality.
			
\subsection{Proof of Theorem \ref{lemma-sqrt-B-regret-bound-time-horizon}.}
Along the same lines as for the case of a single limited resource, we start from inequality \eqref{eq-proof-log-B-start-here-sqrt-B-time-horizon} derived in the proof of Theorem \ref{lemma-log-B-regret-bound-time-horizon} and apply Lemma \ref{lemma-bound-times-non-optimal-pulls-budget-time-horizon} only if $\Delta_x$ is big enough, taking into account the fact that:
$$
	\sum_{x \in \mathcal{B}} \mathbb{E}[n_{x, T}] \leq T.
$$
Specifically, we have: 
\begin{align*}
	R_{B, T} 
		& \leq \sup\limits_{ \substack{(n_x)_{x \in \mathcal{B}} \geq 0 \\ \sum_{x \in \mathcal{B}} n_x \leq T }  } \{ \; \sum_{x \in \mathcal{B} \; | \; \Delta_x > 0} \min( \Delta_x \cdot n_x, 2^9  \frac{\lambda^2}{\epsilon^3} \cdot \frac{\ln(T)}{\Delta_x} +  \frac{ 10 \pi^2 }{3 \epsilon^2} \cdot \Delta_x ) \; \} + O(\frac{K^2 \cdot \sigma }{\epsilon^3} \ln(T)) \\
		& \leq \sup\limits_{ \substack{(n_x)_{x \in \mathcal{B}} \geq 0 \\ \sum_{x \in \mathcal{B}} n_x \leq T }  } \{ \; \sum_{x \in \mathcal{B} \; | \; \Delta_x > 0} \min( \Delta_x \cdot n_x, 2^9 \frac{\lambda^2}{\epsilon^3} \cdot \frac{\ln(T)}{\Delta_x} ) \; \} + O(\frac{K^2 \cdot \sigma }{\epsilon^3} \ln(T)) \\ 
		& \leq \sup\limits_{ \substack{(n_x)_{x \in \mathcal{B}} \geq 0 \\ \sum_{x \in \mathcal{B}} n_x \leq T }  } \{ \; \sum_{x \in \mathcal{B}} \sqrt{2^9  \frac{\lambda^2}{\epsilon^3} \cdot \ln(T)  \cdot  n_x} \; \} + O(\frac{K^2 \cdot \sigma }{\epsilon^3} \ln(T)) \\
		& \leq 2^5 \frac{ \lambda }{\epsilon^{3/2}} \cdot \sqrt{ \ln(  T )  } \cdot \sup\limits_{ \substack{(n_x)_{x \in \mathcal{B}} \geq 0 \\ \sum_{x \in \mathcal{B}} n_x \leq T }  } \{ \; \sum_{x \in \mathcal{B} } \sqrt{n_x} \; \} + O(\frac{K^2 \cdot \sigma }{\epsilon^3} \ln(T)) \\ 
		& \leq 2^5 \frac{ \lambda }{\epsilon^{3/2}} \cdot \sqrt{ |\mathcal{B}| \cdot T \cdot \ln( T )  } + O(\frac{K^2 \cdot \sigma }{\epsilon^3} \ln(T)),
\end{align*}
where we use the fact that $\Delta_x \leq \sum_{k=1}^K \mu^r_k \cdot \xi^{x^*}_k \leq \sum_{k=1}^K \xi^{x^*}_k \leq 1$ for the second inequality, we maximize over each $\Delta_x \geq 0$ to derive the third inequality, and we use Cauchy-Schwartz for the last inequality.

\subsection{Proof of Theorem \ref{lemma-regret-bound-b-small-time-horizon}.}
When $b \leq \epsilon/2$, the analysis almost falls back to the case of a single limited resource. Indeed, we have $\tau^* = \tau(B)$ with high probability given that:
\begin{align*}
	\mathbb{P}[ \tau(B) > T + t ]
		& = \mathbb{P}[ \sum_{\tau=1}^{T+t} c_{a_\tau, \tau} \leq B ] \\
		& \leq \mathbb{P}[ \frac{1}{T+t} \cdot \sum_{t=1}^{T+t} c_{a_t, t} \leq \epsilon - (\epsilon - b) ] \\
		& \leq \exp(- (T+t) \cdot \frac{\epsilon^2}{2}), 
\end{align*}
for any $t \in \mathbb{N}$ using Lemma \ref{lemma-martingale-inequality}. Now observe that, since $b \leq \epsilon/2$, the feasible bases for \eqref{eq-linear-program-general-upperbound-opt-strategy} are exactly the bases $x$ such that $\mathcal{K}_x = \{ k \}$ and $\mathcal{C}_x = \{1\}$ for some $k \in \{1, \cdots, K\}$, which we denote by $(x_k)_{k=1, \cdots, K}$. This shows that $\ropt(B, T) = B \cdot \max_{k=1, \cdots, K} \mu^r_k / \mu^c_k$. Moreover note that Assumption \ref{assumption-simplying-assumption-analysis-budget-and-time-horizon} is automatically satisfied when $b \leq \epsilon/2$. Hence, with a minor modification of the proof of Lemma \ref{lemma-bound-times-infeasible-pulls-budget-time-horizon}, we get:
$$
\mathbb{E}[n_{x, \tau(B)}] \leq \frac{2^{12}}{\epsilon^3} \cdot \mathbb{E}[\ln(\tau(B))] + \frac{40 \pi^2}{3 \epsilon^2},
$$
for any pseudo-basis $x$ involving two arms or any basis $x$ such that $\mathcal{K}_x = \{ k \}$ and $\mathcal{C}_x = \{2\}$ for some arm $k \in \{1, \cdots, K\}$. Similarly, a minor modification of Lemma \ref{lemma-bound-times-non-optimal-pulls-budget-time-horizon} yields:
$$
	\mathbb{E}[n_{x_k, \tau(B)}] \leq 2^{12}	\frac{\lambda^2}{\epsilon^3} \cdot \frac{ \mathbb{E}[\ln(\tau(B))] }{\Delta_{x_k}^2} + \frac{40 \pi^2}{\epsilon^2},
$$
for any $k \in \{1, \cdots, K\}$. What is left is to refine the analysis of Theorem \ref{lemma-log-B-regret-bound-time-horizon} as follows:
	\begin{align*}
				R_{B, T} 
					& \leq B \cdot \max_{k=1, \cdots, K} \frac{\mu^r_k}{\mu^c_k} - \mathbb{E}[\sum_{t=1}^{\tau^*} r_{a_t, t}] + O(1) \\
					& \leq B \cdot \max_{k=1, \cdots, K} \frac{\mu^r_k}{\mu^c_k} - \mathbb{E}[\sum_{t=1}^{\tau(B)} r_{a_t, t}] +  T \cdot \mathbb{P}[ \tau(B) > T ] + \mathbb{E}[(\tau(B) - T)_+] + O(1)  \\					
					& \leq B \cdot \max_{k=1, \cdots, K} \frac{\mu^r_k}{\mu^c_k} - \sum_{k=1}^K \mu^r_k \cdot \mathbb{E}[ n^{x_k}_{k, \tau(B)} ] + O(1) \\
					& \leq \max_{k=1, \cdots, K} \frac{\mu^r_k}{\mu^c_k} \cdot (B - \sum_{k \; | \; \Delta_{x_k} = 0} \mu^c_k \cdot \mathbb{E}[ n^{x_k}_{k, \tau(B)} ]) - \sum_{k \; | \; \Delta_{x_k} > 0} \mu^r_k \cdot \mathbb{E}[ n^{x_k}_{k, \tau(B)} ] + O(1).
	\end{align*} 
	By definition of $\tau(B)$, we have $B \leq \sum_{t=1}^{\tau(B)} c_{a_t, t}$. Taking expectations on both sides yields:
	\begin{align*}
		B 
			& \leq \sum_{k=1}^K \mu^c_k \cdot \mathbb{E}[n_{k, \tau(B)}] \\
			& = \sum_{k \; | \; \Delta_{x_k} = 0} \mu^c_k \cdot \mathbb{E}[ n^{x_k}_{k, \tau(B)} ] + \sum_{k \; | \; \Delta_{x_k} > 0} \mu^c_k \cdot \mathbb{E}[ n^{x_k}_{k, \tau(B)} ] + \sum_{x \notin \mathcal{B}} \mathbb{E}[ n_{x, \tau(B)}].
	\end{align*}
	Plugging this inequality back into the regret bound, we get:
	\begin{align*}
		R_{B, T} 
			& =  \sum_{k \; | \; \Delta_{x_k} > 0} \mu^c_k \cdot \Delta_{x_k} \cdot \mathbb{E}[n^{x_k}_{k, \tau(B)}] + \max_{k=1, \cdots, K} \frac{\mu^r_k}{\mu^c_k} \cdot \sum_{x \notin \mathcal{B}} \mathbb{E}[ n_{x, \tau(B)}] + O(1) \\
			& \leq  \sum_{k \; | \; \Delta_{x_k} > 0} \mu^c_k \cdot \Delta_{x_k} \cdot \mathbb{E}[n^{x_k}_{k, \tau(B)}] + \frac{2^{12} K^2 \cdot \kappa}{\epsilon^3} \cdot \mathbb{E}[\ln(\tau(B))] + O(1) \\
			& \leq 2^{12}	\frac{\lambda^2}{\epsilon^3} \cdot (\sum_{k \; | \; \Delta_{x_k} > 0} \frac{1}{\Delta_{x_k}})  \cdot \ln( \frac{B+1}{\epsilon} ) + \frac{2^{12} K^2 \cdot \kappa}{\epsilon^3} \cdot \ln( \frac{B+1}{\epsilon} ) + O(1),
	\end{align*}
	where we use Assumption \ref{assumption-simplying-assumption-upperboundknown-budget-and-time-horizon} for the second inequality and Lemma \ref{lemma-bound-stopping-time} for the third inequality. A distribution-independent regret bound of order $O(\sqrt{K \cdot \frac{B+1}{\epsilon} \cdot \ln(\frac{B+1}{\epsilon})} + \frac{K^2 \cdot \kappa}{\epsilon^3} \cdot \ln( \frac{B+1}{\epsilon} ) )$ can be derived from the penultimate inequality along the same lines as in Theorem \ref{lemma-sqrt-B-regret-bound-time-horizon}.

\section{Proofs for Section \ref{sec-stochastic-multiple-budget}.}
\label{sec-proofs-general-case}

\subsection{Preliminary work for the proofs of Section \ref{sec-stochastic-multiple-budget}.}
\label{secproofs-preliminarywork-general-case}

\paragraph{Concentration inequality.}
We will use the following inequality repeatedly. For a given round $\tau \geq \tini$ and a basis $x$:
\begin{equation}
	\label{eq-proof-general-case-concentration-mean}
	\begin{aligned}
		\mathbb{P}[ & \exists (k, i) \in \mathcal{K}_x \times \{1, \cdots, C\}, |\bar{c}_{k, \tau}(i) - \mu^c_k(i)| > \maxdevmeangeneralcase ] \\
		& \leq \sum_{s=\tini/K}^T \sum\limits_{ \substack{k \in \mathcal{K}_x \\ i \in \{1, \cdots, C\}}} \mathbb{P}[|\bar{c}_{k, \tau}(i) - \mu^c_k(i)| > \maxdevmeangeneralcase, n_{k, \tau} = s ] \\
		& \leq 2 \cdot C^2 \cdot \sum_{s=\tini/K}^\infty \exp( - s \cdot \frac{\epsilon^6}{2^7 \cdot (C+2)!^4} ) \\
		& \leq \probamaxdevmeangeneralcase,	 
	\end{aligned}
\end{equation}
using Lemma \ref{lemma-martingale-inequality}, the inequality $\exp(-x) \leq 1 - x/2$ for $x \in [0, 1]$, and the fact that we pull each arm $\tini/K \geq \tinivalue \cdot \ln(T)$ times during the initialization phase. 

\paragraph{Useful matrix inequalities.}
For any basis $x$, assume that $\{ |\bar{c}_{k, t}(i) - \mu^c_k(i)| \leq \maxdevmeangeneralcase \}$, for any arm $k \in \mathcal{K}_x$ and resource $i \in \mathcal{C}_x$. We have:
\begin{align*}
	| \det( \bar{A}_{x, t} ) - \det( A_{x} ) |
		& =  |\sum_{\sigma \in S(\mathcal{K}_x, \mathcal{C}_x)} [ \prod_{k \in \mathcal{K}_x} \bar{c}_{k, t}(\sigma(k)) - \prod_{k \in \mathcal{K}_x} \mu^c_{k}(\sigma(k)) ] | \\
		& = |\sum_{\sigma \in S(\mathcal{K}_x, \mathcal{C}_x)} \sum_{l \in \mathcal{K}_x} \prod_{k < l} \bar{c}_{k, t}(\sigma(k)) \cdot [  \bar{c}_{l, t}(\sigma(l)) - \mu^c_{l}(\sigma(l)) ] \cdot \prod_{k > l } \mu^c_{k}(\sigma(k)) | \\
		& \leq \sum_{\sigma \in S(\mathcal{K}_x, \mathcal{C}_x)} \sum_{l \in \mathcal{K}_x} | \bar{c}_{l, t}(\sigma(l)) - \mu^c_{l}(\sigma(l)) | \\
		& \leq \frac{\epsilon^3}{16 \cdot (C+2)!} \\
		& \leq \frac{\epsilon}{2},
\end{align*}
since the amounts of resources consumed at any round are no larger than $1$. This yields:
\begin{equation}
	\label{eq-det-inequality}
	| \det( \bar{A}_{x, t} ) |  \geq | \det( A_{x} ) |  - | \det( \bar{A}_{x, t} ) - \det( A_{x} ) | \geq \frac{\epsilon}{2},
\end{equation}
using Assumption \ref{assumption-simplying-assumption-general case}. Now consider any vector $c$ such that $\norm{c}_\infty \leq 1$, we have:
\begin{align*}
	|c^\tr & \bar{A}_{x, t}^{-1} b_{\mathcal{K}_x}   - c^\tr A_{x}^{-1} b_{\mathcal{K}_x}| \\
		& =  \frac{1}{ |\det(A_x)| \cdot |\det(\bar{A}_{x, t})| }| \cdot | \det(A_x) \cdot c^\tr \adj(\bar{A}_{x, t}) b_{\mathcal{K}_x}  - \det(\bar{A}_{x, t}) \cdot c^\tr \adj(A_{x}) b_{\mathcal{K}_x} | \\
		& \leq \frac{1}{ |\det(\bar{A}_{x, t})| }| \cdot | c^\tr ( \adj(\bar{A}_{x, t}) - \adj(A_{x})) b_{\mathcal{K}_x} |  \\
		& + \frac{1}{ |\det(A_x)| \cdot |\det(\bar{A}_{x, t})| }| \cdot |\det(\bar{A}_{x, t}) - \det(A_x)| \cdot |c^\tr \adj(A_{x}) b_{\mathcal{K}_x} | \\
		& \leq \frac{\epsilon^2}{8} + \frac{\epsilon}{8 \cdot (C+2)!} \cdot \norm{c}_2 \cdot \norm{ \adj(A_x) b_{\mathcal{K}_x}}_2  \\
		& \leq \frac{\epsilon}{4}.
\end{align*} 
The second inequality is obtained using Assumption \ref{assumption-simplying-assumption-general case} and \eqref{eq-det-inequality} by proceeding along the same lines as above to bound the difference between two determinants for each component of $\adj(\bar{A}_{x, t}) - \adj(A_{x})$. The last inequality is obtained using $\norm{c}_2 \leq \sqrt{C}$ and the fact that each component of $A_x$ is smaller than $1$.
If we take $c = e_k$ for $k \in \mathcal{K}_x$, this yields:
\begin{equation}
	\label{eq-proof-non-degeneracy-1}
	|\xi^x_{k, t} - \xi^x_k| \leq \frac{\epsilon}{4}.
\end{equation}
If we take $c = (\mu^c_{k}(i))_{k \in \mathcal{K}_x}$, for any $i \in \{1, \cdots, C\}$, we get:
\begin{equation}
	\label{eq-proof-non-degeneracy-2}
	|\sum_{k \in \mathcal{K}_x} \mu^c_{k}(i) \cdot \xi^x_{k, t} - \sum_{k \in \mathcal{K}_x} \mu^c_{k}(i) \cdot \xi^x_{k}|
		\leq \frac{\epsilon}{4}
\end{equation}
and
\begin{equation}
	\label{eq-proof-non-degeneracy-3}
	\begin{aligned}
		|\sum_{k \in \mathcal{K}_x} \bar{c}_{k, t}(i) \cdot \xi^x_{k, t} - \sum_{k \in \mathcal{K}_x} \mu^c_{k}(i) \cdot \xi^x_{k}|
			& = |\bar{c}^\tr \bar{A}_{x, t}^{-1} b_{\mathcal{K}_x} - c^\tr A_{x}^{-1} b_{\mathcal{K}_x} | \\
			& \leq |(\bar{c} - c)^\tr \bar{A}_{x, t}^{-1} b_{\mathcal{K}_x} | + |c^\tr \bar{A}_{x, t}^{-1} b_{\mathcal{K}_x}   - c^\tr A_{x}^{-1} b_{\mathcal{K}_x}| \\
			& \leq  \norm{ \bar{c} - c }_2 \cdot \frac{1}{| \det(\bar{A}_{x, t}) |} \cdot \norm{ \adj(\bar{A}_{x, t}) b_{\mathcal{K}_x} }_2 + \frac{\epsilon}{4} \\
			& \leq \sqrt{C} \cdot \maxdevmeangeneralcase \cdot \frac{2}{\epsilon} \cdot C!	+ \frac{\epsilon}{4} \\
			& \leq \frac{\epsilon}{2},
	\end{aligned}
\end{equation}
where $\bar{c} = (\bar{c}_{k, t}(i))_{k \in \mathcal{K}_x}$.

\subsection{Proof of Lemma \ref{lemma-bound-times-infeasible-pulls-general-case}.}
First, consider a basis $x \notin \mathcal{B}$. Since $x$ is $\epsilon$-non-degenerate by Assumption \ref{assumption-simplying-assumption-general case}, there must exist $k \in \mathcal{K}_x$ such that $\xi^x_k \leq - \epsilon$ or $i \in \{1, \cdots, C\}$ such that $\sum_{k=1}^K \mu^c_k(i) \cdot \xi^x_k \geq b(i) + \epsilon$. Let us assume that we are in the first situation (the proof is symmetric in the other scenario). Using \eqref{eq-proof-non-degeneracy-1} in the preliminary work of Section \ref{secproofs-preliminarywork-general-case}, we have:
\begin{equation}
	\xi^x_{k,t} = \xi^x_{k} + ( \xi^x_{k,t} - \xi^x_{k} ) \leq -\frac{\epsilon}{2},
\end{equation}
if $\{ |\bar{c}_{k, t}(i) - \mu^c_k(i)| \leq \maxdevmeangeneralcase \}$ for all arms $k \in \mathcal{K}_x$ and resources $i \in \{1, \cdots, C\}$. Hence:
	\begin{align*}
		\mathbb{E}[n_{x, T}] 
			& = \mathbb{E}[ \sum_{t=\tini}^{T} I_{x_t = x}  ] \\
			& \leq  \sum_{t=\tini}^{ T } \mathbb{P}[ \xi^x_{k, t} \geq 0 ] \\
			& \leq  \sum_{t=\tini}^{ T } \sum\limits_{ \substack{ k \in \mathcal{K}_x \\ i \in \{1, \cdots, C\}} } \mathbb{P}[ |\bar{c}_{k, t}(i) - \mu^c_k(i)| > \maxdevmeangeneralcase ] \\
			& \leq \probamaxdevmeangeneralcasewithouttime,
	\end{align*} 
	where the third inequality is derived with \eqref{eq-proof-general-case-concentration-mean}.
	\\
	Second, consider a pseudo-basis $x$ for \eqref{eq-linear-program-general-upperbound-opt-strategy} that is not a basis. Since $\det(A_x) = 0$, either every component of $A_x$ is $0$, in which case $\det(\bar{A}_{x, t}) = 0$ at every round $t$ and $x$ can never be selected, or there exists a basis $\tilde{x}$ for \eqref{eq-linear-program-general-upperbound-opt-strategy} with $\mathcal{K}_{\tilde{x}} \subset \mathcal{K}_x$ and $\mathcal{C}_{\tilde{x}} \subset \mathcal{C}_x$ along with coefficients $(\alpha_{kl})_{k \in \mathcal{K}_x - \mathcal{K}_{\tilde{x}}, l \in \mathcal{K}_{\tilde{x}}}$ such that $\mu^c_k(i) = \sum_{l \in \mathcal{K}_{\tilde{x}}} \alpha_{kl} \cdot \mu^c_l(i)$ for any resource $i \in \mathcal{C}_x$. Assuming we are in the second scenario and since $\tilde{x}$ is $\epsilon$-non-degenerate by Assumption \ref{assumption-simplying-assumption-general case}, we have $\sum_{k=1}^K \mu^c_k(j) \cdot \xi^{\tilde{x}}_k \leq b(j) - \epsilon$ for any $j \notin \mathcal{C}_{\tilde{x}}$. Take $i \in \mathcal{C}_x - \mathcal{C}_{\tilde{x}}$. Suppose that $x$ is feasible for \eqref{eq-algo-general-idea} at round $t$ and assume by contradiction that $\{ |\bar{c}_{k, t}(j) - \mu^c_k(j)| \leq \maxdevmeangeneralcase \}$ for all arms $k \in \mathcal{K}_x$ and resources $j \in \{1, \cdots, C\}$. Using the notations $\tilde{\xi}^x_{k, t} = \xi^x_{k, t} + \sum_{l \in \mathcal{K}_x - \mathcal{K}_{\tilde{x}}} \alpha_{kl} \cdot \xi^x_{l, t}$, we have, for any resource $j \in \mathcal{C}_{\tilde{x}}$:
	\begin{align*}
		b(j) 
		& = \sum_{l \in \mathcal{K}_{x}} \bar{c}_{l, t}(j) \cdot \xi^x_{l, t} \\
		& = \alpha(j) + \sum_{l \in \mathcal{K}_{x}} \mu^c_l(j) \cdot \xi^x_{l, t}  \\
		& = \alpha(j) + \sum_{l \in \mathcal{K}_{\tilde{x}}} \mu^c_l(j) \cdot \tilde{\xi}^x_{l, t},
	\end{align*}
	where $|\alpha(j)| \leq \maxdevmeangeneralcase$ since $\xi_k \in [0, 1] \; \forall k \in \{1, \cdots, K\}$ for any feasible solution to \eqref{eq-algo-general-idea}. We get $A_{\tilde{x}} \tilde{\xi}^x_{\mathcal{K}_{\tilde{x}}, t} = b_{\mathcal{C}_{\tilde{x}}} - \alpha_{\mathcal{C}_{\tilde{x}}}$ while $A_{ \tilde{x} } \xi^{\tilde{x}}_{\mathcal{K}_{\tilde{x}}} = b_{\mathcal{C}_{\tilde{x}}}$. We derive:
	\begin{align*}
		|\sum_{k=1}^K & \bar{c}_{k, t}(i) \cdot \xi^x_{k, t} - \sum_{k=1}^K \mu^c_k(i) \cdot \xi^{\tilde{x}}_{k}| \\
		& \leq |\sum_{k=1}^K \mu^c_k(i) \cdot \xi^x_{k, t} - \sum_{k=1}^K \mu^c_k(i) \cdot \xi^{\tilde{x}}_{k}| + 
\maxdevmeangeneralcase \\
		& \leq |\sum_{k \in \mathcal{K}_{\tilde{x}}} \mu^c_k(i) \cdot (\tilde{\xi}^x_{k, t} - \xi^{\tilde{x}}_{k})| + \frac{\epsilon}{4} \\
		& = \frac{\epsilon}{4} + |\mu^c_{\mathcal{K}_{\tilde{x}}}(i)^\tr A_{\tilde{x}}^{-1} \alpha_{\mathcal{C}_{\tilde{x}}}| \\
		& \leq \frac{\epsilon}{4} + \sqrt{C} \cdot \frac{1}{|\det(A_{\tilde{x}})|} \cdot \norm{ \adj( A_{\tilde{x}} ) \alpha_{\mathcal{C}_{\tilde{x}}} }_2 \\
		& \leq \frac{\epsilon}{4}  + \frac{(C+1)!}{\epsilon} \cdot \maxdevmeangeneralcase \\
		& \leq \frac{\epsilon}{2}.
	\end{align*}
	Thus we obtain:
	\begin{align*}
		\sum_{k=1}^K \bar{c}_{k, t}(i) \cdot \xi^x_{k, t} \leq \sum_{k=1}^K \mu^c_k(i) \cdot \xi^{\tilde{x}}_{k} + \frac{\epsilon}{2} \leq b(i) - \frac{\epsilon}{2} < b(i),
	\end{align*}
	a contradiction since this inequality must be binding by definition if $x$ is selected at round $t$. We finally conclude:
	\begin{align*}
		\mathbb{E}[n_{x, T}] 
			& = \mathbb{E}[ \sum_{t=\tini}^{T} I_{x_t = x}  ] \\
			& \leq  \sum_{t=\tini}^{ T } \sum\limits_{ \substack{ k \in \mathcal{K}_x \\ j \in \{1, \cdots, C\}} } \mathbb{P}[ |\bar{c}_{k, t}(j) - \mu^c_k(j)| > \maxdevmeangeneralcase ] \\
			& \leq \probamaxdevmeangeneralcasewithouttime.
	\end{align*} 
	
\subsection{Proof of Lemma \ref{lemma-load-balance-general-case}.}
\paragraph{Proof of \eqref{eq-load-balance-general-case-eq1}.}
Consider a resource $i \in \mathcal{C}_x$ and $u \geq 1$. We study $\mathbb{P}[b_{x, t}(i) - n_{x, t} \cdot b(i) \geq u]$ but the same technique can be used to bound $\mathbb{P}[b_{x, t}(i) - n_{x, t} \cdot b(i) \leq - u]$. If $b_{x, t}(i) - n_{x, t} \cdot b(i) \geq u$, it must be that $e^x_{i, \tau} = - 1$ for at least $s \geq \floor*{u}$ rounds $\tau = t_1 \leq \cdots \leq t_s \leq t - 1$ where $x$ was selected at Step-Simplex since the last time, denoted by $t_0 < t_1$, where $x$ was selected at Step-Simplex and the budget was below the target, i.e. $b_{x, t_0}(i) \leq n_{x, t_0} \cdot b(i)$ (because the amounts of resources consumed at each round are bounded by $1$). Moreover, we have:
			\begin{align*}
				\sum_{q=1}^s c_{a_{t_q}, t_q}(i)
					& = \sum_{\tau = t_0+1}^{t-1} I_{x_\tau = x} \cdot  c_{a_\tau, \tau}(i) \\
					& = b_{x, t}(i) - b_{x, t_0+1}(i)  \\
					& \geq ( n_{x, t} - n_{x, t_0} ) \cdot b(i) + u - 1 \\
					& \geq  s  \cdot b(i) + u - 1.
			\end{align*}
			Hence:
			\begin{align*}
				\mathbb{P}[ & b_{x, t}(i) - n_{x, t} \cdot b(i) \geq u] \\
					& \leq \sum_{s=\floor*{u}}^t \mathbb{P}[\sum_{q=1}^s c_{a_{t_q}, t_q}(i) \geq s \cdot b(i) + u - 1 \; ; \; e^x_{i, t_q} = - 1 \; \forall q \in \{1, \cdots, s\}] \\
					& \leq \sum_{s=\floor*{u}}^t \mathbb{P}[\sum_{q=1}^s c_{a_{t_q}, t_q}(i) \geq s \cdot b(i) \; ; \; \sum_{k=1}^K \mu^c_k(i) \cdot p^x_{k, t_q} \leq b(i) - \maxdelta \; \forall q \in \{1, \cdots, s\}] \\
					& + \sum_{\tau=\tini}^T \sum\limits_{ \substack{k \in \mathcal{K}_x \\ j \in \{1, \cdots, C\}}} \mathbb{P}[|\bar{c}_{k, \tau}(j) - \mu^c_k(j)| > \maxdevmeangeneralcase ] \\
					& \leq \sum_{s=\floor*{u}}^\infty \exp(- 2 s \cdot (\maxdelta)^2) + \probamaxdevmeangeneralcasewithouttime \cdot \frac{1}{T} \\
					& \leq 16 \frac{(C+1)!^2}{\epsilon^4} \cdot \exp(- u \cdot (\maxdelta)^2) + \probamaxdevmeangeneralcasewithouttime \cdot \frac{1}{T}.
			\end{align*}
The last inequality is obtained using $\exp(-x) \leq 1 - x/2$ for $x \in [0, 1]$. The third inequality is derived using Lemma \ref{lemma-martingale-inequality} for the first term and \eqref{eq-proof-general-case-concentration-mean} for the second term. The second inequality is obtained by observing that if $x$ was selected at time $\tau$ and $|\bar{c}_{k, \tau}(j) - \mu^c_k(j)| \leq \maxdevmeangeneralcase$ for any arm $k \in \mathcal{K}_x$ and resource $j \in \{1, \cdots, C\}$, then we must have $\delta^*_{x, \tau} \geq \maxdelta$. Indeed, using \eqref{eq-det-inequality} and \eqref{eq-proof-non-degeneracy-1}, we have, for $\delta \leq \maxdelta$ and any arm $k \in \mathcal{K}_x$:
\begin{align*}
	c^\tr \bar{A}^{-1}_{x, \tau} (b_{\mathcal{C}_x} + \delta \cdot e^x_\tau)
		& = \xi^x_{k, \tau} + \delta \cdot c^\tr \bar{A}^{-1}_{x, \tau} e^x_\tau \\
		& \geq \xi^{x}_{k} - |\xi^{x}_{k} - \xi^{x}_{k, \tau}| - \delta \cdot |c^\tr \bar{A}^{-1}_{x, \tau} e^x_\tau| \\
		& \geq \frac{\epsilon}{2} - \delta \cdot \norm{\bar{A}^{-1}_{x, \tau} e^x_\tau}_2 \\
		& \geq \frac{\epsilon}{2} - \delta \cdot \frac{1}{ |\det(\bar{A}_{x, \tau})| } \cdot \norm{ \adj(\bar{A}_{x, \tau}) e^x_\tau }_2 \\
		& \geq \frac{\epsilon}{2} - \frac{2 \delta}{\epsilon} \cdot \sqrt{C} \cdot C! \\
		& \geq 0,
\end{align*}
where $c = e_k$ and since $x$ is $\epsilon$-non-degenerate by Assumption \ref{assumption-simplying-assumption-general case}. Similarly, using \eqref{eq-det-inequality} and \eqref{eq-proof-non-degeneracy-3}, we have, for $\delta \leq \maxdelta$ and any resource $j \notin \mathcal{C}_x$:
\begin{align*}
	c^\tr \bar{A}^{-1}_{x, \tau} (b_{\mathcal{C}_x} + \delta \cdot e^x_\tau)
		& = \sum_{k \in \mathcal{K}_x} \bar{c}_{k, \tau}(j) \cdot \xi^x_{k, \tau} + \delta \cdot c^\tr \bar{A}^{-1}_{x, \tau} e^x_\tau \\
		& \leq \sum_{k \in \mathcal{K}_x} \mu^c_{k}(j) \cdot \xi^x_{k, \tau} + | \sum_{k \in \mathcal{K}_x} \bar{c}_{k, \tau}(j) \cdot \xi^x_{k, \tau}  - \sum_{k \in \mathcal{K}_x} \mu^c_{k}(j) \cdot \xi^x_{k, \tau}| + \delta \cdot |c^\tr \bar{A}^{-1}_{x, \tau} e^x_\tau| \\
		& \leq b(j) - \frac{\epsilon}{2} + \delta \cdot \sqrt{C} \cdot \norm{\bar{A}^{-1}_{x, \tau} e^x_\tau}_2 \\
		& \leq b(j) - \frac{\epsilon}{2} + \frac{2 \delta}{\epsilon} \cdot (C+1)! \\
		& \leq b(j),
\end{align*}
where $c = (\bar{c}_{k, \tau}(j))_{k \in \mathcal{K}_x}$ and since $x$ is $\epsilon$-non-degenerate by Assumption \ref{assumption-simplying-assumption-general case}. 

\paragraph{Proof of \eqref{eq-lower-bound-pulls-general-case}.}
	First observe that, using \eqref{eq-load-balance-general-case-eq1}, we have:
	\begin{align*}
		\max_{i \in \mathcal{C}_x} | & \mathbb{E}[b_{x, T}(i)] - \mathbb{E}[n_{x, T}] \cdot b(i) | \\
			& \leq \mathbb{E}[\max_{i \in \mathcal{C}_x} |b_{x, T}(i) - n_{x, T} \cdot b(i)|] \\
			& = \int_0^T \mathbb{P}[ \max_{i \in \mathcal{C}_x} |b_{x, T}(i) - n_{x, T} \cdot b(i) | \geq u ] \mathrm{d}u \\
			& = \sum_{i \in \mathcal{C}_x} \int_0^T \mathbb{P}[  |b_{x, T}(i) - n_{x, T} \cdot b(i) | \geq u ] \mathrm{d}u \\
			& \leq 32 C \cdot \frac{ (C+1)!^2}{\epsilon^4}  \cdot \int_0^T  \exp(- u \cdot (\maxdelta)^2) \mathrm{d}u + C + C \cdot \probamaxdevmeangeneralcasewithouttime \\	
			& = 2^9 C \cdot \frac{ (C+1)!^4 }{\epsilon^8} + C + C \cdot \probamaxdevmeangeneralcasewithouttime  \\
			& \leq 2^{10} C \cdot \frac{(C+3)!^4 }{\epsilon^8}.
	\end{align*}
	Now observe that, for any resource $i \in \mathcal{C}_x$, we have $\mathbb{E}[b_{x, T}(i)] = \sum_{k \in \mathcal{K}_x} \mu^c_k(i) \cdot \mathbb{E}[n^x_{k, T}]$. Hence, defining the vector $p = (\frac{\mathbb{E}[n^x_{k, T}]}{\mathbb{E}[n_{x, T}]})_{k \in \mathcal{K}_x}$, we get:
\begin{align*}
	  \mathbb{E}[n_{x, T}] \cdot \norm{  p - \xi^x  }_2
	  	& = \mathbb{E}[n_{x, T}] \cdot \norm{ A_x^{-1} A_x ( p - \xi^x )  }_2 \\
	  	& \leq \norminduced{ A_x^{-1} }_2 \cdot \norm{  \mathbb{E}[n_{x, T}] \cdot  A_x ( p - \xi^x )  }_2 \\
	  	& = \frac{1}{|\det(A_x)|} \cdot \norminduced{ \adj(A_x) }_2  \cdot \norm{ (\mathbb{E}[b_{x, T}(i)])_{i \in \mathcal{C}_x} - (\mathbb{E}[n_{x, T}] \cdot  b(i))_{i \in \mathcal{C}_x}  }_2 \\
		& \leq \frac{1}{\epsilon} \cdot (C+1)! \cdot \sqrt{C} \cdot 2^{10} C \cdot \frac{(C+3)!^4 }{\epsilon^8} \\	  	
	  	& \leq  2^{10}\frac{ (C+3)!^5 }{\epsilon^9},
\end{align*}
using Assumption \ref{assumption-simplying-assumption-general case}. Finally we obtain:
\begin{align*}
	 \mathbb{E}[n_{x, T}] \cdot \xi^x_k  - \mathbb{E}[n^x_{k, T}]
		& \leq \mathbb{E}[n_{x, T}] \cdot \norm{  p - \xi^x  }_2 \\
		& \leq 2^{10}  \frac{(C+3)!^5 }{\epsilon^9},
\end{align*}
for any arm $k \in \mathcal{K}_x$. 

\paragraph{Proof of \eqref{eq-load-balance-general-case-eq2}.}
Consider a resource $i \notin \mathcal{C}_x$ and assume that $b_{x, t}(i) - n_{x, t} \cdot b(i) \geq 2^8  \frac{(C+3)!^3}{\epsilon^6} \cdot \ln(T)$. By contradiction, suppose that: 
\begin{itemize}
	\item $|b_{x, t}(j) - n_{x, t} \cdot b(j)| \leq 16  \frac{(C+1)!^2}{\epsilon^4} \cdot \ln(T)$ for all resources $j \in \mathcal{C}_x$,
	\item $|\mu^c_k(j) - \tilde{c}_k(j)| \leq \frac{\epsilon^2}{ 8 \cdot (C+2)!} $ for all resources $j \in \{1, \cdots, C\}$ and for all arms $k \in \mathcal{K}_x$ such that $n^x_{k, t} \geq 2^6 \frac{ (C+2)!^2}{\epsilon^4} \cdot \ln(T)$, where $\tilde{c}_k(j)$ denotes the empirical average amount of resource $j$ consumed when selecting basis $x$ and pulling arm $k$, i.e. $\tilde{c}_k(j) = \frac{1}{n^x_{k, t}} \cdot \sum_{\tau=\tini}^{t-1} I_{x_\tau = x} \cdot I_{a_\tau = k} \cdot c_{k, \tau}(j)$. 
\end{itemize}
Observe that if $b_{x, t}(i) - n_{x, t} \cdot b(i) \geq 2^8 \frac{(C+3)!^3}{\epsilon^6} \cdot \ln(T)$, it must be that $x$ has been selected at least $2^8 \frac{(C+3)!^3}{\epsilon^6} \cdot \ln(T)$ times at Step-Simplex since $\tini$, i.e. $n_{x, t} \geq 2^8 \frac{(C+3)!^3}{\epsilon^6} \cdot \ln(T)$. We can partition $\mathcal{K}_x$ into two sets $\mathcal{K}^\mathbbm{1}_x$ and $\mathcal{K}^2_x$ such that $n^x_{k, t} \geq 2^6  \frac{(C+2)!^2}{\epsilon^4} \cdot \ln(T)$ for all $k \in \mathcal{K}^\mathbbm{1}_x$ and $n^x_{k, t} < 2^6  \frac{(C+2)!^2}{\epsilon^4} \cdot \ln(T)$ for all $k \in \mathcal{K}^2_x$. We get, for any $j \in \mathcal{C}_x$:
\begin{align*}
	16 \frac{(C+1)!^2}{\epsilon^4} \cdot \ln(T) 
	& \geq |b_{x, t}(j) - n_{x, t} \cdot b(j)| \\
	& \geq n_{x, t} \cdot | \sum_{k \in \mathcal{K}^\mathbbm{1}_x} \tilde{c}_k(j) \cdot p_k - b(j) | - \sum_{k \in \mathcal{K}^2_x} n^x_{k, t} \\
	& \geq n_{x, t} \cdot | \sum_{k \in \mathcal{K}^\mathbbm{1}_x} \tilde{c}_k(j) \cdot p_k - b(j) |  - 2^6  C \cdot \frac{(C+2)!^2}{\epsilon^4} \cdot \ln(T),
\end{align*}
where $p_k = \frac{n^x_{k, t}}{n_{x, t}}$ for $k \in \mathcal{K}^x_1$ and $p_k = 0$ otherwise. Hence:
\begin{align*}
	| \sum_{k \in \mathcal{K}_x} \mu^c_k(j) \cdot p_k - b(j) | 
		& \leq \max_{k \in \mathcal{K}^\mathbbm{1}_x} |\mu^c_k(j) - \tilde{c}_k(j)| + | \sum_{k \in \mathcal{K}^\mathbbm{1}_x} \tilde{c}_k(j) \cdot p_k - b(j) | \\
		& \leq \frac{\epsilon^2}{ 8 \cdot (C+2)!} +  \frac{(16 \frac{(C+1)!^2}{\epsilon^4} + 2^6 C \cdot \frac{(C+2)!^2}{\epsilon^4}) \cdot \ln(T)}{n_{x, t}} \\
		& \leq \frac{\epsilon^2}{ 4 \cdot (C+2)!},
\end{align*}
where we use the fact that $\sum_{k=1}^K p_k \leq 1$ and $p_k \geq 0$ for any arm $k$ for the first inequality and $n_{x, t} \geq 2^8  \frac{(C+3)!^3}{\epsilon^6} \cdot \ln(T)$ for the last one. We get: 
\begin{align*}
	\norm{ p - \xi^x }_2
	& = \norm{A_x^{-1} A_x ( p - \xi^x )  }_2 \\
	& \leq \norminduced{A_x^{-1}}_2 \cdot \norm{ A_x ( p - \xi^x ) }_2 \\
	& \leq \frac{1}{|\det(A_x)|} \cdot \norminduced{\adj(A_x)}_2 \cdot \sqrt{C} \cdot \frac{\epsilon^2}{ 4 \cdot (C+2)!} \\
	& \leq \frac{1}{\epsilon} \cdot (C+1)! \cdot \sqrt{C} \cdot \frac{\epsilon^2}{ 4 \cdot (C+2)!} \\
	& \leq \frac{\epsilon}{4 \cdot \sqrt{C}},
\end{align*}
using Assumption \ref{assumption-simplying-assumption-general case}. Hence:
\begin{align*}
	2^8  \frac{(C+3)!^3}{\epsilon^6} \cdot \ln(T) 
	& \leq b_{x, t}(i) - n_{x, t} \cdot b(i) \\
	& \leq n_{x, t} \cdot (\sum_{k \in \mathcal{K}^\mathbbm{1}_x} \tilde{c}_k(i) \cdot p_k - b(i))  + \sum_{k \in \mathcal{K}^2_x} n^x_{k, t} \\
	& \leq n_{x, t} \cdot (\sum_{k \in \mathcal{K}^\mathbbm{1}_x} \tilde{c}_k(i) \cdot p_k - b(i))  + 2^6 C \cdot \frac{(C+2)!^2}{\epsilon^4} \cdot \ln(T).
\end{align*}
Using the shorthand notation $c = (\mu^c_k(i))_{k \in \mathcal{K}_x}$, this implies:
\begin{align*}
	0 
	& \leq  \sum_{k \in \mathcal{K}^\mathbbm{1}_x} \tilde{c}_k(i) \cdot p_k - b(i) \\
	& \leq  \sum_{k \in \mathcal{K}^\mathbbm{1}_x} \mu^c_{k}(i) \cdot p_k + \frac{\epsilon^2}{ 8 \cdot (C+2)!} - ( \sum_{k \in \mathcal{K}_x} \mu^c_{k}(i) \cdot \xi^x_k + \epsilon ) \\
	& \leq c^\tr (p - \xi^x) - \frac{\epsilon}{2} \\
	& \leq \sqrt{C} \cdot \norm{p - \xi^x}_2  - \frac{\epsilon}{2} \\
	& < 0,
\end{align*}
a contradiction. Note that we use the fact that $\sum_{k=1}^K p_k \leq 1$ and $p_k \geq 0$ for any arm $k$ and Assumption \ref{assumption-simplying-assumption-general case} for the second inequality. We conclude that:
\begin{align*}
	\mathbb{P}[ & b_{x, t}(i) - n_{x, t} \cdot b(i) \geq 2^8  \frac{(C+3)!^3}{\epsilon^6} \cdot \ln(T)] \\
		& \leq \sum_{j \in \mathcal{C}_x} \mathbb{P}[|b_{x, t}(j) - n_{x, t} \cdot b(j)| \geq 16 \frac{(C+1)!^2}{\epsilon^4} \cdot \ln(T)] \\
		& + \sum\limits_{\substack{k \in \mathcal{K}_x \\ j \in \{1, \cdots, C\} }} \mathbb{P}[|\mu^c_k(j) - \tilde{c}_k(j)| \geq \frac{\epsilon^2}{ 8 \cdot (C+2)!} \; ; \; n^x_{k, t} \geq 2^6 \frac{(C+2)!^2}{\epsilon^4} \cdot \ln(T)] \\
		& \leq  2^5  C \cdot \frac{(C+1)!^2}{\epsilon^4 \cdot T} + C \cdot T \cdot \probamaxdevmeangeneralcase  \\
		& + \sum\limits_{\substack{k \in \mathcal{K}_x \\ j \in \{1, \cdots, C\} }} \sum_{s= 2^6 \cdot (C+2)!^2/\epsilon^4 \cdot \ln(T)} \mathbb{P}[|\mu^c_k(j) - \tilde{c}_k(j)| \geq \frac{\epsilon^2}{ 8 \cdot (C+2)!} \; ; \; n^x_{k, t} = s] \\
		& \leq 2  C \cdot T \cdot \probamaxdevmeangeneralcase + 2^8 C^2 \cdot \frac{ (C+2)!^2}{\epsilon^4 \cdot T^2}  \\
		& \leq 2^{10} \frac{(C+4)!^4}{\epsilon^6 \cdot T},
\end{align*}
where we use \eqref{eq-load-balance-general-case-eq1} for the second inequality and Lemma \ref{lemma-martingale-inequality} for the third inequality.

\subsection{Proof of Lemma \ref{lemma-bound-times-non-optimal-pulls-general-case}.}
		Consider any suboptimal basis $x \in \mathcal{B}$. The proof is along the same lines as for Lemmas \ref{lemma-bound-times-non-optimal-pulls}, \ref{lemma-bound-times-non-optimal-pulls-deterministic}, and \ref{lemma-bound-times-non-optimal-pulls-budget-time-horizon}. We break down the analysis in a series of facts where we emphasize the main differences. We start off with an inequality similar to Fact \ref{fact-wlog-assume-n-big}. We use the shorthand notation $\beta_x = 2^{10} \frac{(C+3)!^3 }{\epsilon^6} \cdot (\frac{\lambda}{\Delta_x})^2$.
		\begin{fact}
			\label{fact-wlog-assume-n-big-budget-generalcase}
			\begin{align}
				\mathbb{E}[n_{x, T}] \leq 
					& 2 \beta_x \cdot \ln( T ) + \probamaxdevmeangeneralcasewithouttime \nonumber \\
					& + \mathbb{E}[\sum_{t=\tini}^{T}  I_{x_t = x} \cdot I_{n_{x, t} \geq \beta_x \ln(t)} \cdot I_{ x^* \in \mathcal{B}_t }]. \label{eq-wlog-assume-n-big-generalcase}
			\end{align}
			\proof{Proof.}
				Similarly as in Fact \ref{fact-wlog-assume-n-big}, we have:
				\begin{align*}
					\mathbb{E}[n_{x, T}] \leq 
						& 2 \beta_x \cdot \ln( T ) + \mathbb{E}[\sum_{t=\tini}^{T}  I_{x_t = x} \cdot I_{n_{x, t} \geq \beta_x \ln(t)} ].
				\end{align*}
				This yields:
				\begin{align*}
					\mathbb{E}[n_{x, T}] 
						& \leq 2 \beta_x \cdot \ln( T ) + \mathbb{E}[\sum_{t=\tini}^{T}  I_{x_t = x} \cdot I_{n_{x, t} \geq \beta_x \ln(t)} \cdot I_{ x^* \in \mathcal{B}_t }] + \mathbb{E}[\sum_{t=\tini}^{T} I_{ x^* \notin \mathcal{B}_t} ].
				\end{align*}
				Using \eqref{eq-proof-non-degeneracy-1}, \eqref{eq-proof-non-degeneracy-3}, and Assumption \ref{assumption-simplying-assumption-general case}, we have:
				\begin{equation*}
					\xi^{x^*}_{k,t} = \xi^{x^*}_{k} - ( \xi^{x^*}_{k} - \xi^{x^*}_{k,t}) \geq \frac{\epsilon}{2} \geq 0,
				\end{equation*}
				for any $k \in \mathcal{K}_{x^*}$ and 
				\begin{align*}
					\sum_{k \in \mathcal{K}_{x^*}} \bar{c}_{k, t}(i) \cdot \xi^{x^*}_{k, t} 
						& = \sum_{k \in \mathcal{K}_{x^*}} \mu^c_{k}(i) \cdot \xi^{x^*}_{k} + ( \sum_{k \in \mathcal{K}_{x^*}} \bar{c}_{k, t}(i) \cdot \xi^{x^*}_{k, t}  - \sum_{k \in \mathcal{K}_{x^*}} \mu^c_{k}(i) \cdot \xi^{x^*}_{k}  ) \\
						& \leq b(i) - \epsilon + \frac{\epsilon}{2} \\
						& \leq b(i) - \frac{\epsilon}{2} \\
						& \leq b(i),
				\end{align*}
	for any resource $i \notin \mathcal{C}_{x^*}$ if $\{ |\bar{c}_{l, t}(j) - \mu^c_l(j)| \leq \maxdevmeangeneralcase \}$ for any arm $l \in \mathcal{K}_{x^*}$ and resource $j \in \{1, \cdots, C\}$. Hence:
				\begin{align*}
					\mathbb{E}[n_{x, T}] 
						& \leq 2 \beta_x \cdot \ln( T ) + \mathbb{E}[\sum_{t=\tini}^{T}  I_{x_t = x} \cdot I_{n_{x, t} \geq \beta_x \ln(t)} \cdot I_{ x^* \in \mathcal{B}_t }]  \\
						& + \sum_{t=\tini}^{ T } \sum\limits_{ \substack{ l \in \mathcal{K}_{x^*} \\ j \in \{1, \cdots, C\}}} \mathbb{P}[ |\bar{c}_{l, t}(j) - \mu^c_l(j)| > \maxdevmeangeneralcase]  \\
						& \leq 2 \beta_x \cdot \ln( T ) + \mathbb{E}[\sum_{t=\tini}^{T}  I_{x_t = x} \cdot I_{n_{x, t} \geq \beta_x \ln(t)} \cdot I_{ x^* \in \mathcal{B}_t }] + \probamaxdevmeangeneralcasewithouttime,
				\end{align*}
				where we bound the third term appearing in the right-hand side using \eqref{eq-proof-general-case-concentration-mean}.
			\endproof				
		\end{fact} \vspace{0.3cm}
		\noindent
		The remainder of this proof is dedicated to show that the last term in \eqref{eq-wlog-assume-n-big-generalcase} can be bounded by a constant. This term can be broken down in three terms similarly as in Lemmas \ref{lemma-bound-times-non-optimal-pulls} and \ref{lemma-bound-times-non-optimal-pulls-deterministic}.
		\begin{align}
		\mathbb{E}[ & \sum_{t=\tini}^{T}  I_{x_t = x} \cdot I_{n_{x, t} \geq \beta_x \ln(t)} \cdot I_{ x^* \in \mathcal{B}_t } ] \nonumber \\ 
			& \leq \mathbb{E}[\sum_{t=\tini}^{T}  I_{ \mathrm{obj}_{x, t} + E_{x, t} \geq \mathrm{obj}_{x^*, t} + E_{x^*, t}} \cdot I_{n_{x, t} \geq \beta_x \ln(t)} \cdot I_{x \in \mathcal{B}_t, x^* \in \mathcal{B}_t} ] \nonumber \\
			& \leq  
				\mathbb{E}[\sum_{t=\tini}^{T} I_{ \mathrm{obj}_{x, t} \geq \mathrm{obj}_{x} + E_{x, t} } \cdot I_{x \in \mathcal{B}_t} ] \label{eq-first-term-upper-bound-generalcase} \\
				& + \mathbb{E}[\sum_{t=\tini}^{T} I_{ \mathrm{obj}_{x^*, t} \leq \mathrm{obj}_{x^*}  - E_{x^*, t} } \cdot I_{x^* \in \mathcal{B}_t}] \label{eq-second-term-upper-bound-generalcase} \\
				& + \mathbb{E}[\sum_{t=\tini}^{T} I_{\mathrm{obj}_{x^*} < \mathrm{obj}_{x} + 2 E_{x, t} } \cdot I_{x \in \mathcal{B}_t} \cdot I_{n_{x, t} \geq \beta_x \ln(t)} ]. \label{eq-third-term-upper-bound-generalcase}
		\end{align}
		
		\begin{fact}
			\label{fact-bound-last-term-is-small-generalcase}
			\begin{align*}
				\mathbb{E}[ & \sum_{t=\tini}^{T} I_{\mathrm{obj}_{x^*} < \mathrm{obj}_{x} + 2 E_{x, t} } \cdot I_{x \in \mathcal{B}_t} \cdot I_{n_{x, t} \geq \beta_x \ln(t)} ] \leq 2^{11} \frac{(C+4)!^4}{\epsilon^6}.
			\end{align*}
		\proof{Proof.}
		Using the shorthand notation $\alpha_x = 8 (\frac{\lambda}{\Delta_x})^2$, we have:
		\begin{align*}
				\mathbb{E}[\sum_{t=\tini}^{T} & I_{\mathrm{obj}_{x^*} < \mathrm{obj}_{x} + 2 E_{x, t} } \cdot I_{x \in \mathcal{B}_t} \cdot I_{n_{x, t} \geq \beta_x \ln(t)} ] \\
					& \leq \mathbb{E}[\sum_{t=\tini}^{T} I_{ \Delta_x < 2 \lambda \cdot \max_{k \in \mathcal{K}_x} \epsilon_{k, t}  } \cdot I_{n_{x, t} \geq \beta_x \ln(t)} ] \\
					& \leq \mathbb{E}[\sum_{t=\tini}^{T} I_{ \min_{k \in \mathcal{K}_x} n_{k, t} \leq \alpha_x \ln(t) } \cdot I_{n_{x, t} \geq \beta_{x} \cdot  \ln(t)} ] \\
					& \leq \sum_{t=\tini}^{T} \sum_{k \in \mathcal{K}_x}  \mathbb{P}[n_{k, t} \leq \alpha_x \ln(t) \; ; \; n_{x, t} \geq \beta_{x} \ln(t)],
			\end{align*} 
			since $\sum_{l=1}^K \xi^x_{l, t} \leq 1$ and $\xi^x_{l, t} \geq 0$ for any arm $l$ when $x$ is feasible for \eqref{eq-algo-general-idea} at time $t$. Consider $k \in \mathcal{K}_x$ and assume that $n_{k, t} \leq \alpha_x \cdot \ln(t)$ and $n_{x, t} \geq \beta_{x} \cdot \ln(t)$. Suppose, by contradiction, that $|b_{x, t}(i) - n_{x, t} \cdot b(i)| \leq 32 \frac{(C+1)!^2}{\epsilon^4} \cdot \ln(t)$ for any resource $i \in \mathcal{C}_x$ and that $|\mu^c_l(i) - \tilde{c}_l(i)| \leq \frac{\epsilon^2}{ 8 \cdot (C+2)!}$ for any resource $i \in \{1, \cdots, C\}$ for all arms $l \in \mathcal{K}_x$ such that $n^x_{l, t} \geq 2^7 \frac{(C+2)!^2}{\epsilon^4} \cdot \ln(t)$, where $\tilde{c}_l(i)$ is the empirical average amount of resource $i$ consumed when selecting basis $x$ and pulling arm $l$, i.e. $\tilde{c}_l(i) = \frac{1}{n^x_{l, t}} \cdot \sum_{\tau=\tini}^{t-1} I_{x_\tau = x} \cdot I_{a_\tau = l} \cdot c_{l, \tau}(i)$. We can partition $\mathcal{K}_x - \{k\}$ into two sets $\mathcal{K}^\mathbbm{1}_x$ and $\mathcal{K}^2_x$ such that $n^x_{l, t} \geq 2^7  \frac{(C+2)!^2}{\epsilon^4} \cdot \ln(t)$ for all $l \in \mathcal{K}^\mathbbm{1}_x$ and $n^x_{l, t} < 2^7 \frac{ (C+2)!^2}{\epsilon^4} \cdot \ln(t)$ for all $l \in \mathcal{K}^2_x$. Similarly as in the proof of Lemma \ref{lemma-load-balance-general-case}, we have, for any resource $i \in \mathcal{C}_x$:
			\begin{align*}
				32  \frac{(C+1)!^2}{\epsilon^4} \cdot \ln(t) 
				& \geq |b_{x, t}(i) - n_{x, t} \cdot b(i)| \\
				& \geq n_{x, t} \cdot | \sum_{l \in \mathcal{K}^\mathbbm{1}_x} \tilde{c}_l(i) \cdot p_l - b(i) | - n_{k, t} - \sum_{l \in \mathcal{K}^2_x} n^x_{l, t} \\
				& \geq n_{x, t} \cdot | \sum_{l \in \mathcal{K}^\mathbbm{1}_x} \tilde{c}_l(i) \cdot p_l - b(i) | - \alpha_x \cdot \ln(t) - C \cdot \frac{2^7 \cdot (C+2)!^2}{\epsilon^4} \cdot \ln(t),
			\end{align*}
			where $p_l = \frac{n^x_{l, t}}{n_{x, t}}$ for $l \in \mathcal{K}^x_1$ and $p_l = 0$ otherwise. Hence:
			\begin{align*}
				| \sum_{l \in \mathcal{K}_x} \mu^c_l(i) \cdot p_l - b(i) | 
					& \leq \max_{l \in \mathcal{K}^\mathbbm{1}_x} |\mu^c_l(i) - \tilde{c}_l(i)| + | \sum_{l \in \mathcal{K}^\mathbbm{1}_x} \tilde{c}_l(i) \cdot p_l - b(i) | \\
					& \leq \frac{\epsilon^2}{ 8 \cdot (C+2)!} +  \frac{(\alpha_x + C \cdot \frac{2^7 \cdot (C+2)!^2}{\epsilon^4}) \cdot \ln(t)}{n_{x, t}} \\
					& \leq \frac{\epsilon^2}{ 4 \cdot (C+2)!}.
			\end{align*}
			To derive the first inequality, we use the fact that $\sum_{l=1}^K p_l \leq 1$ and $p_l \geq 0$ for any arm $l$. For the last inequality, we use $n_{x, t} \geq 2^{10} \frac{(C+3)!^3 }{\epsilon^6} \cdot (\frac{\lambda}{\Delta_x})^2 \cdot \ln(t)$ along with $\lambda \geq 1$ and $\Delta_x \leq \mathrm{obj}_{x^*} = \sum_{k=1}^K \mu^r_k \cdot \xi^{x^*}_{k} \leq \sum_{k=1}^K \xi^{x^*}_{k} \leq 1$ because of the time constraint imposed in \eqref{eq-linear-program-general-upperbound-opt-strategy}. We get: 
			\begin{align*}
				\xi^x_k
				& \leq \norm{ p - \xi^x }_2 \\
				& = \norm{A_x^{-1} A_x ( p - \xi^x )  }_2 \\
				& \leq \norminduced{A_x^{-1}}_2 \cdot \norm{ A_x ( p - \xi^x ) }_2 \\
				& \leq \frac{1}{|\det(A_x)|} \cdot \norminduced{\adj(A_x)}_2 \cdot \sqrt{C} \cdot \frac{\epsilon^2}{ 4 \cdot (C+2)!} \\
				& \leq \frac{1}{\epsilon} \cdot (C+1)! \cdot \sqrt{C} \cdot \frac{\epsilon^2}{ 4 \cdot (C+2)!} \\
				& \leq \frac{\epsilon}{2},
			\end{align*}			
			a contradiction since $x$ is $\epsilon$-non-degenerate by Assumption \ref{assumption-simplying-assumption-general case}. We conclude that:
		\begin{align*}
			\mathbb{E}[\sum_{t=\tini}^{T} & I_{\mathrm{obj}_{x^*} < \mathrm{obj}_{x} + 2 E_{x, t} } \cdot I_{x \in \mathcal{B}_t} \cdot I_{n_{x, t} \geq \beta_x \cdot \ln(t)} ] \\
				& \leq C \cdot \sum_{t=\tini}^{T} \sum_{i \in \mathcal{C}_x} \mathbb{P}[|b_{x, t}(i) - n_{x, t} \cdot b(i)| \geq 32 \frac{(C+1)!^2}{\epsilon^4} \cdot \ln(t)] \\
				& + C \cdot \sum_{t=\tini}^{T} \sum\limits_{\substack{l \in \mathcal{K}_x \\ i \in \mathcal{C}_x }} \mathbb{P}[|\mu^c_l(i) - \tilde{c}_l(i)| \geq \frac{\epsilon^2}{ 8 \cdot (C+2)!} \; ; \; n^x_{l, t} \geq 2^7 \frac{(C+2)!^2}{\epsilon^4} \cdot \ln(t)] \\
				& \leq  32 C^2 \cdot \frac{(C+1)!^2}{\epsilon^4 } \cdot \frac{\pi^2}{6} + C \cdot \probamaxdevmeangeneralcasewithouttime  \\
				& + \sum\limits_{\substack{l \in \mathcal{K}_x \\ j \in \mathcal{C}_x }} \sum_{s= 2^7 \cdot (C+2)!^2/\epsilon^4 \cdot \ln(T)} \mathbb{P}[|\mu^c_l(j) - \tilde{c}_l(j)| \geq \frac{\epsilon^2}{ 8 \cdot (C+2)!} \; ; \; n^x_{l, T} = s] \\
				& \leq 4  C \cdot \probamaxdevmeangeneralcasewithouttime  + 2^7 C^2 \cdot \frac{(C+2)!^2}{\epsilon^4} \cdot \frac{\pi^2}{6}  \\
				& \leq 2^{11} \frac{(C+4)!^4}{\epsilon^6},
		\end{align*}
		where we use \eqref{eq-load-balance-general-case-eq1} for the second inequality and Lemma \ref{lemma-martingale-inequality} for the third inequality.

		\endproof
		\end{fact} \vspace{0.3cm}

		\begin{fact}
			\label{fact-study-bound-first-term-general-case}
			$$
				\mathbb{E}[\sum_{t=\tini}^{T} I_{ \mathrm{obj}_{x, t} \geq \mathrm{obj}_{x} + E_{x, t} } \cdot I_{x \in \mathcal{B}_t}  ] \leq 2^{10}  \frac{(C+3)!^2}{\epsilon^6}.
			$$
			\proof{Proof.}
			First observe that:
			\begin{align*}
				\mathbb{E}[\sum_{t=\tini}^{T} I_{ \mathrm{obj}_{x, t} \geq \mathrm{obj}_{x} + E_{x, t} } \cdot I_{x \in \mathcal{B}_t}  ]  					& \leq \mathbb{E}[\sum_{t=\tini}^{T} I_{ \mathrm{obj}_{x, t} \geq \mathrm{obj}_{x} + E_{x, t} } \cdot I_{x \in \mathcal{B}_t} \cdot I_{ |\det( \bar{A}_{x, t})| \geq \epsilon/2 }  ] \\
					& + \mathbb{E}[\sum_{t=\tini}^{T} I_{ |\det( \bar{A}_{x, t})| < \epsilon/2  } ]  \\
					& \leq \mathbb{E}[\sum_{t=\tini}^{T} I_{ \mathrm{obj}_{x, t} \geq \mathrm{obj}_{x} + E_{x, t} } \cdot I_{x \in \mathcal{B}_t} \cdot I_{ |\det( \bar{A}_{x, t})| \geq \epsilon/2 }  ] \\					
					& + \sum_{t=\tini}^{T}  \sum\limits_{ \substack{ k \in \mathcal{K}_x \\ i \in \mathcal{C}_x} } \mathbb{P}[|\bar{c}_{k, t}(i) - \mu^c_k(i)| > \maxdevmeangeneralcase] \\
					& \leq \mathbb{E}[\sum_{t=\tini}^{T} I_{ \mathrm{obj}_{x, t} \geq \mathrm{obj}_{x} + E_{x, t} } \cdot I_{x \in \mathcal{B}_t} \cdot I_{ |\det( \bar{A}_{x, t})| \geq \epsilon/2  } ] \\
					& + \probamaxdevmeangeneralcasewithouttime,
			\end{align*}				
			where we use the preliminary work of Section \ref{secproofs-preliminarywork-general-case} and in particular \eqref{eq-det-inequality}. The key observation now is that if $\mathrm{obj}_{x, t} \geq \mathrm{obj}_{x} + E_{x, t}$, $x \in \mathcal{B}_t$, and $|\det( \bar{A}_{x, t})| \geq \epsilon/2$, at least one of the following events $\{ \bar{r}_{k, t} \geq \mu^r_k + \epsilon_{k, t} \}$, for $k \in \mathcal{K}_x$, or $\{ |\bar{c}_{k, t}(i) - \mu^c_k(i)| \geq \epsilon_{k, t} \}$, for $k \in \mathcal{K}_x$ and $i \in \{1, \cdots, C\}$, occurs. Otherwise we have:
			
\begin{align*}
	\mathrm{obj}_{x, t} - \mathrm{obj}_{x} 
		& = (\bar{A}^{-1}_{x, t} b_{\mathcal{C}_x})^\tr \bar{r}_{\mathcal{K}_x, t} - (A^{-1}_{x} b_{\mathcal{C}_x})^\tr \mu^r_{\mathcal{K}_x} \\
		& < (\bar{A}^{-1}_{x, t} b_{\mathcal{C}_x})^\tr ( \mu^r_{\mathcal{K}_x} + \epsilon_{\mathcal{K}_x, t} ) - (A^{-1}_{x} b_{\mathcal{C}_x})^\tr \mu^r_{\mathcal{K}_x} \\
		& = \frac{1}{\lambda} \cdot E_{x, t} + ((\bar{A}^{-1}_{x, t} - A^{-1}_{x}) b_{\mathcal{C}_x})^\tr \mu^r_{\mathcal{K}_x},
\end{align*}
where the first inequality is a consequence of the fact that $x$ is feasible for \eqref{eq-algo-general-idea}, i.e. $\bar{A}^{-1}_{x, t} b_{\mathcal{C}_x} \geq 0$ with at least one non-zero coordinate since $\bar{c}_{k, t}(i) \geq \epsilon$ for all arms $k$ and resource $i$ by Assumption \ref{assumption-simplying-assumption-general case}. Writing $\mu^c_k(i) = \bar{c}_{k, t}(i) + u_{i, k} \cdot \epsilon_{k, t}$, with $u_{i, k} \in [-1,1]$ for all $(i, k) \in \mathcal{C}_x \times \mathcal{K}_x$, and defining the matrix $U = (u_{i, k})_{(i, k) \in \mathcal{C}_x \times \mathcal{K}_x}$, we get:
\begin{align*}
	& \mathrm{obj}_{x, t} - \mathrm{obj}_{x} \\
	& < \frac{E_{x, t}}{\lambda} + | (\bar{A}^{-1}_{x, t} - ( \bar{A}_{x, t} + U \diag(\epsilon_{\mathcal{K}_x, t})  )^{-1}  )b_{\mathcal{C}_x})^\tr \mu^r_{\mathcal{K}_x} | \\
	& = \frac{E_{x, t}}{\lambda} +  | (\bar{A}^{-1}_{x, t} U (I + \diag(\epsilon_{\mathcal{K}_x, t}) \bar{A}^{-1}_{x, t} U)^{-1} \diag(\epsilon_{\mathcal{K}_x, t}) \bar{A}^{-1}_{x, t} b_{\mathcal{C}_x})^\tr \mu^r_{\mathcal{K}_x} | \\
	& \leq \frac{E_{x, t}}{\lambda} + \frac{1}{\epsilon} \cdot | \det( \bar{A}_{x, t} + U \diag(\epsilon_{\mathcal{K}_x, t}))| \cdot | (\bar{A}^{-1}_{x, t} U (I + \diag(\epsilon_{\mathcal{K}_x, t}) \bar{A}^{-1}_{x, t} U)^{-1} \diag(\epsilon_{\mathcal{K}_x, t}) \bar{A}^{-1}_{x, t} b_{\mathcal{C}_x})^\tr \mu^r_{\mathcal{K}_x} | \\
	& = \frac{E_{x, t}}{\lambda}+ \frac{1}{\epsilon} \cdot | ( \adj(\bar{A}_{x, t}) U \adj(I + \diag(\epsilon_{\mathcal{K}_x, t}) \bar{A}^{-1}_{x, t} U) \diag(\epsilon_{\mathcal{K}_x, t}) \bar{A}^{-1}_{x, t} b_{\mathcal{C}_x})^\tr \mu^r_{\mathcal{K}_x} | \\
	& \leq \frac{E_{x, t}}{\lambda}+ \frac{1}{\epsilon} \cdot \norminduced{\adj(\bar{A}_{x, t}) U \adj(I + \diag(\epsilon_{\mathcal{K}_x, t}) \bar{A}^{-1}_{x, t} U)}_2 \cdot \norm{\diag(\epsilon_{\mathcal{K}_x, t}) \bar{A}^{-1}_{x, t} b_{\mathcal{C}_x}}_2 \cdot \norm{\mu^r_{\mathcal{K}_x}}_2  \\	
	& \leq \frac{E_{x, t}}{\lambda}+ \frac{1}{\epsilon} \cdot \norminduced{\adj(\bar{A}_{x, t}) U}_2 \cdot \norminduced{\adj(I + \diag(\epsilon_{\mathcal{K}_x, t}) \bar{A}^{-1}_{x, t} U)}_2 \cdot \frac{E_{x, t}}{\lambda} \cdot \sqrt{C} \\	
	& \leq \frac{E_{x, t}}{\lambda}+ \frac{1}{\epsilon} \cdot (C+1)! \cdot \norminduced{\adj(I + \diag(\epsilon_{\mathcal{K}_x, t}) \bar{A}^{-1}_{x, t} U)}_2 \cdot \frac{E_{x, t}}{\lambda}  \\
	& \leq \frac{E_{x, t}}{\lambda}+ \frac{2}{\epsilon} \cdot (C+1)!^2 \cdot \frac{E_{x, t}}{\lambda} \\
	& = E_{x, t}.
\end{align*}
We use the Woodbury matrix identity to derive the first equality and the matrix determinant lemma for the second equality. The second inequality is derived from Assumption \ref{assumption-simplying-assumption-general case} since $\epsilon \leq \det(A_x) = \det( \bar{A}_{x, t} + U \diag(\epsilon_{\mathcal{K}_x, t}) )$ by definition of $U$. The fourth inequality is derived from the observation that $\diag(\epsilon_{\mathcal{K}_x, t}) \bar{A}^{-1}_{x, t} b_{\mathcal{C}_x}$ is the vector $(\epsilon_{k, t} \cdot \xi^x_{k,t})_{k \in \mathcal{K}_x}$. The fifth inequality is obtained by observing that the components of $\bar{A}_{x, t}$ and $U$ are all smaller than $1$ in absolute value. The sixth inequality is obtained by observing that the elements of $\bar{A}^{-1}_{x, t}$ are smaller than $\frac{2}{\epsilon} \cdot (C-1)!$ since $\det(\bar{A}_{x, t}) \geq \frac{\epsilon}{2}$ and that $\epsilon_{k, t} \leq \frac{\epsilon}{2 \cdot C!}$ for all arms $k$ as a result of the initialization phase. We get: 
\begin{align*}
				\mathbb{E}[\sum_{t=\tini}^{T} I_{ \mathrm{obj}_{x, t} \geq \mathrm{obj}_{x} + E_{x, t} } \cdot I_{x \in \mathcal{B}_t}  ]  					& \leq \sum_{t=\tini}^{T} \sum\limits_{ \substack{ k \in \mathcal{K}_x \\ i \in \mathcal{C}_x} } \mathbb{P}[| c_{k, t}(i) - \mu^c_k(i) | \geq \epsilon_{k, t}] \\
				    & + \sum_{t=\tini}^{T} \sum_{k \in \mathcal{K}_x} \mathbb{P}[ r_{k, t}  \geq \mu^r_k + \epsilon_{k, t}] \\
				    & + \probamaxdevmeangeneralcasewithouttime \\
				    & \leq 2 \cdot \probamaxdevmeangeneralcasewithouttime. 
\end{align*}


			\endproof
		\end{fact} \vspace{0.3cm}

		\begin{fact}
			$$
				\mathbb{E}[\sum_{t=\tini}^{T} I_{ \mathrm{obj}_{x^*, t} \leq \mathrm{obj}_{x^*} - E_{x, t} } \cdot I_{x^* \in \mathcal{B}_t}  ] \leq 2^{10} \frac{(C+3)!^2}{\epsilon^6}.
			$$
		\end{fact} \vspace{0.3cm}
		We omit the proof since it is almost identical to the proof of Fact \ref{fact-study-bound-first-term-general-case}.

\subsection{Proof of Theorem \ref{lemma-log-B-regret-bound-general-case}.}
Along the same lines as for Theorem \ref{lemma-log-B-regret-bound-time-horizon}, we build upon \eqref{eq-simplified-general-upper-bound-on-regret}:
\begin{align*}
			R_{B(1), \cdots, B(C-1), T}
				& \leq T \cdot \sum_{k=1}^K \mu^r_k \cdot \xi^{x^*}_k - \mathbb{E}[\sum_{t=1}^{\tau^*} r_{a_t, t}] + O(1) \\
				& = T \cdot \sum_{k=1}^K \mu^r_k \cdot \xi^{x^*}_k - \mathbb{E}[\sum_{t=1}^{T} r_{a_t, t}] + \mathbb{E}[\sum_{t=\tau^*+1}^{T} r_{a_t, t}]  + O(1)  \\					
				& \leq T \cdot \sum_{k=1}^K \mu^r_k \cdot \xi^{x^*}_k - \mathbb{E}[\sum_{t=1}^{T} r_{a_t, t}] + \sigma \cdot \mathbb{E}[ \min\limits_{i=1, \cdots, C} \sum_{t=\tau^*+1}^{T} c_{a_t, t}(i)]  + O(1)  \\
				& \leq T \cdot \sum_{k=1}^K \mu^r_k \cdot \xi^{x^*}_k - \mathbb{E}[\sum_{t=1}^{T} r_{a_t, t}] + \sigma \cdot \sum_{i=1}^C \mathbb{E}[ (\sum_{t=1}^{T} c_{a_t, t}(i) - B(i))_+]  + O(1).
\end{align*}
The second inequality is a direct consequence of Assumption \ref{assumption-cost-bounds-rewards-general-case}. To derive the last inequality, observe that if $\tau^* = T + 1$, we have:
$$
	\sum_{t=\tau^*+1}^{T} c_{a_t, t}(j) = 0 \leq \sum_{i=1}^C \mathbb{E}[ (\sum_{t=1}^{T} c_{a_t, t}(i) - B(i))_+],
$$ 
for any $j \in \{ 1, \cdots, C \}$ while if $\tau^* < T + 1$ we have run out of resources before the end of the game, i.e. there exists $j \in \{1, \cdots, C\}$ such that $\sum_{t=1}^{\tau^*} c_{a_t, t}(j) \geq B(j)$, which implies that:
\begin{align*}
	\min\limits_{i=1, \cdots, C} \sum_{t=\tau^*+1}^{T} c_{a_t, t}(i) 
		& \leq \sum_{t=\tau^*+1}^{T} c_{a_t, t}(j) \\
		& \leq \sum_{t=\tau^*+1}^{T} c_{a_t, t}(j) + \sum_{t=1}^{\tau^*} c_{a_t, t}(j) -  B(j) \\
		& = (\sum_{t=1}^{T} c_{a_t, t}(j) - B(j))_+ \\
		& \leq \sum_{i=1}^C (\sum_{t=1}^{T} c_{a_t, t}(i) - B(i))_+.
\end{align*}
Now observe that, for any resource $i \in \{1, \cdots, C\}$:
\begin{align*}
	\mathbb{E}[ (\sum_{t=1}^{T} c_{a_t, t}(i) - B)_+] 
		& \leq  \mathbb{E}[ (\sum_{t=\tini}^{T} c_{a_t, t}(i) - b(i) )_+] + K \cdot \tinivalue \cdot \ln(T) \\
		& =  \mathbb{E}[ ( \sum_{x \text{ basis for } \eqref{eq-linear-program-general-upperbound-opt-strategy} } \{ b_{x, T}(i) - n_{x, T} \cdot b(i) \} )_+ ] + O( K \cdot \frac{(C+2)!^4}{\epsilon^6} \cdot \ln(T)) \\
		& \leq 	\sum_{x \in \mathcal{B}} \mathbb{E}[ (b_{x, T}(i) - n_{x, T} \cdot b(i) )_+ ] + \sum_{x \notin \mathcal{B}} \mathbb{E}[n_{x, T}] \\
		& + \sum_{ \substack{ x \text{ pseudo-basis for } \eqref{eq-linear-program-general-upperbound-opt-strategy} \\ \text{ with } \det(A_x) = 0}} \mathbb{E}[n_{x, T}] +  O( K \cdot \frac{(C+2)!^4}{\epsilon^6} \cdot \ln(T)) \\
		& \leq  \sum_{x \in \mathcal{B}} \int_0^T \mathbb{P}[ b_{x, T}(i) - n_{x, T} \cdot b(i) \geq u] \mathrm{d}u  +  O( K \cdot \frac{(C+3)!^4}{\epsilon^6} \cdot \ln(T)) \\
		& \leq \sum_{x \in \mathcal{B}} T \cdot \mathbb{P}[ b_{x, T}(i) - n_{x, T} \cdot b(i) \geq 2^8  \frac{(C+3)!^4}{\epsilon^6} \cdot \ln(T)] \\
		& + 2^8 |\mathcal{B}| \cdot \frac{(C+3)!^4}{\epsilon^6} \cdot \ln(T) +  O(K \cdot \frac{(C+3)!^4}{\epsilon^6} \cdot \ln(T)) \\
		& = O(\frac{|\mathcal{B}| \cdot (C+3)!^4 }{\epsilon^6} \cdot \ln(T)),
\end{align*}	
where we use the fact that the amounts of resources consumed at any time period are no larger than $1$ for the first and second inequalities, Lemma \ref{lemma-bound-times-infeasible-pulls-general-case} for the third inequality and inequalities \eqref{eq-load-balance-general-case-eq1} and \eqref{eq-load-balance-general-case-eq2} from Lemma \ref{lemma-load-balance-general-case} along with the fact that there are at least $K$ feasible basis for \eqref{eq-linear-program-general-upperbound-opt-strategy} (corresponding to single-armed strategies) for the last equality. Plugging this back into the regret bound yields:
\begin{align}
	R_{B(1), \cdots, B(C-1), T}
				& \leq T \cdot \sum_{k=1}^K \mu^r_k \cdot \xi^{x^*}_k - \mathbb{E}[\sum_{t=\tini}^{T} r_{a_t, t}] + O(\frac{\sigma \cdot |\mathcal{B}| \cdot (C+3)!^4 }{\epsilon^6} \cdot \ln(T)) \nonumber \\			
				& = T \cdot \sum_{k=1}^K \mu^r_k \cdot \xi^{x^*}_k  - \sum_{x \in \mathcal{B}} \sum_{k=1}^K \mu^r_k \cdot \mathbb{E}[n^x_{k, T} ] + O(\frac{\sigma \cdot |\mathcal{B}| \cdot (C+3)!^4 }{\epsilon^6} \cdot \ln(T)) \nonumber \\
				& \leq T \cdot \sum_{k=1}^K \mu^r_k \cdot \xi^{x^*}_k  - \sum_{x \in \mathcal{B}} (\sum_{k=1}^K \mu^r_k \cdot \xi^x_k) \cdot \mathbb{E}[n_{x, T} ] + O(\frac{\sigma \cdot |\mathcal{B}| \cdot (C+3)!^4 }{\epsilon^6} \cdot \ln(T)) \nonumber \\
				& = \sum_{k=1}^K \mu^r_k \cdot \xi^{x^*}_k \cdot ( T - \sum_{x \in \mathcal{B} \; | \; \Delta_x = 0} \mathbb{E}[n_{x, T}] ) \\
				& - \sum_{x \in \mathcal{B} \; | \; \Delta_x > 0} (\sum_{k=1}^K \mu^r_k \cdot \xi^x_k) \cdot \mathbb{E}[n_{x, T} ] + O(\frac{\sigma \cdot |\mathcal{B}| \cdot (C+3)!^4 }{\epsilon^6} \cdot \ln(T)) \nonumber \\
				& = \sum_{k=1}^K \mu^r_k \cdot \xi^{x^*}_k \cdot ( \tini + \sum_{x \in \mathcal{B} \; | \; \Delta_x > 0} \mathbb{E}[n_{x, T}] + \sum_{x \notin \mathcal{B}}  \mathbb{E}[n_{x, T}] + \sum_{ \substack{ x \text{ pseudo-basis for } \eqref{eq-linear-program-general-upperbound-opt-strategy} \\ \text{ with } \det(A_x) = 0}} \mathbb{E}[n_{x, T}] ) \nonumber \\
				& - \sum_{x \in \mathcal{B} \; | \; \Delta_x > 0} (\sum_{k=1}^K \mu^r_k \cdot \xi^x_k) \cdot \mathbb{E}[n_{x, T} ]  + O(\frac{\sigma \cdot |\mathcal{B}| \cdot (C+3)!^4 }{\epsilon^6} \cdot \ln(T)) \nonumber \\
				& \leq \sum_{x \in \mathcal{B} \; | \; \Delta_x > 0}  \Delta_x \cdot \mathbb{E}[n_{x, T} ] + O(\frac{\sigma \cdot |\mathcal{B}| \cdot (C+3)!^4 }{\epsilon^6} \cdot \ln(T)) \label{eq-proof-log-B-start-here-sqrt-B-general-case} \\	
				& \leq 2^{10} \frac{(C+3)!^3 \cdot \lambda^2}{\epsilon^6} \cdot (\sum_{x \in \mathcal{B} \; | \; \Delta_x > 0} \frac{1}{\Delta_x}) \cdot \ln(T) + O(\frac{\sigma \cdot |\mathcal{B}| \cdot (C+3)!^4 }{\epsilon^6} \cdot \ln(T)), \nonumber 
\end{align}
where we use \eqref{eq-lower-bound-pulls-general-case} from Lemma \ref{lemma-load-balance-general-case}	for the second inequality, Lemma \ref{lemma-bound-times-infeasible-pulls-general-case} for the third inequality. To derive the last inequality, we use Lemma \ref{lemma-bound-times-non-optimal-pulls-general-case} and the fact that $\Delta_x \leq \mathrm{obj}_{x^*} = \sum_{k=1}^K \mu^r_k \cdot \xi^{x^*}_{k} \leq \sum_{k=1}^K \xi^{x^*}_{k} \leq 1$.

\subsection{Proof of Theorem \ref{lemma-sqrt-B-regret-bound-general-case}.}
Along the same lines as for the proof of Theorem \ref{lemma-sqrt-B-regret-bound-time-horizon}, we start from inequality \eqref{eq-proof-log-B-start-here-sqrt-B-general-case} derived in the proof of Theorem \ref{lemma-log-B-regret-bound-general-case} and apply Lemma \ref{lemma-bound-times-non-optimal-pulls-general-case} only if $\Delta_x$ is big enough, taking into account the fact that:
$$
	\sum_{x \in \mathcal{B}} \mathbb{E}[n_{x, T}] \leq T.
$$
Specifically, we have: 
\begin{align*} 
	& R_{B(1), \cdots, B(C-1), T} \\
		& = \sup\limits_{ \substack{(n_x)_{x \in \mathcal{B}} \geq 0 \\ \sum_{x \in \mathcal{B}} n_x \leq T }  } \{ \; \sum_{x \in \mathcal{B} \; | \; \Delta_x > 0} \min( \Delta_x \cdot n_x, 2^{10} \frac{(C+3)!^3 \cdot \lambda^2}{\epsilon^6} \cdot \frac{\ln(T)}{\Delta_x} + 2^{11} \frac{(C+4)!^2}{\epsilon^6} \cdot \Delta_x ) \; \} ) \\
		& + O(\frac{\sigma \cdot |\mathcal{B}| \cdot (C+3)!^4 }{\epsilon^6} \cdot \ln(T)) \\
		& = \sup\limits_{ \substack{(n_x)_{x \in \mathcal{B}} \geq 0 \\ \sum_{x \in \mathcal{B}} n_x \leq T }  } \{ \; \sum_{x \in \mathcal{B} \; | \; \Delta_x > 0} \min( \Delta_x \cdot n_x, 2^{10} \frac{(C+3)!^3 \cdot \lambda^2}{\epsilon^6} \cdot \frac{\ln(T)}{\Delta_x} ) \; \} + O(\frac{\sigma \cdot |\mathcal{B}| \cdot (C+3)!^4 }{\epsilon^6} \cdot \ln(T)) \\
		& \leq \sup\limits_{ \substack{(n_x)_{x \in \mathcal{B}} \geq 0 \\ \sum_{x \in \mathcal{B}} n_x \leq T }  } \{ \; \sum_{x \in \mathcal{B}} \sqrt{ 2^{10} \frac{(C+3)!^3 \cdot \lambda^2}{\epsilon^6} \cdot \ln(T)  \cdot  n_x} \; \} + O(\frac{\sigma \cdot |\mathcal{B}| \cdot (C+3)!^4 }{\epsilon^6} \cdot \ln(T)) \\
		& \leq 2^5 \frac{(C+3)!^{2} \cdot \lambda}{\epsilon^3} \cdot \sqrt{ \ln(  T )  } \cdot \sup\limits_{ \substack{(n_x)_{x \in \mathcal{B}} \geq 0 \\ \sum_{x \in \mathcal{B}} n_x \leq T }  } \{ \; \sum_{x \in \mathcal{B} } \sqrt{n_x} \; \} + O(\frac{\sigma \cdot |\mathcal{B}| \cdot (C+3)!^4 }{\epsilon^6} \cdot \ln(T)) \\
		& \leq 2^5 \frac{(C+3)!^{2} \cdot \lambda}{\epsilon^3} \cdot \sqrt{ \sigma \cdot |\mathcal{B}| \cdot T \cdot \ln( T )  } + O(\frac{\sigma \cdot |\mathcal{B}| \cdot (C+3)!^4 }{\epsilon^6} \cdot \ln(T)),
\end{align*}
where we use the fact that $\Delta_x \leq 1$ (see the proof of Theorem \ref{lemma-log-B-regret-bound-general-case}) for the second equality, we maximize over each $\Delta_x \geq 0$ to derive the first inequality, and we use Cauchy-Schwartz for the last inequality.

\subsection{Proof of Theorem \ref{lemma-regret-bound-b-small-general-case}.}
	Define $\tilde{b}(i) = B(i) / \tilde{T}$ for any $i \in \{1, \cdots, C\}$. If the decision maker stops pulling arms at round $\tilde{T}$ at the latest, all the results derived in Section \ref{sec-stochastic-multiple-budget} hold as long as we substitute $T$ with $\tilde{T}$ and we get:
	\begin{align*}
		\tilde{T} \cdot \tilde{\text{opt}}- \mathbb{E}[\sum_{t=1}^{\min(\tau^*, \tilde{T})} r_{a_t, t}]
		 & \leq X,
	\end{align*}
	where $X$ denotes the right-hand side of the regret bound derived in either Theorem \ref{lemma-log-B-regret-bound-general-case} or Theorem \ref{lemma-sqrt-B-regret-bound-general-case} and $\tilde{\text{opt}}$ denotes the optimal value of \eqref{eq-linear-program-general-upperbound-opt-strategy} when $b(i)$ is substituted with $\tilde{b}(i)$ for any $i \in \{1, \cdots, C\}$. The key observation is that $\tilde{T} \cdot \tilde{\text{opt}} = T \cdot \text{opt}$, where $\text{opt}$ denotes the optimal value of \eqref{eq-linear-program-general-upperbound-opt-strategy}, because the time constraint is redundant in \eqref{eq-linear-program-general-upperbound-opt-strategy} even when $b(i)$ is substituted with $\tilde{b}(i)$ for any $i \in \{1, \cdots, C\}$. This is enough to show the claim as we get:
	\begin{align*}
		X 
			& \geq T \cdot \text{opt} - \mathbb{E}[\sum_{t=1}^{\tau^*} r_{a_t, t}] \\
			& \geq R_{B(1), \cdots, B(C-1), T},
	\end{align*}
	where we use Lemma \ref{lemma-general-bound-optimal-policy} for the last inequality.

\subsection{Proof of Theorem \ref{lemma-relax-assumption-cost-dominate-rewards-general-case}.}
	The only difference with the proofs of Theorems \ref{lemma-log-B-regret-bound-general-case} and \ref{lemma-sqrt-B-regret-bound-general-case} lies in how we bound $\mathbb{E}[\sum_{t=\tau^*+1}^{T} r_{a_t, t}]$. We have:
	\begin{align*}
		\mathbb{E}[ & \sum_{t=\tau^*+1}^{T} r_{a_t, t}] 
			\leq \mathbb{E}[ (T - \tau^*)_+ ] \\
			& = \sum_{t=0}^T \mathbb{P}[ \tau^* \leq T - t ] \\
			& \leq \sum_{i=1}^C \sum_{t=0}^T \mathbb{P}[ \sum_{\tau=1}^{T-t} c_{a_\tau, \tau}(i) \geq B(i) ] \\
			& = \sum_{i=1}^C \sum_{t=0}^T \mathbb{P}[ \sum_{\tau=1}^{T-t} ( c_{a_\tau, \tau}(i) - b(i) ) \geq t \cdot b(i) ] \\
			& = \sum_{i=1}^C \sum_{t=0}^T \mathbb{P}[\tini + \sum_{x \notin \mathcal{B}} n_{x, T} + \sum_{ \substack{ x \text{ pseudo-basis for } \eqref{eq-linear-program-general-upperbound-opt-strategy} \\ \text{ with } \det(A_x) = 0}} n_{x, T} + \sum_{x \in \mathcal{B}} (b_{x, T-t}(i) - n_{x, T-t} \cdot b(i)) \geq t \cdot b(i) ] \\
			& = \sum_{i=1}^C \sum_{t=0}^T \mathbb{P}[\tini + \sum_{x \notin \mathcal{B}} n_{x, T} + \sum_{ \substack{ x \text{ pseudo-basis for } \eqref{eq-linear-program-general-upperbound-opt-strategy} \\ \text{ with } \det(A_x) = 0}} n_{x, T} \geq t \cdot \frac{b(i)}{2} ] \\
			& + \sum_{i=1}^C \sum_{t=0}^T \sum_{x \in \mathcal{B}} \mathbb{P}[b_{x, T-t}(i) - n_{x, T-t} \cdot b(i) \geq t \cdot \frac{b(i)}{2 \cdot |\mathcal{B}|} ] \\
			& \leq 2 \sum_{i=1}^C \sum_{t=1}^T \frac{ \tini + \sum_{x \notin \mathcal{B}} \mathbb{E}[ n_{x, T}] + \sum_{ \substack{ x \text{ pseudo-basis for } \eqref{eq-linear-program-general-upperbound-opt-strategy} \\ \text{ with } \det(A_x) = 0}} \mathbb{E}[n_{x, T}] }{t \cdot b(i)} \\
			& + 2 |\mathcal{B}| \cdot \sum_{i=1}^C \sum_{t=1}^T \sum_{x \in \mathcal{B}} \frac{ \mathbb{E}[|b_{x, T-t}(i) - n_{x, T-t} \cdot b(i)|] }{t \cdot b(i)}  + O(1) \\
			& = O(  \frac{(C+4)!^4 \cdot |\mathcal{B}|^2}{b \cdot \epsilon^6} \cdot  \ln^2(T) ),
	\end{align*}
	where we use Lemma \ref{lemma-bound-times-infeasible-pulls-general-case} and we bound $\mathbb{E}[|b_{x, T-t}(i) - n_{x, T-t} \cdot b(i)|]$ in the same fashion as in the proof of Theorem \ref{lemma-log-B-regret-bound-general-case} using Lemma \ref{lemma-load-balance-general-case}.

\section{Proofs for Section \ref{sec-extensions}.}
\label{sec-proof-extensions}

\subsection{Proof of Lemma \ref{lemma-bound-times-non-optimal-pulls-deterministic-alternaterule}.}
The proof follows the same steps as for Lemma \ref{lemma-bound-times-non-optimal-pulls-deterministic}. We use the shorthand notations $\beta_k = 8  \frac{ \rho \cdot (\sum_{i=1}^C b(i))^2}{ \epsilon^2} \cdot (\frac{1}{\Delta_k})^2$ and $n^{\neq x^*}_{k, t} = \sum_{ x \in \mathcal{B} \; | \; k \in \mathcal{K}_x, x \neq x^* } n^x_{k, t}$. Along the same lines as in Fact \ref{fact-wlog-assume-n-big}, we have:
\begin{equation*}
	\mathbb{E}[n_{k, \tau^*}] \leq 2 \beta_k \cdot \mathbb{E}[\ln( \tau^* )] + \mathbb{E}[\sum_{t=1}^{ \tau^* } I_{x_t \neq x^*} \cdot I_{a_t = k} \cdot I_{n^{\neq x^*}_{k, t} \geq \beta_k \ln(t)} ],
\end{equation*}		
and we can focus on bounding the second term, which can be broken down as follows:
\begin{align*}
	\mathbb{E}[\sum_{t=1}^{ \tau^* } &  I_{x_t \neq x^*} \cdot I_{a_t = k} \cdot I_{n^{\neq x^*}_{k, t} \geq \beta_k \ln(t)} ]  \\
			 & = \mathbb{E}[\sum_{t=1}^{ \tau^* }  I_{ \mathrm{obj}_{x_t, t} + E_{x_t, t} \geq \mathrm{obj}_{x^*, t} + E_{x^*, t}}  \cdot I_{x_t \neq x^*} \cdot I_{a_t = k} \cdot I_{n^{\neq x^*}_{k, t} \geq \beta_k \ln(t)} ] \\
			& \leq  \mathbb{E}[\sum_{t=1}^{ \tau^* } I_{ \mathrm{obj}_{x_t, t} \geq \mathrm{obj}_{x_t} + E_{x_t, t} } ]  \\
				& + \mathbb{E}[\sum_{t=1}^{ \tau^* } I_{ \mathrm{obj}_{x^*, t} \leq \mathrm{obj}_{x^*}  - E_{x^*, t} } ] \\
				& + \mathbb{E}[\sum_{t=1}^{ \tau^* } I_{\mathrm{obj}_{x^*} < \mathrm{obj}_{x_t} + 2 E_{x_t, t} } \cdot I_{x_t \neq x^*} \cdot I_{a_t = k} \cdot I_{n^{\neq x^*}_{k, t} \geq \beta_k \ln(t)} ]. 
\end{align*}
We study each term separately, just like in Lemma \ref{lemma-bound-times-non-optimal-pulls-deterministic}.
		\begin{fact}
			$$
				\mathbb{E}[\sum_{t=1}^{\tau^*} I_{ \mathrm{obj}_{x_t, t} \geq \mathrm{obj}_{x_t} + E_{x_t, t} } ] \leq K \cdot \frac{\pi^2}{6}.
			$$
			\proof{Proof.}
				If $\mathrm{obj}_{x_t, t} \geq \mathrm{obj}_{x_t} + E_{x_t, t}$, there must exist $l \in \mathcal{K}_{x_t}$ such that $\bar{r}_{l, t} \geq \mu^r_l + \epsilon_{l, t}$, otherwise:
				\begin{align*}
					\mathrm{obj}_{x_t, t} - \mathrm{obj}_{x_t} 
						& = \sum_{l \in \mathcal{K}_{x_t}} (\bar{r}_{l, t} - \mu^r_l ) \cdot \xi^{x_t}_l \\
						& < \sum_{l \in \mathcal{K}_{x_t}} \epsilon_{l, t} \cdot \xi^{x_t}_l \\
						& = E_{x_t, t},
				\end{align*}
				where the inequality is strict because there must exist $l \in \mathcal{K}_{x_t}$ such that $\xi^{x_t}_l > 0$ (at least one resource constraint is binding for a feasible basis to \eqref{eq-linear-program-general-upperbound-opt-strategy} aside from the basis $\tilde{x}$ associated with $\mathcal{K}_{\tilde{x}} = \emptyset$). We obtain:
				\begin{align*}
				\mathbb{E}[\sum_{t=1}^{\tau^*}  I_{ \mathrm{obj}_{x_t, t} \geq \mathrm{obj}_{x_t} + E_{x_t, t} } ] 
					& \leq \mathbb{E}[\sum_{t=1}^{\tau^*} \sum_{l=1}^K I_{ \bar{r}_{l, t} \geq \mu^r_l + \epsilon_{l, t}  }	] \\
					& \leq \sum_{l=1}^K \sum_{t=1}^{\infty}  \mathbb{P}[ \bar{r}_{l, t} \geq \mu^r_l + \epsilon_{l, t} ] \\
					& \leq K \cdot \frac{\pi^2}{6},
				\end{align*}
				where the last inequality is derived along the same lines as in the proof of Fact \ref{fact-study-bound-first-term}.
			\endproof					
		\end{fact} \vspace{0.3cm}		
		Similarly, we can show that:
		$$
				\mathbb{E}[\sum_{t=1}^{\tau^*} I_{ \mathrm{obj}_{x^*, t} \leq \mathrm{obj}_{x^*} - E_{x^*, t} } ] \leq K \cdot \frac{\pi^2}{6}.
		$$
		We move on to study the last term.

		\begin{fact}
			$$
				\mathbb{E}[\sum_{t=1}^{ \tau^* } I_{\mathrm{obj}_{x^*} < \mathrm{obj}_{x_t} + 2 E_{x_t, t} } \cdot I_{x_t \neq x^*} \cdot I_{a_t = k} \cdot I_{n^{\neq x^*}_{k, t} \geq \beta_k \ln(t)} ] = 0.
			$$
		\proof{Proof.}
			If $\mathrm{obj}_{x^*} < \mathrm{obj}_{x_t} + 2 E_{x_t, t}$, $x_t \neq x^*$, and $a_t = k$, we have:
			\begin{align*}
				\frac{\Delta_k}{2} 
					& \leq \frac{\Delta_{x_t}}{2}  \\
					& < \sum_{l \in \mathcal{K}_{x_t}} \xi^{x_t}_l \cdot \sqrt{ \frac{2 \ln(t) }{ n_{l, t} } } \\
					& \leq \sum_{l \in \mathcal{K}_{x_t}} \sqrt{ \frac{2 \xi^{x_t}_l \cdot \xi^{x_t}_k \ln(t) }{ n_{k, t}} },
			\end{align*}
			where we use the fact that, by definition of the load balancing algorithm and since $a_t = k$, $\xi^{x_t}_k \neq 0$ (otherwise arm $k$ would not have been selected) and:
			\begin{equation}
				\label{eq-proof-relation-pulls-deterministic-alternaterule}
				n_{l, t} \geq \frac{\xi^{x_t}_l}{\xi^{x_t}_k} n_{k, t},
			\end{equation}
			for all arms $l \in \mathcal{K}_{x_t}$. We get:
			\begin{align*}
				n_{k, t} 
					& < \frac{8}{(\Delta_k)^2}\cdot  \xi^{x_t}_k \cdot ( \sum_{l \in \mathcal{K}_{x_t}} \sqrt{ \xi^{x_t}_l}  )^2 \cdot \ln(t) \\
					& \leq \frac{8}{(\Delta_k)^2} \cdot \xi^{x_t}_k \cdot \rho \cdot \sum_{l \in \mathcal{K}_{x_t}} \xi^{x_t}_l \cdot \ln(t) \\
					& \leq \frac{8}{(\Delta_k)^2} \cdot \rho \cdot (\sum_{l \in \mathcal{K}_{x_t}} \xi^{x_t}_l)^2 \cdot \ln(t),
			\end{align*}
			using the Cauchy$-$Schwarz inequality and the fact that a basis involves at most $\rho$ arms. Now observe that:
			\begin{align*}
				\sum_{l \in \mathcal{K}_{x_t}} \xi^{x_t}_l 
					& \leq \sum_{l \in \mathcal{K}_{x_t}} \frac{ \sum_{i=1}^C c_l(i) }{\epsilon} \cdot \xi^{x_t}_l  \\
					& \leq \frac{ \sum_{i=1}^C b(i) }{ \epsilon }
			\end{align*}
			as $x_t$ is a feasible basis to \eqref{eq-algo-general-idea} and using Assumption \ref{assumption-all-cost-non-zero}. We obtain:
			\begin{align*}
				n^{\neq x^*}_{k, t} 
				& \leq n_{k, t}  \\
				& < 8 \cdot \frac{ \rho \cdot (\sum_{i=1}^C b(i))^2 }{ \epsilon^2 \cdot (\Delta_k)^2 } \cdot \ln(t) \\
				& = \beta_k \cdot \ln(t). 
			\end{align*}
		\endproof
		\end{fact} \vspace{0.3cm}

\subsection{Proof of Lemma \ref{lemma-step-2-well-defined-multiple-budget-alternaterule}.}
We first show \eqref{eq-deterministic-alternaterule-nb-pull-arm-upperbound} by induction on $t$. The base case is straightforward. Suppose that the inequality holds at time $t-1$. There are three cases:
\begin{itemize}
	\item{arm $k$ is not pulled at time $t-1$, in which case the left-hand side of the inequality remains unchanged while the right-hand side can only increase, hence the inequality still holds at time $t$,}
	\item{arm $k$ is pulled at time $t-1$ after selecting $x_{t-1} \neq x^*$, in which case both sides of the inequality increase by one and the inequality still holds at time $t$,}
	\item arm $k$ is pulled at time $t-1$ after selecting $x_{t-1} = x^*$. First observe that there must exist $l \in \mathcal{K}_{x^*}$ such that $n_{l, t-1} \leq ( t-1 ) \cdot \frac{\xi^{x^*}_{l}}{\sum_{r=1}^K \xi^{x^*}_r}$. Suppose otherwise, we have:
	\begin{align*}
	t - 1 
		& = \sum_{l=1}^K n_{l, t} \\
		& \geq \sum_{l \in \mathcal{K}_{x^*}} n_{l, t} \\
		& > \sum_{l \in \mathcal{K}_{x^*}} (t-1) \cdot \frac{\xi^{x^*}_l}{\sum_{r=1}^K \xi^{x^*}_r} \\
		& = t-1,
	\end{align*}
	a contradiction. Suppose now by contradiction that inequality \eqref{eq-deterministic-alternaterule-nb-pull-arm-upperbound} no longer holds at time $t$, we have:
	\begin{align*}
		n_{k, t-1}
			& = n_{k, t} - 1 \\
			& > n_{x^*, t} \cdot \frac{\xi^{x^*}_k}{\sum_{l=1}^K \xi^{x^*}_l} + \sum_{x \in \mathcal{B}, x \neq x^*} n_{x, t} \\
			& \geq ( n_{x^*, t} + \sum_{x \in \mathcal{B}, x \neq x^*} n_{x, t} ) \cdot \frac{\xi^{x^*}_k}{\sum_{l=1}^K \xi^{x^*}_l} \\
			& = ( t-1 ) \cdot \frac{\xi^{x^*}_k}{\sum_{l=1}^K \xi^{x^*}_l},
	\end{align*}
	which implies, using the preliminary remark above, that $\frac{\xi^{x^*}_k}{n_{k, t-1}} < \max\limits_{l \in \mathcal{K}_{x^*}} \frac{\xi^{x^*}_l}{n_{l, t-1}}$, a contradiction given the definition of the load balancing algorithm.
\end{itemize}
We conclude that inequality \eqref{eq-deterministic-alternaterule-nb-pull-arm-upperbound} holds for all times $t$ and arms $k \in \mathcal{K}_{x^*}$. We also derive inequality \eqref{eq-deterministic-alternaterule-nb-pull-arm-lowerbound} as a byproduct, since, at any time $t$ and for any arm $k \in \mathcal{K}_{x^*}$:
\begin{align*}
	n_{k, t} 
		& \geq n_{x^*, t} - \sum_{l \in \mathcal{K}_{x^*}, l \neq k} n_{l, t} \\
		& \geq n_{x^*, t}  \cdot ( 1 - \frac{ \sum_{l \in \mathcal{K}_{x^*}, l \neq k} \xi^{x^*}_l}{\sum_{l=1}^K \xi^{x^*}_l} ) - \rho \cdot (\sum_{x \in \mathcal{B}, x \neq x^*} n_{x, t} + 1) \\
		& = n_{x^*, t} \cdot \frac{\xi^{x^*}_k}{\sum_{l=1}^K \xi^{x^*}_l} - \rho \cdot (\sum_{x \in \mathcal{B}, x \neq x^*} n_{x, t} + 1),
\end{align*}
as a basis involves at most $\rho$ arms.

\subsection{Proof of Theorem \ref{lemma-log-B-regret-bound-deterministic-alternate}.}
	The proof proceeds along the same lines as for Theorem \ref{lemma-log-B-regret-bound-deterministic}. We build upon \eqref{eq-simplified-general-upper-bound-on-regret}:
	\begin{align*}
				R_{B(1), \cdots, B(C)}  
					& \leq B \cdot \sum_{k=1}^K \mu^r_k \cdot \xi^{x^*}_k - \sum_{k=1}^K \mu^r_k \cdot \mathbb{E}[n_{k, \tau^*} ] + O(1) \\
					& \leq B \cdot  \sum_{k=1}^K \mu^r_k \cdot \xi^{x^*}_k - \sum_{k \in \mathcal{K}_{x^*}} \mu^r_k \cdot \mathbb{E}[n_{k, \tau^*} ] + O(1) \\
					& \leq (B -  \frac{\mathbb{E}[n_{x^*, \tau^*} ]}{\sum_{k=1}^K \xi^{x^*}_k }) \cdot \sum_{k=1}^K \mu^r_k \cdot \xi^{x^*}_k + \rho^2 \cdot \sum_{x \in \mathcal{B}, x \neq x^*} \mathbb{E}[n_{x, \tau^*}] + O(1),
	\end{align*}
	where we use \eqref{eq-deterministic-alternaterule-nb-pull-arm-lowerbound} to derive the third inequality. Now observe that, by definition, at least one resource is exhausted at time $\tau^*$. Hence, there exists $i \in  \{1, \cdots, C\}$ such that the following holds almost surely:
	\begin{align*}
	B(i) 
		& \leq \sum_{k=1}^K c_k(i) \cdot n_{k, \tau^*}  \\
		& \leq \sum_{k \notin \mathcal{K}_{x^*}} n_{k, \tau^*} + \sum_{k \in \mathcal{K}_{x^*}} c_k(i) \cdot n_{k, \tau^*}  \\
		& \leq \sum_{x \in \mathcal{B}, x \neq x^*} n_{x, \tau^*} + \sum_{k \in \mathcal{K}_{x^*}} c_k(i) \cdot n_{k, \tau^*}  \\
		& \leq \rho  \cdot (\sum_{x \in \mathcal{B}, x \neq x^*} n_{x, \tau^*} + 2) + n_{x^*, \tau^*} \cdot \sum_{k \in \mathcal{K}_{x^*}} c_k(i) \cdot  \frac{\xi^{x^*}_k}{\sum_{l=1}^K \xi^{x^*}_l} \\
		& \leq \rho \cdot (\sum_{x \in \mathcal{B}, x \neq x^*} n_{x, \tau^*} + 2) +  b(i) \cdot \frac{n_{x^*, \tau^*}}{\sum_{k=1}^K \xi^{x^*}_k}, 
	\end{align*}
	where we use \eqref{eq-deterministic-alternaterule-nb-pull-arm-upperbound} and the fact that $x^*$ is a feasible basis to \eqref{eq-linear-program-general-upperbound-opt-strategy}. Rearranging yields:
	$$
		\frac{n_{x^*, \tau^*}}{\sum_{k=1}^K \xi^{x^*}_k} \geq B - \frac{\rho}{b} \cdot (\sum_{x \in \mathcal{B}, x \neq x^*} n_{x, \tau^*} + 2),
	$$
	almost surely. Plugging this last inequality back into the regret bound, we get:
	\begin{align*}
		R_{B(1), \cdots, B(C)}  
			& \leq \rho \cdot \sum_{x \in \mathcal{B}, x \neq x^*} \mathbb{E}[n_{x, t}] \cdot ( \frac{ \sum_{k=1}^K \mu^r_k \cdot \xi^{x^*}_k}{b}  + \rho) + O(1) \\
			& \leq \rho \cdot \sum_{x \in \mathcal{B}, x \neq x^*} \mathbb{E}[n_{x, t}] \cdot ( \frac{  \sum_{k=1}^K \sum_{i=1}^C c_k(i) \cdot \xi^{x^*}_k}{ \epsilon \cdot b}  + \rho) + O(1) \\
			& \leq  ( \frac{\rho \cdot \sum_{i=1}^C b(i)}{ \epsilon \cdot b} + (\rho)^2) \cdot \sum_{x \in \mathcal{B}, x \neq x^*} \mathbb{E}[n_{x, t}] + O(1) \\
			& = ( \frac{\rho \cdot \sum_{i=1}^C b(i)}{ \epsilon \cdot b} + (\rho)^2) \cdot  \sum_{k=1}^K \mathbb{E}[  \sum_{ x \in \mathcal{B} \; | \; k \in \mathcal{K}_x, x \neq x^* }  n^x_{k, \tau^*}]  + O(1) \\
			& \leq 32 \frac{ \rho^3 \cdot (\sum_{i=1}^C b(i))^3 }{ \epsilon^3 \cdot b} \cdot  (\sum_{k=1}^K \frac{1}{(\Delta_k)^2} ) \cdot \mathbb{E}[ \ln(\tau^*) ]  + O(1) \\
			& \leq 32 \frac{ \rho^3 \cdot (\sum_{i=1}^C b(i))^3 }{ \epsilon^3 \cdot b}  \cdot (\sum_{k=1}^K \frac{1}{(\Delta_k)^2} ) \cdot \ln( \frac{ \sum_{i=1}^C b(i) \cdot B}{\epsilon}  + 1 ) + O(1),
	\end{align*}
	where we use the fact that $x^*$ is a feasible basis to \eqref{eq-linear-program-general-upperbound-opt-strategy} for the third inequality, Lemma \ref{lemma-bound-times-non-optimal-pulls-deterministic-alternaterule} for the fourth inequality, the concavity of the logarithmic function along with Lemma \ref{lemma-bound-stopping-time-deterministic} for the last inequality.

\subsection{Proof of Lemma \ref{lemma-bound-times-non-optimal-pulls-general-case-alternative}.}
We use the shorthand notations $\beta_k = 8 C \cdot (\frac{\lambda}{\Delta_k})^2 $ and, for any round $t$, $n^{\notin \mathcal{O}}_{k, t} = \sum_{ x \in \mathcal{B} \; | \; k \in \mathcal{K}_x, \; x \notin \mathcal{O} } n^x_{k, t}$.  Similarly as in Fact \ref{fact-wlog-assume-n-big-budget-generalcase}, we have:
\begin{align*}
				\mathbb{E}[n^{\notin \mathcal{O}}_{k, T}]
					& \leq 2 \beta_k \cdot \ln( T ) + \probamaxdevmeangeneralcasewithouttime \nonumber \\
					& + \mathbb{E}[\sum_{t=\tini}^{T}  I_{x_t \notin \mathcal{O}} \cdot I_{a_t=k} \cdot I_{n^{\notin \mathcal{O}}_{k, t} \geq \beta_k \ln(t)} \cdot I_{ x^* \in \mathcal{B}_t }],
\end{align*}
and what remains to be done is to bound the second term, which we can break down as follows:
\begin{align*}
		\mathbb{E}[ & \sum_{t=\tini}^{T}  I_{x_t \notin \mathcal{O}} \cdot I_{a_t=k} \cdot I_{n^{\notin \mathcal{O}}_{k, t} \geq \beta_k \ln(t)} \cdot I_{ x^* \in \mathcal{B}_t }] \\ 
			& \leq \mathbb{E}[\sum_{t=\tini}^{T}  I_{ \mathrm{obj}_{x_t, t} + E_{x_t, t} \geq \mathrm{obj}_{x^*, t} + E_{x^*, t}} \cdot I_{x_t \notin \mathcal{O}} \cdot I_{a_t=k} \cdot I_{n^{\notin \mathcal{O}}_{k, t} \geq \beta_k \ln(t)} \cdot I_{x^* \in \mathcal{B}_t} ] \nonumber \\
			& \leq  
				\mathbb{E}[\sum_{t=\tini}^{T} I_{ \mathrm{obj}_{x_t, t} \geq \mathrm{obj}_{x_t} + E_{x_t, t} } \cdot I_{x_t \in \mathcal{B}_t} ]  \\
				& + \mathbb{E}[\sum_{t=\tini}^{T} I_{ \mathrm{obj}_{x^*, t} \leq \mathrm{obj}_{x^*}  - E_{x^*, t} } \cdot I_{x^* \in \mathcal{B}_t}] \\
				& + \mathbb{E}[\sum_{t=\tini}^{T} I_{\mathrm{obj}_{x^*} < \mathrm{obj}_{x_t} + 2 E_{x_t, t} } \cdot I_{x_t \notin \mathcal{O}} \cdot I_{a_t=k} \cdot I_{n^{\notin \mathcal{O}}_{k, t} \geq \beta_k \ln(t)} ]. 
\end{align*}
The study of the second term is the same as in the proof of Lemma \ref{lemma-bound-times-non-optimal-pulls-general-case}. We can also bound the first term in the same fashion as in the proof of Lemma \ref{lemma-bound-times-non-optimal-pulls-general-case} since there is no reference to the load balancing algorithm in the proof of Fact \ref{fact-study-bound-first-term-general-case}. The major difference with the proof of Lemma \ref{lemma-bound-times-non-optimal-pulls-general-case} lies in the study of the last term.
\begin{fact}
			$$
				\mathbb{E}[\sum_{t=\tini}^{T} I_{ \mathrm{obj}_{x_t, t} \geq \mathrm{obj}_{x_t} + E_{x_t, t} } \cdot I_{x \in \mathcal{B}_t}  ] \leq 2^{10}  \frac{K \cdot (C+3)!^2}{\epsilon^6}.
			$$
			\proof{Proof.}
			The only difference with the proof of Fact \ref{fact-study-bound-first-term-general-case} is that the number of arms that belong to $\mathcal{K}_{x}$ for $x$ ranging in $\{\tilde{x} \in \mathcal{B} \; | \; k \in \mathcal{K}_{\tilde{x}}, \; \tilde{x} \notin \mathcal{O}\}$ can be as big as $K$, as opposed to $C$ when we are considering one basis at a time. This increases the bound by a multiplicative factor $K$.
			\endproof
		\end{fact} \vspace{0.3cm}

\begin{fact}
			$$
				\mathbb{E}[\sum_{t=\tini}^{ T } I_{\mathrm{obj}_{x^*} < \mathrm{obj}_{x_t} + 2 E_{x_t, t} } \cdot I_{x_t \notin \mathcal{O}} \cdot I_{a_t = k} \cdot I_{n^{\notin \mathcal{O}}_{k, t} \geq \beta_k \ln(t)} ] = 0.
			$$
		\proof{Proof.}
			 Assume that $\mathrm{obj}_{x^*} < \mathrm{obj}_{x_t} + 2 E_{x_t, t}$, $x_t \notin \mathcal{O}$, and $a_t =k$. We have:
			\begin{align*}
				\frac{\Delta_k}{2} 
					& \leq \frac{\Delta_{x_t}}{2}  \\
					& < \lambda \cdot \sum_{l \in \mathcal{K}_{x_t}} \xi^{x_t}_{l, t} \cdot \sqrt{ \frac{2 \ln(t) }{ n_{l, t} } } \\
					& \leq \lambda \cdot  \sum_{l \in \mathcal{K}_{x_t}} \sqrt{ \frac{2 \xi^{x_t}_{l, t} \cdot \xi^{x_t}_{k, t} \ln(t) }{ n_{k, t}} },
			\end{align*}
			where we use the fact that, by definition of the load balancing algorithm and since $a_t = k$, $\xi^{x_t}_{k,t} \neq 0$ (otherwise arm $k$ would not have been selected) and:
			\begin{equation}
				n_{l, t} \geq \frac{\xi^{x_t}_{l, t}}{\xi^{x_t}_{k, t}} \cdot n_{k, t},
			\end{equation}
			for any arm $l \in \mathcal{K}_{x_t}$. We get:
			\begin{align*}
				n^{\notin \mathcal{O}}_{k, t} 
					& \leq n_{k, t} \\
					& < 8  (\frac{\lambda}{\Delta_k})^2 \cdot  \xi^{x_t}_{k, t} \cdot ( \sum_{l \in \mathcal{K}_{x_t}} \sqrt{ \xi^{x_t}_{l, t}}  )^2 \cdot \ln(t) \\
					& \leq 8 (\frac{\lambda}{\Delta_k})^2 \cdot \xi^{x_t}_{k, t} \cdot C \cdot \sum_{l \in \mathcal{K}_{x_t}} \xi^{x_t}_{l, t} \cdot \ln(t) \\
					& \leq 8 (\frac{\lambda}{\Delta_k})^2 \cdot C \cdot (\sum_{l \in \mathcal{K}_{x_t}} \xi^{x_t}_{l, t})^2 \cdot \ln(t) \\
					& \leq 8 C \cdot (\frac{\lambda}{\Delta_k})^2  \cdot \ln(t),					
			\end{align*}
			using the Cauchy$-$Schwarz inequality, the fact that a basis involves at most $C$ arms, and the fact that $x_t$ is feasible for \eqref{eq-algo-general-idea} whose linear constraints include $\sum_{l=1}^K \xi_{l} \leq 1$ and $\xi_l \geq 0, \forall l \in \{1, \cdots, K\}$. We get $n^{\notin \mathcal{O}}_{k, t} < \beta_k \ln(t)$ by defintion of $\beta_k$.
			\endproof
		\end{fact} \vspace{0.3cm}

\subsection{Proof of Theorem \ref{lemma-log-B-regret-bound-deterministic-when-relaxing-budgetscalinglinearly}.}
Substituting $b(i)$ with $B(i)/B(C)$ for every resource $i \in \{1, \cdots, C\}$, the regret bound obtained in Theorem \ref{lemma-log-B-regret-bound-deterministic} turns into:
\begin{equation}
	\label{eq-proof-lemma-sqrt-B-regret-bound-deterministic-when-relaxing-budgetscalinglinearly}
	R_{B(1), \cdots, B(C)} \leq  16 \frac{\rho }{\epsilon} \cdot \frac{ \sum_{i=1}^C B(i)}{B(C)}\cdot (\sum_{x \in \mathcal{B} \; | \; \Delta_x > 0} \frac{ 1 }{ \Delta_x } ) \cdot \ln( \frac{ \sum_{i=1}^C B(i) }{\epsilon} + 1) + O(1).
\end{equation}
Observe that $\mathcal{B}^\complement$ and $\mathcal{O}^\complement$ are defined by strict inequalities that are linear in the vector $(B(1)/B(C), \cdots, B(C-1)/B(C))$. Hence, for $B(C)$ large enough, $\mathcal{B}^\complement_{\infty} \subset \mathcal{B}^\complement$ and $\mathcal{O}^\complement_{\infty} \subset \mathcal{O}^\complement$ and thus $\mathcal{B} \subset \mathcal{B}_\infty$ and $\mathcal{O} \subset \mathcal{O}_\infty$. We now move on to prove each claim separately.
		\\
		\textbf{First claim}. Suppose that there exists a unique optimal basis to \eqref{eq-linear-program-general-upperbound-opt-strategy}, which we denote by $x^*$. Then, we must have $\mathcal{O} = \{ x^*\} = \mathcal{O}_\infty$ for $B(C)$ large enough. Indeed, using the set inclusion relations shown above, we have $\mathcal{O} \subset \mathcal{O}_\infty = \{ x^* \}$ and $\mathcal{O}$ can never be empty as there exists at least one optimal basis to \eqref{eq-linear-program-general-upperbound-opt-strategy} (this linear program is feasible and bounded). We get $\mathcal{O}^\complement \cap \mathcal{B} \subset \mathcal{O}^\complement_\infty \cap \mathcal{B}_\infty$ for $B(C)$ large enough. Note moreover that, for any $x \in \mathcal{B}$, $\Delta_x$ converges to $\Delta^\infty_x$ (because both the objective value of a feasible basis and the optimal value of a linear program are Lipschitz in the right-hand side of the inequality constraint), which implies that $\Delta_x > \frac{ \Delta^\infty_x }{ 2 } > 0$ when $x \in \mathcal{B} \cap \mathcal{O}^\complement$ for $B(C)$ large enough. We conclude with \eqref{eq-proof-lemma-sqrt-B-regret-bound-deterministic-when-relaxing-budgetscalinglinearly} that:
		\begin{align*} 
			 R_{B(1), \cdots, B(C)}
				& \leq 32 \frac{\rho }{\epsilon} \cdot \frac{ \sum_{i=1}^C B(i)}{B(C)}  \cdot (\sum_{x \in \mathcal{B}_\infty \; | \; \Delta^\infty_x > 0} \frac{ 1 }{ \Delta_x } ) \cdot \ln( \frac{ \sum_{i=1}^C B(i) }{\epsilon} + 1) + O(1),
		\end{align*}
		for $B(C)$ large enough. This yields the result since $B(i)/B(C) \rightarrow b(i) > 0$ for any resource $i=1, \cdots, C-1$. \\
		\textbf{Second claim}. Suppose that $\frac{B(i)}{B(C)} - b(i) = O(\frac{\ln(B(C))}{B(C)})$ for any resource $i \in \{1, \cdots, C-1\}$. Starting from \eqref{eq-proof-log-B-start-here-sqrt-B-deterministic} derived in the proof of Theorem \ref{lemma-log-B-regret-bound-deterministic} and applying Lemma \ref{lemma-bound-times-non-optimal-pulls-deterministic} only if $\Delta_x$ is big enough, we have:
		\begin{align*}		
			& R_{B(1), \cdots, B(C)} \\
			& \leq   \sum_{x \in \mathcal{B} \; | \; \Delta_x > 0} \min( \frac{\Delta_x}{\sum_{k=1}^K \xi^x_k} \cdot \mathbb{E}[n_{x, \tau^*}], 16 \rho \cdot \frac{\sum_{k=1}^K \xi^x_k}{\Delta_x} \cdot \ln( \frac{\sum_{i=1}^C B(i)}{\epsilon} + 1 ) + \frac{\pi^2}{3}  \rho \cdot \frac{\Delta_x}{\sum_{k=1}^K \xi^x_k}) + O(1) \\
			& \leq   \sum_{x \in \mathcal{B} \; | \; \Delta_x > 0} \min(  \Delta_x \cdot \frac{ B(C) }{\min_{i=1, \cdots, C} B(i)} \cdot \frac{\sum_{i=1}^C B(i)}{\epsilon},  16 \frac{\rho \cdot \sum_{i=1}^C B(i)/B(C) }{\epsilon} \cdot \frac{1}{\Delta_x} \cdot \ln( \frac{\sum_{i=1}^C B(i)}{\epsilon}  + 1 )) \\
			& + O(1) \\
			& \leq 16 \frac{\rho \cdot \sum_{i=1}^C B(i)/B(C) }{\epsilon} \cdot (\sum_{x \in \mathcal{B} \cap \mathcal{O}^\complement \cap \mathcal{O}^\complement_\infty } \frac{1}{\Delta_x}) \cdot \ln(\frac{\sum_{i=1}^C B(i)}{\epsilon}  + 1) \\
			& +   (\sum_{x \in \mathcal{B} \cap \mathcal{O}^\complement \cap \mathcal{O}_\infty } \Delta_x) \cdot \frac{ B(C)  }{\min_{i=1, \cdots, C} B(i)} \cdot \frac{\sum_{i=1}^C B(i)}{\epsilon} + O(1),
`		\end{align*} 
	where we use:
	$$
		\sum_{k=1}^K \xi^x_k \in [\min\limits_{i=1, \cdots, C} B(i)/B(C), \frac{\sum_{i=1}^C B(i)/B(C)}{\epsilon}]
	$$ 
	and 
	$$	
		\Delta_x \leq \frac{\sum_{i=1}^C B(i)/B(C)}{\epsilon},
	$$
	as shown in the proof of Theorem \ref{lemma-log-B-regret-bound-deterministic} (substituting $b$ with $\min_{i=1, \cdots, C} B(i)/B(C)$). For $x \in \mathcal{B} \cap \mathcal{O}^\complement \cap \mathcal{O}^\complement_\infty$, we have $x \in \mathcal{B}_\infty$ and $\Delta_x  > \frac{ \Delta^\infty_x }{ 2 } > 0$ for $B(C)$ large enough, as shown for the first claim. For $x \in \mathcal{B} \cap \mathcal{O}^\complement  \cap \mathcal{O}_\infty$, we have $\Delta_x = O( \ln( B(C) )/ B(C) )$ as both the objective value of a feasible basis and the optimal value of a linear program are Lipschitz in the right-hand side of the inequality constraints. We conclude that, for $B(C)$ large enough:
		\begin{align*}		
			& R_{B(1), \cdots, B(C)} \\
			& \leq 32 \frac{\rho \cdot \sum_{i=1}^C B(i)/B(C) }{\epsilon} \cdot (\sum_{x \in \mathcal{B}_\infty \cap \mathcal{O}^\complement_\infty } \frac{1}{\Delta^\infty_x}) \cdot \ln(\frac{\sum_{i=1}^C B(i)}{\epsilon}  + 1) \\
			& + \frac{1}{\epsilon} \cdot \frac{ B(C) }{\min_{i=1, \cdots, C} B(i)} \cdot \sum_{x \in \mathcal{B} \cap \mathcal{O}^\complement \cap \mathcal{O}_\infty } O( \ln(B(C))) + O(1).
		\end{align*}
		This yields the result since $| \mathcal{B} \cap \mathcal{O}^\complement \cap \mathcal{O}_\infty | \leq |\mathcal{O}_\infty|$ and $B(i)/B(C) \rightarrow b(i) > 0$ for any resource $i=1, \cdots, C-1$.

\subsection{Proof of Theorem \ref{lemma-log-B-regret-bound-time-horizon-when-relaxing-budgetscalinglinearly}.}

The proof is along the same lines as for Theorem \ref{lemma-log-B-regret-bound-deterministic-when-relaxing-budgetscalinglinearly}. Specifically, in a first step, we observe that all the proofs of Section \ref{sec-singlebudgettimehorizon} remain valid (up to universal constant factors) for $T$ large enough as long as we substitute $b$ with $B/T$. Indeed, for $T$ large enough, we have $\frac{B}{T} \leq 2$ and $|\mu^c_k -\frac{B}{T}| > \frac{\epsilon}{2}$ for all arms $k \in \{1, \cdots, K\}$ under Assumption \ref{assumption-simplying-assumption-analysis-budget-and-time-horizon}. In a second step, just like in the proof of Theorem \ref{lemma-log-B-regret-bound-deterministic-when-relaxing-budgetscalinglinearly}, we show that we can substitute $\sum_{x \in \mathcal{B} \; | \; \Delta_x > 0} \frac{1}{\Delta_x}$ with $\sum_{x \in \mathcal{B}_\infty \; | \; \Delta^\infty_x > 0} \frac{1}{\Delta^\infty_x}$in the regret bound up to universal constant factors.

\subsection{Proof of Theorem \ref{lemma-log-B-regret-bound-general-case-when-relaxing-budgetscalinglinearly}.}

The proof is along the same lines as for Theorem \ref{lemma-log-B-regret-bound-deterministic-when-relaxing-budgetscalinglinearly}. Specifically, in a first step, we observe that all the proofs of Section \ref{sec-stochastic-multiple-budget} remain valid (up to universal constant factors) for $T$ large enough as long as we substitute $b$ with $\min_{i=1, \cdots, C-1} B(i)/T$. Indeed, for $T$ large enough, we have $\min_{i=1, \cdots, C-1} B(i)/T \leq 2$ and, under Assumption \ref{assumption-simplying-assumption-general case}, any basis to \eqref{eq-linear-program-general-upperbound-opt-strategy-determistic-when-relaxing-budgetscalinglinearly} has determinannt larger than $\epsilon/2$ in absolute value and is $\epsilon/2-$non-degenerate by continuity of linear functions. In a second step, just like in the proof of Theorem \ref{lemma-log-B-regret-bound-deterministic-when-relaxing-budgetscalinglinearly}, we show that we can substitute $\sum_{x \in \mathcal{B} \; | \; \Delta_x > 0} \frac{1}{\Delta_x}$ with $\sum_{x \in \mathcal{B}_\infty \; | \; \Delta^\infty_x > 0} \frac{1}{\Delta^\infty_x}$in the regret bound up to universal constant factors.

\end{APPENDICES}



\bibliographystyle{informs2014} 
\bibliography{biblio_budgeted_multi_arms_bandit} 


\end{document}